%% file: structured-paper.tex
\documentclass[acmlarge,screen,nonacm,10pt]{acmart}

\usepackage{hyperref}
\hypersetup{
    pdftitle={A Comprehensive Evaluation of Four End-to-End AI Autopilots Using CCTest and the Carla Leaderboard},
    pdfauthor={Changwen Li, Joseph Sifakis, Rongjie Yan, Jian Zhang}
}

\usepackage{geometry}
\usepackage{titlesec}
\titleformat{\section}
  {\LARGE\bfseries}{\thesection}{1em}{}
  
\titleformat{\subsection}
  {\Large\bfseries}{\thesubsection}{.5em}{}

\titleformat{\subsubsection}{\large\bfseries}{\thesubsubsection}{.5em}{}

\titleformat{\paragraph}{\normalsize\bfseries}{}{}{}

\author{
\tabcolsep 0.46cm
\MakeTextLowercase{
\begin{tabular}{cccc}
    \MakeTextUppercase{C}hangwen \MakeTextUppercase{L}i$^\dag$ & \MakeTextUppercase{J}oseph \MakeTextUppercase{S}ifakis$^*$ & \MakeTextUppercase{R}ongjie \MakeTextUppercase{Y}an$^\dag$ &
    \MakeTextUppercase{J}ian \MakeTextUppercase{Z}hang$^\dag$\\
    {licw@ios.ac.cn} & {Joseph.Sifakis@univ-grenoble-alpes.fr} & {yrj@ios.ac.cn} & {zj@ios.ac.cn}
\end{tabular}}
}

\authorsaddresses{$\dag$ Key Laboratory of System Software (Chinese Academy of Sciences), Institute of Software, Chinese Academy of Sciences. \\
$*$ VERIMAG, Univ. Grenoble Alpes, CNRS, Grenoble INP,  and SUSTECH/RITAS, Shenzhen.}

\usepackage{multicol}
\usepackage{amsmath}
\usepackage{graphicx}
\usepackage{subfigure}
\usepackage{algorithm2e}
\usepackage{algorithmic}
\usepackage{multirow}
\usepackage{makecell}
\usepackage{tabularx}
\usepackage{multicol}
\usepackage{nicematrix}
\usepackage{array}
\usepackage{tabularray}
\usepackage{enumitem}
\setlist[itemize]{leftmargin=*}
\usepackage{adjustbox}
\usepackage{marginnote}
\usepackage{amsthm}
\usepackage{bm}
\usepackage{diagbox}
\usepackage{longtable}
\usepackage{caption}
\usepackage{subcaption}
\usepackage{mathtools,leftindex,leftidx,tensor,mhchem}
\setcopyright{none}

\newcommand{\removable}{\color{black}}

\newcommand{\hidden}[1]{}
\definecolor{ao(english)}{rgb}{0.0, 0.5, 0.0}

\usepackage{relsize}

\newcolumntype{Y}{>{\hsize=\hsize\centering\arraybackslash}X}
\newcolumntype{Z}{>{\hsize=1.5\hsize\centering\arraybackslash}X}

\setlist[itemize]{itemsep=0pt, topsep=0pt, parsep=0pt, left=0pt}
\setlist[enumerate]{itemsep=0pt, topsep=0pt, parsep=0pt, left=0pt}

\usepackage{xcolor,colortbl}
\newcommand{\ps}[1]{\cellcolor{green!25}#1}
\newcommand{\pu}[1]{\cellcolor{olive!35}#1}
\newcommand{\ac}[1]{\cellcolor{red!25}#1}
\newcommand{\cs}[1]{\cellcolor{cyan!25}#1}
\newcommand{\cu}[1]{\cellcolor{violet!25}#1}
\newcommand{\co}[1]{\cellcolor{violet!25}#1}

\newcommand{\ro}[1]{\cellcolor{orange!25}#1}
\newcommand{\blk}[1]{\cellcolor{gray!25}#1}

\titlespacing*{\paragraph}{0em}{10pt}{-7pt}

\titleformat{\paragraph}{\normalsize\bfseries}{}{}{}

\newcommand{\ct}[1]{{\bf\color{red}#1}}

\title[A comprehensive  evaluation of four E2E AI autopilots using CCTest and the Carla Leaderboard]{\huge A Comprehensive  Evaluation of Four End-to-End AI Autopilots Using CCTest and the Carla Leaderboard}

\begin{abstract}

\textbf{Abstract.}~
End-to-end AI autopilots for autonomous driving systems have emerged as a promising alternative to traditional modular autopilots, offering the potential to reduce development costs and mitigate defects arising from module composition. However, they suffer from the well-known problems of AI systems such as non-determinism, non-explainability, and anomalies. This naturally raises the question of their evaluation and, in particular, their comparison with existing modular solutions. 

This work extends a study of the critical configuration testing (CCTest) approach that has been applied to four open modular autopilots. This approach differs from others in that it generates test cases ensuring safe control policies are possible for the tested autopilots. This enables an accurate assessment of the ability to drive safely in critical situations, as any incident observed in the simulation involves the failure of a tested autopilot. The contribution of this paper is twofold. 

Firstly, we apply the CCTest approach to four end-to-end open autopilots, InterFuser, MILE, Transfuser, and LMDrive, and compare their test results with those of the four modular open autopilots previously tested with the same approach implemented in the Carla simulation environment. This comparison identifies both differences and similarities in the failures of the two autopilot types in critical configurations.

Secondly, we compare the evaluations of the four autopilots carried out in the Carla Leaderboard with the CCTest results. This comparison reveals significant discrepancies, reflecting differences in test case generation criteria and risk assessment methods. It underlines the need to work towards the development of objective assessment methods combining qualitative and quantitative criteria.
\end{abstract}

\begin{document}

\maketitle

\section{Introduction}

Recently, interest has shifted from traditional modular autopilots to end-to-end AI autopilots for autonomous driving systems (ADS) for two reasons. Firstly, development costs are lower for end-to-end solutions. Secondly, the adoption of a holistic development minimizes the risks inherent in the non-composability of functionality in modular architectures. However, end-to-end AI autopilots are less deterministic and suffer from the well-known problems of AI systems, such as non-explainability and anomalies~\cite{storey2022explainable, kurakin2018adversarial}.

Applying end-to-end AI solutions to ADS is a bold step and poses two challenges. The first is knowing to what extent these solutions have the same weaknesses as those identified for modular autopilots, and how their characteristic differences may be reflected in their respective evaluation results. The second challenge is to choose an effective approach to testing and evaluating end-to-end AI solutions, so that the risks induced by autopilots are carefully uncovered and properly reflected in the results of their evaluation.

Scenario-based testing is the dominant approach to testing ADS in a simulated validation environment. Its effective application depends on two interrelated issues. 

The first is the choice of the method used to generate test cases, based on various criteria such as risk, degree of autonomy, degree of coverage and representativeness, and complexity~\cite{nalic2020scenario,lehmann2019use}. In making this choice, we need to bear in mind that accidents and all kinds of hazards are rare events that are difficult to reproduce. Furthermore, we need to cover a wide variety of situations, including different traffic densities, road geometries or topological conditions, urban, regional, or national road traffic, and so on. These requirements involve covering a wide variety of average and extreme situations, making the ADS testing problem immensely complex. 

The second issue concerns the selection of evaluation methods, which are not independent of test generation methods. As far as possible, the evaluation should cover normal conditions, with representative benchmarks involving fairly long scenarios. In this case, a quantitative analysis using statistics is required to determine the extent to which the benchmarks reflect the wide variety of real-life situations. In addition, the evaluation should focus on rare and particular critical situations. In this case, a qualitative analysis is necessary to identify the risks associated with the vehicles tested and their causes. 

A test case, in its most general form, involves two sets of vehicles: autonomous vehicles and non-player character (NPC) vehicles.
\begin{itemize}
    \item Autonomous vehicles are driven by autopilots following their routes. Typically, there is only one autonomous vehicle under test, the \textit{ego vehicle}, whose observed behavior is evaluated by an oracle. The other autonomous vehicles, called \textit{background vehicles}, act as moving obstacles in the ego vehicle's environment.
    \item NPC vehicles are controlled to bring the vehicles under test to into specific situations. The behavior of an NPC is defined by a sequence of positions, with corresponding speeds or timestamps that may be given at the outset or can be generated online. This implies that the NPCs do not drive autonomously, but they move so as to respect timing or speed constraints.
\end{itemize}

\hidden{An important requirement for scenario generation is that the scenarios are realistic and involve situations that are faithful to real-world situations. In particular, the dynamic behavior of a vehicle must not be incompatible with the applicable physical constraints, e.g. acceleration or deceleration rates that are too high for an autonomous vehicle or an NPC vehicle. Furthermore, a vehicle's speed must not reach extremely high values or be limited by too low a ceiling. We require all scenarios to be realistic. Applying unrealistic scenarios can alter the quality of test results.}

\vspace{0.5em}

The application of a test case results in a scenario, the sequence of states of the vehicles involved, on which the validation of safety properties is performed. The generation of test cases must take into account two interdependent factors: the \textit{criticality} of the test cases for the autopilots under test, and the \textit{degree of risk} for the resulting scenario. Intuitively, one test case is more critical than another if the success of the scenario generated implies the existence of a successful scenario for the other test case. The degree of risk of a scenario can be characterized by various indicators \cite{mahmud2017application} that estimate the likelihood of a hazardous outcome and its severity, based on ego vehicle's behavior \cite{riedmaier2020survey,karimi2022automatic,xia2017automatic}.

For natural driving data, risk thresholds are used to distinguish between safe driving, dangerous situations, and accidents~\cite{nalic2020scenario,lehmann2019use,mahmud2017application}. This distinction is important because accidents have a low probability, but their impact can be very significant.

Based on the criticality of test cases and their resulting risks, we
adopt the following classification:

\begin{itemize}
\item \textit{Normal test cases}  do not deviate significantly from commonly accepted patterns and reflect ordinary real-life situations, usually requiring straightforward driving policies.

\item \textit{Critical test cases}~\cite{nalic2020scenario,lehmann2019use} are generated by maximizing their criticality or the risk of the resulting scenarios while taking into account nominal operational design domains (ODDs)~\cite{thorn2018framework}. They are calculated to push the tested vehicles to their limits while ensuring that they are not involved in unmanageable situations that inevitably lead to accidents \cite{riedmaier2020survey,lehmann2019use,hallerbach2018simulation}. 

\item \textit{Accidental test cases}~\cite{nalic2020scenario,lehmann2019use} differ from critical ones in that they intentionally involve abnormal situations that could lead to accidents. These test cases are used to assess which situations beyond ODDs can be managed by the autopilots to minimize the risk.
\end{itemize}

\vspace{3pt}
We use the term \textit{adversarial test cases} to denote critical or accidental test cases ~\cite{10328462}.
\hidden{Scenario controllability is the subject of active research, notably using diffusion techniques~\cite{NEURIPS2023Pro,KDD24Zhu,ICRA24Lu}.}

\vspace{5pt}
An essential part of a test method is the evaluation of the scenarios. Clearly, the evaluation method should take into account not only various risks detected but also the possibility of avoiding them in the generated test cases. It is very easy to create serious accidents that are impossible to manage safely, such as a sudden cut-in by an NPC that does not give the ego vehicle enough room to brake, or the intentional disregard of a red light. Accounting for the unavoidable risks may result in a biased evaluation of the true risks posed by the tested autopilot. In line with these considerations, we distinguish two risk evaluation approaches:

\begin{itemize}
\item 
\textit{Global quantitative evaluation} is used to estimate the risk induced by benchmarks involving a mixture of normal, critical, and accidental test cases. In this approach, the representativeness of the benchmarks in relation to a population of real-world situations is essential~\cite{lehmann2019use}. Evaluation methods calculate the associated risk using different criteria. Some apply the usual definition of risk as the weighted sum of terms that are the product of the probability of occurrence of a hazard and a corresponding severity coefficient~\cite{carla2020leaderboard}. Others adopt criteria involving time to collision, distance, or deceleration~\cite{mahmud2017application}. 
\item \textit{Qualitative evaluation} involves analyzing test results on a case-by-case basis and, in particular, determining the extent to which each observed hazard could have been avoided. This type of assessment is essential for high-risk scenarios and naturally complements the global quantitative approaches.
\end{itemize}

\vspace{5pt}

We recently studied the critical configuration test approach (CCTest) that we applied to four modular autopilots: Apollo, Autoware, and built-in autopilots in Carla, and LGSVL simulators \cite{li2024rigorous}. This approach focuses on generating critical test cases from initial configurations of autonomous vehicles. Test cases are potentially safe by construction, meaning that each vehicle involved can have proven safety policies based on a predictive analysis that takes into account, in addition to the initial kinetic attributes of the vehicles, their dynamic characteristics, including their braking and acceleration capabilities. The inability to drive safely from a given potentially safe test case, directly calls into question the safety of the autopilot. The associated evaluation method is therefore qualitative and focuses on specific types of safety violations.

CCTest is not based on representativeness and coverage criteria. Instead of testing safety for a wide variety of arbitrarily long scenarios, it focuses on the circumstances in which accident-prone conflicts can occur. It considers that the ability to drive safely boils down to the ability to drive safely in a small number of driving situations determined by analysis and compositional reasoning \cite{bozga2024safe}. These driving situations arise in a limited number of conflictual contexts, such as intersections, overtaking, merging, etc. A further simplification comes from the fact that, for each context, the safe driving of the vehicle under test depends on its interaction with a small number of vehicles: the vehicle in front of it and a vehicle that may be arriving in a path that interferes with its route. In this way, critical configuration testing involves only a few vehicles in conflicting situations on specific route patterns, greatly simplifying the calculation and validation of test cases.

This paper investigates the application of critical configuration testing to four open end-to-end AI autopilots: Transfuser \cite{chitta2022transfuser}, InterFuser \cite{shao2023safety}, MILE \cite{hu2022model}, and LMDrive \cite{shao2024lmdrive}. This allows identifying the main issues in the four AI autopilots and comparing them with previous critical configuration testing results of four modular autopilots \cite{li2024rigorous}, which revealed numerous problems, including accidents, software failures, blockages, and traffic rule violations. In addition, the four AI autopilots have been integrated and tested in the Carla Leaderboard test environment \cite{carla2020leaderboard} that provides detailed quantitative evaluation results. It is therefore appropriate to compare the results obtained for these four autopilots using two very different testing approaches, one general and quantitative, the other qualitative and focusing on critical test cases. In conclusion, the paper addresses the following research questions:

\begin{itemize}
    \item \textbf{RQ1.1}: To what extent do the four AI autopilots withstand the critical test cases generated by CCTest?
    \begin{itemize}
        \item \textbf{RQ1.1}: What are the main issues discovered for the four AI autopilots using CCTest?
        \item \textbf{RQ1.2}: How do the evaluation results of the four AI autopilots compare to the results of the four modular autopilots tested in \cite{li2024rigorous}, both obtained using CCTest?
    \end{itemize}
    
    \item \textbf{RQ2}: To what extent do the quantitative evaluation of the four end-to-end autopilots in the Carla Leaderboard and their qualitative evaluation of CCTest in the same framework, produce comparable or complementary results?
    \begin{itemize}
        \item \textbf{RQ2.1}: 
        How well do the Carla Leaderboard test method and its results allow for a faithful estimation of the safety and performance capabilities of ADS?
        \item \textbf{RQ2.2}: What does the comparison of the two test methods teach us about the strengths and weaknesses of each, and in particular the possibility of closing the gap between quantitative and qualitative methods?
    \end{itemize}
\end{itemize}

\vspace{5pt}

This work makes three contributions to the field of evaluating the safety of ADS through testing:
\begin{itemize}
\item It presents an implementation of the CCTest framework in the Carla Leaderboard environment,  carrying out the necessary adaptation work to enable the application of   CCTest to the four end-to-end AI autopilots.
\item It presents an in-depth evaluation of the four AI autopilots, InterFuser, MILE, Transfuser, and LMDrive, with critical test cases generated by CCTest. The results show safety issues for all four autopilots, including accidents, traffic violations, unintended route changes, and road deviations.
\item It carries out a complete evaluation of the four end-to-end  AI autopilots using the Carla Leaderboard. A detailed analysis shows its inability to distinguish faulty scenarios where problems could have been avoided according to the capabilities of the vehicles tested and others where danger is unavoidable. The evaluation results also show the limitation of Carla Leaderboard in detecting hazards in critical configurations, which is highlighted by comparison with the CCTest results.
\end{itemize}

\vspace{0.5em}
{\removable The paper is structured as follows. In Section~\ref{se:cctest}, we present the critical configuration testing approach, its theoretical foundations and applications. In Section~\ref{se:carla}, we describe the Carla Simulator and its architecture. Then, we present Carla's Leaderboard test environment, as well as the CCTest environment that we implemented in Carla. In Section \ref{se:analysis}, we present a comparative analysis of our results with the results of these autopilots in the Carla ranking, providing answers to the research questions posed. Section \ref{se:relatedwork} discusses related work. Section \ref{se:discussion} summarizes the main results and draws conclusions on the suitability of existing test methods for the provision of well-founded, scalable, and reproducible evaluations of ADS.}

\section{Critical Configuration Testing of the Autopilots}\label{se:cctest}

\subsection{The CCTest approach}

CCTest focuses on the generation of critical test cases and their simulation-based evaluation \cite{li2024rigorous}. Its implementation in the Carla Simulator is presented in Section~\ref{se:testenv}. Instead of analyzing long scenarios for maximum coverage, CCTest aims to identify minimal and critical configurations by applying the following compositional principles:
\begin{itemize}
    \item Locality of context:  
 The traffic infrastructure can be seen as the composition of a finite number of patterns comprising different types of roads and junctions with their signaling equipment. ADS safety strongly depends on the context in which vehicles operate. We can therefore imagine that a vehicle's safety policy is the composition of elementary policies, each of which is used to drive safely according to the corresponding basic patterns.
 \item Locality of knowledge: A vehicle's driving policy is based only on local knowledge of the state of the obstacles due to limited visibility. 
 \item Rights-based responsibility: Each vehicle is responsible for driving safely with respect to the first front vehicle on its route, and the first vehicle arriving from a direction that may cross its route.
\end{itemize}

\vspace{3pt}

With these considerations, CCTest simplifies configurations involving a minimal number of vehicles that determine the safe policies of the vehicle under test, known as the ego vehicle. It then focuses on critical configurations where conflicts can arise when the ego vehicle has an obligation to apply traffic rules to avoid risks. In particular, it addresses two cases of potential conflict.

The first case occurs when the trajectory of the ego vehicle intersects that of arriving vehicles, and the ego vehicle must yield the right of way. This can happen when the ego vehicle is driving on a road that merges onto a main road, or on a road crossing a higher priority road, or when the ego vehicle changes lanes on a highway. In these cases, the ego vehicle must perform a maneuver, applying successively two different control policies: 1) a caution policy, when the vehicle ends up moderating its speed when approaching a critical zone where a collision with arriving vehicles of higher priority is possible; 2) a progress policy when the ego vehicle decides to cross the critical zone, possibly accelerating while avoiding a collision with the vehicles in front.

The second case occurs when the route of the ego vehicle crosses an intersection protected by traffic lights. In this case as well, the ego vehicle applies successively two different control policies. A caution policy during which the ego vehicle reduces its speed as it approaches the intersection to ensure that it can stop safely if the traffic lights turn red. Then, a progress policy, when the ego vehicle is close enough to the intersection and decides to cross it, possibly accelerating. This requires the ego vehicle to estimate crossing time, and make sure that it enters the critical zone before the light turns red and gets out of the critical zone before the lights of a cross street turn green. In addition, in this maneuver, the ego vehicle should take into account the position and speed of vehicles in front of it to avoid a collision. 

CCTest generates critical test cases for the identified situations, seeking in particular to explore the transition between caution and progress policies. It is based on an analysis that takes two important factors into account.

\begin{itemize}
    \item The distances from the critical zone and the speeds of the ego vehicle and of an arriving vehicle heading towards the critical zone, and a vehicle in front moving away from it. These quantities define a configuration, which can be specified as a triplet $cf=\langle q_e,q_a,q_f\rangle$, consisting of the kinematic states of the three vehicles in their environment modeled by a map. More precisely, $q_e=\langle x_e,v_e\rangle, q_a=\langle x_a,v_a\rangle, q_f=\langle x_f,v_f\rangle$, where the first component is the distance of the corresponding vehicle from the critical zone and $v_e$ is its speed.
    \item The ability of the vehicles to accelerate and decelerate by modifying their speed according to the needs of their autopilot’s control policies.

\end{itemize}

For a given configuration, $cf=\langle q_e,q_a,q_f\rangle$, a test case for the ego vehicle is a triplet $tc=\langle q_e, Q_a(t), Q_f(t)\rangle$ where $q_e$ is the initial state of the ego vehicle, and $Q_a(t)$, $Q_f(t)$ are time functions specifying the evolution of the kinematic states of the arriving vehicle and the front vehicle on the map representing their environment. The application of a test case in the test environment generates a scenario in which the autopilot of the ego vehicle generates a control policy that adjusts its speed in order to avoid collisions and to respect the applicable traffic rules. Formally, a control policy is a function $cp$ that for a given test case $tc$, gives the evolution of the state of the ego vehicle $Q_e(t)$. In other words, $cp(tc)=Q_e(t)$, which defines a scenario $sc=\langle Q_e(t),Q_a(t),Q_f(t)\rangle$ from the initial configuration $cf=\langle Q_e(0), Q_a(0), Q_f(0)\rangle$. A control policy is safe if the corresponding scenario avoids collision and respects traffic rules. 

CCTest allows the generation of test cases $tc$ that meet the following two requirements.
\begin{itemize}
    \item The test cases generated from a given configuration are the most critical according to a criticality order between test cases defined as follows. A test case $tc_1$ is more critical than a test case $tc_2$, if the existence of a safe policy for $tc_1$ implies the existence of a safe policy for $tc_2$.
    \item The generated test cases are potentially safe. A test case is potentially safe if it is theoretically possible for the ego vehicle to generate a safe control policy for this test case. Deciding this property requires the use of predictive analysis techniques, which in general is a hard problem that requires taking into account ego vehicle dynamics, in particular its acceleration and deceleration capabilities. However, potential safety is a necessary condition for considering that an observed risk calls into question the safety of the ego vehicle. 
\end{itemize}

As explained, the generation of test cases meeting these two requirements has to take into account vehicle dynamics characterized by the following three functions, which we call A/D (acceleration/deceleration) functions, for short \cite{li2024rigorous,li2021estimation}:

 (1) The braking function $B(v)$ that gives the distance needed to brake from speed $v$ to speed $0$.

(2) The acceleration time function $AT(v,x)$ that gives the time needed to cover distance $x$ by accelerating from speed $v$.

(3) The acceleration speed function $AV(v,x)$ that gives the speed reached from speed $v$ after accelerating for distance $x$.

In Section \ref{sec:estimatedyn}, we show how these functions can be estimated experimentally.

Below, we present the foundations of the CCTest test case generation method, which is based on the analysis of constraints on configuration parameters. The method guarantees both high criticality and potential safety of the generated test cases. All technical details are available in \cite{li2024rigorous}.

The basis of the critical test case calculation method is to consider test cases $tc= \langle x_e, v_e, X_a(t), V_a(t), X_f(t), V_f(t)\rangle$ characterizing the following worst-case conditions, for given $\langle x_e,v_e\rangle$.
The speed of the arriving vehicle is constant and equal to the maximum authorized speed $vl$ on the road it is traveling on. Thus, $X_a(t)= X_a(0)-vl*t$. Additionally, the speed of the preceding vehicle is equal to $0$ and, consequently, $X_f(t)=X_f(0)$. 

Clearly, $tc$ is more critical than any test case $tc'=\langle x_e, v_e, X'_a(t),$ $V'_a(t), X'_f(t), V'_f(t)\rangle$ such that $V'_a(t)$ is bounded by $vl$.
Critical test cases of this form have the advantage of being able to decide on their potential safety by analyzing a set of constraints on a tuple of values $\tilde{tc}=\langle x_e, v_e, x_a, x_f\rangle$. This is the initial configuration from which the test case is applied for speeds $vl$ and $0$, respectively, of the arriving vehicle and the front vehicle. 
We provide the method for calculating initial configurations for the type of critical test case considered for two configurations: 1) priority-protected meeting of routes; and 2) traffic-light-protected junction. These configurations specify precisely the states in which transitions take place between a cautious ego vehicle policy that allows it to stop safely before a critical zone and a policy of progression based on the estimation that there is no risk of collision.

\subsubsection{Priority-protected meeting of routes}

\begin{figure}[h]
\centering
\includegraphics[width=0.25\linewidth]{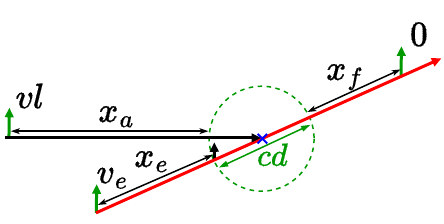}
\vspace{-10pt}
\caption{Priority-protected meeting of routes}\label{fig:vistapara}
\vspace{-5pt}
\end{figure}

Fig. \ref{fig:vistapara} shows the configuration for a test case $\tilde{tc} = \langle x_e,v_e,x_a, x_f\rangle$, in which the route of the ego vehicle, in red, meets the high-priority route of an arriving vehicle. In addition, there is a vehicle in front, after the critical zone, on the route of the ego vehicle.  The ego vehicle must give way to the arriving vehicle, while maintaining a safe distance from the front vehicle. We assume that when the two routes intersect, there is a critical zone covering the ego vehicle’s route for a distance $cd$.

This form of test cases covers the situations where the ego vehicle merges onto a main road, changes lanes, or crosses an intersection protected by yield signs.
As explained, we consider critical test cases where: 1) the arriving vehicle travels at constant speed $vl$ that is the maximal allowed speed on its route; 2) the front vehicle is static, i.e., $v_f = 0$. 

In this test case, the ego vehicle can adopt a cautious policy and stop before the critical zone, waiting for the arriving vehicle to cross.
This policy should respect the constraint $B(v_e)\leq x_e$, for braking safely before the critical zone. An alternative is that the ego vehicle progresses without waiting for the arriving vehicle. In that case, it will accelerate for distance $x_e + cd$ exiting the critical zone, at speed $AV(v_e,x_e+cd)$, then to avoid collision with the front vehicle, it should have enough distance to brake. As the front vehicle is static, the constraint $B (AV(v_e,x_e+cd))\leq x_f$  should be met.   

For the arriving vehicle, the worst case is that it drives at the maximal speed limit $vl$. So, if the ego vehicle decides to progress by exiting the critical zone during time $AT(v_e, x_e+cd)$, the arriving vehicle will travel a distance $vl * AT(v_e, x_e+cd)$. Then, the constraints on safe progress can be determined in two cases:
\begin{itemize}
\item If the routes of the arriving vehicle and the ego vehicle merge, the distance of the arriving vehicle should be large enough such that it can brake to avoid collision with the ego vehicle after the ego vehicle merges onto the arriving vehicle's route, i.e., $B(vl) + vl* AT(v_e, x_e+cd) \leq  x_a$. 
\item If the routes of the arriving vehicle and the ego vehicle intersect in the critical zone, the arriving vehicle's initial distance to the critical zone should be larger than the maximal traveled distance at speed $vl$ before the ego vehicle exits the zone, which gives $vl * AT(v_e, x_e + cd) \le x_a$.
\end{itemize}

\vspace{3pt}
For the ego vehicle progressing under this type of test case, both the constraints on $x_f$ and $x_a$ should be satisfied at the same time. A critical test case for given $x_e$ and $v_e$ is characterized by the minimal values of $x_a$ and $x_f$ that allow the ego vehicle to progress.

\subsubsection{Junction protected by traffic lights}

\begin{figure}[h]

\centering
\vspace{-5pt}
\includegraphics[width=0.35\linewidth]{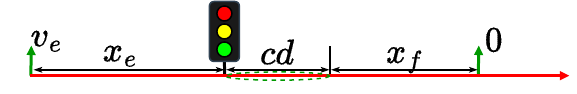}
\vspace{-10pt}
\caption{test case with traffic light}\label{fig:trafficlight}
\vspace{-5pt}
\end{figure}

Fig.~\ref{fig:trafficlight} shows the configuration for a test case  $\tilde{tc}=\langle x_e, v_e, x_f \rangle$ with a traffic light where the arriving vehicles are irrelevant. It involves the ego vehicle approaching at speed $v_e$ and at distance $x_e$ from a traffic-light protected intersection of width $cd$, and a static front vehicle located at the ego vehicle's route at a distance $x_f$ after the exit of the critical zone. 
The traffic light has a state variable taking values ``red'', ``yellow'', and ``green''. We assume that we know the duration $t_y$ of the yellow light. Furthermore, the lights of the intersection have an ``all red'' phase of duration $t_{ar}$ where all the lights are red before the lights of the intersecting road pass from red to green. These constants are important for safety regulations requiring that the ego vehicle enters the critical zone only when the lights are either green or yellow and exits before the lights of intersecting roads turn green.

In this test case, when the ego vehicle approaches the traffic lights, the constraint applied for a cautious policy of stopping before the intersection is $B(v_e) \leq  x_e$.

The constraint for progress is that the ego vehicle must not enter the critical zone when it sees a red light. This means that even if the light switches to yellow right after the decision to cross is taken, the ego vehicle will reach the entrance to the intersection before the lights turn to red, i.e., $AT(v_e, x_e) \leq  t_y$.

Additionally, the ego vehicle should leave the critical zone before the end of the all-red phase. That is, the time to travel distance $x_e+cd$ should be less than $t_y+t_{ar}$, i.e., $AT(v_e, x_e+cd) \leq  t_y+t_{ar}$. 

Moreover, the constraint induced by the presence of a front obstacle at $x_f$ from the intersection is $B(AV(v_e, x_e+cd))\leq  x_f$.

For given $x_e, v_e$ and contextual parameters $t_y, t_{ar}, cd$, the conditions for safe progress are: $AT(v_e, x_e)\le t_y$, $AT(v_e, x_e + cd) \le t_y + t_{ar}$, and $B(AV(v_e, x_e + cd)) \le x_f$. A critical test case is determined for safe values of $x_e$ and $x_f$ such that the time constraints are met and the distance $x_f$ is minimal.

\subsection{Estimating vehicle dynamics}\label{sec:estimatedyn}

We show how we can obtain specifications of the A/D functions for a given end-to-end autopilot. Unlike most modular autopilots, which have explicitly defined functions for acceleration and deceleration that allow for deriving A/D functions analytically \cite{li2024rigorous}, end-to-end autopilots lack such specifications. Therefore, we estimate the A/D functions empirically, as explained below.

To estimate the function $B(v)$, we consider scenarios with an initial speed $v$ and a front obstacle in distance $x$. The simulation runs until the vehicle either stops safely or collides with the obstacle. We determine the value of $B(v)$ as the minimal distance $x$ where the vehicle stops without collision. This minimal distance is obtained by progressively increasing $x$ from 0 until the value at which a safe stop occurs.

To estimate the functions $AT(v, x)$ and $AV(v, x)$, we consider scenarios where a vehicle starts at speed $v$ on a clear road. The simulation runs until the vehicle travels for distance $x$. We estimate $AT(v, x)$ as the time taken by the ego vehicle to travel distance $x$, and $AV(v, x)$ as the final speed of the ego vehicle.

\section{Testing End-to-end Autopilots in the Carla Simulator}\label{se:carla}

{
\subsection{The Carla Simulator}

The CARLA Simulator is designed to mimic a real-world driving environment. Its Runtime, shown in Fig.~\ref{fig:carla-arch}, supplies sensory data to the autopilots, and processes the control signals it receives from them to update the state of the world. It controls the execution of four key functions:
\begin{itemize}
    \item \textbf{World state update}: This function maintains a world state, including the positions and speeds of vehicles and obstacles on a map. It provides an API to set the initial world state.
    \item \textbf{Route Planner}: Given an initial position and a destination, this function plans a route on the map, specified as a sequence of waypoints, and used by the world state update function. 
    \item \textbf{Sensor simulation}: This function generates simulated raw sensor data, such as camera images or LiDAR scans, corresponding to the current world state.
    \item \textbf{Dynamics simulation}: Based on the control signals generated by the autopilot, this function simulates vehicle movements, updating the world state at each cycle to reflect the vehicle's new position and speed.
\end{itemize}

\vspace{3pt}
The Carla Simulator simulates the behavior of a set of vehicles from an initial state of the world and their routes by alternating the execution of their autopilots and the calculation of the resulting state of the world by the Runtime. At the beginning of each cycle, the Runtime provides sensory data from the Sensor Simulation function to the autopilot. The autopilot then computes control signals for a fixed time period. The cycle ends by updating the world with the vehicles’ kinetic state, based on the received control signals, using the Dynamic Simulation function.


\begin{figure}[h]
    \centering 
    \includegraphics[width=0.53\linewidth]{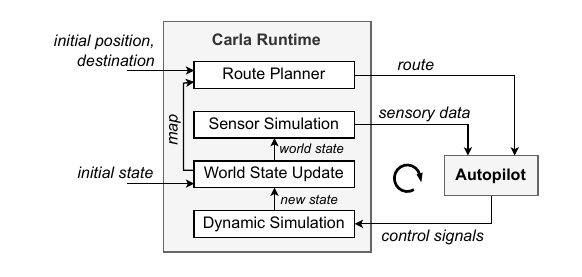}
    \vspace{-15pt}
    \caption{Simulating one vehicle in the Carla Simulator}
    \label{fig:carla-arch}
\end{figure}
}

\subsection{The Carla Leaderboard Test Framework}

The Carla Leaderboard \cite{carla2020leaderboard} is a testing and evaluation environment using the Carla Simulator. It includes benchmarks comprising sequences of test cases applied sequentially along the ego vehicle's predefined route on a given map so as to faithfully cover real-life situations with a mix of normal, critical, and accidental test cases. In addition, it proposes a quantitative evaluation method that calculates scores measuring driving proficiency. It serves as an open platform for the comparison between different approaches \cite{al2024end}.
We present below the Carla Leaderboard 1.0, its framework, and its evaluation method. The choice of Leaderboard 1.0 is due to the fact that the four autopilots studied were evaluated with this version in~\cite{shao2023safety, chitta2022transfuser, hu2022model, shao2024lmdrive}.

\begin{figure}[h]
    \vspace{10pt}
    \centering
    \includegraphics[width=0.55\linewidth]{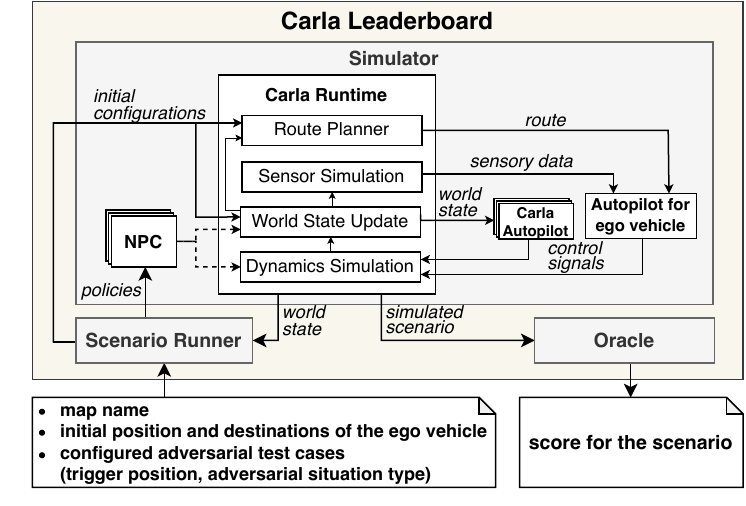}
    \vspace{-10pt}
    \caption{The Carla Leaderboard test framework}
    \label{fig:lederboard-arch}
    \vspace{-10pt}
\end{figure}

The Leaderboard framework is shown in Fig.~\ref{fig:lederboard-arch}, including the Carla Simulator, a Scenario Runner, and an Oracle. Its input data are benchmarks made up of test case sequences. A test case sequence involves the ego vehicle, background vehicles, and NPCs. It is linked to a map in the Carla Simulator and a predefined route of the ego vehicle. Along this route, adversarial test cases in the sequence are configured as pairs of trigger positions and their corresponding types~\cite{thorn2018framework}. The ego vehicle is controlled by the autopilot under test starting at zero speed. The background vehicles are controlled by Carla's built-in autopilot to create normal test cases. The NPCs are planned to create adversarial test cases. The execution for adversarial test cases is coordinated during the simulation, by the Scenario Runner. The Oracle performs a quantitative evaluation of the simulated scenario and assigns a score.



To generate normal test cases, the Leaderboard spawns 100–200 background vehicles, depending on the map size. These vehicles are randomly placed at lane centers, initialized at zero speed. Their routes are randomly generated at runtime, adapting to the map layout and allowing the vehicles to take any direction at junctions.   

Adversarial test cases of the Leaderboard are instances of 10 types C1-C10 selected from the NHTSA's 37-pre-crash typology \cite{najm2007pre}. Their names are listed in Tab. \ref{tab:adversarial-scenarios}, which also provides the informal description of these test cases obtained from Python code used by the Scenario Runner of the Leaderboard. Each type of adversarial test case corresponds to a specific road or junction context, and may also define parameters that determine NPC behavior or contextual factors (e.g., traffic light policies or road damages) to create critical or abnormal situations conducive to accidents. 

While the ego vehicle is traveling, as soon it reaches a trigger position, the corresponding adversarial test case is activated, controlling NPCs or modifying contextual parameters. When the ego vehicle leaves the relevant area or is blocked, the adversarial test case ends, the NPCs are removed, and the contextual parameters are reset to their default values, allowing the ego vehicle to continue on its way.

\begin{table*}[t]
    \centering
    \caption{Adversarial test cases types defined in Carla Leaderboard 1.0}
    \label{tab:adversarial-scenarios}
    \vspace{-5pt}
    \setlength{\aboverulesep}{0pt}
    \setlength{\belowrulesep}{0pt}
    \small
    \begin{tabular}{c|m{3.5cm}|m{12cm}}
    \toprule
         \textbf{Type ID} & \textbf{Name}& \textbf{Description}\\
    \midrule
         C1& Control loss without previous action & Introduce three damage areas at distances 14, 48, and 74 meters in front of the ego vehicle. \\
    \hline
         C2& Longitudinal control after leading vehicle's brake & Spawn a front vehicle at a distance of 25 meters. Make it keep a constant speed of 10 m/s until reaching 20 m before the next intersection and then brake at the maximum deceleration. \\
    \hline
         C3& Obstacle avoidance without prior action & The ego vehicle is driving on the road with $N$ lanes and has $M$ lanes to the right shoulder. Spawn a pedestrian at a distance of $12 + 1.5 * M$ m and cross the road with a speed of $3 + 0.4 * N$ m/s.  \\
    \hline
         C4 & Obstacle avoidance with prior action & Spawn a bicycle on the right shoulder of a road connected to a junction exit. Assume the ego vehicle is driving in a lane that is a lateral distance $L$ from the shoulder. When the distance between the ego vehicle and the bicycle becomes less than $13+L$ m, the bicycle will start crossing the road at a speed of 10 m/s. \\
    \hline
         C5& Lane changing to evade the slow leading vehicle & Spawn a vehicle on the inner lane at 35 m in front of the ego vehicle that starts to brake when the ego vehicle is within 55 m. Another vehicle (with length $l$) on the outer lane at 36 m front drives and will start to brake when detecting a vehicle ahead (with length $l'$) at $3 * (l + l')$. \\
    \hline
         C6& Vehicle passing dealing with oncoming traffic & Spawn a static vehicle at 50 meters front. Another vehicle (with length $l$) on the outer lane is driving from the opposite direction with a speed of 5.56 m/s and will start to brake when detecting a vehicle ahead (with length $l'$) at $3*(l + l')$.  \\
    \hline
         C7& Crossing traffic running a red light at an intersection & Set both the front and side traffic lights for the ego vehicle to green, allowing vehicles controlled by the Carla autopilot approaching from lateral directions to cross the junction. This simulates a situation where lateral vehicles run red lights. \\
    \hline
         C8& Unprotected left turn at an intersection with oncoming traffic & Adjust the traffic light colors to permit the ego vehicle to make a left turn while allowing the oncoming vehicles in the opposite direction to proceed straight. \\
    \hline
         C9& Right turn at junctions with crossing traffic & If the ego vehicle's route is turning right, adjust the traffic light colors to permit the ego vehicle to make a right turn while allowing the oncoming vehicles on the left to proceed straight. \\
    \hline
         C10& Crossing negotiation at an unsignalized intersection & The ego vehicle is crossing an unsignalized intersection. The Leaderboard only specifies an expected duration (180 seconds) for passing the intersection without other operations.\\
    \bottomrule
    \end{tabular}
\end{table*}

Through our further classification, five of the ten types of adversarial test cases are potentially safe, including C2, C5, C6, C8, and C10. These test cases either involve only autonomous vehicles starting at zero speed under normal conditions (C8, C10), or feature only NPCs at higher priority, leaving the ego vehicle responsible for managing potential risks (C2, C5, C6).

These five types are considered critical instead of normal because they involve situations where accidents are more frequent, according to NHTSA~\cite{thorn2018framework}. However, the Leaderboard does not distinguish the varying levels of criticality for each type of test case. The test cases are generated by background vehicles with randomly determined initial positions and routes, all starting from zero speed, while NPCs, where they exist, have fixed speed policy parameters. In particular, C10, which involves crossing negotiation at an unsignalized intersection, uses default context settings and does not involve NPCs. Thus, it is generated in the same way as a normal test case. 

The other adversarial test cases are accidental, involving NPCs breaking traffic regulations (C3, C4) or abnormal traffic control mechanisms (C1, C7, C9). For these types of {test cases}, it is important to know whether an accident could have been avoided, which requires analyses of the control capabilities of the vehicles involved. However, the Leaderboard does not provide this type of information.

The Leaderboard adopts a global quantitative evaluation approach by computing scores for the simulated scenarios. These depend on the route completion rate of the ego vehicle and the number of different incident types that occurred during the route completion. More precisely, the score $Sc$ is defined as $Sc=R*P$, where $R$ is the percentage of the route distance completed by the ego vehicle, $P$ is the incident penalty rate. Route completion can fail for a variety of reasons, including untimely route changes, inability to complete the route in the allotted time, or stalling without progress. The penalty rate $P$ is further defined as $P=\prod_{i_\in I} p_i^{n_i}$, where $p_i$ is the penalty rate for an incident type $i$ from a given set of incident types $I$, and $n_i$ is the number of occurrences of this type.
The Leaderboard sets the penalty rate to 0.50 for an accident with a pedestrian, 0.60 for an accident with a vehicle, 0.65 for an accident with a static layout, 0.7 for running a red light, and 0.8 for running a stop sign.

It should be noted that the evaluation approach does not take into account the potential safety of the test case and the implied responsibility for each incident. It also fails to distinguish between the incidents in normal and critical test cases, and those in accidental test cases that may be unavoidable, which can significantly distort the calculated score of the ego vehicle.

\subsection{The CCTest Environment}\label{se:testenv}

\begin{figure}[h]
    \centering
    \includegraphics[width=0.48\linewidth]{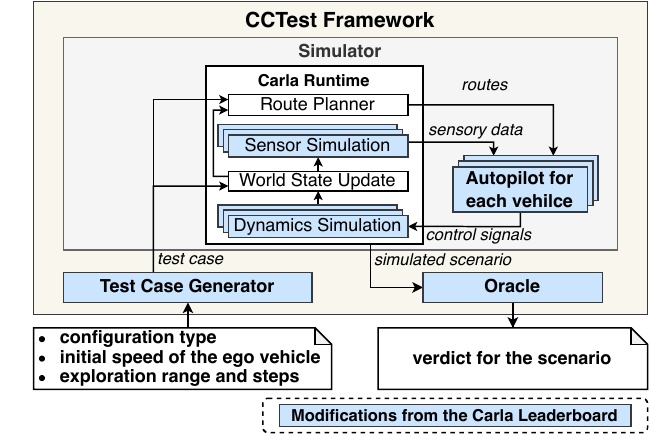}
    \vspace{-5pt}
    \caption{The architecture of the CCTest Framework}
    \label{fig:cctest}
\end{figure}

To apply CCTest to the four autopilots, we modified the Carla Leaderboard, as shown in Fig.~\ref{fig:cctest}.

The Simulator was modified to support the simulation of several vehicles controlled by their own autopilots, driving autonomously along their own routes. They therefore differ from NPCs in that they do not intentionally violate safety rules, and also from Leaderboard background vehicles, which make random decisions by changing their route at runtime. In addition, we modified the Simulator to allow a vehicle's initial speed to be set to values greater than zero.

Unlike the Leaderboard, which requires benchmark test cases, CCTest uses a Test Case Generator to automatically create test cases. The Test Case Generator receives inputs specifying the configuration type, the initial speed of the ego vehicle, and the range and step for exploring the parameters around the critical values. It first generates the test case with critical values based on the analysis of the vehicle dynamics and context parameters accessed from the Simulator. Then it explores the parameter values following the given range and step, and finally refines the intervals where the verdict given by the Oracle changes from caution to progress.

We envisage four types of configurations for testing, including:
\begin{itemize}
    \item \textbf{Merging}: The ego vehicle is on a road that merges onto a main road protected by a yield sign on which there is an arriving vehicle and a static front vehicle.
    \item \textbf{Lane change}: On a two-lane straight road, the ego vehicle changes from the inside lane to the outside lane with a travel distance of $x_e=13.5$ m to overtake a static front vehicle at a distance of $x_{f'}$. On the outside lane, there is an arriving vehicle and a static front vehicle.
    \item \textbf{Crossing with yield sign}: At an intersection where the critical zone covers 20 meters of the ego vehicle's route, the ego vehicle's entry is controlled by a yield sign. A vehicle arrives to the left of the ego vehicle, and a vehicle in front of it is outside the critical zone on the ego vehicle's route.
    \item  \textbf{Crossing with traffic lights}: 
    The ego vehicle approaches an intersection protected by traffic lights. After the critical zone, which covers 20 meters of its route, a static front vehicle is positioned. The ego vehicle sees first a green
light, then immediately the light changes to yellow at the next time step. The initial state of the ego vehicle is therefore exactly the state where the light changes from green to yellow. We use the default traffic light policy in the Carla Simulator, where the duration of the yellow light is 3 seconds, and the duration of the all-red phase is 2 seconds.
\end{itemize}

\vspace{5pt}

The CCTest Oracle analyzes the behavior of the ego vehicle in the simulated scenario to detect caution or progress. It also checks safety issues, including accidents and violations of the safety properties listed in Tab. \ref{tab:safety-properties}. It delivers the following types of verdicts:
\begin{itemize}
\item \texttt{PS} and \texttt{PU}$p$, respectively for safe and unsafe progress violating property $p$. 
\item \texttt{CS}, \texttt{CU}$p$, respectively for safe caution and unsafe caution violating property $p$.
\item \texttt{Ae} and \texttt{Aa} respectively, for accidents in which the ego vehicle collides with the arriving vehicle and the arriving vehicle collides with the ego vehicle.  \texttt{Af} for accidents in which the ego vehicle collides with the front vehicle.
\item \texttt{Blk} when the ego vehicle runs aground midway and blocks the path of the arriving vehicle.
\item \texttt{DRe} and \texttt{DRa}, respectively, denote the deviation of the ego and the arriving vehicle from lanes. \texttt{CRe} and \texttt{CRa}, respectively, represent the fact that the ego vehicle and the arriving vehicle change their specified route instead of following it. In particular, \texttt{CDR} denotes the fact that a vehicle changes its route and then deviates from the road.
\end{itemize}


\begin{table}[h]
    \centering
    \caption{Safety properties}
    \label{tab:safety-properties}
    \vspace{-10pt}
    \adjustbox{max width=0.7\linewidth}{
    \small
    \begin{tabular}{l}
    \hline
         \textbf{For crossing context (with yield signs or traffic lights)}: \\
         $p_1$:  Two vehicles must not be in the critical zone at the same time. \\
         $p_2$: The ego vehicle must not stop inside the critical zone.\\
         \textbf{For crossing with traffic lights}:\\
         $p_3$: The ego vehicle must not enter the critical zone when the light is red.\\
         $p_4$: The ego vehicle must not be in the critical zone when a side light is green.\\
    \hline
    \end{tabular}
    }
\end{table}


\section{Test results analysis and evaluation}\label{se:analysis}

\subsection{The Four Evaluated End-to-end Autopilots}

\begin{figure}[h]
    \centering
    \vspace{20pt}
    \includegraphics[width=\linewidth]{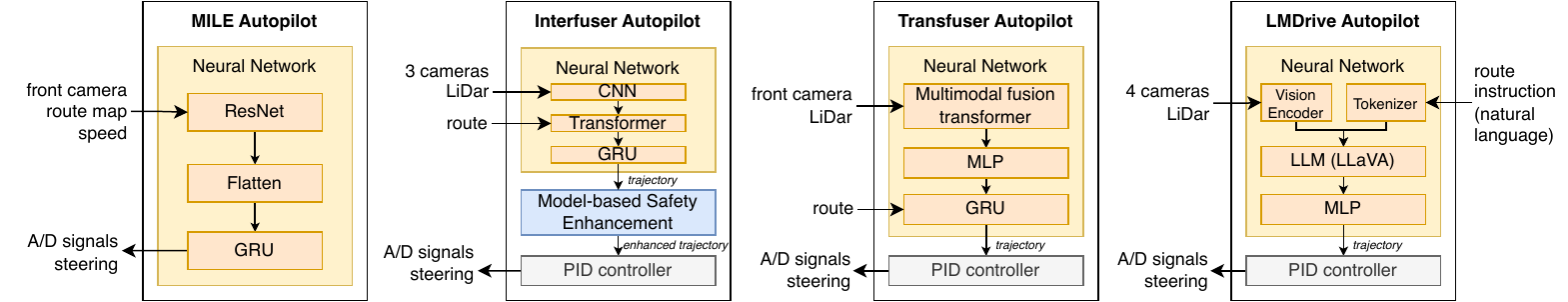}
    \vspace{-15pt}
    \caption{Architecture of the four autopilots}
    \label{fig:autopilots}
    \vspace{-5pt}
\end{figure}

\noindent\textbf{Architectures.~} In Fig. \ref{fig:autopilots}, we present the relevant characteristics and architectures of the four end-to-end open autopilots that were evaluated in the Carla Leaderboard.

MILE employs a neural network to directly convert sensory input into control signals, including acceleration/deceleration (A/D) and steering. In contrast, the other three autopilots generate trajectories using neural networks, which are then converted into control signals via a PID controller.
Additionally, InterFuser incorporates a model-based safety module that enhances the neural network's trajectory output. This module adjusts the trajectory's speed using a linear programming model, taking into account nearby vehicles, obstacles, and their potential movements.

All autopilots, except MILE, rely on fusion-based methods that combine inputs from cameras and LiDAR. Transfuser and InterFuser use a Transformer network, while LMDrive employs the large language model LLaVA. LLaVA \cite{liu2023llava} is a multimodal LLM that adapts to various input types, not limited to natural languages, such as images and other data formats, with proper adaption. For trajectory or control signal generation, MILE, Transfuser, and InterFuser use a Gated Recurrent Unit (GRU) to process the fused input. In contrast, LMDrive uses a multilayer perceptron (MLP) to extract future trajectories from the LLM. 




\vspace{5pt}
\noindent\textbf{Estimating autopilot dynamics.~} For the four autopilots, we experimentally estimated their maximum speeds $v_{max}$ by letting vehicles drive long enough on freeways with no obstacles in sight. We also estimate their A/D functions using the method described in Section \ref{sec:estimatedyn}.

The maximum speeds of the four autopilots are generally low. We found that LMDrive and MILE have a maximum speed of 6.5 m/s (23.4 km/h), InterFuser is limited to 5.0 m/s (18 km/h), and Transfuser to 4.0 m/s (14.4 km/h). It should be noted that the low maximum speeds of the autopilots can mask design flaws that only become apparent at higher speeds. 

InterFuser, Transfuser, and LMDrive share the same A/D functions as shown in Tab. \ref{tab:bv-value}(a) and Tab. \ref{tab:av-value}(a) when their speeds do not exceed the maximum, as they use PID controllers with the same acceleration and deceleration parameters. MILE has different A/D functions, shown in Tab. \ref{tab:bv-value}(b) and Tab. \ref{tab:av-value}(b), as it relies on a neural network to determine acceleration and deceleration, which distinguishes it from the other autopilots.

\begin{table}[h]
    \centering
    \caption{$B(v)$ for the four autopilots}
    \label{tab:bv-value}
    \vspace{-21pt}
\end{table}

\begin{minipage}{0.49\linewidth}
\begin{center}
    \small{(a) \textbf{For InterFuser, Transfuser, and LMDrive}}
    
    {\footnotesize
    \begin{tabular}{|c|c|c|c|c|c|c|c|}
    \hline
         $\bm{v}$ & 0.0 & 1.0 & 2.0 & 3.0 & 4.0 & 5.0 \\
    \hline
         $\bm{B(v)}$ & 0.0 & 0.1 & 0.4 & 1.0 & 1.9 & 2.3 \\
    \hline
    \end{tabular}
    }
\end{center}
\end{minipage}
\begin{minipage}{0.49\linewidth}
\begin{center}
    \small{(b) \textbf{For MILE}}\\
    
    {\footnotesize
    \begin{tabular}{|c|c|c|c|c|c|c|c|}
    \hline
         $\bm{v}$ & 0.0 & 1.0 & 2.0 & 3.0 & 4.0 & 5.0 \\
    \hline
         $\bm{B(v)}$ & 0.0 & 0.1 & 0.6 & 1.5 & 2.4 & 2.4 \\
    \hline
    \end{tabular}
    }
\end{center}
\end{minipage}

\vspace{10pt}

\begin{table}[h]
    \centering
    \caption{$AV(v, x), AT(v, x)$ for the four autopilots}
    \label{tab:av-value}
    \vspace{-21pt}
\end{table}

\begin{minipage}{0.49\linewidth}
\begin{center}
\small{(a) \textbf{For Interufser, Transfuser, and LMDrive}}

\adjustbox{max width=\linewidth}{\footnotesize
    \begin{tabular}{|c|c|c|c|c|c|c|c|}
        \hline
        \diagbox[width=3em,height=2em]{~$\bm v$}{$\bm x$~} 
             & 0.0 & 1.0 & 2.0 & 3.0 & 4.0 & 5.0 & 6.0 \\
        \hline
         0.0& 0.0, 0.0& 2.3, 0.9 & 3.7, 1.2 & 4.7, 1.4 & 5.1, 1.6 & 5.2, 1.8 & 5.4, 2.0 \\
          \hline
         1.0& 1.0, 0.0& 2.4, 0.6 & 3.8, 0.9 & 4.7, 1.2 & 5.1, 1.4 & 5.2, 1.6 & 5.4, 1.8 \\
         \hline
         2.0& 2.0, 0.0& 2.9, 0.4 & 4.1, 0.7 & 5.0, 1.0 & 5.1, 1.1 & 5.2, 1.3 & 5.4, 1.5 \\
         \hline
         3.0& 3.0, 0.0& 3.7, 0.3 & 4.7, 0.5 & 5.2, 0.7 & 5.2, 0.9 & 5.7, 1.1 & 6.5, 1.3 \\
         \hline
         4.0& 4.0, 0.0& 4.6, 0.2 & 5.2, 0.4 & 5.2, 0.6 & 5.8, 0.8 & 6.2, 1.0 & 6.5, 1.1 \\
        \hline
    \end{tabular}
}
\end{center}
\end{minipage}
\begin{minipage}{0.49\linewidth}
\begin{center}
\small{(b) \textbf{For MILE}}
\adjustbox{max width=\linewidth}{\footnotesize
    \begin{tabular}{|c|c|c|c|c|c|c|c|}
        \hline
        \diagbox[width=3em,height=2em]{~$\bm v$}{$\bm x$~} 
            & 0.0 & 1.0 & 2.0 & 3.0 & 4.0 & 5.0 & 6.0 \\
        \hline
        0.0  & 0.0, 0.0& 3.2, 0.7 & 4.9, 0.9 & 5.4, 1.1 & 5.9, 1.3 & 6.3, 1.5 & 6.5, 1.6 \\
        \hline
        1.0  & 1.0, 0.0& 3.3, 0.5 & 5.0, 0.8 & 5.4, 1.0 & 5.9, 1.2 & 6.3, 1.3 & 6.5, 1.5 \\
        \hline
        2.0  & 2.0, 0.0& 3.5, 0.4 & 5.1, 0.6 & 5.4, 0.8 & 5.9, 1.0 & 6.3, 1.2 & 6.5, 1.3 \\
        \hline
        3.0  & 3.0, 0.0& 4.1, 0.3 & 5.3, 0.5 & 5.4, 0.7 & 5.9, 0.9 & 6.3, 1.0 & 6.5, 1.2 \\
        \hline
        4.0  & 4.0, 0.0& 4.9, 0.2 & 5.3, 0.4 & 5.5, 0.6 & 6.1, 0.8 & 6.3, 0.9 & 6.5, 1.1 \\
        \hline
    \end{tabular}
}
\end{center}
\end{minipage}

\vspace{20pt}

\hidden{
\vspace{-2pt}
\noindent\textbf{Determinism.~} While non-determinism in the Carla Runtime has been identified, we further examine the determinism of the autopilots by verifying the reproducibility of observed behavior with identical sensory data. The results show that LMDrive exhibited non-deterministic behavior, whereas the other autopilots are deterministic. A detailed analysis of LMDrive's implementation revealed two sources of its non-determinism, both aiming to improve computational performance. First, during LiDAR data processing, LMDrive reduces the data size by randomly sampling points from the raw input. Second, during neural network inference, LMDrive uses parallel operators that lead to floating-point rounding variations due to different computation orders between threads.
}

\subsection{RQ1: CCTest result analysis}

\subsubsection{Experimental settings}

In the experiments, we apply test cases whose $x_f$ and $x_a$ values are around the critical ones, $\hat x_a$ and $\hat x_f$. 

Given the low maximum speeds of the four autopilots, we set $x_a$ between 0.0 and 40.0 m, considering that a vehicle arriving from a greater distance does not present a significant hazard. However, if the ego vehicle consistently exhibits cautious behavior in this range, we extend $x_a$ to 520.0 m to examine whether the ego vehicle can progress in the given type of test case. For $x_f$, we evaluate over a broader range of 0.0 to 520.0 m. Finally, for merging, passing yield signs and traffic lights, we take $v_e$ to be 0.0 m/s and 4.0 m/s with $x_e=B(v_e)$; for lane changes, we take $v_e=4.0$ m/s and $x_f'=B(v_e)$.

Additionally, we observe that the Carla Runtime can produce different sensory data for the same system state, hindering the test result reproducibility. To address this, we execute the same test case five times and record the number of times each verdict is issued.

\subsubsection{Test results for InterFuser}

\paragraph{1) InterFuser Autopilot: Analysis of merging scenarios}

\begin{minipage}{\linewidth}
\begin{table}[H]
    \centering
    \caption{InterFuser: Verdicts for merging scenarios}
    \label{tab:interfuser-merging}
    \vspace{-20pt}
\end{table}

\begin{center}
Table \ref{tab:interfuser-merging} (a). InterFuser: Verdicts for merging scenarios with $v_e=0.0$ m/s

\adjustbox{max width=0.9\linewidth}{
\input{tables/interfuser-merging-0}
}
\end{center}
\end{minipage}

\vspace{10pt}
\noindent
\begin{minipage}{\linewidth}
\begin{center}
Table \ref{tab:interfuser-merging} (b). InterFuser: Verdicts for merging scenarios with $v_e=4.0$ m/s
\adjustbox{max width=0.9\linewidth}{
    \input{tables/interfuser-merging-4}
}
\end{center}
\end{minipage}

~

Table \ref{tab:interfuser-merging} presents verdicts for InterFuser in merging scenarios generated by application of test cases defined by the parameters $x_a$ and $x_f$, whose critical values are indicated in red. 

In Table \ref{tab:interfuser-merging}(a) where $v_e=0.0$, accidents are mainly observed during the transition from caution to progress when $x_a = 11.3$. Additionally, the ego vehicle has significantly non-deterministic behavior observed for this value of $x_a$. In particular, four different verdicts, \texttt{Ae}, \texttt{Ae}, \texttt{PS}, \texttt{CS}, are produced among five runs for $x_f=400.0$.  

For $\langle x_a, x_f \rangle = \langle 15.0, 240.0 \rangle, \langle 15.0, 440.0 \rangle, \langle 20.0, 480.0 \rangle, \langle 25.0, 240.0 \rangle$, four accidents are observed abnormally because their test cases are less critical than the ones with smaller $x_f$, in which only safe progress is observed.  In these accidents, the arriving vehicle fails to stop safely when it is following the ego vehicle after the ego vehicle changes the lane and stops before the front vehicle.

In Table \ref{tab:interfuser-merging}(b) where $v_e=4.0$, we observe similar patterns for $x_a = 4.2, 5.0, 7.5$ around $\hat x_a=4.2$ with accidents in the transition from caution to progress. Accidents that occur when $x_a=4.2, 5.0$ are mainly due to the progress of the ego vehicle when the arriving vehicle reaches the merge point. For $x_a=7.5$, accidents are mostly reproducible when $x_f\ge 40.0$, where the ego vehicle merges onto the main road leaving insufficient space for the arriving vehicle to brake. 

Anomalies occur again for $\langle x_a, x_f \rangle = \langle 25.0, 280.0 \rangle$, where we observe safe progress for the configurations with the same $x_a$ and smaller values of $x_f$. In this abnormal accident, the arriving vehicle also fails to stop safely when it is following the ego vehicle after the ego vehicle changes the lane and stops before the front vehicle.

\paragraph{2) InterFuser Autopilot: Analysis of lane change scenarios}

\begin{minipage}{\textwidth}
\begin{table}[H]
    \centering
    \caption{InterFuser: Verdicts for lane change scenarios with $v_e=4.0$ m/s}
    \label{tab:interfuser-lane_change}
    \vspace{-15pt}
\end{table}

\begin{center}

\adjustbox{max width=0.9\linewidth}{
\input{tables/interfuser-lane_change-4}
}
\end{center}
\end{minipage}

~

Table \ref{tab:interfuser-lane_change} presents verdicts for InterFuser in lane change scenarios generated by application of test cases defined by the parameters $x_a$ and $x_f$, whose critical values are indicated in red, while the value of $v_e$ is 4.0. We observe accidents in the transition from caution to progress for a wide range of $x_a \in [0.0, 25.0]$, while most accidents occur when $x_a = 15.0$ and independently of the value of $x_f$. Abnormal accidents are also observed for $x_a=0.0, 5.0, 10.0, 19.2, 20.0$, while safe progress is possible for smaller values of $x_f$. These accidents mainly occur after the ego vehicle has safely changed lanes, either because the arriving vehicle fails to follow the ego vehicle safely or because the ego vehicle fails to follow the arriving vehicle safely.

\paragraph{3) InterFuser Autopilot: Analysis of crossing with yield sign scenarios}

\begin{minipage}{\linewidth}
\begin{table}[H]
    \centering
    \caption{InterFuser: Verdicts for crossing with yield sign scenarios}
    \label{tab:interfuser-crossing}
    \vspace{-15pt}
\end{table}

\begin{center}
Table \ref{tab:interfuser-crossing} (a). InterFuser: Verdicts for crossing with yield sign scenarios with $v_e=0.0$ m/s
\end{center}
\adjustbox{max width=\linewidth}{
\input{tables/interfuser-crossing-0}
}
\end{minipage}

\vspace{5pt}
\noindent
\begin{minipage}{\linewidth}
\begin{center}
Table \ref{tab:interfuser-crossing} (b). InterFuser: Verdicts for crossing with yield sign scenarios with $v_e=4.0$ m/s
\end{center}
\adjustbox{max width=\linewidth}{
\input{tables/interfuser-crossing-4}
}
\end{minipage}

~

Table \ref{tab:interfuser-crossing} presents verdicts for InterFuser in crossing with yield sign  scenarios generated by the application of test cases defined by the parameters $x_a$ and $x_f$, whose critical values are indicated in red. 

In Table \ref{tab:interfuser-crossing}(a) where $v_e=0.0$, for $x_a=0.0, 2.5$, we observe failures due to the ego vehicle constantly moving into the critical zone despite the presence of the arriving vehicle. In particular, no safe caution is observed. The identified failures include many accidents and, in some cases, blockages, violations of the mutual exclusion rule $p_1$, or stopping inside the junction violating $p_2$. For values of $5.0 \le x_a \le 25.0$, we observe violations of traffic rules when $x_f\ge 40.0$ and blockages where the ego vehicle stops before the junction exit when $x_f=0.0$.

In Table \ref{tab:interfuser-crossing}(b) where $v_e=4.0$, we observe similar patterns for $x_a=0.0$ with accidents, blockages, and unsafe cautious behavior violating mutual exclusion rule $p_1$ and violating $p_2$ by stopping inside the junction. For $2.5 \le x_a \le 15.0$, we observe uniform violations of traffic rules $p_1$ and $p_2$ when $x_f\ge 40.0$ and blockages for $x_f=0.0$. Consistently safe progress is achieved for values of $x_a\ge30.0$ and $x_f\ge 2.3$, while for $20.0\le x_a \le 25.0$, non-deterministic verdicts switching between safe and unsafe progress are observed. 

\vspace{10pt}
\paragraph{4) InterFuser Autopilot: Analysis of  crossing with traffic light scenarios}

\begin{minipage}{\linewidth}
\begin{table}[H]
    \centering
    \caption{InterFuser: Verdicts for crossing with traffic light scenarios}
    \label{tab:interfuser-traffic_light}
    \vspace{-15pt}
\end{table}
\begin{center}
    Table \ref{tab:interfuser-traffic_light} (a). InterFuser: Verdicts for crossing with traffic light scenarios with $v_e=0.0$ m/s

    \adjustbox{max width=\linewidth}{
        \input{tables/interfuser-traffic_light-0}
    }
\end{center}
\end{minipage}

\vspace{10pt}

\noindent
\begin{minipage}{\linewidth}
\begin{center}
    {Table \ref{tab:interfuser-traffic_light} (b). InterFuser: Verdicts for crossing with traffic light scenarios with $v_e=4.0$ m/s}
    \adjustbox{max width=\linewidth}{
        \input{tables/interfuser-traffic_light-4}
    }   
\end{center}
\end{minipage}

\vspace{5pt}

Table \ref{tab:interfuser-traffic_light} presents verdicts for InterFuser in crossing traffic lights  scenarios generated by application of test cases defined by the parameters $x_a$ and $x_f$, whose critical values are indicated in red. 

For both speeds $v_e=0.0$ and 4.0 m/s,  we observe safe progress with the exception of the borderline cases for $v_e=0.0$ and $x_f=0.0, 2.8$, where the ego vehicle does not leave the critical zone within the allotted time. 

Note that a closer look at the observed behavior reveals some limitations of CCTest, which focuses on critical situations.  Indeed, we found that for $v_e = 0.0$ and $v_e = 4.0$, when the ego vehicle faces a red light, it decides to cross in clear violation of traffic rules. This surprising behavior would have been discovered using normal test cases. The almost correct behavior observed in our experiments may therefore be the result of a poor control policy that simply does not respect red lights.

\subsubsection{Test results for MILE}

\paragraph{1) MILE Autopilot: Analysis of  merging scenarios}

\begin{minipage}{\textwidth}
\begin{table}[H]
    \centering
    \caption{MILE: Verdicts for merging scenarios}
    \label{tab:mile-merging}
    \vspace{-15pt}
\end{table}
\begin{center}
    Table \ref{tab:mile-merging} (a). MILE: Verdicts for merging scenarios with $v_e=0.0$ m/s
\end{center}
\adjustbox{max width=\linewidth}{
    \input{tables/mile-merging-0}
}
\vspace{10pt}
\noindent
\begin{center}
    {Table \ref{tab:mile-merging} (b). MILE: Verdicts for merging scenarios with $v_e=4.0$ m/s}
\end{center}
\adjustbox{max width=\linewidth}{
    \input{tables/mile-merging-4}
}
\end{minipage}

~

Table \ref{tab:mile-merging} presents verdicts for MILE  in merging scenarios generated by application of test cases defined by the parameters $x_a$ and $x_f$, whose critical values are indicated in red. 

In Table \ref{tab:mile-merging}(a), we observe accidents in the transition from caution to progress for $x_a$ equal to 10 m  for all the values of $x_f$ different from 0.0. For values not less than $x_a=12.5$, we observe uniformly safe behavior. 

In Table \ref{tab:mile-merging}(b), we observe similar patterns with the difference that accidents occur for $x_a =7.5$ and $x_f\ge 40$. For all other configurations, we observe uniformly safe behavior.  It should be noted that the behavior observed in all merging scenarios does not vary when the same test case is applied 5 times.

\paragraph{2) MILE Autopilot: Analysis of lane change scenarios}

\begin{minipage}{\textwidth}
\begin{table}[H]
    \centering
    \caption{MILE: Verdicts for lane change scenarios with $v_e=4.0$ m/s}
    \label{tab:mile-lane_change}
    \vspace{-15pt}
\end{table}

\adjustbox{max width=\linewidth}{
    \input{tables/mile-lane_change-4}
}   
\end{minipage}

~

Table \ref{tab:mile-lane_change} presents verdicts for MILE  in lane  change scenarios generated by application of test cases defined by the parameters $x_a$ and $x_f$, whose critical values are indicated in red. The observed behavior is safe cautious. We do not observe progress even when $x_a$ and $x_f$ take values for which safe progress is possible. This is indicated on the table by the pink area with uniformly safe overcautious verdict.

Further analysis shows that MILE's over-cautiousness can be attributed to its inability to overtake a static front vehicle or follow a planned route involving a lane change. This is supported by two extra experiments: 1) In scenarios with only the ego vehicle and a stationary front vehicle, the ego vehicle consistently fails to overtake, regardless of the distance to the front vehicle, whether 10, 20, 30, ..., or 100 meters; {2) In scenarios involving only the ego vehicle, even when its route is set to use the outer lane, the ego vehicle is unable to make the required lane change.}

\paragraph{3) MILE Autopilot: Analysis of  crossing with yield sign scenarios}

\begin{minipage}{\textwidth}
\begin{table}[H]
    \centering
    \caption{MILE: Verdicts for crossing with yield sign scenarios}
    \label{tab:mile-crossing}
    \vspace{-15pt}
\end{table}
\begin{center}
    Table \ref{tab:mile-crossing} (a). MILE: Verdicts for crossing with yield sign scenarios with $v_e=0.0$ m/s
\end{center}
\adjustbox{max width=\linewidth}{
    \input{tables/mile-crossing-0}
}
\end{minipage}

\vspace{5pt}

\noindent
\begin{center}
    Table \ref{tab:mile-crossing} (b). MILE: Verdicts for crossing with yield sign scenarios with $v_e=4.0$ m/s
\end{center}
\adjustbox{max width=\linewidth}{
    \input{tables/mile-crossing-4}
}   

~

Table \ref{tab:mile-crossing} presents verdicts for MILE  in crossing with yield sign  scenarios generated by application of test cases defined by the parameters $x_a$ and $x_f$, whose critical values are indicated in red. 

In Table \ref{tab:mile-crossing}(a), the observed verdicts are not sensitive to changes of $x_f$. For $x_a=0.0$, we observe violations of traffic rules, while for $x_a=2.5$, we observe accidents. Then for values of $x_a$ between 5.0 and 24.5 we observe again violations of the mutual exclusion property ($p_1$). For values of $x_a$ not less than 29.1, we observe safe progress. 

In Table \ref{tab:mile-crossing}(b), we observe similar patterns  with the difference that accidents occur for $x_a=0.0$ and violations of the mutual exclusion occur for values of $x_a$ between 2.5 and 20.0. For larger values of $x_a$, and $x_f$ not less than 3.6 safe progress is uniformly observed.

\paragraph{4) MILE Autopilot: Analysis of crossing with traffic light scenarios}

\begin{minipage}{\linewidth}
\begin{table}[H]
    \centering
    \caption{MILE: Verdicts for crossing with traffic light scenarios}
    \label{tab:mile-traffic_light}
    \vspace{-15pt}
\end{table}
\begin{center}
    Table \ref{tab:mile-traffic_light} (a). MILE: Verdicts for crossing with traffic light scenarios with $v_e=0.0$ m/s

    \adjustbox{max width=\linewidth}{
        \input{tables/mile-traffic_light-0}
    }
\end{center}
\end{minipage}

\vspace{10pt}

\noindent
\begin{minipage}{\linewidth}
\begin{center}
    {Table \ref{tab:mile-traffic_light} (b). MILE: Verdicts for crossing with traffic light scenarios with $v_e=4.0$ m/s}
    \adjustbox{max width=\linewidth}{
        \input{tables/mile-traffic_light-4}
    }   
\end{center}
\end{minipage}

\vspace{5pt}

Table \ref{tab:mile-traffic_light} presents verdicts for MILE  in crossing traffic lights scenarios generated by the application of test cases defined by the parameters $x_a$ and $x_f$, whose critical values are indicated in red. 

For Table \ref{tab:mile-traffic_light}(a),  the behavior involves a blockage where the ego vehicle remains stopped before the exit of the critical zone for $x_f=0$. For all other values of $x_f$, we have in most cases, in a non-deterministic way, either safe progress, or an inability to leave the critical zone in the allotted time.

For Table \ref{tab:mile-traffic_light}(b), we have safe progress for all test cases, except for $x_f=0$, where a blockage is observed.

As with the InterFuser, a closer look at the observed behavior reveals the limitations of our approach, which focuses on critical situations. Indeed, we found that for $v_e = 0.0$ and $v_e = 4.0$, when the ego vehicle faces a red light, it decides to cross in clear violation of traffic rules.  The almost correct behavior observed may therefore be the result of a poor control policy that simply does not respect red lights.

\subsubsection{Test results for Transfuser}

\paragraph{1) Transfuser Autopilot: Analysis of  merging scenarios}

\begin{minipage}{\linewidth}
\begin{table}[H]
    \centering
    \caption{Transfuser: Verdicts for merging scenarios}
    \label{tab:transfuser-merging}
    \vspace{-15pt}
\end{table}
\begin{center}
    Table \ref{tab:transfuser-merging} (a). Transfuser: Verdicts for merging scenarios with $v_e=0.0$ m/s
\end{center}
\adjustbox{max width=\linewidth}{
\input{tables/transfuser-merging-0}
}

\vspace{10pt}
\noindent
\end{minipage}
\begin{minipage}{\linewidth}
\begin{center}
    {Table \ref{tab:transfuser-merging} (b). Transfuser: Verdicts for merging scenarios with $v_e=4.0$ m/s}
\end{center}
\adjustbox{max width=\linewidth}{
\input{tables/transfuser-merging-4}
}
\end{minipage}

\vspace{5pt}

Table \ref{tab:transfuser-merging} presents verdicts for Transfuser  in merging scenarios generated by application of test cases defined by the parameters $x_a$ and $x_f$, whose critical values are indicated in red. 

In Table \ref{tab:transfuser-merging}(a) where $v_e=0.0$, we observe accidents in the transition from caution to progress for $x_a = 10.0$ and all the values of $x_f$ different from 0.0, with the possibility of safe progress and safe caution due to non-determinism. 

In addition, we have abnormal accidents for $x_f=440.0$ and $0 \le x_a \le 30.0$, which occur after the ego vehicle has safely merged onto the main road but either the arriving vehicle fails to follow the ego vehicle safely, causing \texttt{Aa}, or the ego vehicle fails to follow the arriving vehicle safely, causing \texttt{Ae}. 

Finally, four road deviations occur with low probabilities, for the ego vehicle whose verdicts are marked \texttt{DRe}.

In Table \ref{tab:transfuser-merging}(b) where $v_e=4.0$, we observe similar patterns of abnormal accidents for $x_a$ =7.5 and $x_f\ge 40$. Furthermore, when $x_f=440.0$ and $0.0 \le x_a \le 30.0$, there are abnormal accidents.

\paragraph{2) Transfuser Autopilot: Analysis of  lane change scenarios}

\begin{minipage}{\linewidth}
\begin{table}[H]
    \centering
    \caption{Transfuser: Verdicts for lane change scenarios with $v_e=4.0$ m/s}
    \label{tab:transfuser-lane_change}
    \vspace{-15pt}
\end{table}

\adjustbox{max width=\linewidth}{
\input{tables/transfuser-lane_change-4}
}
\end{minipage}

~

Table \ref{tab:transfuser-lane_change} presents verdicts for MILE  in lane change scenarios generated by the application of test cases defined by the parameters $x_a$ and $x_f$, whose critical values are indicated in red. As for the MILE autopilot, the observed behavior is safe cautious. We do not observe progress even when $x_a$ and $x_f$ take values for which safe progress is possible. This is indicated on the table by the pink area with uniformly safe overcautious verdicts.

Further experimental analysis of Transfuser's over-cautiousness shows that it cannot perform the lane change maneuver for distances less than 45 meters from the front stationary vehicle on the inside lane. 
However, the observed over-cautiousness can be explained by the fact that the front vehicle on the inside lane is at the critical braking distance $\hat{x_f} = B(v_e) = B(4) = 1.9$. This explains the observed over-cautiousness.

\vspace{15pt}

\paragraph{3) Transfuser Autopilot: Analysis of crossing with yield sign scenarios}

\begin{minipage}{\linewidth}
\begin{table}[H]
    \centering
    \caption{Transfuser: Verdicts for crossing with yield sign scenarios}
    \label{tab:transfuser-crossing}
    \vspace{-15pt}
\end{table}
\begin{center}
    Table \ref{tab:transfuser-crossing} (a). Transfuser: Verdicts for crossing with yield sign scenarios with $v_e=0.0$ m/s
\end{center}
\adjustbox{max width=\linewidth}{
\input{tables/transfuser-crossing-0}
}
\end{minipage}

\vspace{10pt}

\noindent
\begin{minipage}{\linewidth}
\begin{center}
    {Table \ref{tab:transfuser-crossing} (b). Transfuser: Verdicts for crossing with yield sign scenarios with $v_e=4.0$ m/s}
    \adjustbox{max width=\linewidth}{
        \input{tables/transfuser-crossing-4}
    }   
\end{center}
\end{minipage}

~

Table \ref{tab:transfuser-crossing} presents verdicts for Transfuser in crossing with yield sign scenarios generated by application of test cases defined by the parameters $x_a$ and $x_f$, whose critical values are indicated in red. 

In Table \ref{tab:transfuser-crossing}(a) where $v_e=0.0$, the observed verdicts are not sensitive to changes of $x_f$. For $x_a=0.0$, we observe non-deterministic verdicts, including accidents, violations of the mutual exclusion rule ($p_1$), and stopping inside the junction ($p_2$), as well as blockages. For $x_a=10.0$ and  $20.0$, we uniformly observe violations of the mutual exclusion rule ($p_1$). For $x_a\ge 30.0$ and all the values of $x_f$, we observe safe progress. 

In Table \ref{tab:transfuser-crossing}(b) where $v_e=4.0$, similar patterns are observed and the verdicts are independent of the value of $x_f$. For $x_a\le 25.0$,  we observe unsafe progress violating mutual exclusion rules ($p_1$). For greater values, we have safe progress, for all $x_f$ values less than 40.0. For smaller values of $x_f$, the behavior can give either safe progress or violations of traffic rules. 

\vspace{10pt}
\paragraph{4) Transfuser Autopilot: Analysis of crossing with traffic light scenarios}

\begin{minipage}{\linewidth}
\begin{table}[H]
    \centering
    \caption{Transfuser: Verdicts for crossing with traffic light scenarios}
    \label{tab:transfuser-traffic_light}
    \vspace{-15pt}
\end{table}
\begin{center}
    Table \ref{tab:transfuser-traffic_light} (a). Transfuser: Verdicts for crossing with traffic light scenarios with $v_e=0.0$ m/s

    \adjustbox{max width=\linewidth}{
        \input{tables/transfuser-traffic_light-0}
    }
\end{center}
\end{minipage}

\vspace{10pt}

\noindent
\begin{minipage}{\linewidth}
\begin{center}
    {Table \ref{tab:transfuser-traffic_light} (b). Transfuser: Verdicts for crossing with traffic light scenarios with $v_e=4.0$ m/s}
    \adjustbox{max width=\linewidth}{
        \input{tables/transfuser-traffic_light-4}
    }   
\end{center}
\end{minipage}

\vspace{5pt}

Table \ref{tab:transfuser-traffic_light} presents verdicts for Transfuser  in crossing traffic lights  scenarios generated by application of test cases defined by the parameters $x_a$ and $x_f$.

For both $v_e= 0.0$ and $v_e=4.0$, the observed behavior is uniformly cautious. This cautious behavior can be explained by the fact that the Transfuser's maximum speed is 4.0 m/s, which is not enough to cross the 20-meter-long critical zone in the allotted time of 5 seconds.

\vspace{20pt}
\subsubsection{Test results for LMDrive}

\paragraph{1) LMDrive Autopilot: Analysis of  merging scenarios}

\begin{minipage}{\linewidth}
\begin{table}[H]
    \centering
    \caption{LMDrive: Verdicts for merging scenarios}
    \label{tab:lmdrive-merging}
    \vspace{-15pt}
\end{table}
\begin{center}
    Table \ref{tab:lmdrive-merging} (a). LMDrive: Verdicts for merging scenarios with $v_e=0.0$ m/s
\end{center}
\adjustbox{max width=\linewidth}{
\input{tables/lmdrive-merging-0}
}
\end{minipage}

\vspace{10pt}

\noindent
\begin{minipage}{\linewidth}
\begin{center}
    {Table \ref{tab:lmdrive-merging} (b). LMDrive: Verdicts for merging scenarios with $v_e=4.0$ m/s}
    \adjustbox{max width=\linewidth}{
        \input{tables/lmdrive-merging-4}
    }   
\end{center}
\end{minipage}

\vspace{10pt}

Table \ref{tab:lmdrive-merging} presents verdicts for LMDrive in merging scenarios generated by application of test cases defined by the parameters $x_a$ and $x_f$, whose critical values are indicated in red. 

In Table \ref{tab:lmdrive-merging}(a) where $v_e=0.0$, many road deviation behaviors are observed for the ego vehicle whose verdicts are marked as DRe. These deviations mainly occur after the ego vehicle merges onto the main road but fails to keep on the road while approaching the front vehicle. In addition, accidents are observed during the transition from caution to progress for $x_a = 25.0, 30.0$ and $x_f \ge 40.0$.

In Table \ref{tab:lmdrive-merging}(b), we observe similar road deviation behaviors occurring mainly for large $x_f$ values. There is a safe transition from caution to progress when $x_a=10.0$ and $x_f\ge 40.0$. However, an accident occurs where the ego vehicle hits the front vehicle stopping at the merging point for $x_a=25.0$ and $x_f=0.0$. 

\paragraph{2) LMDrive Autopilot: Analysis of lane change scenarios}

\begin{minipage}{\textwidth}
\begin{table}[H]
    \centering
    \caption{LMDrive: Verdicts for lane change scenarios}
    \label{tab:lmdrive-lane_change}
    \vspace{-15pt}
\end{table}

\begin{center}
    Table \ref{tab:lmdrive-lane_change} (a). LMDrive: Verdicts for lane change scenarios with $v_e=4.0$ m/s
\end{center}
\adjustbox{max width=\linewidth}{
    \input{tables/lmdrive-lane_change-4}
}
\end{minipage}

~

Table \ref{tab:lmdrive-lane_change} presents verdicts for LMDrive  in lane change scenarios generated by application of test cases defined by the parameters $x_a$ and $x_f$, whose critical values are indicated in red. As for the MILE and Transfuser autopilots, the observed behaviors are all safely cautious. We do not observe progress even when $x_a$ and $x_f$ take values for which safe progress is possible. This is indicated on the table by the pink area with uniformly safe overcautious verdict.

Further analysis shows that LMDrive's over-cautiousness can be attributed to its inability to overtake a static front vehicle or follow a planned route involving a lane change. This is supported by two extra experiments: 1) In scenarios with only an ego vehicle and a stationary front vehicle, the ego vehicle consistently fails to overtake, regardless of the distance to the front vehicle, whether 10, 20, 30, ..., or 100 meters; {2) In scenarios involving only the ego vehicle, even when the route is set to pass via the outer lane, the ego vehicle is unable to perform the required lane change.}

\paragraph{3) LMDrive Autopilot: Analysis of crossing with yield sign scenarios}

\noindent
\begin{minipage}{\textwidth}
\begin{table}[H]
    \centering
    \caption{LMDrive: Verdicts for crossing with yield sign scenarios}
    \label{tab:lmdrive-crossing}
    \vspace{-15pt}
\end{table}
\begin{center}
    Table \ref{tab:lmdrive-crossing} (a). LMDrive: Verdicts for crossing with yield sign scenarios with $v_e=0.0$ m/s
\end{center}

\adjustbox{max width=\linewidth}{
    \input{tables/lmdrive-crossing-0}
} 
\end{minipage}

\vspace{10pt}
\noindent
\begin{minipage}{\linewidth}
\begin{center}
    Table \ref{tab:mile-crossing} (b). LMDrive: Verdicts for crossing with yield sign scenarios with $v_e=4.0$ m/s
\end{center}
\adjustbox{max width=\linewidth}{
    \input{tables/lmdrive-crossing-4}
}   
\end{minipage}

~

Table \ref{tab:lmdrive-crossing} presents verdicts for LMDrive in crossing with yield sign  scenarios generated by application of test cases defined by the parameters $x_a$ and $x_f$, whose critical values are indicated in red. 

For both speeds $v_e = 0.0$ and $v_e = 4.0$, the observed behavior is flawed with exceptions of safe progress occurring under 4 test cases.

For low values of $x_a$ and $x_f$ , the main issues are blockages, where both the ego vehicle and the arriving vehicle enter the critical zone, leading to a deadlock. 

In addition, we observe a large number of arbitrary changes of route denoted \texttt{CR} either by the ego vehicle (\texttt{CRe}) or by the arriving vehicle (\texttt{CRa}) or both (\texttt{CRea}). These surprising changes of route consist of making a right turn at the intersection instead of proceeding straight ahead as specified by their routes. In some cases, the change of route is followed by a deviation of the road and a passage on the sidewalk, indicated by \texttt{CDRe}, \texttt{CDRa}, \texttt{CDRea} respectively for the ego, the arriving vehicle or both. 

Finally, for $v_e = 0.0$, there
are 3 accidents where the ego vehicle hits the arriving vehicle for $x_a = 25.0$, and unsafe progress violating $p_1$, as the two vehicles are in the critical zone at the same time, for $x_a = 20.0, 25.0, 27.0$.

\paragraph{4) LMDrive Autopilot: Analysis of crossing with traffic light scenarios}

\begin{minipage}{\linewidth}
\begin{table}[H]
    \centering
    \caption{LMDrive: Verdicts for crossing with traffic light scenarios}
    \label{tab:lmdrive-traffic_light}
    \vspace{-15pt}
\end{table}
\begin{center}
    Table \ref{tab:lmdrive-traffic_light} (a). LMDrive: Verdicts for crossing with traffic light scenarios with $v_e=0.0$ m/s

    \adjustbox{max width=\linewidth}{
        \input{tables/lmdrive-traffic_light-0}
    }
\end{center}
\end{minipage}

\vspace{10pt}

\noindent
\begin{minipage}{\linewidth}
\begin{center}
    {Table \ref{tab:lmdrive-traffic_light} (b). LMDrive: Verdicts for crossing with traffic light scenarios with $v_e=4.0$ m/s}
    \adjustbox{max width=\linewidth}{
        \input{tables/lmdrive-traffic_light-4}
    }   
\end{center}
\end{minipage}

\vspace{5pt}

Table \ref{tab:lmdrive-traffic_light} presents verdicts for LMDrive in crossing with traffic light  scenarios generated by application of test cases defined by the parameters $x_a$ and $x_f$, whose critical values are indicated in red. 

For both speeds $v_e=0.0$ and $v_e= 4.0$, the observed behaviors are flawed. 

For $v_e=0.0$ and for all values of $x_f$ other than 0.0, the ego vehicle fails to leave the critical zone in the allotted time. For $x_f= 0.0$ we observe blockages.

For $v_e = 4.0$, as for the yield sign crossing scenario, the ego vehicle changes route (\texttt{CRe}) by turning right instead of continuing straight on its route. 

Further analysis with test cases where only the ego vehicle is confronted with a constant red light reveals significant shortcomings. For $v_e = 0.0$, the ego vehicle ignores the red light and crosses the critical zone. For $v_e = 4.0$, the ego vehicle ignores the red light and also changes the specified route (\texttt{CRe}) by turning right instead of going straight ahead.

\begin{figure*}[h]
    \begin{minipage}{0.48\linewidth}
        \centering
        \includegraphics[width=\linewidth]{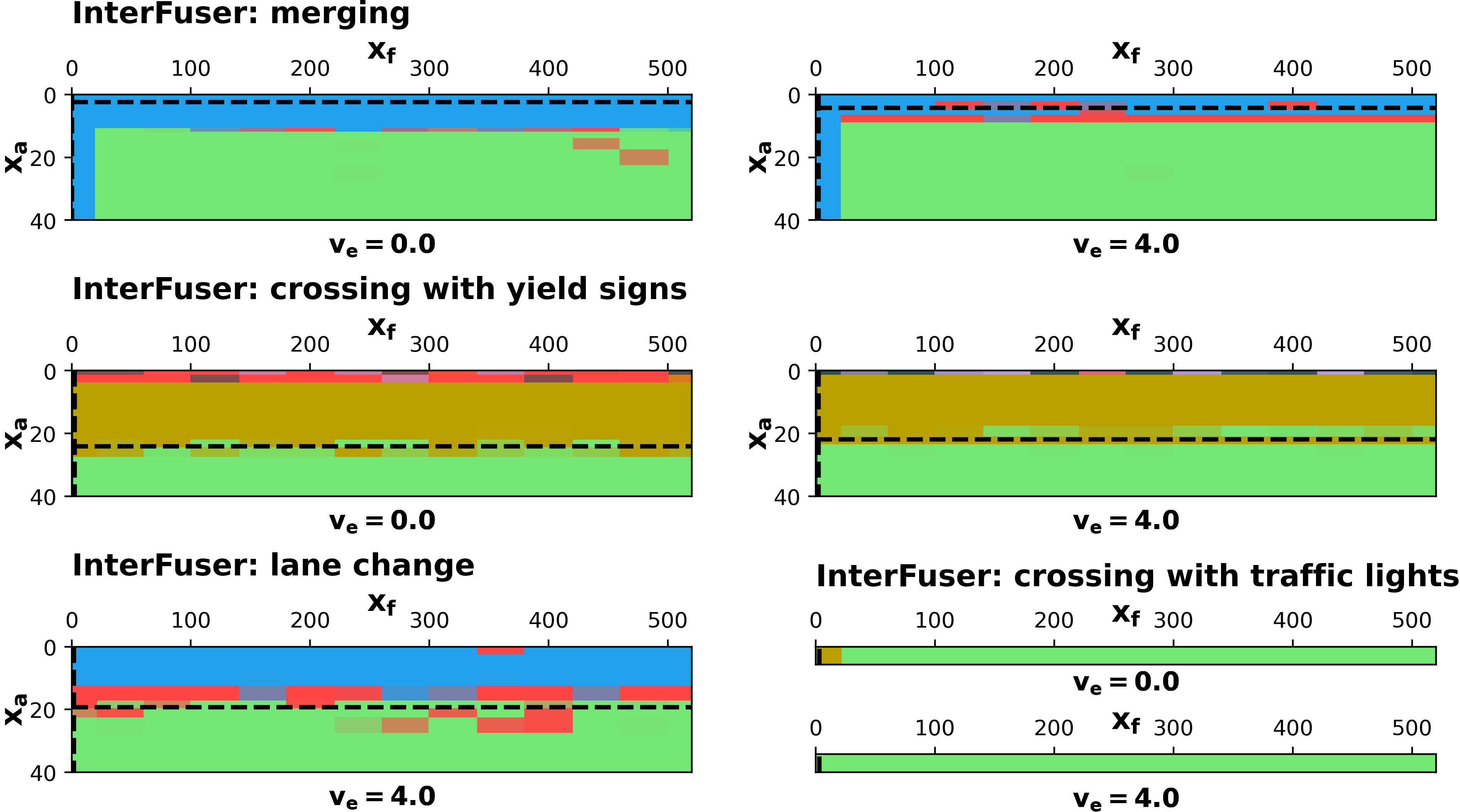}
        
        \textbf{\small (a) verdicts for the InterFuser autopilot}
        \label{fig:interfuser-result}
    \end{minipage}
    \hspace{1em}
    \begin{minipage}{0.48\linewidth}
        \centering
        \includegraphics[width=\linewidth]{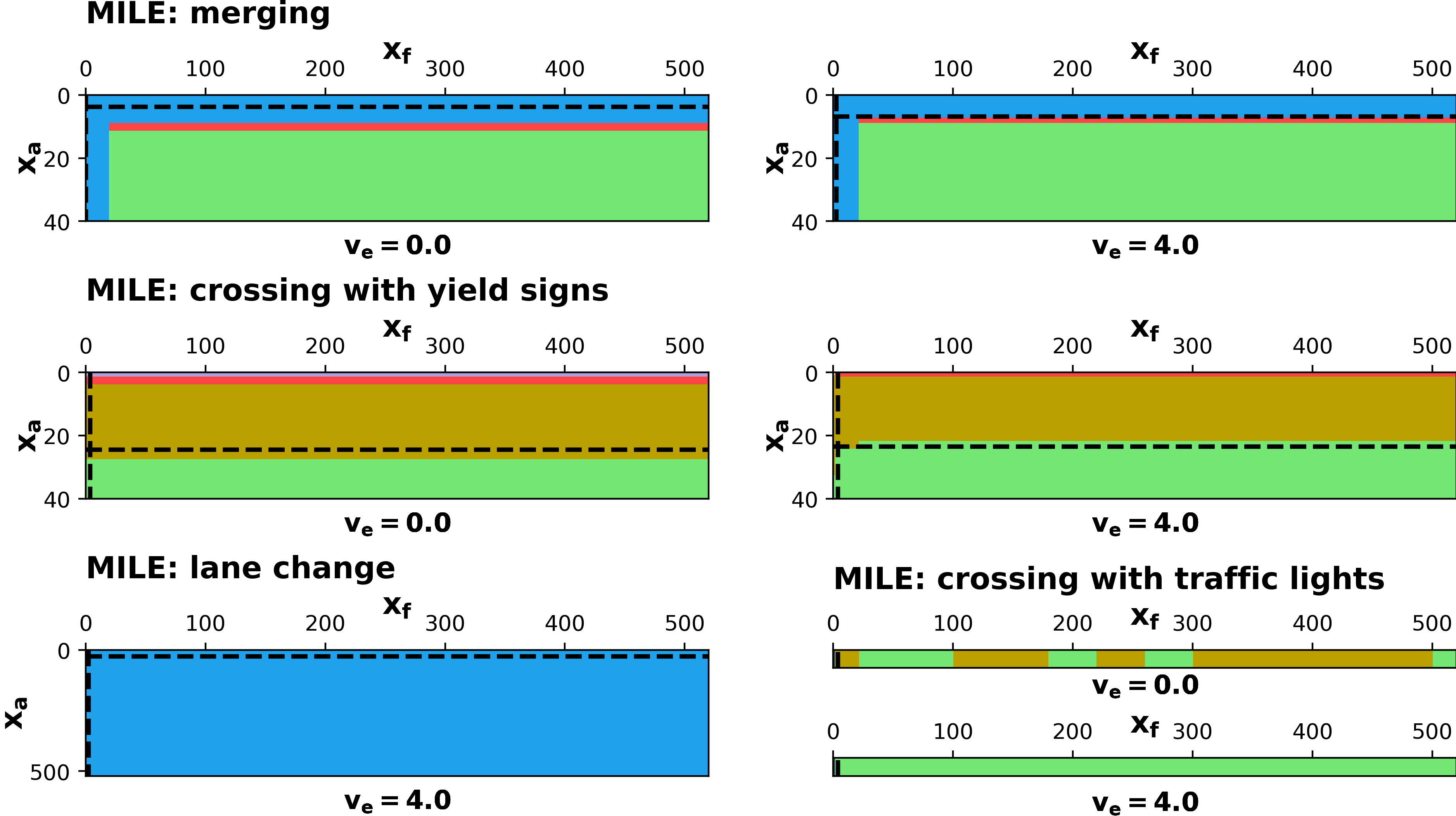}
        
        \textbf{\small (b) verdicts for the MILE autopilot}
        \label{fig:mile-result}
    \end{minipage}
    
    \vspace{0.5em}
    
    \begin{minipage}[t]{0.48\linewidth}
        \centering
        \includegraphics[width=\linewidth]{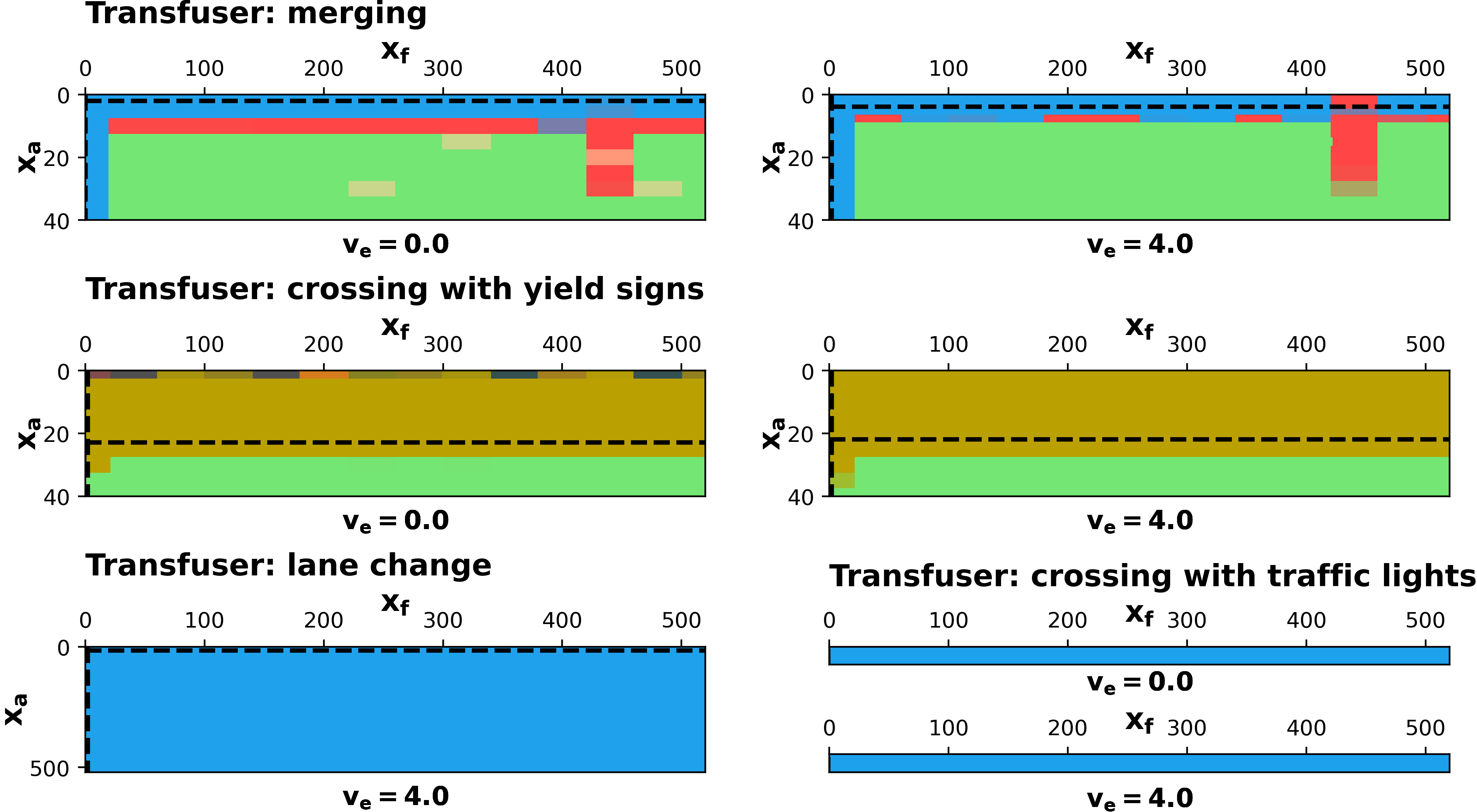}
        
        \textbf{\small (c) verdicts for the Transfuser autopilot}
        \label{fig:transfuser-result}
    \end{minipage}
    \hspace{1em}
    \begin{minipage}[t]{0.48\linewidth}
        \centering
        \includegraphics[width=\linewidth]{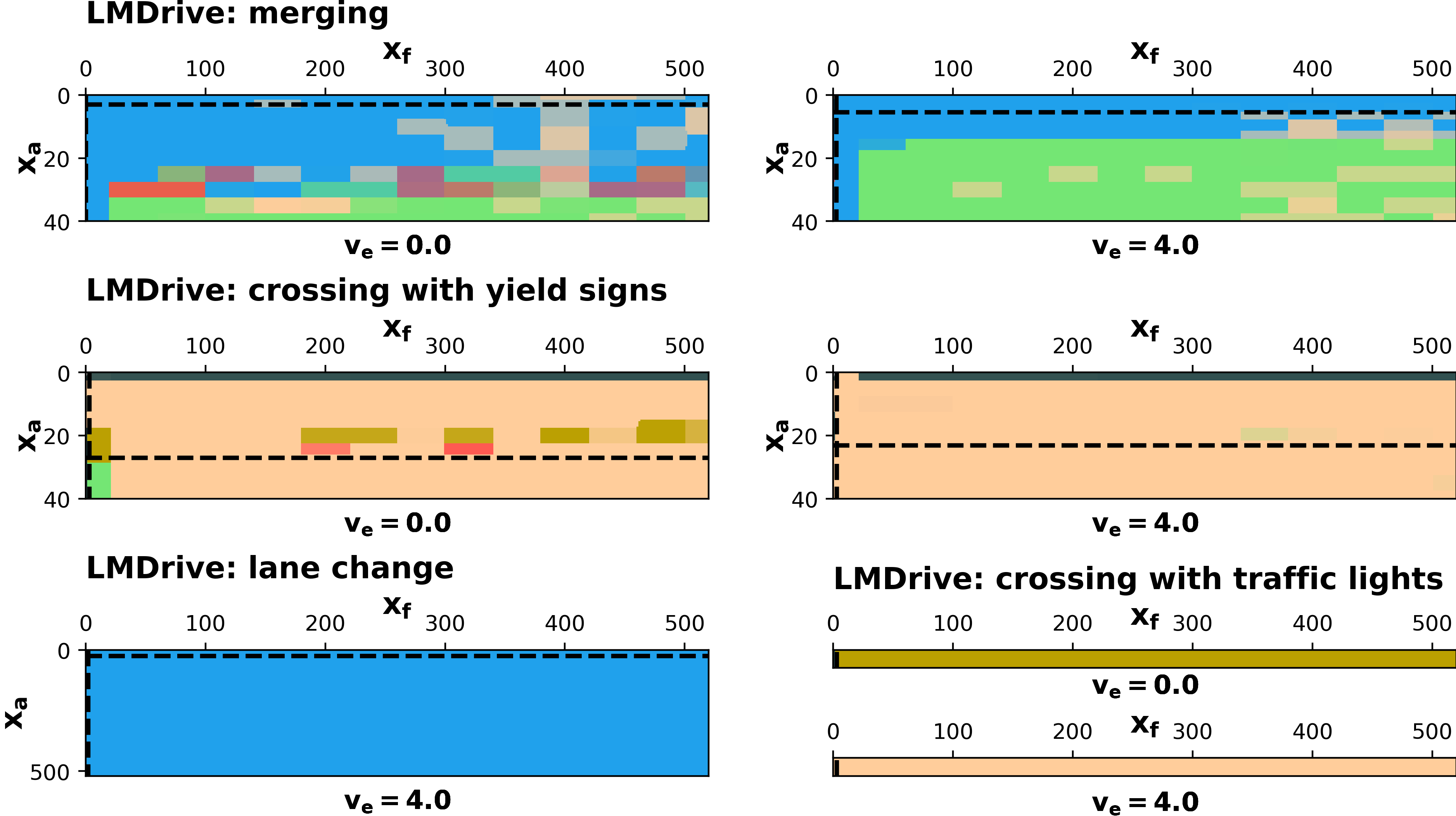}
        
        \textbf{\small (d) verdicts for the LMDrive autopilot}
        \label{fig:lmdrive-result}
    \end{minipage}
    \begin{minipage}{\linewidth}
    
    \vspace{0.2em}
    
    \hfill
    \includegraphics[width=0.97\linewidth]{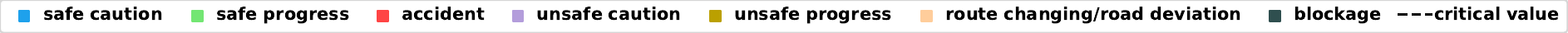}    
    \end{minipage}
    
    \vspace{-0.5em}
    
    \caption{Verdicts for each autopilot and configuration}
    \label{fig:results}
    
    \vspace{-1em}
\end{figure*}

\subsubsection{Summary of CCTest results}
\label{sec:summary}

Fig. \ref{fig:results} shows the partition of verdicts for each autopilot in each context, with distinct colors representing different verdict categories. Mixed colors indicate different verdicts observed under the same test case due to the non-determinism. The verdicts include safe caution and progress, as well as safety issues such as traffic rule violations leading to unsafe caution or progress, accidents, and route changes/road deviations.

The global view of the verdicts highlights some cases in which the autopilot is generally flawed regardless of the values of $x_a$ and $x_f$. These include the overcautious behavior of Transfuser, MILE and LMDrive in lane change scenarios, as well as modifications of the given route or deviation from the road caused by LMDrive in crossing with yield sign scenarios and crossing with traffic light scenarios.

Except for these cases with general flaws, problematic scenarios mainly occur during the transition from caution to progress as $x_a$ and $x_f$ increase. Additionally, the non-determinism of the test results is most pronounced during this transition, while the verdicts are more stable when $x_a$ and $x_f$ are both small or large enough.

In addition to the transition from caution to progress, we observe that all four autopilots behave abnormally in at least one context: they fail for a test case but succeed for a more critical one. This failure could be avoided by applying existing safe policies of the more critical test case. Such abnormal behaviors are observed in the following situations: InterFuser in merging scenarios with $v_e = 0.0$, MILE in crossings with traffic lights at $v_e = 0.0$, as well as Transfuser and LMDrive in merging scenarios with $v_e = 0.0$ and $v_e = 4.0$.

In crossing scenarios with traffic lights, both InterFuser and MILE predominantly exhibit high frequencies of safe progress and only unsafe progress violating $p_4$ (the ego vehicle should leave the critical zone when the lateral light turns green) occurs with low frequency. However, a closer look reveals that even when InterFuser and MILE autopilots face a red light, they decide to cross in clear violation of traffic rules. The almost correct behavior observed in our experiments may therefore be the result of a poor control policy that simply does not respect red lights. This shows a limitation of CCTest, which focuses only on critical configurations and may not expose defects in simple policies for non-critical configurations.

Transfuser is always cautious under the crossing with traffic lights test cases generated by CCTest. This is explained by the fact that, according to the CCTest analysis, there is no safe progress policy for Transfuser to pass the critical zone during the duration of the yellow light and the all-red phase, due to its lowest maximum speed of 4 m/s among the four autopilots.

\begin{center}
\fcolorbox{black}{gray!10}{\parbox{\linewidth}{\textbf{Answer to RQ1.1}: For each autopilot across the four contexts considered by CCTest, we observed either general failures, where the autopilot lacks safe progress policies under almost all generated test cases, or accidents and other hazards in critical situations where the transition from caution to progress occurs as $x_a$ and $x_f$ increase. Additionally, all four autopilots exhibited abnormal behavior in at least one of the four contexts, where an autopilot fails in a test case but succeeds in a more critical test case. This indicates that, beyond critical test cases, the four AI autopilots present safety problems in less critical, normal situations.}}
\end{center}

\begin{figure}[h]
    \centering
    \begin{minipage}{0.3\linewidth}
        \centering
        \includegraphics[width=\linewidth]{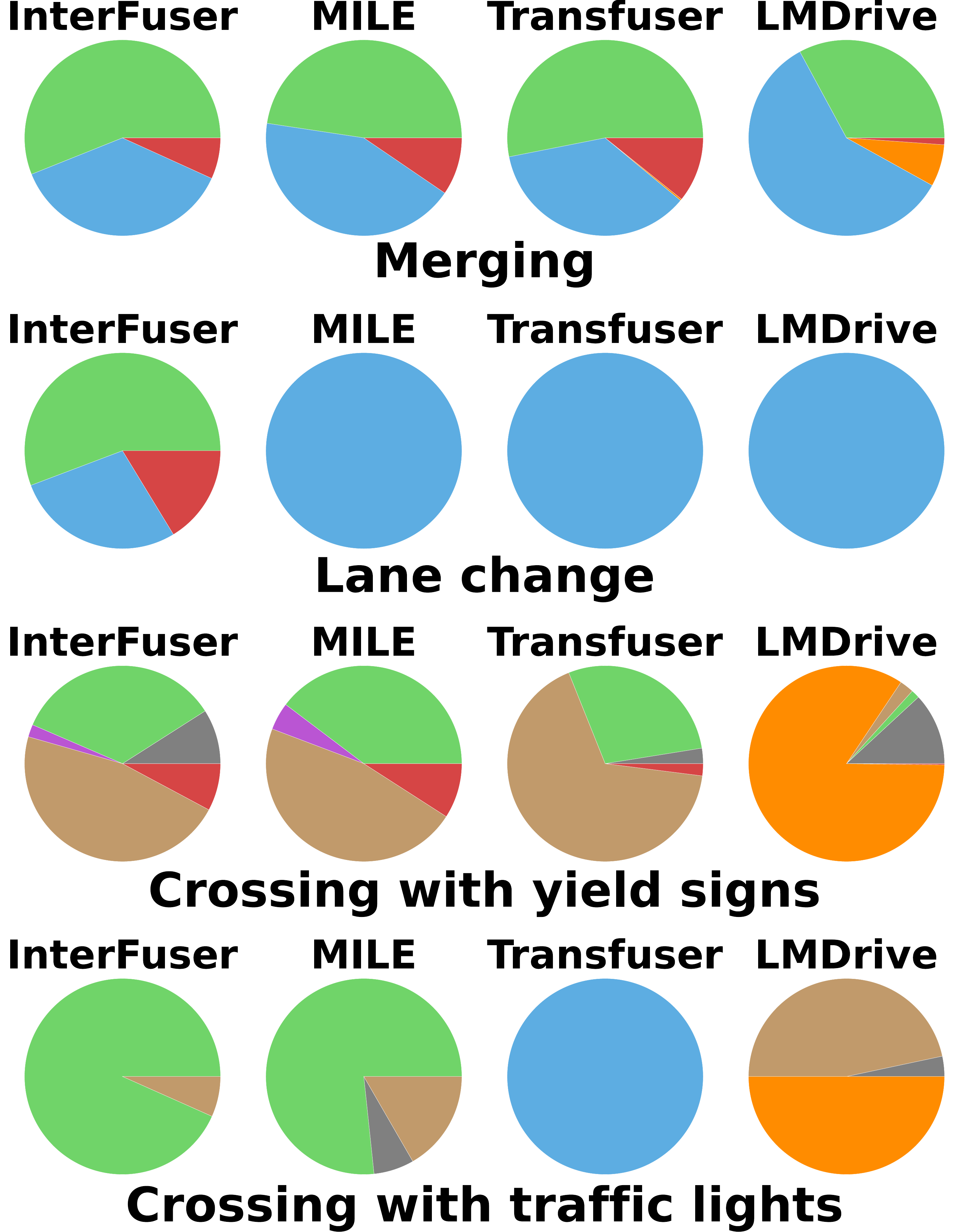}
        \textbf{\small{(a) AI autopilots}}
        \label{fig:stat-ai}
    \end{minipage}%
    \hspace{5em}
    \begin{minipage}{0.3\linewidth}
        \centering
        \includegraphics[width=\linewidth]{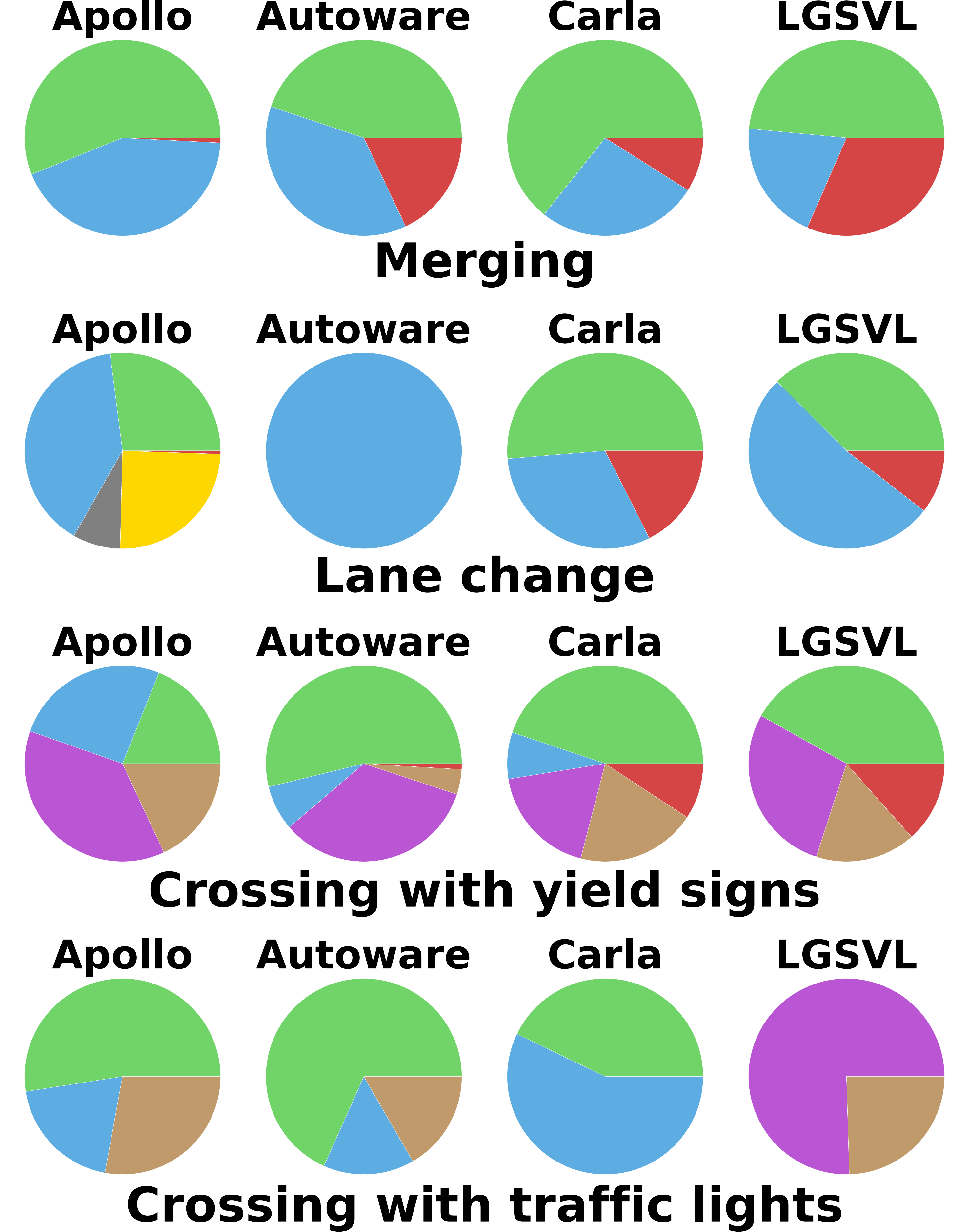}
        \textbf{\small{(b) Modular autopilots}}
    \end{minipage}

    \vspace{3pt}
    \includegraphics[width=0.6\linewidth]{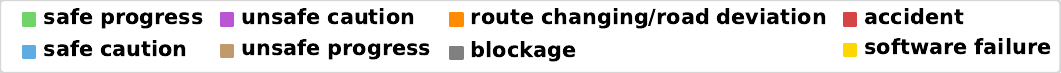}

    \vspace{-10pt}
    
    \caption{Synthesis of results for AI and modular autopilots}
    \label{fig:synthesis}

    \vspace{-10pt}
\end{figure}

\subsubsection{Comparison with modular autopilots}

Fig. \ref{fig:synthesis} (a) and Fig. \ref{fig:synthesis} (b) respectively show the synthesis of results for the four end-to-end AI autopilots, as well as existing results for four traditional modular autopilots \cite{li2024rigorous}, Apollo, Autoware, Carla, and LGSVL. 

A comparison of the results for identical configurations \cite{li2024rigorous} reveals similarities in the types of hazards observed, with the exception that AI autopilots encounter additional hazards, including route changes and road deviations.

Merging is the most successful maneuver for both types of autopilot, where all the tested autopilot had both safe cautious and
safe progress policies. However, two of the AI autopilots, Transfuser and LMDrive, have additional hazards of deviating roads, which is not observed for any of the four modular autopilots.

For lane changes, three AI autopilots adopted a policy of excessive caution, even when the maneuver was possible without risk. In contrast, only one modular autopilot, Autoware, is overcautious in lane change.

For crossings with yield signs, both AI and modular autopilots exhibited a high percentage of traffic violations. None of the four AI autopilots has safe cautious behavior, indicating their general defects in strictly respecting traffic rules. In contrast, three of the four modular autopilots had safe cautious behavior in this context. Additionally, LMDrive suffered consistent failures, which is not the case for modular autopilots.

For crossings with traffic lights, both AI and modular autopilots had high frequencies of safe policies with LMDrive and LGSVL being the only two exceptions. However, Transfuser adopted cautious policies across all test cases, whereas InterFuser and MILE failed to adopt a cautious policy in any test case. In contrast, three of the four modular autopilots were able to apply both safe cautious and safe progress policies.

Finally, our global analysis presented in Section \ref{sec:summary} reveals that all four end-to-end AI autopilots exhibit abnormal policies in at least one context. In contrast, abnormal policies are not generally observed for modular autopilots \cite{li2024rigorous}, with Apollo being the only exception.

\begin{center}
\fcolorbox{black}{gray!10}{\parbox{\linewidth}{\textbf{Answer to RQ1.2}: 
Both AI and modular autopilots exhibit failures
in the critical test cases generated by CCTest, while AI autopilots
have additional issues in certain contexts, including 1) failures
for almost all $x_a$ and $x_f$ values when fixing $v_e$, including excessive caution in lane changes and safety violations in crossings; 2) prevalent failures to strictly adhere to traffic rules at crossings with yield signs or traffic lights; and 3) more defects in executing simple driving policies, such as keeping within road boundaries or respecting a red light.}}
\end{center}

\subsection{RQ2: Leaderboard test result analysis}

\subsubsection{Experimental settings}

We evaluate each autopilot against four open benchmarks designed for the Carla Leaderboard 1.0: Longest6, Town05 Long, NEAT, and Langauto. Their main characteristics are summarized in Tab. \ref{tab:synthesis_benchmark}, including the maps used, the number of test case sequences, the average length of the ego vehicle’s route for each sequence, and the number of adversarial test cases of each type in the benchmark. It should be noted that the benchmarks do not cover all types of adversarial test cases. In particular, Langauto covers only three out of the ten types (C1, C3, C4), and none of the benchmarks include C6.

\begin{table}[htbp]
\vspace{-5pt}
\centering
\caption{Synthesis of information for the benchmarks}
\label{tab:synthesis_benchmark}
\vspace{-10pt}
\setlength{\aboverulesep}{0pt}  
\setlength{\belowrulesep}{0pt} 
\adjustbox{max width=\linewidth}{
\small
\begin{tabular}{c|c|c|c|ccccc|ccccc}
\toprule
\multirow{3}{*}{\textbf{Benchmark}} & \multirow{3}{*}{\shortstack{\textbf{Maps}\\\textbf{(Town)}}} & \multirow{3}{*}{\shortstack{\textbf{\# Test}\\\textbf{case}\\\textbf{sequence}}} & \multirow{3}{*}{\shortstack{\textbf{Average}\\\textbf{length}}} & \multicolumn{10}{c}{\# \textbf{Adversarial} \textbf{test cases}}\\
\cline{5-14}
& & & & \multicolumn{5}{c|}{\textbf{Critical}} & \multicolumn{5}{c}{\textbf{Accidental}} \\
\cline{5-14}
 &  &  &  & \textbf{C2}&\textbf{C5}&\textbf{C6}&\textbf{C8}&\textbf{C10}&\textbf{C1}&\textbf{C3}&\textbf{C4}&\textbf{C7}&\textbf{C9} \\ 
\midrule
Town05 Long & 05 & 10 & 1.5 km & 0&0&0&28&130&151&167&62&111&39 \\
Longest6 & 01-06 & 36 & 1.6 km &0&0&0&43&48&80&116&2&112&77 \\
NEAT & 01-06 & 42 & 765 m &6&3&0&9&6&93&24&33&30&9 \\
Langauto & 01-06, 10 & 32 & 636 m &0&0&0&0&0&80&12&29&0&0 \\
\bottomrule
\end{tabular}
}
\vspace{-10pt}
\end{table}

\begin{table}[h]
    \centering
    \setlength{\arraycolsep}{0pt}
    \setlength{\tabcolsep}{1pt}
    \caption{Synthesis of Carla Leaderboard evaluation results}
    \label{fig:carla-evaluation}
    \vspace{-10pt}
    \setlength{\aboverulesep}{0pt}
    \setlength{\belowrulesep}{0pt}
    \adjustbox{max width=\linewidth}{
    \small
    \begin{tabular}{c||c|c|c|c||c|c|c|c||c|c|c|c||c|c|c|c} \toprule 
         &  \multicolumn{4}{c||}{\textbf{Town05 Long}}&  \multicolumn{4}{c||}{\textbf{Longest6}}&  \multicolumn{4}{c||}{\textbf{NEAT}}& \multicolumn{4}{c}{\textbf{Longauto}}\\ \hline 
 & \textbf{I}& \textbf{M}& \textbf{T}& \textbf{L}& \textbf{I}& \textbf{M}& \textbf{T}& \textbf{L}& \textbf{I}& \textbf{M}& \textbf{T}& \textbf{L}& \textbf{I}& \textbf{M}& \textbf{T}&\textbf{L}\\ \midrule
 \textbf{Sc}& \textbf{57.6} & 10.0 & 49.1 & 27.1 & \textbf{55.8} & 19.3 & 51.2 & 19.5 & \textbf{79.5} & 65.9 & 76.1 & 41.5 & \textbf{76.9} & 57.0 & 73.7 &31.3 \\ \hline 
 \textbf{P}& 0.68 & 0.10 & 0.52 & 0.94 & 0.70 & 0.21 & 0.57 & 0.81 & 0.88 & 0.67 & 0.88 & 0.77 & 0.84 & 0.63 & 0.83 &0.75 \\ \hline
 \textbf{R}& 85\%& 100\%& 94\%& 29\%& 80\%& 91\%& 90\%& 24\%& 91\%& 99\%& 87\%& 54\%& 92\%& 90\%& 89\%&42\%\\
 \hline\hline
 \textbf{RC}& 0& 0& 0& 6& 0& 0& 0& 18& 0& 1& 0& 21& 1& 1& 1&16
\\ \hline
 \textbf{Blk}& 0& 0& 1& 3& 3& 6& 5& 17& 2& 0& 8& 6& 0& 4& 2&12
\\ \hline
 \textbf{RT}& 5& 0& 0& 0& 12& 1& 0& 0& 7& 0& 0& 5& 7& 0& 3&2\\
 \bottomrule
    \end{tabular}
    }

    \hfill\footnotesize{InterFuser (I)\quad MILE (M)\quad Transfuser (T)\quad LMDrive (L)}\hspace{100pt}
    
    \hfill\footnotesize{Score (Sc)\quad Penalty rate (P)\quad Route completion (R)}\hspace{100pt}
    
    \hfill\footnotesize{Route change (RC)\quad Blk (blockage)\quad Route timeout (RT)}\hspace{100pt}

\end{table}

\begin{figure}[h]
    \centering
    \includegraphics[width=0.45\linewidth]{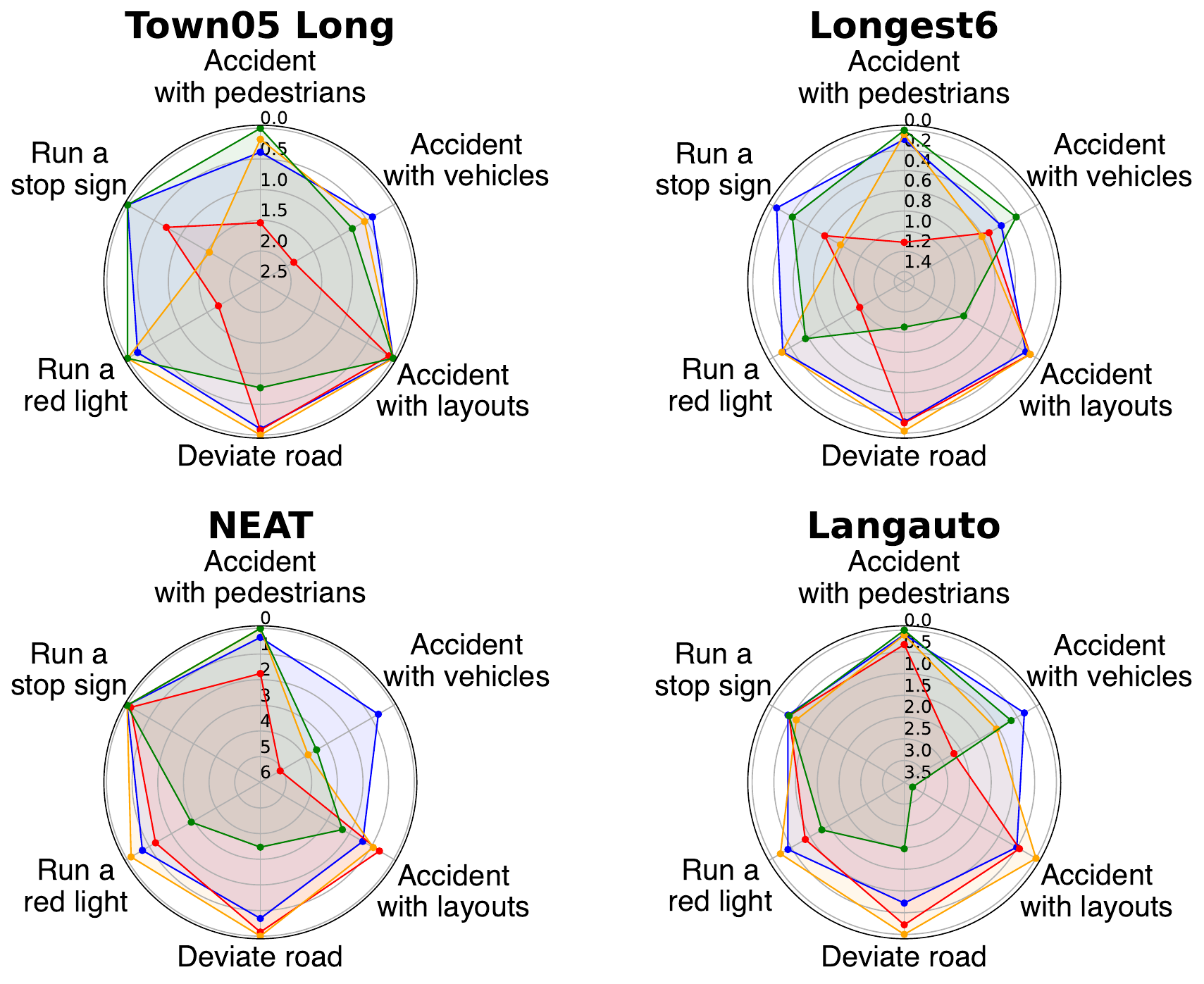}
    
    \includegraphics[width=0.45\linewidth]{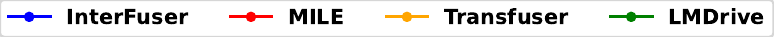}\hspace{8pt}
    \vspace{-10pt}
    \caption{Occurrence of safety violations per kilometer}
    \label{fig:occurence-leaderboard}
\end{figure}

\subsubsection{Overview of Carla Leaderboard test results}

Tab. \ref{fig:carla-evaluation} presents the evaluation results of the four autopilots tested on the Leaderboard for the four benchmarks. The Leaderboard provides the average score (\textbf{Sc}) for all the scenarios generated from a benchmark. It also provides metrics that contribute to the score, including the penalty rate (\textbf{P}) and the route completion rate (\textbf{R}). Recall that three types of incidents can prevent full route completion, including route change (\textbf{RC}), blockage (\textbf{Blk}), and route timeout (\textbf{RT}), whose occurrence counts are also shown in the table. Additionally, Fig.~\ref{fig:occurence-leaderboard} illustrates the safety violations contributing to the penalty rate, normalized as occurrences per kilometer.

The global results show that InterFuser achieved the highest scores across all four benchmarks. Transfuser ranked second, with scores 3 to 5 points lower than InterFuser in three benchmarks and 8.47 points lower in the Town05 Long benchmark. MILE is placed third in three benchmarks, falling behind Transfuser by over 10 points, and scored a very low 10.0 in the Town05 Long benchmark. LMDrive, on the other hand, performed poorly in all benchmarks. Its highest score was 41.5 in the NEAT benchmark.

To gain a deeper understanding of the test results, we manually conducted a qualitative analysis of each safety violation type and performance issue reported by the leaderboard below.

\noindent\textbf{1) Analysis of accidents and road deviations}

All four autopilots experienced accidents with pedestrians, vehicles, layout elements, and road deviations, though the distribution of these incidents varied. LMDrive had a high frequency of accidents with layout elements and road deviations. MILE, in contrast, showed a significantly higher rate of accidents with pedestrians and vehicles. Meanwhile, InterFuser and Transfuser exhibited relatively lower rates across all accident types and road deviations.

Accidents involving layout elements and road deviations for the four autopilots mainly occurred on sharp turns, where the ego vehicle failed to stay within its lane. Among them, LMDrive demonstrated a higher tendency to fail in lane-keeping during turns, leading to a greater frequency of such incidents.

For accidents with pedestrians and vehicles, the Leaderboard reports the observed incidents without distinguishing between normal and adversarial test cases or assessing their potential safety and the responsibility of the participants. We further analyze the reported accidents based on their occurrence in normal or adversarial test cases, as shown in Tab. \ref{fig:carla-analysis}. 

For normal test cases, we classify the accidents based on the context for the ego vehicle, including following a vehicle, changing lanes, merging onto a main road, crossing a traffic light junction, or crossing a priority-protected junction (e.g., stop/yield sign junction and unprotected junction). In addition, we analyze responsibility for each accident, identifying whether the fault lies with the ego vehicle or with background vehicles, respectively denoted as \textbf{EF} and \textbf{BF}. Since normal test cases only involve autonomous vehicles starting at zero speed, we consider that they are all potentially safe.

In adversarial test cases, we distinguish between accidents occurring in 10 test case types C1-C10 described in Section 3.2. Recall that five of them (C2, C5, C6, C8, C10) are critical test cases, while the others are accidental test cases. However, since crossing negotiation at an unsignalized intersection (C10) is generated in the same way as normal test cases, we group the accidents observed in C10 with those of the normal test cases at intersections protected by a right-of-way.

We also noticed that if two vehicles remain in collision for a long period, the Leaderboard may record the same accident several times. In Tab. \ref{fig:carla-analysis}, we count only distinct accidents to avoid double-counting.

\begin{table}[h]
    \centering
    \caption{Classification and statistics of accidents with pedestrians and vehicles reported by Carla Leaderboard}
    \label{fig:carla-analysis}
    \vspace{-10pt}
    \includegraphics[width=\linewidth]{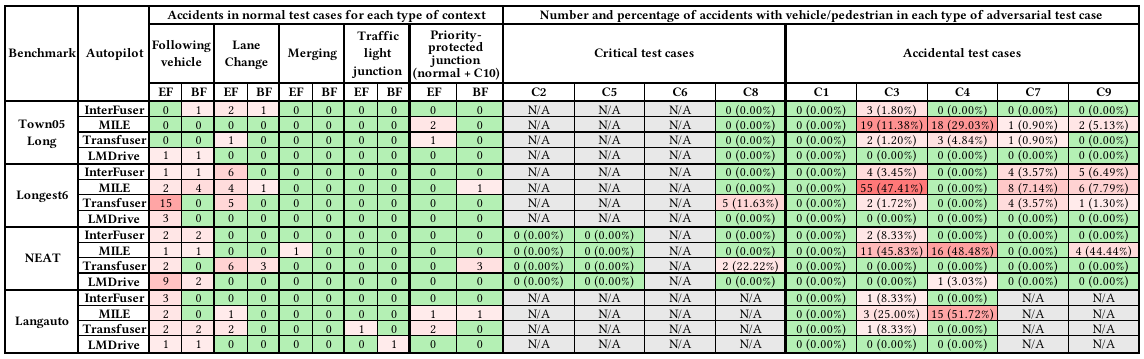}
    \vspace{-15pt}
\end{table}

In normal test cases, Transfuser has the largest number of defects. In particular, it causes 15 accidents when following lanes in the Longest6 benchmark, significantly higher than InterFuser, which comes second and causes only 3 accidents. Transfuser also causes 6 accidents during lane changes in the NEAT benchmark, whereas no such accidents are observed with the other autopilots.

In critical test cases, the Leaderboard reports a low overall number of accidents for the four benchmarks. This can be attributed to the limited number of such scenarios included in the benchmarks, as well as to the fact that the Leaderboard's scenario generation does not create sufficiently critical test cases taking into account participant positions and speeds. Among the accidents, Transfuser causes the highest number of them with 5 in Longest6 and 1 in NEAT. InterFuser ranks second, with 2 accidents in Longest6. The other autopilots do not cause any accidents.

Accidental test cases generally result in higher accident counts, with the exception of C1 involving road damage. Although C1 occurs 404 times across the four benchmarks, it does not result in any accident.

In C3 and C4 involving pedestrians or bicycles crossing roads, MILE exhibits an exceptionally high accident rate. Especially in Longest6, NEAT, and Langauto, nearly half of the test cases of these types result in accidents for MILE. In contrast, the other three autopilots have lower accident rates in these test cases, typically below 10\%. However, such accidents are difficult to fully avoid, as pedestrians or bicycles are NPCs moving at constant speeds and can actively collide with the side of the ego vehicle.

For C7 and C9 involving abnormal traffic light settings, none of the autopilots managed to achieve zero accidents across all benchmarks. Similar to C3 and C4, some accidents are difficult to avoid, such as a vehicle running a red light actively colliding with the ego vehicle's side.

\noindent\textbf{2) Analysis of traffic rule violations }

As shown in Fig. \ref{fig:occurence-leaderboard}, MILE and LMDrive exhibit high frequencies of traffic light violations. MILE has up to 1.71 traffic light violations per kilometer in Town05 Long, while LMDrive reaches 2.89 traffic light violations per kilometer in NEAT. In contrast, Transfuser demonstrates consistently lower rates of traffic light violations, ranging from 0.0 to 0.21 violations per kilometer across the four benchmarks. InterFuser shows slightly higher rates, ranging from 0.11 to 0.69 violations per kilometer. Most traffic light violations occur when the ego vehicle passes the intersection, and the light has been red for a while. 

As far as stop sign violations are concerned, MILE and Transfuser exhibit significantly higher frequencies of violations compared to InterFuser and LMDrive across the four benchmarks. In particular, Transfuser has the highest frequency of 1.54 violations per kilometer in Town05 Long, followed by MILE with 0.73 violations per km, while InterFuser and LMDrive have no such violations in Town05 Long.

\noindent\textbf{3) Analysis of performance issues} 

With regard to performance issues such as route changes, blockages, and route timeouts, we observe the following.  LMDrive primarily makes the mistake of changing routes, when it fails to strictly follow the route  assigned to it. In contrast, the other autopilots generally perform well in this respect, with at most one exception per benchmark.

All four autopilots experience blockages, but a closer look reveals that most blockages occur after accidents, where the ego vehicle is unable to move forward due to obstruction from the collided vehicle or other layout elements. In real-world conditions, blockages are more often the result of traffic congestion, road obstructions, or vehicle breakdowns rather than the after-effects of accidents. Additionally, some blockages are caused by background vehicles becoming stuck due to their own deadlocks or accidents.

InterFuser experiences a higher number of route timeouts compared to the other autopilots. A closer analysis shows that this is caused by its overly cautious behavior at traffic lights. Specifically, InterFuser usually halts in the middle of the road when facing a red light, instead of stopping at the junction entrance. This inefficient behavior often causes it to wait through multiple traffic light cycles when approaching a traffic light junction, resulting in timeouts.

\begin{center}
\fcolorbox{black}{gray!10}{\parbox{\linewidth}{\textbf{Answer to RQ2.1}: The Leaderboard detected accidents for the four AI autopilots in diverse accidental test cases, as well as failures for simple policies such as following front vehicles or stopping at stop signs and red lights. However, it found few accidents in critical contexts, such as merging and crossings. In addition, its evaluation method does not analyze the responsibility and possibility of avoiding the hazards, resulting in a bias. For example, Transfuser ranked second, despite having the highest number of accidents in normal test cases due to its own faults.}}
\end{center}

\subsubsection{Comparison between Leaderboard and CCTest results}

Interestingly, the Leaderboard and CCTest results concur in their assessment of certain safety violations, such as accidents, running red lights and road deviations, as well as performance issues, such as route changes and blockages. In this section, we compare the safety and performance issues reported by CCTest and the Leaderboard to show the complementarity of two approaches in evaluating autopilots.

\noindent\textbf{1) Analysis of accidents and road deviations}

When comparing accidents detected by the Leaderboard and CCTest, although the Leaderboard identified more diverse accidents in accidental test cases, it proved less effective at detecting accidents in critical test cases. Specifically, no accidents were detected at merging for InterFuser, Transfuser, or LMDrive, nor at priority-protected junctions for InterFuser or LMDrive. In contrast, CCTest consistently discovered accidents for the four autopilots in both merging and priority-protected crossing contexts.

Furthermore, the Leaderboard requires background vehicles to be managed by the Carla autopilot. Our analysis shows that, out of a total of 104 accidents in normal scenarios, 25 (24.0\%) of them were caused by background vehicles. However, all accidents are treated equivalently in the Leaderboard's quantitative evaluation. In contrast, CCTest generates critical scenarios where each moving vehicle is equipped with the autopilot under test and has a safe policy. This ensures that any detected violations can be attributed to the autopilot under test.

\noindent\textbf{2) Analysis of traffic rule violations}

For traffic light violations, the issues identified by the Leaderboard align with the findings from our analysis of CCTest results that the ego vehicle can cross when it is facing a red light. However, CCTest does not detect these problems directly, as they occur in non-critical cases involving simple control policies. Similarly, CCTest does not consider test cases involving stopping before stop signs.

\noindent\textbf{3) Analysis of performance issues} 

The issue of route changes for LMDrive is identified by both the Leaderboard and CCTest. In the Leaderboard, LMDrive modified the assigned route in more than half of the routes in each benchmark. On the other hand, CCTest reveals that LMDrive exhibits route changes across a wide range of configuration parameters. The mutually confirmed results revealed inherent flaws in LMDrive.

Blockages identified by the Leaderboard are primarily the result of secondary effects from accidents or issues caused by background vehicles. In contrast, CCTest detects blockages for Transfuser and LMDrive under situations with safe progress policies. Such blockages reflect real defects in the autopilot.

The overly cautious behavior of InterFuser at traffic lights is not detected by CCTest, as this behavior occurs exclusively from non-critical configurations, whereas CCTest is designed to focus on critical configurations.

\begin{center}
\fcolorbox{black}{gray!10}{\parbox{\linewidth}{\textbf{Answer to RQ2.2}: Compared to the Leaderboard that covers both normal and adversarial test cases, CCTest has narrower coverage but demonstrated significantly higher effectiveness in detecting accidents under potentially safe and critical test cases. It systematically detected accidents involving all four autopilots in critical contexts, something the Leaderboard failed to do. This highlights CCTest’s essential role in identifying accidents under extreme conditions, which are hard to uncover without predicting and adjusting vehicle behavior to create critical situations. In addition, CCTest generates simple test cases that enable automatic qualitative evaluation, allowing precise risk analysis free from biases introduced by Leaderboard’s global quantitative evaluation.}}
\end{center}

\section{Related work}
\label{se:relatedwork}
There exists a very rich literature on testing ADS using scenario generation techniques. These techniques range from knowledge-based generation, e.g. \cite{nalic2020scenario,ding2023survey}, 
to model-based generation \cite{riedmaier2020survey,karimi2022automatic,xia2017automatic} and data-driven generation, e.g. \cite{riedmaier2020survey,nalic2020scenario,watanabe2019methodology}, to generation from natural driving data, e.g. \cite{lehmann2019use}. 

CCTest is model-based and focuses on calculating the critical test cases that are the most difficult to manage, taking into account the driving capabilities of the vehicles tested and parameters such as their relative position and speed. It differs from other model-based methods that pursue different objectives or account for various factors, such as searching for scenarios that violate traffic rules by monitoring temporal logic specifications \cite{zhou2023specification,li2023simulation, li2024viohawk}, or creating NPC trajectories that conflict with the ego vehicle's route using SMT solvers \cite{tian2022generating,kim2019test}. It also differs from complexity-based generation relying on influence factor analysis~\cite{karimi2022automatic,xia2017automatic}, or applying combinatorial testing methods to generate scenarios with high coverage rates.~\cite{gao2019test,li2022comopt}.

Many other methods apply search or fuzz techniques to explore model parameters with objectives such as maximizing coverage or risk metrics \cite{2020AV-fuzzer,kim2022drivefuzz,gorbunov2010autofuzz,huai2023sceno,huai2023doppelganger,cheng2023behavexplor,wang2024dance,lin2024tm}. CCTest is much more effective in detecting real defects in autopilots than these techniques, which lack rigorous criteria for selecting test cases that are both critical and potentially safe.

Data-driven approaches reconstruct or generalize scenarios from real-world driving data. The Carla Leaderboard applies this kind of approach to create adversarial situations. Recent works leverage AI techniques, such as Large Language Models (LLMs) and diffusion-based techniques, to generate realistic controllable scenarios meeting a given set of properties. LLMs are used to generate parameterized scenarios from accident records while generalizing them towards diversity or risk using appropriate prompts \cite{miao2024dashcam,tang2024legend,chang2024llmscenario}. Diffusion-based techniques are trained on large sets of traffic data and using guided sampling so that the scenarios involve scenes that exhibit the desired characteristics~\cite{NEURIPS2023Pro,KDD24Zhu,ICRA24Lu}.

As far as evaluation methods are concerned, our distinction between quantitative and qualitative evaluation is reflected in numerous works. For example, \cite{riedmaier2020survey} distinguishes between macroscopic and microscopic evaluation, the former being mainly quantitative, focusing on the overall impact of ADS on traffic and using large amounts of data. Various indicators have been proposed to estimate the risk for a scenario \cite{nalic2020scenario,lehmann2019use}. These include temporal measures \cite{nalic2020scenario}, such as time to collision (TTC), worst TTC, time to braking, time to steering, time to accident, and distance-related measures \cite{nalic2020scenario}.  A detailed list and comparison of safety indicators is provided in \cite{mahmud2017application}, including a summary of the main guiding principles, as well as their most important features, such as critical value or thresholds for indicators. 

\section{Discussion}
\label{se:discussion}
This paper makes multiple contributions to the study of the ADS testing problem. The starting point is the growing interest in end-to-end AI autopilots, which seem to be a promising alternative to traditional modular autopilots. Recently, four such modular open autopilots have been tested using CCTest, demonstrating an impressive number of faults of all types. So, it was tempting to compare end-to-end AI autopilots with modular autopilots by applying the same test method in the Carla Simulator. We identified four such open autopilots, namely InterFuser, MILE, Transfuser, and LMDrive, for which evaluations in the framework of the Carla Leaderboard have been published. 

We tested these autopilots, after having carried out the necessary implementation work in the Carla Simulator to apply the test method. This enabled us to obtain results in two interesting directions.

First, we evaluated the four end-to-end AI autopilots and compared the results with existing results for the four open modular autopilots. Using the same testing environment provided a common basis for comparison. We have found that the two types of autopilot have comparable overall failure rates, and are therefore indistinguishable in terms of their ability to handle critical situations. In addition, the two types of autopilots show the same kind of weaknesses for each type of configuration tested: relatively good scores for merging, over-cautiousness for lane changes, and a large number of traffic rule violations for yield-protected crossings. Finally, for junctions with traffic lights, the modular autopilots obtained better scores overall, but the results were very uneven between autopilots of the same type.

Secondly, the analysis of the Leaderboard evaluation method led us to reflect on the complementary roles of quantitative evaluation methods and CCTest's qualitative method. The Leaderboard benchmarks offer broader but less focused coverage of situations, with sequences of test cases involving adversarial test cases derived from the NHTSA typology. The first question that arose was how effective the test method was compared with the CCTest, which is based on much shorter, more focused scenarios. The second question was the extent to which Leaderboard's quantitative evaluation method provides a scaled comparison of each autopilot's proficiency. Our detailed analysis of the verdicts reveals problems concerning both the effectiveness of the test generation methodology and the suitability of the Leaderboard evaluation method. 
 
One problem is that test cases often involve numerous dangerous accidental situations whose parameters are chosen on an ad hoc basis, and can lead to unavoidable accidents. So they do not call into question the safety of autopilot control policies. In these cases, the inability of a vehicle to escape should not be taken into account when estimating the score of its autopilot. 
 
The second problem is that the normal test cases involve only one vehicle equipped with the autopilot under test, the ego vehicle, whereas the background vehicles are driven by different autopilots, Carla's built-in autopilots. In this case, accidents caused by background vehicles are irrelevant and should not be taken into account in the calculated score. Overall, CCTest performs very well for critical situations it is designed to test. However, it does not detect faults that occur in non-conflict situations, and which need to be detected by adequate coverage of normal situations. 

In summary, this comparison suggests that a test methodology for ADS should adequately cover both average and extreme cases. Average cases are useful for ensuring that autopilots can drive autonomously in everyday situations, safely and efficiently. Extreme cases are experiences that must be carefully designed using test cases that orchestrate NPCs or place an ADS on critical initial configurations from the start. In both cases, however, the behavior of the vehicles concerned must be predicted with a certain degree of accuracy, to avoid potentially unsafe situations that exceed the capabilities of the vehicles under test.

We hope this work will contribute to a better understanding of the characteristics of end-to-end AI autopilots and the associated testing methodologies, while advancing the development of a unified evaluation framework with sufficiently objective criteria.

\section{Data Availability}

The code and detailed results for this paper are available at \url{https://github.com/LIIHWF/testing-end-to-end-ai-autopilots}.

\bibliographystyle{abbrv}
\bibliography{reference}
\end{document}

%% file: tables/interfuser-merging-0.tex
\begin{tabular}{|c| *{14}{c|}} 
\hline 
\diagbox{$\bm{x_a}$}{$\bm{x_f}$} & \ct{0.0} & 40.0 & 80.0 & 120.0 & 160.0 & 200.0 & 240.0 & 280.0 & 320.0 & 360.0 & 400.0 & 440.0 & 480.0 & 520.0 \\ 
\hline 
0.0 & - & \cs{\textsf{CS} (5/5)} & \cs{\textsf{CS} (5/5)} & \cs{\textsf{CS} (5/5)} & \cs{\textsf{CS} (5/5)} & \cs{\textsf{CS} (5/5)} & \cs{\textsf{CS} (5/5)} & \cs{\textsf{CS} (5/5)} & \cs{\textsf{CS} (5/5)} & \cs{\textsf{CS} (5/5)} & \cs{\textsf{CS} (5/5)} & \cs{\textsf{CS} (5/5)} & \cs{\textsf{CS} (5/5)} & \cs{\textsf{CS} (5/5)} \\ 
\hline 
\ct{2.3} & \cs{\textsf{CS} (5/5)} & \cs{\textsf{CS} (5/5)} & \cs{\textsf{CS} (5/5)} & \cs{\textsf{CS} (5/5)} & \cs{\textsf{CS} (5/5)} & \cs{\textsf{CS} (5/5)} & \cs{\textsf{CS} (5/5)} & \cs{\textsf{CS} (5/5)} & \cs{\textsf{CS} (5/5)} & \cs{\textsf{CS} (5/5)} & \cs{\textsf{CS} (5/5)} & \cs{\textsf{CS} (5/5)} & \cs{\textsf{CS} (5/5)} & \cs{\textsf{CS} (5/5)} \\ 
\hline 
5.0 & \cs{\textsf{CS} (5/5)} & \cs{\textsf{CS} (5/5)} & \cs{\textsf{CS} (5/5)} & \cs{\textsf{CS} (5/5)} & \cs{\textsf{CS} (5/5)} & \cs{\textsf{CS} (5/5)} & \cs{\textsf{CS} (5/5)} & \cs{\textsf{CS} (5/5)} & \cs{\textsf{CS} (5/5)} & \cs{\textsf{CS} (5/5)} & \cs{\textsf{CS} (5/5)} & \cs{\textsf{CS} (5/5)} & \cs{\textsf{CS} (5/5)} & \cs{\textsf{CS} (5/5)} \\ 
\hline 
10.0 & \cs{\textsf{CS} (5/5)} & \cs{\textsf{CS} (5/5)} & \cs{\textsf{CS} (5/5)} & \cs{\textsf{CS} (5/5)} & \cs{\textsf{CS} (5/5)} & \cs{\textsf{CS} (5/5)} & \cs{\textsf{CS} (5/5)} & \cs{\textsf{CS} (5/5)} & \cs{\textsf{CS} (5/5)} & \cs{\textsf{CS} (5/5)} & \cs{\textsf{CS} (5/5)} & \cs{\textsf{CS} (5/5)} & \cs{\textsf{CS} (5/5)} & \cs{\textsf{CS} (5/5)} \\ 
\hline 
11.3 & \cs{\textsf{CS} (5/5)} & \ps{\textsf{PS} (5/5)}& \begin{tabular}{@{}c@{}}\begin{minipage}[c][1.0cm][c]{1.2cm}\centering \ac{\textsf{Aa} (2/5)}\ \end{minipage} \\\begin{minipage}[c][1.0cm][c]{1.2cm}\centering \ps{\textsf{PS} (3/5)}\ \end{minipage} \\\end{tabular}& \begin{tabular}{@{}c@{}}\begin{minipage}[c][0.6666666666666666cm][c]{1.2cm}\centering \ps{\textsf{PS} (3/5)}\ \end{minipage} \\\begin{minipage}[c][0.6666666666666666cm][c]{1.2cm}\centering \ac{\textsf{Aa} (1/5)}\ \end{minipage} \\\begin{minipage}[c][0.6666666666666666cm][c]{1.2cm}\centering \cs{\textsf{CS} (1/5)}\ \end{minipage} \\\end{tabular}& \begin{tabular}{@{}c@{}}\begin{minipage}[c][0.6666666666666666cm][c]{1.2cm}\centering \ps{\textsf{PS} (2/5)}\ \end{minipage} \\\begin{minipage}[c][0.6666666666666666cm][c]{1.2cm}\centering \cs{\textsf{CS} (2/5)}\ \end{minipage} \\\begin{minipage}[c][0.6666666666666666cm][c]{1.2cm}\centering \ac{\textsf{Aa} (1/5)}\ \end{minipage} \\\end{tabular}& \begin{tabular}{@{}c@{}}\begin{minipage}[c][0.6666666666666666cm][c]{1.2cm}\centering \cs{\textsf{CS} (1/5)}\ \end{minipage} \\\begin{minipage}[c][0.6666666666666666cm][c]{1.2cm}\centering \ps{\textsf{PS} (1/5)}\ \end{minipage} \\\begin{minipage}[c][0.6666666666666666cm][c]{1.2cm}\centering \ac{\textsf{Aa} (3/5)}\ \end{minipage} \\\end{tabular}& \begin{tabular}{@{}c@{}}\begin{minipage}[c][1.0cm][c]{1.2cm}\centering \ac{\textsf{Aa} (1/5)}\ \end{minipage} \\\begin{minipage}[c][1.0cm][c]{1.2cm}\centering \cs{\textsf{CS} (4/5)}\ \end{minipage} \\\end{tabular}& \begin{tabular}{@{}c@{}}\begin{minipage}[c][0.6666666666666666cm][c]{1.2cm}\centering \ps{\textsf{PS} (1/5)}\ \end{minipage} \\\begin{minipage}[c][0.6666666666666666cm][c]{1.2cm}\centering \cs{\textsf{CS} (3/5)}\ \end{minipage} \\\begin{minipage}[c][0.6666666666666666cm][c]{1.2cm}\centering \ac{\textsf{Aa} (1/5)}\ \end{minipage} \\\end{tabular}& \begin{tabular}{@{}c@{}}\begin{minipage}[c][0.6666666666666666cm][c]{1.2cm}\centering \cs{\textsf{CS} (3/5)}\ \end{minipage} \\\begin{minipage}[c][0.6666666666666666cm][c]{1.2cm}\centering \ps{\textsf{PS} (1/5)}\ \end{minipage} \\\begin{minipage}[c][0.6666666666666666cm][c]{1.2cm}\centering \ac{\textsf{Ae} (1/5)}\ \end{minipage} \\\end{tabular}& \begin{tabular}{@{}c@{}}\begin{minipage}[c][1.0cm][c]{1.2cm}\centering \ac{\textsf{Aa} (4/5)}\ \end{minipage} \\\begin{minipage}[c][1.0cm][c]{1.2cm}\centering \cs{\textsf{CS} (1/5)}\ \end{minipage} \\\end{tabular}& \begin{tabular}{@{}c@{}}\begin{minipage}[c][0.5cm][c]{1.2cm}\centering \ps{\textsf{PS} (2/5)}\ \end{minipage} \\\begin{minipage}[c][0.5cm][c]{1.2cm}\centering \ac{\textsf{Ae} (1/5)}\ \end{minipage} \\\begin{minipage}[c][0.5cm][c]{1.2cm}\centering \cs{\textsf{CS} (1/5)}\ \end{minipage} \\\begin{minipage}[c][0.5cm][c]{1.2cm}\centering \ac{\textsf{Aa} (1/5)}\ \end{minipage} \\\end{tabular}& \begin{tabular}{@{}c@{}}\begin{minipage}[c][1.0cm][c]{1.2cm}\centering \cs{\textsf{CS} (2/5)}\ \end{minipage} \\\begin{minipage}[c][1.0cm][c]{1.2cm}\centering \ac{\textsf{Ae} (3/5)}\ \end{minipage} \\\end{tabular}& \begin{tabular}{@{}c@{}}\begin{minipage}[c][1.0cm][c]{1.2cm}\centering \cs{\textsf{CS} (3/5)}\ \end{minipage} \\\begin{minipage}[c][1.0cm][c]{1.2cm}\centering \ps{\textsf{PS} (2/5)}\ \end{minipage} \\\end{tabular}& \begin{tabular}{@{}c@{}}\begin{minipage}[c][1.0cm][c]{1.2cm}\centering \cs{\textsf{CS} (4/5)}\ \end{minipage} \\\begin{minipage}[c][1.0cm][c]{1.2cm}\centering \ps{\textsf{PS} (1/5)}\ \end{minipage} \\\end{tabular} \\ 
\hline 
12.5 & \cs{\textsf{CS} (5/5)} & \ps{\textsf{PS} (5/5)} & \ps{\textsf{PS} (5/5)} & \ps{\textsf{PS} (5/5)} & \ps{\textsf{PS} (5/5)}& \begin{tabular}{@{}c@{}}\begin{minipage}[c][0.5cm][c]{1.2cm}\centering \ac{\textsf{Aa} (1/5)}\ \end{minipage} \\\begin{minipage}[c][0.5cm][c]{1.2cm}\centering \ps{\textsf{PS} (4/5)}\ \end{minipage} \\\end{tabular} & \ps{\textsf{PS} (5/5)} & \ps{\textsf{PS} (5/5)} & \ps{\textsf{PS} (5/5)} & \ps{\textsf{PS} (5/5)} & \ps{\textsf{PS} (5/5)} & \ps{\textsf{PS} (5/5)} & \ps{\textsf{PS} (5/5)} & \ps{\textsf{PS} (5/5)} \\ 
\hline 
15.0 & \cs{\textsf{CS} (5/5)} & \ps{\textsf{PS} (5/5)} & \ps{\textsf{PS} (5/5)} & \ps{\textsf{PS} (5/5)} & \ps{\textsf{PS} (5/5)} & \ps{\textsf{PS} (5/5)}& \begin{tabular}{@{}c@{}}\begin{minipage}[c][0.5cm][c]{1.2cm}\centering \ac{\textsf{Aa} (1/5)}\ \end{minipage} \\\begin{minipage}[c][0.5cm][c]{1.2cm}\centering \ps{\textsf{PS} (4/5)}\ \end{minipage} \\\end{tabular} & \ps{\textsf{PS} (5/5)} & \ps{\textsf{PS} (5/5)} & \ps{\textsf{PS} (5/5)} & \ps{\textsf{PS} (5/5)}& \begin{tabular}{@{}c@{}}\begin{minipage}[c][0.5cm][c]{1.2cm}\centering \ps{\textsf{PS} (4/5)}\ \end{minipage} \\\begin{minipage}[c][0.5cm][c]{1.2cm}\centering \ac{\textsf{Aa} (1/5)}\ \end{minipage} \\\end{tabular} & \ps{\textsf{PS} (5/5)} & \ps{\textsf{PS} (5/5)} \\ 
\hline 
20.0 & \cs{\textsf{CS} (5/5)} & \ps{\textsf{PS} (5/5)} & \ps{\textsf{PS} (5/5)} & \ps{\textsf{PS} (5/5)} & \ps{\textsf{PS} (5/5)} & \ps{\textsf{PS} (5/5)} & \ps{\textsf{PS} (5/5)} & \ps{\textsf{PS} (5/5)} & \ps{\textsf{PS} (5/5)} & \ps{\textsf{PS} (5/5)} & \ps{\textsf{PS} (5/5)} & \ps{\textsf{PS} (5/5)}& \begin{tabular}{@{}c@{}}\begin{minipage}[c][0.5cm][c]{1.2cm}\centering \ps{\textsf{PS} (4/5)}\ \end{minipage} \\\begin{minipage}[c][0.5cm][c]{1.2cm}\centering \ac{\textsf{Aa} (1/5)}\ \end{minipage} \\\end{tabular} & \ps{\textsf{PS} (5/5)} \\ 
\hline 
25.0 & \cs{\textsf{CS} (5/5)} & \ps{\textsf{PS} (5/5)} & \ps{\textsf{PS} (5/5)} & \ps{\textsf{PS} (5/5)} & \ps{\textsf{PS} (5/5)} & \ps{\textsf{PS} (5/5)}& \begin{tabular}{@{}c@{}}\begin{minipage}[c][0.5cm][c]{1.2cm}\centering \ac{\textsf{Aa} (1/5)}\ \end{minipage} \\\begin{minipage}[c][0.5cm][c]{1.2cm}\centering \ps{\textsf{PS} (4/5)}\ \end{minipage} \\\end{tabular} & \ps{\textsf{PS} (5/5)} & \ps{\textsf{PS} (5/5)} & \ps{\textsf{PS} (5/5)} & \ps{\textsf{PS} (5/5)} & \ps{\textsf{PS} (5/5)} & \ps{\textsf{PS} (5/5)} & \ps{\textsf{PS} (5/5)} \\ 
\hline 
30.0 & \cs{\textsf{CS} (5/5)} & \ps{\textsf{PS} (5/5)} & \ps{\textsf{PS} (5/5)} & \ps{\textsf{PS} (5/5)} & \ps{\textsf{PS} (5/5)} & \ps{\textsf{PS} (5/5)} & \ps{\textsf{PS} (5/5)} & \ps{\textsf{PS} (5/5)} & \ps{\textsf{PS} (5/5)} & \ps{\textsf{PS} (5/5)} & \ps{\textsf{PS} (5/5)} & \ps{\textsf{PS} (5/5)} & \ps{\textsf{PS} (5/5)} & \ps{\textsf{PS} (5/5)} \\ 
\hline 
35.0 & \cs{\textsf{CS} (5/5)} & \ps{\textsf{PS} (5/5)} & \ps{\textsf{PS} (5/5)} & \ps{\textsf{PS} (5/5)} & \ps{\textsf{PS} (5/5)} & \ps{\textsf{PS} (5/5)} & \ps{\textsf{PS} (5/5)} & \ps{\textsf{PS} (5/5)} & \ps{\textsf{PS} (5/5)} & \ps{\textsf{PS} (5/5)} & \ps{\textsf{PS} (5/5)} & \ps{\textsf{PS} (5/5)} & \ps{\textsf{PS} (5/5)} & \ps{\textsf{PS} (5/5)} \\ 
\hline 
40.0 & \cs{\textsf{CS} (5/5)} & \ps{\textsf{PS} (5/5)} & \ps{\textsf{PS} (5/5)} & \ps{\textsf{PS} (5/5)} & \ps{\textsf{PS} (5/5)} & \ps{\textsf{PS} (5/5)} & \ps{\textsf{PS} (5/5)} & \ps{\textsf{PS} (5/5)} & \ps{\textsf{PS} (5/5)} & \ps{\textsf{PS} (5/5)} & \ps{\textsf{PS} (5/5)} & \ps{\textsf{PS} (5/5)} & \ps{\textsf{PS} (5/5)} & \ps{\textsf{PS} (5/5)} \\ 
\hline 
\end{tabular}

%% file: tables/interfuser-merging-4.tex
\begin{tabular}{|c| *{15}{c|}} 
\hline 
\diagbox{$\bm{x_a}$}{$\bm{x_f}$} & 0.0 & \ct{2.3} & 40.0 & 80.0 & 120.0 & 160.0 & 200.0 & 240.0 & 280.0 & 320.0 & 360.0 & 400.0 & 440.0 & 480.0 & 520.0 \\ 
\hline 
0.0 & - & - & \cs{\textsf{CS} (5/5)} & \cs{\textsf{CS} (5/5)} & \cs{\textsf{CS} (5/5)} & \cs{\textsf{CS} (5/5)} & \cs{\textsf{CS} (5/5)} & \cs{\textsf{CS} (5/5)} & \cs{\textsf{CS} (5/5)} & \cs{\textsf{CS} (5/5)} & \cs{\textsf{CS} (5/5)} & \cs{\textsf{CS} (5/5)} & \cs{\textsf{CS} (5/5)} & \cs{\textsf{CS} (5/5)} & \cs{\textsf{CS} (5/5)} \\ 
\hline 
\ct{4.2} & \cs{\textsf{CS} (5/5)} & \cs{\textsf{CS} (5/5)} & \cs{\textsf{CS} (5/5)} & \cs{\textsf{CS} (5/5)}& \begin{tabular}{@{}c@{}}\begin{minipage}[c][0.5cm][c]{1.2cm}\centering \cs{\textsf{CS} (2/5)}\ \end{minipage} \\\begin{minipage}[c][0.5cm][c]{1.2cm}\centering \ac{\textsf{Ae} (3/5)}\ \end{minipage} \\\end{tabular}& \begin{tabular}{@{}c@{}}\begin{minipage}[c][0.5cm][c]{1.2cm}\centering \ac{\textsf{Ae} (4/5)}\ \end{minipage} \\\begin{minipage}[c][0.5cm][c]{1.2cm}\centering \cs{\textsf{CS} (1/5)}\ \end{minipage} \\\end{tabular} & \ac{\textsf{Ae} (5/5)}& \begin{tabular}{@{}c@{}}\begin{minipage}[c][0.5cm][c]{1.2cm}\centering \ac{\textsf{Ae} (4/5)}\ \end{minipage} \\\begin{minipage}[c][0.5cm][c]{1.2cm}\centering \cs{\textsf{CS} (1/5)}\ \end{minipage} \\\end{tabular} & \cs{\textsf{CS} (5/5)} & \cs{\textsf{CS} (5/5)} & \cs{\textsf{CS} (5/5)}& \begin{tabular}{@{}c@{}}\begin{minipage}[c][0.5cm][c]{1.2cm}\centering \cs{\textsf{CS} (1/5)}\ \end{minipage} \\\begin{minipage}[c][0.5cm][c]{1.2cm}\centering \ac{\textsf{Ae} (4/5)}\ \end{minipage} \\\end{tabular} & \cs{\textsf{CS} (5/5)} & \cs{\textsf{CS} (5/5)} & \cs{\textsf{CS} (5/5)} \\ 
\hline
5.0 & \cs{\textsf{CS} (5/5)} & \cs{\textsf{CS} (5/5)} & \cs{\textsf{CS} (5/5)} & \cs{\textsf{CS} (5/5)} & \cs{\textsf{CS} (5/5)} & \cs{\textsf{CS} (5/5)} & \cs{\textsf{CS} (5/5)}& \begin{tabular}{@{}c@{}}\begin{minipage}[c][0.5cm][c]{1.2cm}\centering \cs{\textsf{CS} (3/5)}\ \end{minipage} \\\begin{minipage}[c][0.5cm][c]{1.2cm}\centering \ac{\textsf{Ae} (2/5)}\ \end{minipage} \\\end{tabular} & \cs{\textsf{CS} (5/5)} & \cs{\textsf{CS} (5/5)} & \cs{\textsf{CS} (5/5)}& \begin{tabular}{@{}c@{}}\begin{minipage}[c][0.5cm][c]{1.2cm}\centering \ac{\textsf{Ae} (1/5)}\ \end{minipage} \\\begin{minipage}[c][0.5cm][c]{1.2cm}\centering \cs{\textsf{CS} (4/5)}\ \end{minipage} \\\end{tabular} & \cs{\textsf{CS} (5/5)} & \cs{\textsf{CS} (5/5)} & \cs{\textsf{CS} (5/5)} \\ 
\hline 
7.5 & \cs{\textsf{CS} (5/5)} & \cs{\textsf{CS} (5/5)} & \ac{\textsf{Aa} (5/5)} & \ac{\textsf{Aa} (5/5)} & \ac{\textsf{Aa} (5/5)}& \begin{tabular}{@{}c@{}}\begin{minipage}[c][0.5cm][c]{1.2cm}\centering \ac{\textsf{Aa} (4/5)}\ \end{minipage} \\\begin{minipage}[c][0.5cm][c]{1.2cm}\centering \cs{\textsf{CS} (1/5)}\ \end{minipage} \\\end{tabular} & \ac{\textsf{Aa} (5/5)} & \ac{\textsf{Aa} (5/5)} & \ac{\textsf{Aa} (5/5)} & \ac{\textsf{Aa} (5/5)} & \ac{\textsf{Aa} (5/5)} & \ac{\textsf{Aa} (5/5)} & \ac{\textsf{Aa} (5/5)} & \ac{\textsf{Aa} (5/5)} & \ac{\textsf{Aa} (5/5)} \\ 
\hline 
10.0 & \cs{\textsf{CS} (5/5)} & \cs{\textsf{CS} (5/5)} & \ps{\textsf{PS} (5/5)} & \ps{\textsf{PS} (5/5)} & \ps{\textsf{PS} (5/5)} & \ps{\textsf{PS} (5/5)} & \ps{\textsf{PS} (5/5)} & \ps{\textsf{PS} (5/5)} & \ps{\textsf{PS} (5/5)} & \ps{\textsf{PS} (5/5)} & \ps{\textsf{PS} (5/5)} & \ps{\textsf{PS} (5/5)} & \ps{\textsf{PS} (5/5)} & \ps{\textsf{PS} (5/5)} & \ps{\textsf{PS} (5/5)} \\ 
\hline 
15.0 & \cs{\textsf{CS} (5/5)} & \cs{\textsf{CS} (5/5)} & \ps{\textsf{PS} (5/5)} & \ps{\textsf{PS} (5/5)} & \ps{\textsf{PS} (5/5)} & \ps{\textsf{PS} (5/5)} & \ps{\textsf{PS} (5/5)} & \ps{\textsf{PS} (5/5)} & \ps{\textsf{PS} (5/5)} & \ps{\textsf{PS} (5/5)} & \ps{\textsf{PS} (5/5)} & \ps{\textsf{PS} (5/5)} & \ps{\textsf{PS} (5/5)} & \ps{\textsf{PS} (5/5)} & \ps{\textsf{PS} (5/5)} \\ 
\hline 
20.0 & \cs{\textsf{CS} (5/5)} & \cs{\textsf{CS} (5/5)} & \ps{\textsf{PS} (5/5)} & \ps{\textsf{PS} (5/5)} & \ps{\textsf{PS} (5/5)} & \ps{\textsf{PS} (5/5)} & \ps{\textsf{PS} (5/5)} & \ps{\textsf{PS} (5/5)} & \ps{\textsf{PS} (5/5)} & \ps{\textsf{PS} (5/5)} & \ps{\textsf{PS} (5/5)} & \ps{\textsf{PS} (5/5)} & \ps{\textsf{PS} (5/5)} & \ps{\textsf{PS} (5/5)} & \ps{\textsf{PS} (5/5)} \\ 
\hline 
25.0 & \cs{\textsf{CS} (5/5)} & \cs{\textsf{CS} (5/5)} & \ps{\textsf{PS} (5/5)} & \ps{\textsf{PS} (5/5)} & \ps{\textsf{PS} (5/5)} & \ps{\textsf{PS} (5/5)} & \ps{\textsf{PS} (5/5)} & \ps{\textsf{PS} (5/5)}& \begin{tabular}{@{}c@{}}\begin{minipage}[c][0.5cm][c]{1.2cm}\centering \ac{\textsf{Aa} (1/5)}\ \end{minipage} \\\begin{minipage}[c][0.5cm][c]{1.2cm}\centering \ps{\textsf{PS} (4/5)}\ \end{minipage} \\\end{tabular} & \ps{\textsf{PS} (5/5)} & \ps{\textsf{PS} (5/5)} & \ps{\textsf{PS} (5/5)} & \ps{\textsf{PS} (5/5)} & \ps{\textsf{PS} (5/5)} & \ps{\textsf{PS} (5/5)} \\ 
\hline 
30.0 & \cs{\textsf{CS} (5/5)} & \cs{\textsf{CS} (5/5)} & \ps{\textsf{PS} (5/5)} & \ps{\textsf{PS} (5/5)} & \ps{\textsf{PS} (5/5)} & \ps{\textsf{PS} (5/5)} & \ps{\textsf{PS} (5/5)} & \ps{\textsf{PS} (5/5)} & \ps{\textsf{PS} (5/5)} & \ps{\textsf{PS} (5/5)} & \ps{\textsf{PS} (5/5)} & \ps{\textsf{PS} (5/5)} & \ps{\textsf{PS} (5/5)} & \ps{\textsf{PS} (5/5)} & \ps{\textsf{PS} (5/5)} \\ 
\hline 
35.0 & \cs{\textsf{CS} (5/5)} & \cs{\textsf{CS} (5/5)} & \ps{\textsf{PS} (5/5)} & \ps{\textsf{PS} (5/5)} & \ps{\textsf{PS} (5/5)} & \ps{\textsf{PS} (5/5)} & \ps{\textsf{PS} (5/5)} & \ps{\textsf{PS} (5/5)} & \ps{\textsf{PS} (5/5)} & \ps{\textsf{PS} (5/5)} & \ps{\textsf{PS} (5/5)} & \ps{\textsf{PS} (5/5)} & \ps{\textsf{PS} (5/5)} & \ps{\textsf{PS} (5/5)} & \ps{\textsf{PS} (5/5)} \\ 
\hline 
40.0 & \cs{\textsf{CS} (5/5)} & \cs{\textsf{CS} (5/5)} & \ps{\textsf{PS} (5/5)} & \ps{\textsf{PS} (5/5)} & \ps{\textsf{PS} (5/5)} & \ps{\textsf{PS} (5/5)} & \ps{\textsf{PS} (5/5)} & \ps{\textsf{PS} (5/5)} & \ps{\textsf{PS} (5/5)} & \ps{\textsf{PS} (5/5)} & \ps{\textsf{PS} (5/5)} & \ps{\textsf{PS} (5/5)} & \ps{\textsf{PS} (5/5)} & \ps{\textsf{PS} (5/5)} & \ps{\textsf{PS} (5/5)} \\ 
\hline 
\end{tabular} 

%% file: tables/interfuser-lane_change-4.tex
\begin{tabular}{|c| *{15}{c|}} 
\hline 
\diagbox{$\bm{x_a}$}{$\bm{x_f}$} & 0.0 & \ct{1.9} & 40.0 & 80.0 & 120.0 & 160.0 & 200.0 & 240.0 & 280.0 & 320.0 & 360.0 & 400.0 & 440.0 & 480.0 & 520.0 \\ 
\hline 
0.0 & - & - & \cs{\textsf{CS} (5/5)} & \cs{\textsf{CS} (5/5)} & \cs{\textsf{CS} (5/5)} & \cs{\textsf{CS} (5/5)} & \cs{\textsf{CS} (5/5)} & \cs{\textsf{CS} (5/5)} & \cs{\textsf{CS} (5/5)} & \cs{\textsf{CS} (5/5)} & \ac{\textsf{Ae} (5/5)} & \cs{\textsf{CS} (5/5)} & \cs{\textsf{CS} (5/5)} & \cs{\textsf{CS} (5/5)}& \begin{tabular}{@{}c@{}}\begin{minipage}[c][0.5cm][c]{1.2cm}\centering \ac{\textsf{Ae} (1/5)}\ \end{minipage} \\\begin{minipage}[c][0.5cm][c]{1.2cm}\centering \cs{\textsf{CS} (4/5)}\ \end{minipage} \\\end{tabular} \\ 
\hline 
5.0 & - & - & \cs{\textsf{CS} (5/5)} & \cs{\textsf{CS} (5/5)} & \cs{\textsf{CS} (5/5)} & \cs{\textsf{CS} (5/5)} & \cs{\textsf{CS} (5/5)} & \cs{\textsf{CS} (5/5)} & \cs{\textsf{CS} (5/5)} & \cs{\textsf{CS} (5/5)} & \cs{\textsf{CS} (5/5)} & \cs{\textsf{CS} (5/5)} & \cs{\textsf{CS} (5/5)} & \cs{\textsf{CS} (5/5)} & \cs{\textsf{CS} (5/5)} \\ 
\hline 
10.0 & \cs{\textsf{CS} (5/5)} & \cs{\textsf{CS} (5/5)} & \cs{\textsf{CS} (5/5)} & \cs{\textsf{CS} (5/5)} & \cs{\textsf{CS} (5/5)}& \begin{tabular}{@{}c@{}}\begin{minipage}[c][0.5cm][c]{1.2cm}\centering \ac{\textsf{Ae} (1/5)}\ \end{minipage} \\\begin{minipage}[c][0.5cm][c]{1.2cm}\centering \cs{\textsf{CS} (4/5)}\ \end{minipage} \\\end{tabular} & \cs{\textsf{CS} (5/5)} & \cs{\textsf{CS} (5/5)} & \cs{\textsf{CS} (5/5)} & \cs{\textsf{CS} (5/5)}& \begin{tabular}{@{}c@{}}\begin{minipage}[c][0.5cm][c]{1.2cm}\centering \ac{\textsf{Ae} (1/5)}\ \end{minipage} \\\begin{minipage}[c][0.5cm][c]{1.2cm}\centering \cs{\textsf{CS} (4/5)}\ \end{minipage} \\\end{tabular} & \cs{\textsf{CS} (5/5)} & \cs{\textsf{CS} (5/5)} & \cs{\textsf{CS} (5/5)} & \cs{\textsf{CS} (5/5)} \\ 
\hline 
15.0 & \ac{\textsf{Aa} (5/5)} & \ac{\textsf{Aa} (5/5)} & \ac{\textsf{Aa} (5/5)} & \ac{\textsf{Aa} (5/5)} & \ac{\textsf{Aa} (5/5)}& \begin{tabular}{@{}c@{}}\begin{minipage}[c][0.5cm][c]{1.2cm}\centering \ac{\textsf{Aa} (3/5)}\ \end{minipage} \\\begin{minipage}[c][0.5cm][c]{1.2cm}\centering \ac{\textsf{Ae} (1/5)}\ \end{minipage} \\\begin{minipage}[c][0.5cm][c]{1.2cm}\centering \cs{\textsf{CS} (1/5)}\ \end{minipage} \\\end{tabular} & \ac{\textsf{Aa} (5/5)}& \begin{tabular}{@{}c@{}}\begin{minipage}[c][0.75cm][c]{1.2cm}\centering \cs{\textsf{CS} (2/5)}\ \end{minipage} \\\begin{minipage}[c][0.75cm][c]{1.2cm}\centering \ac{\textsf{Aa} (3/5)}\ \end{minipage} \\\end{tabular}& \begin{tabular}{@{}c@{}}\begin{minipage}[c][0.75cm][c]{1.2cm}\centering \ac{\textsf{Aa} (3/5)}\ \end{minipage} \\\begin{minipage}[c][0.75cm][c]{1.2cm}\centering \cs{\textsf{CS} (2/5)}\ \end{minipage} \\\end{tabular}& \begin{tabular}{@{}c@{}}\begin{minipage}[c][0.75cm][c]{1.2cm}\centering \ac{\textsf{Aa} (4/5)}\ \end{minipage} \\\begin{minipage}[c][0.75cm][c]{1.2cm}\centering \cs{\textsf{CS} (1/5)}\ \end{minipage} \\\end{tabular} & \ac{\textsf{Aa} (5/5)} & \ac{\textsf{Aa} (5/5)}& \begin{tabular}{@{}c@{}}\begin{minipage}[c][0.75cm][c]{1.2cm}\centering \ac{\textsf{Aa} (4/5)}\ \end{minipage} \\\begin{minipage}[c][0.75cm][c]{1.2cm}\centering \cs{\textsf{CS} (1/5)}\ \end{minipage} \\\end{tabular} & \ac{\textsf{Aa} (5/5)}& \begin{tabular}{@{}c@{}}\begin{minipage}[c][0.75cm][c]{1.2cm}\centering \ac{\textsf{Aa} (4/5)}\ \end{minipage} \\\begin{minipage}[c][0.75cm][c]{1.2cm}\centering \ac{\textsf{Ae} (1/5)}\ \end{minipage} \\\end{tabular} \\ 
\hline 
\ct{19.2} & \ac{\textsf{Aa} (5/5)} & \ac{\textsf{Aa} (5/5)}& \begin{tabular}{@{}c@{}}\begin{minipage}[c][0.5cm][c]{1.2cm}\centering \ac{\textsf{Aa} (3/5)}\ \end{minipage} \\\begin{minipage}[c][0.5cm][c]{1.2cm}\centering \ps{\textsf{PS} (2/5)}\ \end{minipage} \\\end{tabular}& \begin{tabular}{@{}c@{}}\begin{minipage}[c][0.5cm][c]{1.2cm}\centering \ps{\textsf{PS} (4/5)}\ \end{minipage} \\\begin{minipage}[c][0.5cm][c]{1.2cm}\centering \ac{\textsf{Aa} (1/5)}\ \end{minipage} \\\end{tabular} & \ps{\textsf{PS} (5/5)} & \ps{\textsf{PS} (5/5)}& \begin{tabular}{@{}c@{}}\begin{minipage}[c][0.5cm][c]{1.2cm}\centering \ps{\textsf{PS} (2/5)}\ \end{minipage} \\\begin{minipage}[c][0.5cm][c]{1.2cm}\centering \ac{\textsf{Aa} (3/5)}\ \end{minipage} \\\end{tabular} & \ps{\textsf{PS} (5/5)} & \ps{\textsf{PS} (5/5)} & \ps{\textsf{PS} (5/5)} & \ps{\textsf{PS} (5/5)}& \begin{tabular}{@{}c@{}}\begin{minipage}[c][0.5cm][c]{1.2cm}\centering \ac{\textsf{Aa} (4/5)}\ \end{minipage} \\\begin{minipage}[c][0.5cm][c]{1.2cm}\centering \ps{\textsf{PS} (1/5)}\ \end{minipage} \\\end{tabular} & \ps{\textsf{PS} (5/5)}& \begin{tabular}{@{}c@{}}\begin{minipage}[c][0.5cm][c]{1.2cm}\centering \ac{\textsf{Aa} (2/5)}\ \end{minipage} \\\begin{minipage}[c][0.5cm][c]{1.2cm}\centering \ps{\textsf{PS} (3/5)}\ \end{minipage} \\\end{tabular} & \ps{\textsf{PS} (5/5)} \\ 
\hline 
20.0 & \ps{\textsf{PS} (5/5)}& \begin{tabular}{@{}c@{}}\begin{minipage}[c][0.5cm][c]{1.2cm}\centering \ps{\textsf{PS} (4/5)}\ \end{minipage} \\\begin{minipage}[c][0.5cm][c]{1.2cm}\centering \ac{\textsf{Aa} (1/5)}\ \end{minipage} \\\end{tabular}& \begin{tabular}{@{}c@{}}\begin{minipage}[c][0.5cm][c]{1.2cm}\centering \ps{\textsf{PS} (2/5)}\ \end{minipage} \\\begin{minipage}[c][0.5cm][c]{1.2cm}\centering \ac{\textsf{Aa} (3/5)}\ \end{minipage} \\\end{tabular} & \ps{\textsf{PS} (5/5)} & \ps{\textsf{PS} (5/5)} & \ps{\textsf{PS} (5/5)} & \ps{\textsf{PS} (5/5)} & \ps{\textsf{PS} (5/5)} & \ps{\textsf{PS} (5/5)}& \begin{tabular}{@{}c@{}}\begin{minipage}[c][0.5cm][c]{1.2cm}\centering \ps{\textsf{PS} (3/5)}\ \end{minipage} \\\begin{minipage}[c][0.5cm][c]{1.2cm}\centering \ac{\textsf{Aa} (2/5)}\ \end{minipage} \\\end{tabular} & \ps{\textsf{PS} (5/5)}& \begin{tabular}{@{}c@{}}\begin{minipage}[c][0.5cm][c]{1.2cm}\centering \ps{\textsf{PS} (2/5)}\ \end{minipage} \\\begin{minipage}[c][0.5cm][c]{1.2cm}\centering \ac{\textsf{Aa} (3/5)}\ \end{minipage} \\\end{tabular} & \ps{\textsf{PS} (5/5)} & \ps{\textsf{PS} (5/5)} & \ps{\textsf{PS} (5/5)} \\ 
\hline 
25.0 & \ps{\textsf{PS} (5/5)} & \ps{\textsf{PS} (5/5)}& \begin{tabular}{@{}c@{}}\begin{minipage}[c][0.5cm][c]{1.2cm}\centering \ac{\textsf{Aa} (1/5)}\ \end{minipage} \\\begin{minipage}[c][0.5cm][c]{1.2cm}\centering \ps{\textsf{PS} (4/5)}\ \end{minipage} \\\end{tabular} & \ps{\textsf{PS} (5/5)} & \ps{\textsf{PS} (5/5)} & \ps{\textsf{PS} (5/5)} & \ps{\textsf{PS} (5/5)}& \begin{tabular}{@{}c@{}}\begin{minipage}[c][0.5cm][c]{1.2cm}\centering \ac{\textsf{Aa} (3/5)}\ \end{minipage} \\\begin{minipage}[c][0.5cm][c]{1.2cm}\centering \ps{\textsf{PS} (2/5)}\ \end{minipage} \\\end{tabular}& \begin{tabular}{@{}c@{}}\begin{minipage}[c][0.5cm][c]{1.2cm}\centering \ps{\textsf{PS} (4/5)}\ \end{minipage} \\\begin{minipage}[c][0.5cm][c]{1.2cm}\centering \ac{\textsf{Aa} (1/5)}\ \end{minipage} \\\end{tabular} & \ps{\textsf{PS} (5/5)}& \begin{tabular}{@{}c@{}}\begin{minipage}[c][0.5cm][c]{1.2cm}\centering \ps{\textsf{PS} (3/5)}\ \end{minipage} \\\begin{minipage}[c][0.5cm][c]{1.2cm}\centering \ac{\textsf{Aa} (2/5)}\ \end{minipage} \\\end{tabular}& \begin{tabular}{@{}c@{}}\begin{minipage}[c][0.5cm][c]{1.2cm}\centering \ps{\textsf{PS} (2/5)}\ \end{minipage} \\\begin{minipage}[c][0.5cm][c]{1.2cm}\centering \ac{\textsf{Aa} (3/5)}\ \end{minipage} \\\end{tabular} & \ps{\textsf{PS} (5/5)}& \begin{tabular}{@{}c@{}}\begin{minipage}[c][0.5cm][c]{1.2cm}\centering \ac{\textsf{Aa} (1/5)}\ \end{minipage} \\\begin{minipage}[c][0.5cm][c]{1.2cm}\centering \ps{\textsf{PS} (4/5)}\ \end{minipage} \\\end{tabular} & \ps{\textsf{PS} (5/5)} \\ 
\hline 
30.0 & \ps{\textsf{PS} (5/5)} & \ps{\textsf{PS} (5/5)} & \ps{\textsf{PS} (5/5)} & \ps{\textsf{PS} (5/5)} & \ps{\textsf{PS} (5/5)} & \ps{\textsf{PS} (5/5)} & \ps{\textsf{PS} (5/5)} & \ps{\textsf{PS} (5/5)} & \ps{\textsf{PS} (5/5)} & \ps{\textsf{PS} (5/5)} & \ps{\textsf{PS} (5/5)} & \ps{\textsf{PS} (5/5)} & \ps{\textsf{PS} (5/5)} & \ps{\textsf{PS} (5/5)} & \ps{\textsf{PS} (5/5)} \\ 
\hline 
35.0 & \ps{\textsf{PS} (5/5)} & \ps{\textsf{PS} (5/5)} & \ps{\textsf{PS} (5/5)} & \ps{\textsf{PS} (5/5)} & \ps{\textsf{PS} (5/5)} & \ps{\textsf{PS} (5/5)} & \ps{\textsf{PS} (5/5)} & \ps{\textsf{PS} (5/5)} & \ps{\textsf{PS} (5/5)} & \ps{\textsf{PS} (5/5)} & \ps{\textsf{PS} (5/5)} & \ps{\textsf{PS} (5/5)} & \ps{\textsf{PS} (5/5)} & \ps{\textsf{PS} (5/5)} & \ps{\textsf{PS} (5/5)} \\ 
\hline 
40.0 & \ps{\textsf{PS} (5/5)} & \ps{\textsf{PS} (5/5)} & \ps{\textsf{PS} (5/5)} & \ps{\textsf{PS} (5/5)} & \ps{\textsf{PS} (5/5)} & \ps{\textsf{PS} (5/5)} & \ps{\textsf{PS} (5/5)} & \ps{\textsf{PS} (5/5)} & \ps{\textsf{PS} (5/5)} & \ps{\textsf{PS} (5/5)} & \ps{\textsf{PS} (5/5)} & \ps{\textsf{PS} (5/5)} & \ps{\textsf{PS} (5/5)} & \ps{\textsf{PS} (5/5)} & \ps{\textsf{PS} (5/5)} \\ 
\hline 
\end{tabular} 

%% file: tables/interfuser-crossing-0.tex
\begin{tabular}{|c| *{15}{c|}} 
\hline 
\diagbox{$\bm{x_a}$}{$\bm{x_f}$} & 0.0 & \ct{2.3} & 40.0 & 80.0 & 120.0 & 160.0 & 200.0 & 240.0 & 280.0 & 320.0 & 360.0 & 400.0 & 440.0 & 480.0 & 520.0 \\ 
\hline 
0.0& \begin{tabular}{@{}c@{}}\begin{minipage}[c][0.75cm][c]{2.0cm}\centering \ac{\textsf{Ae} (4/5)}\ \end{minipage} \\\begin{minipage}[c][0.75cm][c]{2.0cm}\centering \blk{\textsf{Blk} (1/5)}\ \end{minipage} \\\end{tabular}& \begin{tabular}{@{}c@{}}\begin{minipage}[c][0.75cm][c]{2.0cm}\centering \ac{\textsf{Ae} (4/5)}\ \end{minipage} \\\begin{minipage}[c][0.75cm][c]{2.0cm}\centering \blk{\textsf{Blk} (1/5)}\ \end{minipage} \\\end{tabular}& \begin{tabular}{@{}c@{}}\begin{minipage}[c][0.75cm][c]{2.0cm}\centering \ac{\textsf{Ae} (4/5)}\ \end{minipage} \\\begin{minipage}[c][0.75cm][c]{2.0cm}\centering \blk{\textsf{Blk} (1/5)}\ \end{minipage} \\\end{tabular}& \begin{tabular}{@{}c@{}}\begin{minipage}[c][0.75cm][c]{2.0cm}\centering \cu{\textsf{CU$p_1p_2$} (2/5)}\ \end{minipage} \\\begin{minipage}[c][0.75cm][c]{2.0cm}\centering \ac{\textsf{Ae} (3/5)}\ \end{minipage} \\\end{tabular}& \begin{tabular}{@{}c@{}}\begin{minipage}[c][0.75cm][c]{2.0cm}\centering \cu{\textsf{CU$p_1p_2$} (1/5)}\ \end{minipage} \\\begin{minipage}[c][0.75cm][c]{2.0cm}\centering \ac{\textsf{Ae} (4/5)}\ \end{minipage} \\\end{tabular}& \begin{tabular}{@{}c@{}}\begin{minipage}[c][0.75cm][c]{2.0cm}\centering \ac{\textsf{Ae} (4/5)}\ \end{minipage} \\\begin{minipage}[c][0.75cm][c]{2.0cm}\centering \cu{\textsf{CU$p_1p_2$} (1/5)}\ \end{minipage} \\\end{tabular} & \ac{\textsf{Ae} (5/5)}& \begin{tabular}{@{}c@{}}\begin{minipage}[c][0.5cm][c]{2.0cm}\centering \blk{\textsf{Blk} (2/5)}\ \end{minipage} \\\begin{minipage}[c][0.5cm][c]{2.0cm}\centering \ac{\textsf{Ae} (2/5)}\ \end{minipage} \\\begin{minipage}[c][0.5cm][c]{2.0cm}\centering \cu{\textsf{CU$p_1p_2$} (1/5)}\ \end{minipage} \\\end{tabular}& \begin{tabular}{@{}c@{}}\begin{minipage}[c][0.75cm][c]{2.0cm}\centering \ac{\textsf{Ae} (4/5)}\ \end{minipage} \\\begin{minipage}[c][0.75cm][c]{2.0cm}\centering \blk{\textsf{Blk} (1/5)}\ \end{minipage} \\\end{tabular}& \begin{tabular}{@{}c@{}}\begin{minipage}[c][0.5cm][c]{2.0cm}\centering \blk{\textsf{Blk} (1/5)}\ \end{minipage} \\\begin{minipage}[c][0.5cm][c]{2.0cm}\centering \pu{\textsf{PU$p_1$} (2/5)}\ \end{minipage} \\\begin{minipage}[c][0.5cm][c]{2.0cm}\centering \ac{\textsf{Ae} (2/5)}\ \end{minipage} \\\end{tabular}& \begin{tabular}{@{}c@{}}\begin{minipage}[c][0.5cm][c]{2.0cm}\centering \blk{\textsf{Blk} (2/5)}\ \end{minipage} \\\begin{minipage}[c][0.5cm][c]{2.0cm}\centering \ac{\textsf{Ae} (2/5)}\ \end{minipage} \\\begin{minipage}[c][0.5cm][c]{2.0cm}\centering \cu{\textsf{CU$p_1p_2$} (1/5)}\ \end{minipage} \\\end{tabular}& \begin{tabular}{@{}c@{}}\begin{minipage}[c][0.75cm][c]{2.0cm}\centering \blk{\textsf{Blk} (1/5)}\ \end{minipage} \\\begin{minipage}[c][0.75cm][c]{2.0cm}\centering \ac{\textsf{Ae} (4/5)}\ \end{minipage} \\\end{tabular}& \begin{tabular}{@{}c@{}}\begin{minipage}[c][0.5cm][c]{2.0cm}\centering \pu{\textsf{PU$p_1$} (2/5)}\ \end{minipage} \\\begin{minipage}[c][0.5cm][c]{2.0cm}\centering \blk{\textsf{Blk} (1/5)}\ \end{minipage} \\\begin{minipage}[c][0.5cm][c]{2.0cm}\centering \ac{\textsf{Ae} (2/5)}\ \end{minipage} \\\end{tabular}& \begin{tabular}{@{}c@{}}\begin{minipage}[c][0.75cm][c]{2.0cm}\centering \blk{\textsf{Blk} (1/5)}\ \end{minipage} \\\begin{minipage}[c][0.75cm][c]{2.0cm}\centering \ac{\textsf{Ae} (4/5)}\ \end{minipage} \\\end{tabular}& \begin{tabular}{@{}c@{}}\begin{minipage}[c][0.75cm][c]{2.0cm}\centering \ac{\textsf{Ae} (3/5)}\ \end{minipage} \\\begin{minipage}[c][0.75cm][c]{2.0cm}\centering \blk{\textsf{Blk} (2/5)}\ \end{minipage} \\\end{tabular} \\ 
\hline 
2.5 & \ac{\textsf{Ae} (5/5)} & \ac{\textsf{Ae} (5/5)}& \begin{tabular}{@{}c@{}}\begin{minipage}[c][0.75cm][c]{2.0cm}\centering \ac{\textsf{Ae} (4/5)}\ \end{minipage} \\\begin{minipage}[c][0.75cm][c]{2.0cm}\centering \ac{\textsf{Aa} (1/5)}\ \end{minipage} \\\end{tabular} & \ac{\textsf{Ae} (5/5)}& \begin{tabular}{@{}c@{}}\begin{minipage}[c][0.75cm][c]{2.0cm}\centering \ac{\textsf{Ae} (4/5)}\ \end{minipage} \\\begin{minipage}[c][0.75cm][c]{2.0cm}\centering \blk{\textsf{Blk} (1/5)}\ \end{minipage} \\\end{tabular}& \begin{tabular}{@{}c@{}}\begin{minipage}[c][0.75cm][c]{2.0cm}\centering \ac{\textsf{Aa} (2/5)}\ \end{minipage} \\\begin{minipage}[c][0.75cm][c]{2.0cm}\centering \ac{\textsf{Ae} (3/5)}\ \end{minipage} \\\end{tabular} & \ac{\textsf{Ae} (5/5)}& \begin{tabular}{@{}c@{}}\begin{minipage}[c][0.75cm][c]{2.0cm}\centering \ac{\textsf{Aa} (2/5)}\ \end{minipage} \\\begin{minipage}[c][0.75cm][c]{2.0cm}\centering \ac{\textsf{Ae} (3/5)}\ \end{minipage} \\\end{tabular}& \begin{tabular}{@{}c@{}}\begin{minipage}[c][0.75cm][c]{2.0cm}\centering \ac{\textsf{Aa} (4/5)}\ \end{minipage} \\\begin{minipage}[c][0.75cm][c]{2.0cm}\centering \cu{\textsf{CU$p_1p_2$} (1/5)}\ \end{minipage} \\\end{tabular} & \ac{\textsf{Ae} (5/5)} & \ac{\textsf{Ae} (5/5)}& \begin{tabular}{@{}c@{}}\begin{minipage}[c][0.5cm][c]{2.0cm}\centering \ac{\textsf{Ae} (3/5)}\ \end{minipage} \\\begin{minipage}[c][0.5cm][c]{2.0cm}\centering \ac{\textsf{Aa} (1/5)}\ \end{minipage} \\\begin{minipage}[c][0.5cm][c]{2.0cm}\centering \blk{\textsf{Blk} (1/5)}\ \end{minipage} \\\end{tabular}& \begin{tabular}{@{}c@{}}\begin{minipage}[c][0.75cm][c]{2.0cm}\centering \ac{\textsf{Ae} (3/5)}\ \end{minipage} \\\begin{minipage}[c][0.75cm][c]{2.0cm}\centering \ac{\textsf{Aa} (2/5)}\ \end{minipage} \\\end{tabular}& \begin{tabular}{@{}c@{}}\begin{minipage}[c][0.75cm][c]{2.0cm}\centering \ac{\textsf{Ae} (4/5)}\ \end{minipage} \\\begin{minipage}[c][0.75cm][c]{2.0cm}\centering \ac{\textsf{Aa} (1/5)}\ \end{minipage} \\\end{tabular}& \begin{tabular}{@{}c@{}}\begin{minipage}[c][0.5cm][c]{2.0cm}\centering \ac{\textsf{Aa} (2/5)}\ \end{minipage} \\\begin{minipage}[c][0.5cm][c]{2.0cm}\centering \ac{\textsf{Ae} (2/5)}\ \end{minipage} \\\begin{minipage}[c][0.5cm][c]{2.0cm}\centering \pu{\textsf{PU$p_1$} (1/5)}\ \end{minipage} \\\end{tabular} \\ 
\hline 
5.0 & \blk{\textsf{Blk} (5/5)} & \pu{\textsf{PU$p_1$} (5/5)} & \pu{\textsf{PU$p_1$} (5/5)} & \pu{\textsf{PU$p_1$} (5/5)} & \pu{\textsf{PU$p_1$} (5/5)} & \pu{\textsf{PU$p_1$} (5/5)} & \pu{\textsf{PU$p_1$} (5/5)} & \pu{\textsf{PU$p_1$} (5/5)} & \pu{\textsf{PU$p_1$} (5/5)} & \pu{\textsf{PU$p_1$} (5/5)} & \pu{\textsf{PU$p_1$} (5/5)} & \pu{\textsf{PU$p_1$} (5/5)} & \pu{\textsf{PU$p_1$} (5/5)} & \pu{\textsf{PU$p_1$} (5/5)} & \pu{\textsf{PU$p_1$} (5/5)} \\ 
\hline
10.0 & \blk{\textsf{Blk} (5/5)} & \pu{\textsf{PU$p_1$} (5/5)} & \pu{\textsf{PU$p_1$} (5/5)} & \pu{\textsf{PU$p_1$} (5/5)} & \pu{\textsf{PU$p_1$} (5/5)} & \pu{\textsf{PU$p_1$} (5/5)} & \pu{\textsf{PU$p_1$} (5/5)} & \pu{\textsf{PU$p_1$} (5/5)} & \pu{\textsf{PU$p_1$} (5/5)} & \pu{\textsf{PU$p_1$} (5/5)} & \pu{\textsf{PU$p_1$} (5/5)} & \pu{\textsf{PU$p_1$} (5/5)} & \pu{\textsf{PU$p_1$} (5/5)} & \pu{\textsf{PU$p_1$} (5/5)} & \pu{\textsf{PU$p_1$} (5/5)} \\ 
\hline 
15.0 & \blk{\textsf{Blk} (5/5)} & \pu{\textsf{PU$p_1$} (5/5)} & \pu{\textsf{PU$p_1$} (5/5)} & \pu{\textsf{PU$p_1$} (5/5)} & \pu{\textsf{PU$p_1$} (5/5)} & \pu{\textsf{PU$p_1$} (5/5)} & \pu{\textsf{PU$p_1$} (5/5)} & \pu{\textsf{PU$p_1$} (5/5)} & \pu{\textsf{PU$p_1$} (5/5)} & \pu{\textsf{PU$p_1$} (5/5)} & \pu{\textsf{PU$p_1$} (5/5)} & \pu{\textsf{PU$p_1$} (5/5)} & \pu{\textsf{PU$p_1$} (5/5)} & \pu{\textsf{PU$p_1$} (5/5)} & \pu{\textsf{PU$p_1$} (5/5)} \\ 
\hline 
20.0 & \blk{\textsf{Blk} (5/5)} & \pu{\textsf{PU$p_1$} (5/5)} & \pu{\textsf{PU$p_1$} (5/5)} & \pu{\textsf{PU$p_1$} (5/5)} & \pu{\textsf{PU$p_1$} (5/5)} & \pu{\textsf{PU$p_1$} (5/5)} & \pu{\textsf{PU$p_1$} (5/5)} & \pu{\textsf{PU$p_1$} (5/5)} & \pu{\textsf{PU$p_1$} (5/5)} & \pu{\textsf{PU$p_1$} (5/5)}& \begin{tabular}{@{}c@{}}\begin{minipage}[c][0.5cm][c]{2.0cm}\centering \ps{\textsf{PS} (1/5)}\ \end{minipage} \\\begin{minipage}[c][0.5cm][c]{2.0cm}\centering \pu{\textsf{PU$p_1$} (4/5)}\ \end{minipage} \\\end{tabular}& \begin{tabular}{@{}c@{}}\begin{minipage}[c][0.5cm][c]{2.0cm}\centering \ps{\textsf{PS} (1/5)}\ \end{minipage} \\\begin{minipage}[c][0.5cm][c]{2.0cm}\centering \pu{\textsf{PU$p_1$} (4/5)}\ \end{minipage} \\\end{tabular} & \pu{\textsf{PU$p_1$} (5/5)} & \pu{\textsf{PU$p_1$} (5/5)} & \pu{\textsf{PU$p_1$} (5/5)} \\ 
\hline 
\ct{24.0} & \blk{\textsf{Blk} (5/5)} & \pu{\textsf{PU$p_1$} (5/5)}& \begin{tabular}{@{}c@{}}\begin{minipage}[c][0.5cm][c]{2.0cm}\centering \ps{\textsf{PS} (3/5)}\ \end{minipage} \\\begin{minipage}[c][0.5cm][c]{2.0cm}\centering \pu{\textsf{PU$p_1$} (2/5)}\ \end{minipage} \\\end{tabular} & \pu{\textsf{PU$p_1$} (5/5)}& \begin{tabular}{@{}c@{}}\begin{minipage}[c][0.5cm][c]{2.0cm}\centering \pu{\textsf{PU$p_1$} (2/5)}\ \end{minipage} \\\begin{minipage}[c][0.5cm][c]{2.0cm}\centering \ps{\textsf{PS} (3/5)}\ \end{minipage} \\\end{tabular}& \begin{tabular}{@{}c@{}}\begin{minipage}[c][0.5cm][c]{2.0cm}\centering \ps{\textsf{PS} (3/5)}\ \end{minipage} \\\begin{minipage}[c][0.5cm][c]{2.0cm}\centering \pu{\textsf{PU$p_1$} (2/5)}\ \end{minipage} \\\end{tabular} & \pu{\textsf{PU$p_1$} (5/5)} & \ps{\textsf{PS} (5/5)} & \ps{\textsf{PS} (5/5)} & \pu{\textsf{PU$p_1$} (5/5)}& \begin{tabular}{@{}c@{}}\begin{minipage}[c][0.5cm][c]{2.0cm}\centering \pu{\textsf{PU$p_1$} (4/5)}\ \end{minipage} \\\begin{minipage}[c][0.5cm][c]{2.0cm}\centering \ps{\textsf{PS} (1/5)}\ \end{minipage} \\\end{tabular} & \pu{\textsf{PU$p_1$} (5/5)} & \ps{\textsf{PS} (5/5)}& \begin{tabular}{@{}c@{}}\begin{minipage}[c][0.5cm][c]{2.0cm}\centering \ps{\textsf{PS} (2/5)}\ \end{minipage} \\\begin{minipage}[c][0.5cm][c]{2.0cm}\centering \pu{\textsf{PU$p_1$} (3/5)}\ \end{minipage} \\\end{tabular} & \pu{\textsf{PU$p_1$} (5/5)} \\ 
\hline 
25.0 & \blk{\textsf{Blk} (5/5)} & \pu{\textsf{PU$p_1$} (5/5)}& \begin{tabular}{@{}c@{}}\begin{minipage}[c][0.5cm][c]{2.0cm}\centering \ps{\textsf{PS} (3/5)}\ \end{minipage} \\\begin{minipage}[c][0.5cm][c]{2.0cm}\centering \pu{\textsf{PU$p_1$} (2/5)}\ \end{minipage} \\\end{tabular}& \begin{tabular}{@{}c@{}}\begin{minipage}[c][0.5cm][c]{2.0cm}\centering \pu{\textsf{PU$p_1$} (2/5)}\ \end{minipage} \\\begin{minipage}[c][0.5cm][c]{2.0cm}\centering \ps{\textsf{PS} (3/5)}\ \end{minipage} \\\end{tabular}& \begin{tabular}{@{}c@{}}\begin{minipage}[c][0.5cm][c]{2.0cm}\centering \ps{\textsf{PS} (4/5)}\ \end{minipage} \\\begin{minipage}[c][0.5cm][c]{2.0cm}\centering \pu{\textsf{PU$p_1$} (1/5)}\ \end{minipage} \\\end{tabular}& \begin{tabular}{@{}c@{}}\begin{minipage}[c][0.5cm][c]{2.0cm}\centering \pu{\textsf{PU$p_1$} (3/5)}\ \end{minipage} \\\begin{minipage}[c][0.5cm][c]{2.0cm}\centering \ps{\textsf{PS} (2/5)}\ \end{minipage} \\\end{tabular}& \begin{tabular}{@{}c@{}}\begin{minipage}[c][0.5cm][c]{2.0cm}\centering \pu{\textsf{PU$p_1$} (3/5)}\ \end{minipage} \\\begin{minipage}[c][0.5cm][c]{2.0cm}\centering \ps{\textsf{PS} (2/5)}\ \end{minipage} \\\end{tabular} & \pu{\textsf{PU$p_1$} (5/5)}& \begin{tabular}{@{}c@{}}\begin{minipage}[c][0.5cm][c]{2.0cm}\centering \pu{\textsf{PU$p_1$} (4/5)}\ \end{minipage} \\\begin{minipage}[c][0.5cm][c]{2.0cm}\centering \ps{\textsf{PS} (1/5)}\ \end{minipage} \\\end{tabular}& \begin{tabular}{@{}c@{}}\begin{minipage}[c][0.5cm][c]{2.0cm}\centering \ps{\textsf{PS} (3/5)}\ \end{minipage} \\\begin{minipage}[c][0.5cm][c]{2.0cm}\centering \pu{\textsf{PU$p_1$} (2/5)}\ \end{minipage} \\\end{tabular}& \begin{tabular}{@{}c@{}}\begin{minipage}[c][0.5cm][c]{2.0cm}\centering \pu{\textsf{PU$p_1$} (4/5)}\ \end{minipage} \\\begin{minipage}[c][0.5cm][c]{2.0cm}\centering \ps{\textsf{PS} (1/5)}\ \end{minipage} \\\end{tabular}& \begin{tabular}{@{}c@{}}\begin{minipage}[c][0.5cm][c]{2.0cm}\centering \pu{\textsf{PU$p_1$} (3/5)}\ \end{minipage} \\\begin{minipage}[c][0.5cm][c]{2.0cm}\centering \ps{\textsf{PS} (2/5)}\ \end{minipage} \\\end{tabular}& \begin{tabular}{@{}c@{}}\begin{minipage}[c][0.5cm][c]{2.0cm}\centering \pu{\textsf{PU$p_1$} (4/5)}\ \end{minipage} \\\begin{minipage}[c][0.5cm][c]{2.0cm}\centering \ps{\textsf{PS} (1/5)}\ \end{minipage} \\\end{tabular} & \pu{\textsf{PU$p_1$} (5/5)}& \begin{tabular}{@{}c@{}}\begin{minipage}[c][0.5cm][c]{2.0cm}\centering \ps{\textsf{PS} (3/5)}\ \end{minipage} \\\begin{minipage}[c][0.5cm][c]{2.0cm}\centering \pu{\textsf{PU$p_1$} (2/5)}\ \end{minipage} \\\end{tabular} \\ 
\hline 
30.0 & \blk{\textsf{Blk} (5/5)} & \ps{\textsf{PS} (5/5)} & \ps{\textsf{PS} (5/5)} & \ps{\textsf{PS} (5/5)} & \ps{\textsf{PS} (5/5)} & \ps{\textsf{PS} (5/5)} & \ps{\textsf{PS} (5/5)} & \ps{\textsf{PS} (5/5)} & \ps{\textsf{PS} (5/5)} & \ps{\textsf{PS} (5/5)} & \ps{\textsf{PS} (5/5)} & \ps{\textsf{PS} (5/5)} & \ps{\textsf{PS} (5/5)} & \ps{\textsf{PS} (5/5)} & \ps{\textsf{PS} (5/5)} \\ 
\hline 
35.0 & \blk{\textsf{Blk} (5/5)} & \ps{\textsf{PS} (5/5)} & \ps{\textsf{PS} (5/5)} & \ps{\textsf{PS} (5/5)} & \ps{\textsf{PS} (5/5)} & \ps{\textsf{PS} (5/5)} & \ps{\textsf{PS} (5/5)} & \ps{\textsf{PS} (5/5)} & \ps{\textsf{PS} (5/5)} & \ps{\textsf{PS} (5/5)} & \ps{\textsf{PS} (5/5)} & \ps{\textsf{PS} (5/5)} & \ps{\textsf{PS} (5/5)} & \ps{\textsf{PS} (5/5)} & \ps{\textsf{PS} (5/5)} \\ 
\hline 
40.0 & \blk{\textsf{Blk} (5/5)} & \ps{\textsf{PS} (5/5)} & \ps{\textsf{PS} (5/5)} & \ps{\textsf{PS} (5/5)} & \ps{\textsf{PS} (5/5)} & \ps{\textsf{PS} (5/5)} & \ps{\textsf{PS} (5/5)} & \ps{\textsf{PS} (5/5)} & \ps{\textsf{PS} (5/5)} & \ps{\textsf{PS} (5/5)} & \ps{\textsf{PS} (5/5)} & \ps{\textsf{PS} (5/5)} & \ps{\textsf{PS} (5/5)} & \ps{\textsf{PS} (5/5)} & \ps{\textsf{PS} (5/5)} \\ 
\hline 
\end{tabular} 

%% file: tables/interfuser-crossing-4.tex
\begin{tabular}{|c| *{15}{c|}} 
\hline 
\diagbox{$\bm{x_a}$}{$\bm{x_f}$} & 0.0 & \ct{2.3} & 40.0 & 80.0 & 120.0 & 160.0 & 200.0 & 240.0 & 280.0 & 320.0 & 360.0 & 400.0 & 440.0 & 480.0 & 520.0 \\ 
\hline 
0.0& \begin{tabular}{@{}c@{}}\begin{minipage}[c][1.0cm][c]{2.0cm}\centering \ac{\textsf{Ae} (1/5)}\ \end{minipage} \\\begin{minipage}[c][1.0cm][c]{2.0cm}\centering \blk{\textsf{Blk} (4/5)}\ \end{minipage} \\\end{tabular}& \begin{tabular}{@{}c@{}}\begin{minipage}[c][1.0cm][c]{2.0cm}\centering \ac{\textsf{Ae} (1/5)}\ \end{minipage} \\\begin{minipage}[c][1.0cm][c]{2.0cm}\centering \blk{\textsf{Blk} (4/5)}\ \end{minipage} \\\end{tabular}& \begin{tabular}{@{}c@{}}\begin{minipage}[c][0.6666666666666666cm][c]{2.0cm}\centering \pu{\textsf{PU$p_1$} (1/5)}\ \end{minipage} \\\begin{minipage}[c][0.6666666666666666cm][c]{2.0cm}\centering \blk{\textsf{Blk} (3/5)}\ \end{minipage} \\\begin{minipage}[c][0.6666666666666666cm][c]{2.0cm}\centering \cu{\textsf{CU$p_1p_2$} (1/5)}\ \end{minipage} \\\end{tabular}& \begin{tabular}{@{}c@{}}\begin{minipage}[c][1.0cm][c]{2.0cm}\centering \ac{\textsf{Ae} (1/5)}\ \end{minipage} \\\begin{minipage}[c][1.0cm][c]{2.0cm}\centering \blk{\textsf{Blk} (4/5)}\ \end{minipage} \\\end{tabular}& \begin{tabular}{@{}c@{}}\begin{minipage}[c][1.0cm][c]{2.0cm}\centering \blk{\textsf{Blk} (3/5)}\ \end{minipage} \\\begin{minipage}[c][1.0cm][c]{2.0cm}\centering \cu{\textsf{CU$p_1p_2$} (2/5)}\ \end{minipage} \\\end{tabular}& \begin{tabular}{@{}c@{}}\begin{minipage}[c][1.0cm][c]{2.0cm}\centering \pu{\textsf{PU$p_1$} (1/5)}\ \end{minipage} \\\begin{minipage}[c][1.0cm][c]{2.0cm}\centering \cu{\textsf{CU$p_1p_2$} (4/5)}\ \end{minipage} \\\end{tabular}& \begin{tabular}{@{}c@{}}\begin{minipage}[c][1.0cm][c]{2.0cm}\centering \ac{\textsf{Ae} (2/5)}\ \end{minipage} \\\begin{minipage}[c][1.0cm][c]{2.0cm}\centering \blk{\textsf{Blk} (3/5)}\ \end{minipage} \\\end{tabular}& \begin{tabular}{@{}c@{}}\begin{minipage}[c][0.5cm][c]{2.0cm}\centering \pu{\textsf{PU$p_1$} (1/5)}\ \end{minipage} \\\begin{minipage}[c][0.5cm][c]{2.0cm}\centering \blk{\textsf{Blk} (1/5)}\ \end{minipage} \\\begin{minipage}[c][0.5cm][c]{2.0cm}\centering \cu{\textsf{CU$p_1p_2$} (2/5)}\ \end{minipage} \\\begin{minipage}[c][0.5cm][c]{2.0cm}\centering \ac{\textsf{Ae} (1/5)}\ \end{minipage} \\\end{tabular}& \begin{tabular}{@{}c@{}}\begin{minipage}[c][1.0cm][c]{2.0cm}\centering \cu{\textsf{CU$p_1p_2$} (2/5)}\ \end{minipage} \\\begin{minipage}[c][1.0cm][c]{2.0cm}\centering \blk{\textsf{Blk} (3/5)}\ \end{minipage} \\\end{tabular}& \begin{tabular}{@{}c@{}}\begin{minipage}[c][1.0cm][c]{2.0cm}\centering \blk{\textsf{Blk} (2/5)}\ \end{minipage} \\\begin{minipage}[c][1.0cm][c]{2.0cm}\centering \cu{\textsf{CU$p_1p_2$} (3/5)}\ \end{minipage} \\\end{tabular}& \begin{tabular}{@{}c@{}}\begin{minipage}[c][1.0cm][c]{2.0cm}\centering \cu{\textsf{CU$p_1p_2$} (3/5)}\ \end{minipage} \\\begin{minipage}[c][1.0cm][c]{2.0cm}\centering \blk{\textsf{Blk} (2/5)}\ \end{minipage} \\\end{tabular}& \begin{tabular}{@{}c@{}}\begin{minipage}[c][1.0cm][c]{2.0cm}\centering \cu{\textsf{CU$p_1p_2$} (2/5)}\ \end{minipage} \\\begin{minipage}[c][1.0cm][c]{2.0cm}\centering \blk{\textsf{Blk} (3/5)}\ \end{minipage} \\\end{tabular}& \begin{tabular}{@{}c@{}}\begin{minipage}[c][1.0cm][c]{2.0cm}\centering \ac{\textsf{Aa} (1/5)}\ \end{minipage} \\\begin{minipage}[c][1.0cm][c]{2.0cm}\centering \cu{\textsf{CU$p_1p_2$} (4/5)}\ \end{minipage} \\\end{tabular}& \begin{tabular}{@{}c@{}}\begin{minipage}[c][1.0cm][c]{2.0cm}\centering \cu{\textsf{CU$p_1p_2$} (3/5)}\ \end{minipage} \\\begin{minipage}[c][1.0cm][c]{2.0cm}\centering \blk{\textsf{Blk} (2/5)}\ \end{minipage} \\\end{tabular}& \begin{tabular}{@{}c@{}}\begin{minipage}[c][1.0cm][c]{2.0cm}\centering \cu{\textsf{CU$p_1p_2$} (1/5)}\ \end{minipage} \\\begin{minipage}[c][1.0cm][c]{2.0cm}\centering \blk{\textsf{Blk} (4/5)}\ \end{minipage} \\\end{tabular} \\ 
\hline 
2.5 & \blk{\textsf{Blk} (5/5)} & \pu{\textsf{PU$p_1$} (5/5)} & \pu{\textsf{PU$p_1$} (5/5)} & \pu{\textsf{PU$p_1$} (5/5)} & \pu{\textsf{PU$p_1$} (5/5)} & \pu{\textsf{PU$p_1$} (5/5)} & \pu{\textsf{PU$p_1$} (5/5)} & \pu{\textsf{PU$p_1$} (5/5)} & \pu{\textsf{PU$p_1$} (5/5)} & \pu{\textsf{PU$p_1$} (5/5)} & \pu{\textsf{PU$p_1$} (5/5)} & \pu{\textsf{PU$p_1$} (5/5)} & \pu{\textsf{PU$p_1$} (5/5)} & \pu{\textsf{PU$p_1$} (5/5)} & \pu{\textsf{PU$p_1$} (5/5)} \\ 
\hline 
5.0 & \blk{\textsf{Blk} (5/5)} & \pu{\textsf{PU$p_1$} (5/5)} & \pu{\textsf{PU$p_1$} (5/5)} & \pu{\textsf{PU$p_1$} (5/5)} & \pu{\textsf{PU$p_1$} (5/5)} & \pu{\textsf{PU$p_1$} (5/5)} & \pu{\textsf{PU$p_1$} (5/5)} & \pu{\textsf{PU$p_1$} (5/5)} & \pu{\textsf{PU$p_1$} (5/5)} & \pu{\textsf{PU$p_1$} (5/5)} & \pu{\textsf{PU$p_1$} (5/5)} & \pu{\textsf{PU$p_1$} (5/5)} & \pu{\textsf{PU$p_1$} (5/5)} & \pu{\textsf{PU$p_1$} (5/5)} & \pu{\textsf{PU$p_1$} (5/5)} \\ 
\hline 
10.0 & \blk{\textsf{Blk} (5/5)} & \pu{\textsf{PU$p_1$} (5/5)} & \pu{\textsf{PU$p_1$} (5/5)} & \pu{\textsf{PU$p_1$} (5/5)} & \pu{\textsf{PU$p_1$} (5/5)} & \pu{\textsf{PU$p_1$} (5/5)} & \pu{\textsf{PU$p_1$} (5/5)} & \pu{\textsf{PU$p_1$} (5/5)} & \pu{\textsf{PU$p_1$} (5/5)} & \pu{\textsf{PU$p_1$} (5/5)} & \pu{\textsf{PU$p_1$} (5/5)} & \pu{\textsf{PU$p_1$} (5/5)} & \pu{\textsf{PU$p_1$} (5/5)} & \pu{\textsf{PU$p_1$} (5/5)} & \pu{\textsf{PU$p_1$} (5/5)} \\ 
\hline 
15.0 & \blk{\textsf{Blk} (5/5)} & \pu{\textsf{PU$p_1$} (5/5)} & \pu{\textsf{PU$p_1$} (5/5)} & \pu{\textsf{PU$p_1$} (5/5)} & \pu{\textsf{PU$p_1$} (5/5)} & \pu{\textsf{PU$p_1$} (5/5)} & \pu{\textsf{PU$p_1$} (5/5)} & \pu{\textsf{PU$p_1$} (5/5)} & \pu{\textsf{PU$p_1$} (5/5)} & \pu{\textsf{PU$p_1$} (5/5)} & \pu{\textsf{PU$p_1$} (5/5)} & \pu{\textsf{PU$p_1$} (5/5)} & \pu{\textsf{PU$p_1$} (5/5)} & \pu{\textsf{PU$p_1$} (5/5)} & \pu{\textsf{PU$p_1$} (5/5)} \\ 
\hline 
20.0 & \blk{\textsf{Blk} (5/5)} & \pu{\textsf{PU$p_1$} (5/5)}& \begin{tabular}{@{}c@{}}\begin{minipage}[c][0.5cm][c]{2.0cm}\centering \ps{\textsf{PS} (4/5)}\ \end{minipage} \\\begin{minipage}[c][0.5cm][c]{2.0cm}\centering \pu{\textsf{PU$p_1$} (1/5)}\ \end{minipage} \\\end{tabular} & \pu{\textsf{PU$p_1$} (5/5)}& \begin{tabular}{@{}c@{}}\begin{minipage}[c][0.5cm][c]{2.0cm}\centering \ps{\textsf{PS} (2/5)}\ \end{minipage} \\\begin{minipage}[c][0.5cm][c]{2.0cm}\centering \pu{\textsf{PU$p_1$} (3/5)}\ \end{minipage} \\\end{tabular}& \begin{tabular}{@{}c@{}}\begin{minipage}[c][0.5cm][c]{2.0cm}\centering \pu{\textsf{PU$p_1$} (3/5)}\ \end{minipage} \\\begin{minipage}[c][0.5cm][c]{2.0cm}\centering \ps{\textsf{PS} (2/5)}\ \end{minipage} \\\end{tabular}& \begin{tabular}{@{}c@{}}\begin{minipage}[c][0.5cm][c]{2.0cm}\centering \pu{\textsf{PU$p_1$} (4/5)}\ \end{minipage} \\\begin{minipage}[c][0.5cm][c]{2.0cm}\centering \ps{\textsf{PS} (1/5)}\ \end{minipage} \\\end{tabular}& \begin{tabular}{@{}c@{}}\begin{minipage}[c][0.5cm][c]{2.0cm}\centering \ps{\textsf{PS} (3/5)}\ \end{minipage} \\\begin{minipage}[c][0.5cm][c]{2.0cm}\centering \pu{\textsf{PU$p_1$} (2/5)}\ \end{minipage} \\\end{tabular}& \begin{tabular}{@{}c@{}}\begin{minipage}[c][0.5cm][c]{2.0cm}\centering \ps{\textsf{PS} (3/5)}\ \end{minipage} \\\begin{minipage}[c][0.5cm][c]{2.0cm}\centering \pu{\textsf{PU$p_1$} (2/5)}\ \end{minipage} \\\end{tabular}& \begin{tabular}{@{}c@{}}\begin{minipage}[c][0.5cm][c]{2.0cm}\centering \pu{\textsf{PU$p_1$} (4/5)}\ \end{minipage} \\\begin{minipage}[c][0.5cm][c]{2.0cm}\centering \ps{\textsf{PS} (1/5)}\ \end{minipage} \\\end{tabular}& \begin{tabular}{@{}c@{}}\begin{minipage}[c][0.5cm][c]{2.0cm}\centering \pu{\textsf{PU$p_1$} (1/5)}\ \end{minipage} \\\begin{minipage}[c][0.5cm][c]{2.0cm}\centering \ps{\textsf{PS} (4/5)}\ \end{minipage} \\\end{tabular}& \begin{tabular}{@{}c@{}}\begin{minipage}[c][0.5cm][c]{2.0cm}\centering \pu{\textsf{PU$p_1$} (2/5)}\ \end{minipage} \\\begin{minipage}[c][0.5cm][c]{2.0cm}\centering \ps{\textsf{PS} (3/5)}\ \end{minipage} \\\end{tabular}& \begin{tabular}{@{}c@{}}\begin{minipage}[c][0.5cm][c]{2.0cm}\centering \pu{\textsf{PU$p_1$} (3/5)}\ \end{minipage} \\\begin{minipage}[c][0.5cm][c]{2.0cm}\centering \ps{\textsf{PS} (2/5)}\ \end{minipage} \\\end{tabular}& \begin{tabular}{@{}c@{}}\begin{minipage}[c][0.5cm][c]{2.0cm}\centering \pu{\textsf{PU$p_1$} (4/5)}\ \end{minipage} \\\begin{minipage}[c][0.5cm][c]{2.0cm}\centering \ps{\textsf{PS} (1/5)}\ \end{minipage} \\\end{tabular}& \begin{tabular}{@{}c@{}}\begin{minipage}[c][0.5cm][c]{2.0cm}\centering \pu{\textsf{PU$p_1$} (3/5)}\ \end{minipage} \\\begin{minipage}[c][0.5cm][c]{2.0cm}\centering \ps{\textsf{PS} (2/5)}\ \end{minipage} \\\end{tabular} \\ 
\hline 
21.9 & \blk{\textsf{Blk} (5/5)} & \pu{\textsf{PU$p_1$} (5/5)}& \begin{tabular}{@{}c@{}}\begin{minipage}[c][0.5cm][c]{2.0cm}\centering \ps{\textsf{PS} (1/5)}\ \end{minipage} \\\begin{minipage}[c][0.5cm][c]{2.0cm}\centering \pu{\textsf{PU$p_1$} (4/5)}\ \end{minipage} \\\end{tabular} & \pu{\textsf{PU$p_1$} (5/5)} & \pu{\textsf{PU$p_1$} (5/5)} & \pu{\textsf{PU$p_1$} (5/5)} & \pu{\textsf{PU$p_1$} (5/5)} & \pu{\textsf{PU$p_1$} (5/5)} & \pu{\textsf{PU$p_1$} (5/5)} & \pu{\textsf{PU$p_1$} (5/5)} & \pu{\textsf{PU$p_1$} (5/5)} & \pu{\textsf{PU$p_1$} (5/5)} & \pu{\textsf{PU$p_1$} (5/5)} & \pu{\textsf{PU$p_1$} (5/5)} & \pu{\textsf{PU$p_1$} (5/5)} \\ 
\hline 
25.0 & \blk{\textsf{Blk} (5/5)} & \ps{\textsf{PS} (5/5)} & \ps{\textsf{PS} (5/5)}& \begin{tabular}{@{}c@{}}\begin{minipage}[c][0.5cm][c]{2.0cm}\centering \pu{\textsf{PU$p_1$} (1/5)}\ \end{minipage} \\\begin{minipage}[c][0.5cm][c]{2.0cm}\centering \ps{\textsf{PS} (4/5)}\ \end{minipage} \\\end{tabular} & \ps{\textsf{PS} (5/5)} & \ps{\textsf{PS} (5/5)}& \begin{tabular}{@{}c@{}}\begin{minipage}[c][0.5cm][c]{2.0cm}\centering \pu{\textsf{PU$p_1$} (1/5)}\ \end{minipage} \\\begin{minipage}[c][0.5cm][c]{2.0cm}\centering \ps{\textsf{PS} (4/5)}\ \end{minipage} \\\end{tabular} & \ps{\textsf{PS} (5/5)}& \begin{tabular}{@{}c@{}}\begin{minipage}[c][0.5cm][c]{2.0cm}\centering \pu{\textsf{PU$p_1$} (1/5)}\ \end{minipage} \\\begin{minipage}[c][0.5cm][c]{2.0cm}\centering \ps{\textsf{PS} (4/5)}\ \end{minipage} \\\end{tabular} & \ps{\textsf{PS} (5/5)} & \ps{\textsf{PS} (5/5)} & \ps{\textsf{PS} (5/5)}& \begin{tabular}{@{}c@{}}\begin{minipage}[c][0.5cm][c]{2.0cm}\centering \pu{\textsf{PU$p_1$} (1/5)}\ \end{minipage} \\\begin{minipage}[c][0.5cm][c]{2.0cm}\centering \ps{\textsf{PS} (4/5)}\ \end{minipage} \\\end{tabular} & \ps{\textsf{PS} (5/5)} & \ps{\textsf{PS} (5/5)} \\ 
\hline 
30.0 & \blk{\textsf{Blk} (5/5)} & \ps{\textsf{PS} (5/5)} & \ps{\textsf{PS} (5/5)} & \ps{\textsf{PS} (5/5)} & \ps{\textsf{PS} (5/5)} & \ps{\textsf{PS} (5/5)} & \ps{\textsf{PS} (5/5)} & \ps{\textsf{PS} (5/5)} & \ps{\textsf{PS} (5/5)} & \ps{\textsf{PS} (5/5)} & \ps{\textsf{PS} (5/5)} & \ps{\textsf{PS} (5/5)} & \ps{\textsf{PS} (5/5)} & \ps{\textsf{PS} (5/5)} & \ps{\textsf{PS} (5/5)} \\ 
\hline 
35.0 & \blk{\textsf{Blk} (5/5)} & \ps{\textsf{PS} (5/5)} & \ps{\textsf{PS} (5/5)} & \ps{\textsf{PS} (5/5)} & \ps{\textsf{PS} (5/5)} & \ps{\textsf{PS} (5/5)} & \ps{\textsf{PS} (5/5)} & \ps{\textsf{PS} (5/5)} & \ps{\textsf{PS} (5/5)} & \ps{\textsf{PS} (5/5)} & \ps{\textsf{PS} (5/5)} & \ps{\textsf{PS} (5/5)} & \ps{\textsf{PS} (5/5)} & \ps{\textsf{PS} (5/5)} & \ps{\textsf{PS} (5/5)} \\ 
\hline 
40.0 & \blk{\textsf{Blk} (5/5)} & \ps{\textsf{PS} (5/5)} & \ps{\textsf{PS} (5/5)} & \ps{\textsf{PS} (5/5)} & \ps{\textsf{PS} (5/5)} & \ps{\textsf{PS} (5/5)} & \ps{\textsf{PS} (5/5)} & \ps{\textsf{PS} (5/5)} & \ps{\textsf{PS} (5/5)} & \ps{\textsf{PS} (5/5)} & \ps{\textsf{PS} (5/5)} & \ps{\textsf{PS} (5/5)} & \ps{\textsf{PS} (5/5)} & \ps{\textsf{PS} (5/5)} & \ps{\textsf{PS} (5/5)} \\ 
\hline 
\end{tabular} 

%% file: tables/interfuser-traffic_light-0.tex
\begin{tabular}{|c|c|c|c|c|c|c|c|c|c|c|c|c|c|c|}
\hline
\multicolumn{15}{|c|}{$\bm{xf}$}\\
\hline
0.0 & \ct{2.8} & 40.0 & 80.0 & 120.0 & 160.0 & 200.0 & 240.0 & 280.0 & 320.0 & 360.0 & 400.0 & 440.0 & 480.0 & 520.0 \\
\hline
\pu{\textsf{PU$p_4$ (5/5)}} & \pu{\textsf{PU$p_4$ (5/5)}} & \ps{\textsf{PS (5/5)}} & \ps{\textsf{PS (5/5)}} & \ps{\textsf{PS (5/5)}} & \ps{\textsf{PS (5/5)}} & \ps{\textsf{PS (5/5)}} & \ps{\textsf{PS (5/5)}} & \ps{\textsf{PS (5/5)}} & \ps{\textsf{PS (5/5)}} & \ps{\textsf{PS (5/5)}} & \ps{\textsf{PS (5/5)}} & \ps{\textsf{PS (5/5)}} & \ps{\textsf{PS (5/5)}} & \ps{\textsf{PS (5/5)}} \\
\hline
\end{tabular}

%% file: tables/interfuser-traffic_light-4.tex
\begin{tabular}{|c|c|c|c|c|c|c|c|c|c|c|c|c|c|c|}
\hline
\multicolumn{15}{|c|}{$\bm{xf}$}\\
\hline
0.0 & \ct{2.8} & 40.0 & 80.0 & 120.0 & 160.0 & 200.0 & 240.0 & 280.0 & 320.0 & 360.0 & 400.0 & 440.0 & 480.0 & 520.0 \\
\hline
\ps{\textsf{PS (5/5)}} & \ps{\textsf{PS (5/5)}} & \ps{\textsf{PS (5/5)}} & \ps{\textsf{PS (5/5)}} & \ps{\textsf{PS (5/5)}} & \ps{\textsf{PS (5/5)}} & \ps{\textsf{PS (5/5)}} & \ps{\textsf{PS (5/5)}} & \ps{\textsf{PS (5/5)}} & \ps{\textsf{PS (5/5)}} & \ps{\textsf{PS (5/5)}} & \ps{\textsf{PS (5/5)}} & \ps{\textsf{PS (5/5)}} & \ps{\textsf{PS (5/5)}} & \ps{\textsf{PS (5/5)}} \\
\hline
\end{tabular}

%% file: tables/mile-merging-0.tex
\begin{tabular}{|c| *{14}{c|}} 
\hline 
\diagbox{$\bm{x_a}$}{$\bm{x_f}$} & \ct{0.0} & 40.0 & 80.0 & 120.0 & 160.0 & 200.0 & 240.0 & 280.0 & 320.0 & 360.0 & 400.0 & 440.0 & 480.0 & 520.0 \\ 
\hline 
0.0 & - & \cs{\textsf{CS} (5/5)} & \cs{\textsf{CS} (5/5)} & \cs{\textsf{CS} (5/5)} & \cs{\textsf{CS} (5/5)} & \cs{\textsf{CS} (5/5)} & \cs{\textsf{CS} (5/5)} & \cs{\textsf{CS} (5/5)} & \cs{\textsf{CS} (5/5)} & \cs{\textsf{CS} (5/5)} & \cs{\textsf{CS} (5/5)} & \cs{\textsf{CS} (5/5)} & \cs{\textsf{CS} (5/5)} & \cs{\textsf{CS} (5/5)} \\ 
\hline 
\ct{3.6} & \cs{\textsf{CS} (5/5)} & \cs{\textsf{CS} (5/5)} & \cs{\textsf{CS} (5/5)} & \cs{\textsf{CS} (5/5)} & \cs{\textsf{CS} (5/5)} & \cs{\textsf{CS} (5/5)} & \cs{\textsf{CS} (5/5)} & \cs{\textsf{CS} (5/5)} & \cs{\textsf{CS} (5/5)} & \cs{\textsf{CS} (5/5)} & \cs{\textsf{CS} (5/5)} & \cs{\textsf{CS} (5/5)} & \cs{\textsf{CS} (5/5)} & \cs{\textsf{CS} (5/5)} \\ 
\hline 
5.0 & \cs{\textsf{CS} (5/5)} & \cs{\textsf{CS} (5/5)} & \cs{\textsf{CS} (5/5)} & \cs{\textsf{CS} (5/5)} & \cs{\textsf{CS} (5/5)} & \cs{\textsf{CS} (5/5)} & \cs{\textsf{CS} (5/5)} & \cs{\textsf{CS} (5/5)} & \cs{\textsf{CS} (5/5)} & \cs{\textsf{CS} (5/5)} & \cs{\textsf{CS} (5/5)} & \cs{\textsf{CS} (5/5)} & \cs{\textsf{CS} (5/5)} & \cs{\textsf{CS} (5/5)} \\ 
\hline
7.5 & \cs{\textsf{CS} (5/5)} & \cs{\textsf{CS} (5/5)} & \cs{\textsf{CS} (5/5)} & \cs{\textsf{CS} (5/5)} & \cs{\textsf{CS} (5/5)} & \cs{\textsf{CS} (5/5)} & \cs{\textsf{CS} (5/5)} & \cs{\textsf{CS} (5/5)} & \cs{\textsf{CS} (5/5)} & \cs{\textsf{CS} (5/5)} & \cs{\textsf{CS} (5/5)} & \cs{\textsf{CS} (5/5)} & \cs{\textsf{CS} (5/5)} & \cs{\textsf{CS} (5/5)} \\ 
\hline 
10.0 & \cs{\textsf{CS} (5/5)} & \ac{\textsf{Ae} (5/5)} & \ac{\textsf{Ae} (5/5)} & \ac{\textsf{Ae} (5/5)} & \ac{\textsf{Ae} (5/5)} & \ac{\textsf{Ae} (5/5)} & \ac{\textsf{Ae} (5/5)} & \ac{\textsf{Ae} (5/5)} & \ac{\textsf{Ae} (5/5)} & \ac{\textsf{Ae} (5/5)} & \ac{\textsf{Ae} (5/5)} & \ac{\textsf{Ae} (5/5)} & \ac{\textsf{Ae} (5/5)} & \ac{\textsf{Ae} (5/5)} \\ 
\hline 
12.5 & \cs{\textsf{CS} (5/5)} & \ps{\textsf{PS} (5/5)} & \ps{\textsf{PS} (5/5)} & \ps{\textsf{PS} (5/5)} & \ps{\textsf{PS} (5/5)} & \ps{\textsf{PS} (5/5)} & \ps{\textsf{PS} (5/5)} & \ps{\textsf{PS} (5/5)} & \ps{\textsf{PS} (5/5)} & \ps{\textsf{PS} (5/5)} & \ps{\textsf{PS} (5/5)} & \ps{\textsf{PS} (5/5)} & \ps{\textsf{PS} (5/5)} & \ps{\textsf{PS} (5/5)} \\ 
\hline
15.0 & \cs{\textsf{CS} (5/5)} & \ps{\textsf{PS} (5/5)} & \ps{\textsf{PS} (5/5)} & \ps{\textsf{PS} (5/5)} & \ps{\textsf{PS} (5/5)} & \ps{\textsf{PS} (5/5)} & \ps{\textsf{PS} (5/5)} & \ps{\textsf{PS} (5/5)} & \ps{\textsf{PS} (5/5)} & \ps{\textsf{PS} (5/5)} & \ps{\textsf{PS} (5/5)} & \ps{\textsf{PS} (5/5)} & \ps{\textsf{PS} (5/5)} & \ps{\textsf{PS} (5/5)} \\ 
\hline 
20.0 & \cs{\textsf{CS} (5/5)} & \ps{\textsf{PS} (5/5)} & \ps{\textsf{PS} (5/5)} & \ps{\textsf{PS} (5/5)} & \ps{\textsf{PS} (5/5)} & \ps{\textsf{PS} (5/5)} & \ps{\textsf{PS} (5/5)} & \ps{\textsf{PS} (5/5)} & \ps{\textsf{PS} (5/5)} & \ps{\textsf{PS} (5/5)} & \ps{\textsf{PS} (5/5)} & \ps{\textsf{PS} (5/5)} & \ps{\textsf{PS} (5/5)} & \ps{\textsf{PS} (5/5)} \\ 
\hline 
22.9 & \cs{\textsf{CS} (5/5)} & \ps{\textsf{PS} (5/5)} & \ps{\textsf{PS} (5/5)} & \ps{\textsf{PS} (5/5)} & \ps{\textsf{PS} (5/5)} & \ps{\textsf{PS} (5/5)} & \ps{\textsf{PS} (5/5)} & \ps{\textsf{PS} (5/5)} & \ps{\textsf{PS} (5/5)} & \ps{\textsf{PS} (5/5)} & \ps{\textsf{PS} (5/5)} & \ps{\textsf{PS} (5/5)} & \ps{\textsf{PS} (5/5)} & \ps{\textsf{PS} (5/5)} \\ 
\hline 
\end{tabular} 

%% file: tables/mile-merging-4.tex
\begin{tabular}{|c| *{15}{c|}} 
\hline 
\diagbox{$\bm{x_a}$}{$\bm{x_f}$} & 0.0 & \ct{2.6} & 40.0 & 80.0 & 120.0 & 160.0 & 200.0 & 240.0 & 280.0 & 320.0 & 360.0 & 400.0 & 440.0 & 480.0 & 520.0 \\ 
\hline 
0.0 & - & - & \cs{\textsf{CS} (5/5)} & \cs{\textsf{CS} (5/5)} & \cs{\textsf{CS} (5/5)} & \cs{\textsf{CS} (5/5)} & \cs{\textsf{CS} (5/5)} & \cs{\textsf{CS} (5/5)} & \cs{\textsf{CS} (5/5)} & \cs{\textsf{CS} (5/5)} & \cs{\textsf{CS} (5/5)} & \cs{\textsf{CS} (5/5)} & \cs{\textsf{CS} (5/5)} & \cs{\textsf{CS} (5/5)} & \cs{\textsf{CS} (5/5)} \\ 
\hline 
5.0 & \cs{\textsf{CS} (5/5)} & \cs{\textsf{CS} (5/5)} & \cs{\textsf{CS} (5/5)} & \cs{\textsf{CS} (5/5)} & \cs{\textsf{CS} (5/5)} & \cs{\textsf{CS} (5/5)} & \cs{\textsf{CS} (5/5)} & \cs{\textsf{CS} (5/5)} & \cs{\textsf{CS} (5/5)} & \cs{\textsf{CS} (5/5)} & \cs{\textsf{CS} (5/5)} & \cs{\textsf{CS} (5/5)} & \cs{\textsf{CS} (5/5)} & \cs{\textsf{CS} (5/5)} & \cs{\textsf{CS} (5/5)} \\ 
\hline 
\ct{6.7} & \cs{\textsf{CS} (5/5)} & \cs{\textsf{CS} (5/5)} & \cs{\textsf{CS} (5/5)} & \cs{\textsf{CS} (5/5)} & \cs{\textsf{CS} (5/5)} & \cs{\textsf{CS} (5/5)} & \cs{\textsf{CS} (5/5)} & \cs{\textsf{CS} (5/5)} & \cs{\textsf{CS} (5/5)} & \cs{\textsf{CS} (5/5)} & \cs{\textsf{CS} (5/5)} & \cs{\textsf{CS} (5/5)} & \cs{\textsf{CS} (5/5)} & \cs{\textsf{CS} (5/5)} & \cs{\textsf{CS} (5/5)} \\ 
\hline 
7.5 & \cs{\textsf{CS} (5/5)} & \cs{\textsf{CS} (5/5)} & \ac{\textsf{Ae} (5/5)} & \ac{\textsf{Ae} (5/5)} & \ac{\textsf{Ae} (5/5)} & \ac{\textsf{Ae} (5/5)} & \ac{\textsf{Ae} (5/5)} & \ac{\textsf{Ae} (5/5)} & \ac{\textsf{Ae} (5/5)} & \ac{\textsf{Ae} (5/5)} & \ac{\textsf{Ae} (5/5)} & \ac{\textsf{Ae} (5/5)} & \ac{\textsf{Ae} (5/5)} & \ac{\textsf{Ae} (5/5)} & \ac{\textsf{Ae} (5/5)} \\ 
\hline 
10.0 & \cs{\textsf{CS} (5/5)} & \cs{\textsf{CS} (5/5)} & \ps{\textsf{PS} (5/5)} & \ps{\textsf{PS} (5/5)} & \ps{\textsf{PS} (5/5)} & \ps{\textsf{PS} (5/5)} & \ps{\textsf{PS} (5/5)} & \ps{\textsf{PS} (5/5)} & \ps{\textsf{PS} (5/5)} & \ps{\textsf{PS} (5/5)} & \ps{\textsf{PS} (5/5)} & \ps{\textsf{PS} (5/5)} & \ps{\textsf{PS} (5/5)} & \ps{\textsf{PS} (5/5)} & \ps{\textsf{PS} (5/5)} \\ 
\hline 
20.0 & \cs{\textsf{CS} (5/5)} & \cs{\textsf{CS} (5/5)} & \ps{\textsf{PS} (5/5)} & \ps{\textsf{PS} (5/5)} & \ps{\textsf{PS} (5/5)} & \ps{\textsf{PS} (5/5)} & \ps{\textsf{PS} (5/5)} & \ps{\textsf{PS} (5/5)} & \ps{\textsf{PS} (5/5)} & \ps{\textsf{PS} (5/5)} & \ps{\textsf{PS} (5/5)} & \ps{\textsf{PS} (5/5)} & \ps{\textsf{PS} (5/5)} & \ps{\textsf{PS} (5/5)} & \ps{\textsf{PS} (5/5)} \\ 
\hline 
25.0 & \cs{\textsf{CS} (5/5)} & \cs{\textsf{CS} (5/5)} & \ps{\textsf{PS} (5/5)} & \ps{\textsf{PS} (5/5)} & \ps{\textsf{PS} (5/5)} & \ps{\textsf{PS} (5/5)} & \ps{\textsf{PS} (5/5)} & \ps{\textsf{PS} (5/5)} & \ps{\textsf{PS} (5/5)} & \ps{\textsf{PS} (5/5)} & \ps{\textsf{PS} (5/5)} & \ps{\textsf{PS} (5/5)} & \ps{\textsf{PS} (5/5)} & \ps{\textsf{PS} (5/5)} & \ps{\textsf{PS} (5/5)} \\ 
\hline 
30.0 & \cs{\textsf{CS} (5/5)} & \cs{\textsf{CS} (5/5)} & \ps{\textsf{PS} (5/5)} & \ps{\textsf{PS} (5/5)} & \ps{\textsf{PS} (5/5)} & \ps{\textsf{PS} (5/5)} & \ps{\textsf{PS} (5/5)} & \ps{\textsf{PS} (5/5)} & \ps{\textsf{PS} (5/5)} & \ps{\textsf{PS} (5/5)} & \ps{\textsf{PS} (5/5)} & \ps{\textsf{PS} (5/5)} & \ps{\textsf{PS} (5/5)} & \ps{\textsf{PS} (5/5)} & \ps{\textsf{PS} (5/5)} \\ 
\hline 
35.0 & \cs{\textsf{CS} (5/5)} & \cs{\textsf{CS} (5/5)} & \ps{\textsf{PS} (5/5)} & \ps{\textsf{PS} (5/5)} & \ps{\textsf{PS} (5/5)} & \ps{\textsf{PS} (5/5)} & \ps{\textsf{PS} (5/5)} & \ps{\textsf{PS} (5/5)} & \ps{\textsf{PS} (5/5)} & \ps{\textsf{PS} (5/5)} & \ps{\textsf{PS} (5/5)} & \ps{\textsf{PS} (5/5)} & \ps{\textsf{PS} (5/5)} & \ps{\textsf{PS} (5/5)} & \ps{\textsf{PS} (5/5)} \\ 
\hline 
40.0 & \cs{\textsf{CS} (5/5)} & \cs{\textsf{CS} (5/5)} & \ps{\textsf{PS} (5/5)} & \ps{\textsf{PS} (5/5)} & \ps{\textsf{PS} (5/5)} & \ps{\textsf{PS} (5/5)} & \ps{\textsf{PS} (5/5)} & \ps{\textsf{PS} (5/5)} & \ps{\textsf{PS} (5/5)} & \ps{\textsf{PS} (5/5)} & \ps{\textsf{PS} (5/5)} & \ps{\textsf{PS} (5/5)} & \ps{\textsf{PS} (5/5)} & \ps{\textsf{PS} (5/5)} & \ps{\textsf{PS} (5/5)} \\ 
\hline 
\end{tabular} 

%% file: tables/mile-lane_change-4.tex
\begin{tabular}{|c| *{15}{c|}}
\hline
\diagbox{$\bm{x_a}$}{$\bm{x_f}$} & 0.0 & \ct{2.4} & 40.0 & 80.0 & 120.0 & 160.0 & 200.0 & 240.0 & 280.0 & 320.0 & 360.0 & 400.0 & 440.0 & 480.0 & 520.0 \\
\hline
0.0 & - & \cs{\textsf{CS}} (5/5) & \cs{\textsf{CS}} (5/5) & \cs{\textsf{CS}} (5/5) & \cs{\textsf{CS}} (5/5) & \cs{\textsf{CS}} (5/5) & \cs{\textsf{CS}} (5/5) & \cs{\textsf{CS}} (5/5) & \cs{\textsf{CS}} (5/5) & \cs{\textsf{CS}} (5/5) & \cs{\textsf{CS}} (5/5) & \cs{\textsf{CS}} (5/5) & \cs{\textsf{CS}} (5/5) & \cs{\textsf{CS}} (5/5) & \cs{\textsf{CS}} (5/5) \\
\hline
\ct{25.6} & \cs{\textsf{CS}} (5/5) & \cs{\textsf{CS}} (5/5) & \cs{\textsf{CS}} (5/5) & \cs{\textsf{CS}} (5/5) & \cs{\textsf{CS}} (5/5) & \cs{\textsf{CS}} (5/5) & \cs{\textsf{CS}} (5/5) & \cs{\textsf{CS}} (5/5) & \cs{\textsf{CS}} (5/5) & \cs{\textsf{CS}} (5/5) & \cs{\textsf{CS}} (5/5) & \cs{\textsf{CS}} (5/5) & \cs{\textsf{CS}} (5/5) & \cs{\textsf{CS}} (5/5) & \cs{\textsf{CS}} (5/5)  \\
\hline
40.0 & \cs{\textsf{CS}} (5/5) & \cs{\textsf{CS}} (5/5) & \co{\textsf{CO}} (5/5) & \co{\textsf{CO}} (5/5) & \co{\textsf{CO}} (5/5) & \co{\textsf{CO}} (5/5) & \co{\textsf{CO}} (5/5) & \co{\textsf{CO}} (5/5) & \co{\textsf{CO}} (5/5) & \co{\textsf{CO}} (5/5) & \co{\textsf{CO}} (5/5) & \co{\textsf{CO}} (5/5) & \co{\textsf{CO}} (5/5) & \co{\textsf{CO}} (5/5) & \co{\textsf{CO}} (5/5) \\
\hline
80.0 & \cs{\textsf{CS}} (5/5) & \cs{\textsf{CS}} (5/5) & \co{\textsf{CO}} (5/5) & \co{\textsf{CO}} (5/5) & \co{\textsf{CO}} (5/5) & \co{\textsf{CO}} (5/5) & \co{\textsf{CO}} (5/5) & \co{\textsf{CO}} (5/5) & \co{\textsf{CO}} (5/5) & \co{\textsf{CO}} (5/5) & \co{\textsf{CO}} (5/5) & \co{\textsf{CO}} (5/5) & \co{\textsf{CO}} (5/5) & \co{\textsf{CO}} (5/5) & \co{\textsf{CO}} (5/5) \\
\hline
120.0 & \cs{\textsf{CS}} (5/5) & \cs{\textsf{CS}} (5/5) & \co{\textsf{CO}} (5/5) & \co{\textsf{CO}} (5/5) & \co{\textsf{CO}} (5/5) & \co{\textsf{CO}} (5/5) & \co{\textsf{CO}} (5/5) & \co{\textsf{CO}} (5/5) & \co{\textsf{CO}} (5/5) & \co{\textsf{CO}} (5/5) & \co{\textsf{CO}} (5/5) & \co{\textsf{CO}} (5/5) & \co{\textsf{CO}} (5/5) & \co{\textsf{CO}} (5/5) & \co{\textsf{CO}} (5/5) \\
\hline
160.0 & \cs{\textsf{CS}} (5/5) & \cs{\textsf{CS}} (5/5) & \co{\textsf{CO}} (5/5) & \co{\textsf{CO}} (5/5) & \co{\textsf{CO}} (5/5) & \co{\textsf{CO}} (5/5) & \co{\textsf{CO}} (5/5) & \co{\textsf{CO}} (5/5) & \co{\textsf{CO}} (5/5) & \co{\textsf{CO}} (5/5) & \co{\textsf{CO}} (5/5) & \co{\textsf{CO}} (5/5) & \co{\textsf{CO}} (5/5) & \co{\textsf{CO}} (5/5) & \co{\textsf{CO}} (5/5) \\
\hline
200.0 & \cs{\textsf{CS}} (5/5) & \cs{\textsf{CS}} (5/5) & \co{\textsf{CO}} (5/5) & \co{\textsf{CO}} (5/5) & \co{\textsf{CO}} (5/5) & \co{\textsf{CO}} (5/5) & \co{\textsf{CO}} (5/5) & \co{\textsf{CO}} (5/5) & \co{\textsf{CO}} (5/5) & \co{\textsf{CO}} (5/5) & \co{\textsf{CO}} (5/5) & \co{\textsf{CO}} (5/5) & \co{\textsf{CO}} (5/5) & \co{\textsf{CO}} (5/5) & \co{\textsf{CO}} (5/5) \\
\hline
240.0 & \cs{\textsf{CS}} (5/5) & \cs{\textsf{CS}} (5/5) & \co{\textsf{CO}} (5/5) & \co{\textsf{CO}} (5/5) & \co{\textsf{CO}} (5/5) & \co{\textsf{CO}} (5/5) & \co{\textsf{CO}} (5/5) & \co{\textsf{CO}} (5/5) & \co{\textsf{CO}} (5/5) & \co{\textsf{CO}} (5/5) & \co{\textsf{CO}} (5/5) & \co{\textsf{CO}} (5/5) & \co{\textsf{CO}} (5/5) & \co{\textsf{CO}} (5/5) & \co{\textsf{CO}} (5/5) \\
\hline
280.0 & \cs{\textsf{CS}} (5/5) & \cs{\textsf{CS}} (5/5) & \co{\textsf{CO}} (5/5) & \co{\textsf{CO}} (5/5) & \co{\textsf{CO}} (5/5) & \co{\textsf{CO}} (5/5) & \co{\textsf{CO}} (5/5) & \co{\textsf{CO}} (5/5) & \co{\textsf{CO}} (5/5) & \co{\textsf{CO}} (5/5) & \co{\textsf{CO}} (5/5) & \co{\textsf{CO}} (5/5) & \co{\textsf{CO}} (5/5) & \co{\textsf{CO}} (5/5) & \co{\textsf{CO}} (5/5) \\
\hline
320.0 & \cs{\textsf{CS}} (5/5) & \cs{\textsf{CS}} (5/5) & \co{\textsf{CO}} (5/5) & \co{\textsf{CO}} (5/5) & \co{\textsf{CO}} (5/5) & \co{\textsf{CO}} (5/5) & \co{\textsf{CO}} (5/5) & \co{\textsf{CO}} (5/5) & \co{\textsf{CO}} (5/5) & \co{\textsf{CO}} (5/5) & \co{\textsf{CO}} (5/5) & \co{\textsf{CO}} (5/5) & \co{\textsf{CO}} (5/5) & \co{\textsf{CO}} (5/5) & \co{\textsf{CO}} (5/5) \\
\hline
360.0 & \cs{\textsf{CS}} (5/5) & \cs{\textsf{CS}} (5/5) & \co{\textsf{CO}} (5/5) & \co{\textsf{CO}} (5/5) & \co{\textsf{CO}} (5/5) & \co{\textsf{CO}} (5/5) & \co{\textsf{CO}} (5/5) & \co{\textsf{CO}} (5/5) & \co{\textsf{CO}} (5/5) & \co{\textsf{CO}} (5/5) & \co{\textsf{CO}} (5/5) & \co{\textsf{CO}} (5/5) & \co{\textsf{CO}} (5/5) & \co{\textsf{CO}} (5/5) & \co{\textsf{CO}} (5/5) \\
\hline
400.0 & \cs{\textsf{CS}} (5/5) & \cs{\textsf{CS}} (5/5) & \co{\textsf{CO}} (5/5) & \co{\textsf{CO}} (5/5) & \co{\textsf{CO}} (5/5) & \co{\textsf{CO}} (5/5) & \co{\textsf{CO}} (5/5) & \co{\textsf{CO}} (5/5) & \co{\textsf{CO}} (5/5) & \co{\textsf{CO}} (5/5) & \co{\textsf{CO}} (5/5) & \co{\textsf{CO}} (5/5) & \co{\textsf{CO}} (5/5) & \co{\textsf{CO}} (5/5) & \co{\textsf{CO}} (5/5) \\
\hline
440.0 & \cs{\textsf{CS}} (5/5) & \cs{\textsf{CS}} (5/5) & \co{\textsf{CO}} (5/5) & \co{\textsf{CO}} (5/5) & \co{\textsf{CO}} (5/5) & \co{\textsf{CO}} (5/5) & \co{\textsf{CO}} (5/5) & \co{\textsf{CO}} (5/5) & \co{\textsf{CO}} (5/5) & \co{\textsf{CO}} (5/5) & \co{\textsf{CO}} (5/5) & \co{\textsf{CO}} (5/5) & \co{\textsf{CO}} (5/5) & \co{\textsf{CO}} (5/5) & \co{\textsf{CO}} (5/5) \\
\hline
480.0 & \cs{\textsf{CS}} (5/5) & \cs{\textsf{CS}} (5/5) & \co{\textsf{CO}} (5/5) & \co{\textsf{CO}} (5/5) & \co{\textsf{CO}} (5/5) & \co{\textsf{CO}} (5/5) & \co{\textsf{CO}} (5/5) & \co{\textsf{CO}} (5/5) & \co{\textsf{CO}} (5/5) & \co{\textsf{CO}} (5/5) & \co{\textsf{CO}} (5/5) & \co{\textsf{CO}} (5/5) & \co{\textsf{CO}} (5/5) & \co{\textsf{CO}} (5/5) & \co{\textsf{CO}} (5/5) \\
\hline
520.0 & \cs{\textsf{CS}} (5/5) & \cs{\textsf{CS}} (5/5) & \co{\textsf{CO}} (5/5) & \co{\textsf{CO}} (5/5) & \co{\textsf{CO}} (5/5) & \co{\textsf{CO}} (5/5) & \co{\textsf{CO}} (5/5) & \co{\textsf{CO}} (5/5) & \co{\textsf{CO}} (5/5) & \co{\textsf{CO}} (5/5) & \co{\textsf{CO}} (5/5) & \co{\textsf{CO}} (5/5) & \co{\textsf{CO}} (5/5) & \co{\textsf{CO}} (5/5) & \co{\textsf{CO}} (5/5) \\
\hline
\end{tabular}

%% file: tables/mile-crossing-0.tex
\begin{tabular}{|c| *{15}{c|}}
\hline
\diagbox{$\bm{x_a}$}{$\bm{x_f}$} & 0.0 & \ct{3.6} & 40.0 & 80.0 & 120.0 & 160.0 & 200.0 & 240.0 & 280.0 & 320.0 & 360.0 & 400.0 & 440.0 & 480.0 & 520.0 \\
\hline
0.0 & \begin{tabular}{@{}c@{}}\cu{\textsf{CU}}$p_1$ (1/5)\\ \cu{\textsf{CU}}$p_1p_2$ (4/5)\end{tabular} & \begin{tabular}{@{}c@{}}\cu{\textsf{CU}}$p_1$ (1/5)\\ \cu{\textsf{CU}}$p_1p_2$ (4/5)\end{tabular} & \cu{\textsf{CU}}$p_1p_2$ (5/5) & \begin{tabular}{@{}c@{}}\cu{\textsf{CU}}$p_1$ (1/5)\\ \cu{\textsf{CU}}$p_1p_2$ (4/5)\end{tabular} & \begin{tabular}{@{}c@{}}\cu{\textsf{CU}}$p_1$ (1/5)\\ \cu{\textsf{CU}}$p_1p_2$ (4/5)\end{tabular} & \cu{\textsf{CU}}$p_1p_2$ (5/5) & \begin{tabular}{@{}c@{}}\cu{\textsf{CU}}$p_1$ (1/5)\\ \cu{\textsf{CU}}$p_1p_2$ (4/5)\end{tabular} & \cu{\textsf{CU}}$p_1p_2$ (5/5) & \cu{\textsf{CU}}$p_1p_2$ (5/5) & \cu{\textsf{CU}}$p_1p_2$ (5/5) & \cu{\textsf{CU}}$p_1p_2$ (5/5) & \cu{\textsf{CU}}$p_1p_2$ (5/5) & \cu{\textsf{CU}}$p_1p_2$ (5/5) & \cu{\textsf{CU}}$p_1p_2$ (5/5) & \cu{\textsf{CU}}$p_1p_2$ (5/5) \\
\hline
2.5 & \ac{\textsf{Ae}} (5/5) & \ac{\textsf{Ae}} (5/5) & \ac{\textsf{Ae}} (5/5) & \ac{\textsf{Ae}} (5/5) & \ac{\textsf{Ae}} (5/5) & \ac{\textsf{Ae}} (5/5) & \ac{\textsf{Ae}} (5/5) & \ac{\textsf{Ae}} (5/5) & \ac{\textsf{Ae}} (5/5) & \ac{\textsf{Ae}} (5/5) & \ac{\textsf{Ae}} (5/5) & \ac{\textsf{Ae}} (5/5) & \ac{\textsf{Ae}} (5/5) & \ac{\textsf{Ae}} (5/5) & \ac{\textsf{Ae}} (5/5) \\
\hline
5.0 & \pu{\textsf{PU}}$p_1$ (5/5) & \pu{\textsf{PU}}$p_1$ (5/5) & \pu{\textsf{PU}}$p_1$ (5/5) & \pu{\textsf{PU}}$p_1$ (5/5) & \pu{\textsf{PU}}$p_1$ (5/5) & \pu{\textsf{PU}}$p_1$ (5/5) & \pu{\textsf{PU}}$p_1$ (5/5) & \pu{\textsf{PU}}$p_1$ (5/5) & \pu{\textsf{PU}}$p_1$ (5/5) & \pu{\textsf{PU}}$p_1$ (5/5) & \pu{\textsf{PU}}$p_1$ (5/5) & \pu{\textsf{PU}}$p_1$ (5/5) & \pu{\textsf{PU}}$p_1$ (5/5) & \pu{\textsf{PU}}$p_1$ (5/5) & \pu{\textsf{PU}}$p_1$ (5/5) \\
\hline
10.0 & \pu{\textsf{PU}}$p_1$ (5/5) & \pu{\textsf{PU}}$p_1$ (5/5) & \pu{\textsf{PU}}$p_1$ (5/5) & \pu{\textsf{PU}}$p_1$ (5/5) & \pu{\textsf{PU}}$p_1$ (5/5) & \pu{\textsf{PU}}$p_1$ (5/5) & \pu{\textsf{PU}}$p_1$ (5/5) & \pu{\textsf{PU}}$p_1$ (5/5) & \pu{\textsf{PU}}$p_1$ (5/5) & \pu{\textsf{PU}}$p_1$ (5/5) & \pu{\textsf{PU}}$p_1$ (5/5) & \pu{\textsf{PU}}$p_1$ (5/5) & \pu{\textsf{PU}}$p_1$ (5/5) & \pu{\textsf{PU}}$p_1$ (5/5) & \pu{\textsf{PU}}$p_1$ (5/5) \\
\hline
20.0 & \pu{\textsf{PU}}$p_1$ (5/5) & \pu{\textsf{PU}}$p_1$ (5/5) & \pu{\textsf{PU}}$p_1$ (5/5) & \pu{\textsf{PU}}$p_1$ (5/5) & \pu{\textsf{PU}}$p_1$ (5/5) & \pu{\textsf{PU}}$p_1$ (5/5) & \pu{\textsf{PU}}$p_1$ (5/5) & \pu{\textsf{PU}}$p_1$ (5/5) & \pu{\textsf{PU}}$p_1$ (5/5) & \pu{\textsf{PU}}$p_1$ (5/5) & \pu{\textsf{PU}}$p_1$ (5/5) & \pu{\textsf{PU}}$p_1$ (5/5) & \pu{\textsf{PU}}$p_1$ (5/5) & \pu{\textsf{PU}}$p_1$ (5/5) & \pu{\textsf{PU}}$p_1$ (5/5) \\
\hline
\ct{24.4} & \pu{\textsf{PU}}$p_1$ (5/5) & \pu{\textsf{PU}}$p_1$ (5/5) & \pu{\textsf{PU}}$p_1$ (5/5) & \pu{\textsf{PU}}$p_1$ (5/5) & \pu{\textsf{PU}}$p_1$ (5/5) & \pu{\textsf{PU}}$p_1$ (5/5) & \pu{\textsf{PU}}$p_1$ (5/5) & \pu{\textsf{PU}}$p_1$ (5/5) & \pu{\textsf{PU}}$p_1$ (5/5) & \pu{\textsf{PU}}$p_1$ (5/5) & \pu{\textsf{PU}}$p_1$ (5/5) & \pu{\textsf{PU}}$p_1$ (5/5) & \pu{\textsf{PU}}$p_1$ (5/5) & \pu{\textsf{PU}}$p_1$ (5/5) & \pu{\textsf{PU}}$p_1$ (5/5) \\
\hline
25.0 & \pu{\textsf{PU}}$p_1$ (5/5) & \pu{\textsf{PU}}$p_1$ (5/5) & \pu{\textsf{PU}}$p_1$ (5/5) & \pu{\textsf{PU}}$p_1$ (5/5) & \pu{\textsf{PU}}$p_1$ (5/5) & \pu{\textsf{PU}}$p_1$ (5/5) & \pu{\textsf{PU}}$p_1$ (5/5) & \pu{\textsf{PU}}$p_1$ (5/5) & \pu{\textsf{PU}}$p_1$ (5/5) & \pu{\textsf{PU}}$p_1$ (5/5) & \pu{\textsf{PU}}$p_1$ (5/5) & \pu{\textsf{PU}}$p_1$ (5/5) & \pu{\textsf{PU}}$p_1$ (5/5) & \pu{\textsf{PU}}$p_1$ (5/5) & \pu{\textsf{PU}}$p_1$ (5/5) \\
\hline
30.0 & \ps{\textsf{PS}} (5/5) & \ps{\textsf{PS}} (5/5) & \ps{\textsf{PS}} (5/5) & \ps{\textsf{PS}} (5/5) & \ps{\textsf{PS}} (5/5) & \ps{\textsf{PS}} (5/5) & \ps{\textsf{PS}} (5/5) & \ps{\textsf{PS}} (5/5) & \ps{\textsf{PS}} (5/5) & \ps{\textsf{PS}} (5/5) & \ps{\textsf{PS}} (5/5) & \ps{\textsf{PS}} (5/5) & \ps{\textsf{PS}} (5/5) & \ps{\textsf{PS}} (5/5) & \ps{\textsf{PS}} (5/5) \\
\hline
40.0 & \ps{\textsf{PS}} (5/5) & \ps{\textsf{PS}} (5/5) & \ps{\textsf{PS}} (5/5) & \ps{\textsf{PS}} (5/5) & \ps{\textsf{PS}} (5/5) & \ps{\textsf{PS}} (5/5) & \ps{\textsf{PS}} (5/5) & \ps{\textsf{PS}} (5/5) & \ps{\textsf{PS}} (5/5) & \ps{\textsf{PS}} (5/5) & \ps{\textsf{PS}} (5/5) & \ps{\textsf{PS}} (5/5) & \ps{\textsf{PS}} (5/5) & \ps{\textsf{PS}} (5/5) & \ps{\textsf{PS}} (5/5) \\
\hline
49.1 & \ps{\textsf{PS}} (5/5) & \ps{\textsf{PS}} (5/5) & \ps{\textsf{PS}} (5/5) & \ps{\textsf{PS}} (5/5) & \ps{\textsf{PS}} (5/5) & \ps{\textsf{PS}} (5/5) & \ps{\textsf{PS}} (5/5) & \ps{\textsf{PS}} (5/5) & \ps{\textsf{PS}} (5/5) & \ps{\textsf{PS}} (5/5) & \ps{\textsf{PS}} (5/5) & \ps{\textsf{PS}} (5/5) & \ps{\textsf{PS}} (5/5) & \ps{\textsf{PS}} (5/5) & \ps{\textsf{PS}} (5/5) \\
\hline
\end{tabular}

%% file: tables/mile-crossing-4.tex
\begin{tabular}{|c| *{15}{c|}} 
\hline 
\diagbox{$\bm{x_a}$}{$\bm{x_f}$} & 0.0 & \ct{3.6} & 40.0 & 80.0 & 120.0 & 160.0 & 200.0 & 240.0 & 280.0 & 320.0 & 360.0 & 400.0 & 440.0 & 480.0 & 520.0 \\ 
\hline 
0.0 & \ac{\textsf{Ae} (5/5)} & \ac{\textsf{Ae} (5/5)} & \ac{\textsf{Ae} (5/5)} & \ac{\textsf{Ae} (5/5)} & \ac{\textsf{Ae} (5/5)} & \ac{\textsf{Ae} (5/5)} & \ac{\textsf{Ae} (5/5)} & \ac{\textsf{Ae} (5/5)} & \ac{\textsf{Ae} (5/5)} & \ac{\textsf{Ae} (5/5)} & \ac{\textsf{Ae} (5/5)} & \ac{\textsf{Ae} (5/5)} & \ac{\textsf{Ae} (5/5)} & \ac{\textsf{Ae} (5/5)} & \ac{\textsf{Ae} (5/5)} \\ 
\hline 
2.5 & \pu{\textsf{PU$p_1$} (5/5)} & \pu{\textsf{PU$p_1$} (5/5)} & \pu{\textsf{PU$p_1$} (5/5)} & \pu{\textsf{PU$p_1$} (5/5)} & \pu{\textsf{PU$p_1$} (5/5)} & \pu{\textsf{PU$p_1$} (5/5)} & \pu{\textsf{PU$p_1$} (5/5)} & \pu{\textsf{PU$p_1$} (5/5)} & \pu{\textsf{PU$p_1$} (5/5)} & \pu{\textsf{PU$p_1$} (5/5)} & \pu{\textsf{PU$p_1$} (5/5)} & \pu{\textsf{PU$p_1$} (5/5)} & \pu{\textsf{PU$p_1$} (5/5)} & \pu{\textsf{PU$p_1$} (5/5)} & \pu{\textsf{PU$p_1$} (5/5)} \\ 
\hline 
5.0 & \pu{\textsf{PU$p_1$} (5/5)} & \pu{\textsf{PU$p_1$} (5/5)} & \pu{\textsf{PU$p_1$} (5/5)} & \pu{\textsf{PU$p_1$} (5/5)} & \pu{\textsf{PU$p_1$} (5/5)} & \pu{\textsf{PU$p_1$} (5/5)} & \pu{\textsf{PU$p_1$} (5/5)} & \pu{\textsf{PU$p_1$} (5/5)} & \pu{\textsf{PU$p_1$} (5/5)} & \pu{\textsf{PU$p_1$} (5/5)} & \pu{\textsf{PU$p_1$} (5/5)} & \pu{\textsf{PU$p_1$} (5/5)} & \pu{\textsf{PU$p_1$} (5/5)} & \pu{\textsf{PU$p_1$} (5/5)} & \pu{\textsf{PU$p_1$} (5/5)} \\ 
\hline 
10.0 & \pu{\textsf{PU$p_1$} (5/5)} & \pu{\textsf{PU$p_1$} (5/5)} & \pu{\textsf{PU$p_1$} (5/5)} & \pu{\textsf{PU$p_1$} (5/5)} & \pu{\textsf{PU$p_1$} (5/5)} & \pu{\textsf{PU$p_1$} (5/5)} & \pu{\textsf{PU$p_1$} (5/5)} & \pu{\textsf{PU$p_1$} (5/5)} & \pu{\textsf{PU$p_1$} (5/5)} & \pu{\textsf{PU$p_1$} (5/5)} & \pu{\textsf{PU$p_1$} (5/5)} & \pu{\textsf{PU$p_1$} (5/5)} & \pu{\textsf{PU$p_1$} (5/5)} & \pu{\textsf{PU$p_1$} (5/5)} & \pu{\textsf{PU$p_1$} (5/5)} \\ 
\hline 
15.0 & \pu{\textsf{PU$p_1$} (5/5)} & \pu{\textsf{PU$p_1$} (5/5)} & \pu{\textsf{PU$p_1$} (5/5)} & \pu{\textsf{PU$p_1$} (5/5)} & \pu{\textsf{PU$p_1$} (5/5)} & \pu{\textsf{PU$p_1$} (5/5)} & \pu{\textsf{PU$p_1$} (5/5)} & \pu{\textsf{PU$p_1$} (5/5)} & \pu{\textsf{PU$p_1$} (5/5)} & \pu{\textsf{PU$p_1$} (5/5)} & \pu{\textsf{PU$p_1$} (5/5)} & \pu{\textsf{PU$p_1$} (5/5)} & \pu{\textsf{PU$p_1$} (5/5)} & \pu{\textsf{PU$p_1$} (5/5)} & \pu{\textsf{PU$p_1$} (5/5)} \\ 
\hline 
20.0 & \pu{\textsf{PU$p_1$} (5/5)} & \pu{\textsf{PU$p_1$} (5/5)} & \pu{\textsf{PU$p_1$} (5/5)} & \pu{\textsf{PU$p_1$} (5/5)} & \pu{\textsf{PU$p_1$} (5/5)} & \pu{\textsf{PU$p_1$} (5/5)} & \pu{\textsf{PU$p_1$} (5/5)} & \pu{\textsf{PU$p_1$} (5/5)} & \pu{\textsf{PU$p_1$} (5/5)} & \pu{\textsf{PU$p_1$} (5/5)} & \pu{\textsf{PU$p_1$} (5/5)} & \pu{\textsf{PU$p_1$} (5/5)} & \pu{\textsf{PU$p_1$} (5/5)} & \pu{\textsf{PU$p_1$} (5/5)} & \pu{\textsf{PU$p_1$} (5/5)} \\ 
\hline 
\ct{23.5} & \pu{\textsf{PU$p_1$} (5/5)} & \pu{\textsf{PU$p_1$} (5/5)} & \ps{\textsf{PS} (5/5)} & \ps{\textsf{PS} (5/5)} & \ps{\textsf{PS} (5/5)} & \ps{\textsf{PS} (5/5)} & \ps{\textsf{PS} (5/5)} & \ps{\textsf{PS} (5/5)} & \ps{\textsf{PS} (5/5)} & \ps{\textsf{PS} (5/5)} & \ps{\textsf{PS} (5/5)} & \ps{\textsf{PS} (5/5)} & \ps{\textsf{PS} (5/5)} & \ps{\textsf{PS} (5/5)} & \ps{\textsf{PS} (5/5)} \\ 
\hline 
25.0 & \pu{\textsf{PU$p_1$} (5/5)} & \ps{\textsf{PS} (5/5)} & \ps{\textsf{PS} (5/5)} & \ps{\textsf{PS} (5/5)} & \ps{\textsf{PS} (5/5)} & \ps{\textsf{PS} (5/5)} & \ps{\textsf{PS} (5/5)} & \ps{\textsf{PS} (5/5)} & \ps{\textsf{PS} (5/5)} & \ps{\textsf{PS} (5/5)} & \ps{\textsf{PS} (5/5)} & \ps{\textsf{PS} (5/5)} & \ps{\textsf{PS} (5/5)} & \ps{\textsf{PS} (5/5)} & \ps{\textsf{PS} (5/5)} \\ 
\hline 
30.0 & \pu{\textsf{PU$p_1$} (5/5)} & \ps{\textsf{PS} (5/5)} & \ps{\textsf{PS} (5/5)} & \ps{\textsf{PS} (5/5)} & \ps{\textsf{PS} (5/5)} & \ps{\textsf{PS} (5/5)} & \ps{\textsf{PS} (5/5)} & \ps{\textsf{PS} (5/5)} & \ps{\textsf{PS} (5/5)} & \ps{\textsf{PS} (5/5)} & \ps{\textsf{PS} (5/5)} & \ps{\textsf{PS} (5/5)} & \ps{\textsf{PS} (5/5)} & \ps{\textsf{PS} (5/5)} & \ps{\textsf{PS} (5/5)} \\ 
\hline 
35.0 & \ps{\textsf{PS} (5/5)} & \ps{\textsf{PS} (5/5)} & \ps{\textsf{PS} (5/5)} & \ps{\textsf{PS} (5/5)} & \ps{\textsf{PS} (5/5)} & \ps{\textsf{PS} (5/5)} & \ps{\textsf{PS} (5/5)} & \ps{\textsf{PS} (5/5)} & \ps{\textsf{PS} (5/5)} & \ps{\textsf{PS} (5/5)} & \ps{\textsf{PS} (5/5)} & \ps{\textsf{PS} (5/5)} & \ps{\textsf{PS} (5/5)} & \ps{\textsf{PS} (5/5)} & \ps{\textsf{PS} (5/5)} \\ 
\hline 
40.0 & \ps{\textsf{PS} (5/5)} & \ps{\textsf{PS} (5/5)} & \ps{\textsf{PS} (5/5)} & \ps{\textsf{PS} (5/5)} & \ps{\textsf{PS} (5/5)} & \ps{\textsf{PS} (5/5)} & \ps{\textsf{PS} (5/5)} & \ps{\textsf{PS} (5/5)} & \ps{\textsf{PS} (5/5)} & \ps{\textsf{PS} (5/5)} & \ps{\textsf{PS} (5/5)} & \ps{\textsf{PS} (5/5)} & \ps{\textsf{PS} (5/5)} & \ps{\textsf{PS} (5/5)} & \ps{\textsf{PS} (5/5)} \\ 
\hline 
46.8 & \ps{\textsf{PS} (5/5)} & \ps{\textsf{PS} (5/5)} & \ps{\textsf{PS} (5/5)} & \ps{\textsf{PS} (5/5)} & \ps{\textsf{PS} (5/5)} & \ps{\textsf{PS} (5/5)} & \ps{\textsf{PS} (5/5)} & \ps{\textsf{PS} (5/5)} & \ps{\textsf{PS} (5/5)} & \ps{\textsf{PS} (5/5)} & \ps{\textsf{PS} (5/5)} & \ps{\textsf{PS} (5/5)} & \ps{\textsf{PS} (5/5)} & \ps{\textsf{PS} (5/5)} & \ps{\textsf{PS} (5/5)} \\ 
\hline 
\end{tabular}

%% file: tables/mile-traffic_light-0.tex
\begin{tabular}{|c| *{14}{c|}} 
\hline
\multicolumn{15}{|c|}{$\bm{xf}$} \\
\hline
0.0 & \ct{3.6} & 40.0 & 80.0 & 120.0 & 160.0 & 200.0 & 240.0 & 280.0 & 320.0 & 360.0 & 400.0 & 440.0 & 480.0 & 520.0 \\
\hline
\blk{\textsf{Blk} (5/5)} & \pu{\textsf{PU$p_4$} (5/5)}& \begin{tabular}{@{}c@{}}\begin{minipage}[c][0.5cm][c]{1.8cm}\centering \pu{\textsf{PU$p_4$} (3/5)}\ \end{minipage} \\\begin{minipage}[c][0.5cm][c]{1.8cm}\centering \ps{\textsf{PS} (2/5)}\ \end{minipage} \\\end{tabular}& \begin{tabular}{@{}c@{}}\begin{minipage}[c][0.5cm][c]{1.8cm}\centering \pu{\textsf{PU$p_4$} (3/5)}\ \end{minipage} \\\begin{minipage}[c][0.5cm][c]{1.8cm}\centering \ps{\textsf{PS} (2/5)}\ \end{minipage} \\\end{tabular}& \begin{tabular}{@{}c@{}}\begin{minipage}[c][0.5cm][c]{1.8cm}\centering \ps{\textsf{PS} (3/5)}\ \end{minipage} \\\begin{minipage}[c][0.5cm][c]{1.8cm}\centering \pu{\textsf{PU$p_4$} (2/5)}\ \end{minipage} \\\end{tabular}& \begin{tabular}{@{}c@{}}\begin{minipage}[c][0.5cm][c]{1.8cm}\centering \ps{\textsf{PS} (4/5)}\ \end{minipage} \\\begin{minipage}[c][0.5cm][c]{1.8cm}\centering \pu{\textsf{PU$p_4$} (1/5)}\ \end{minipage} \\\end{tabular} & \ps{\textsf{PS} (5/5)}& \begin{tabular}{@{}c@{}}\begin{minipage}[c][0.5cm][c]{1.8cm}\centering \ps{\textsf{PS} (4/5)}\ \end{minipage} \\\begin{minipage}[c][0.5cm][c]{1.8cm}\centering \pu{\textsf{PU$p_4$} (1/5)}\ \end{minipage} \\\end{tabular} & \ps{\textsf{PS} (5/5)}& \begin{tabular}{@{}c@{}}\begin{minipage}[c][0.5cm][c]{1.8cm}\centering \ps{\textsf{PS} (4/5)}\ \end{minipage} \\\begin{minipage}[c][0.5cm][c]{1.8cm}\centering \pu{\textsf{PU$p_4$} (1/5)}\ \end{minipage} \\\end{tabular}& \begin{tabular}{@{}c@{}}\begin{minipage}[c][0.5cm][c]{1.8cm}\centering \ps{\textsf{PS} (4/5)}\ \end{minipage} \\\begin{minipage}[c][0.5cm][c]{1.8cm}\centering \pu{\textsf{PU$p_4$} (1/5)}\ \end{minipage} \\\end{tabular}& \begin{tabular}{@{}c@{}}\begin{minipage}[c][0.5cm][c]{1.8cm}\centering \ps{\textsf{PS} (3/5)}\ \end{minipage} \\\begin{minipage}[c][0.5cm][c]{1.8cm}\centering \pu{\textsf{PU$p_4$} (2/5)}\ \end{minipage} \\\end{tabular}& \begin{tabular}{@{}c@{}}\begin{minipage}[c][0.5cm][c]{1.8cm}\centering \ps{\textsf{PS} (3/5)}\ \end{minipage} \\\begin{minipage}[c][0.5cm][c]{1.8cm}\centering \pu{\textsf{PU$p_4$} (2/5)}\ \end{minipage} \\\end{tabular}& \begin{tabular}{@{}c@{}}\begin{minipage}[c][0.5cm][c]{1.8cm}\centering \ps{\textsf{PS} (3/5)}\ \end{minipage} \\\begin{minipage}[c][0.5cm][c]{1.8cm}\centering \pu{\textsf{PU$p_4$} (2/5)}\ \end{minipage} \\\end{tabular}& \begin{tabular}{@{}c@{}}\begin{minipage}[c][0.5cm][c]{1.8cm}\centering \pu{\textsf{PU$p_4$} (2/5)}\ \end{minipage} \\\begin{minipage}[c][0.5cm][c]{1.8cm}\centering \ps{\textsf{PS} (3/5)}\ \end{minipage} \\\end{tabular} \\
\hline
\end{tabular}

%% file: tables/mile-traffic_light-4.tex
\begin{tabular}{|c|c|c|c|c|c|c|c|c|c|c|c|c|c|c|}
\hline
\multicolumn{15}{|c|}{$\bm{xf}$}\\
\hline
0.0 & \ct{3.6} & 40.0 & 80.0 & 120.0 & 160.0 & 200.0 & 240.0 & 280.0 & 320.0 & 360.0 & 400.0 & 440.0 & 480.0 & 520.0 \\
\hline
\blk{\textsf{Blk (5/5)}} & \ps{\textsf{PS (5/5)}} & \ps{\textsf{PS (5/5)}} & \ps{\textsf{PS (5/5)}} & \ps{\textsf{PS (5/5)}} & \ps{\textsf{PS (5/5)}} & \ps{\textsf{PS (5/5)}} & \ps{\textsf{PS (5/5)}} & \ps{\textsf{PS (5/5)}} & \ps{\textsf{PS (5/5)}} & \ps{\textsf{PS (5/5)}} & \ps{\textsf{PS (5/5)}} & \ps{\textsf{PS (5/5)}} & \ps{\textsf{PS (5/5)}} & \ps{\textsf{PS (5/5)}} \\
\hline
\end{tabular}

%% file: tables/transfuser-merging-0.tex
\begin{tabular}{|c| *{14}{c|}} 
\hline 
\diagbox{$\bm{x_a}$}{$\bm{x_f}$} & \ct{0.0} & 40.0 & 80.0 & 120.0 & 160.0 & 200.0 & 240.0 & 280.0 & 320.0 & 360.0 & 400.0 & 440.0 & 480.0 & 520.0 \\ 
\hline 
0.0 & - & \cs{\textsf{CS} (5/5)} & \cs{\textsf{CS} (5/5)} & \cs{\textsf{CS} (5/5)} & \cs{\textsf{CS} (5/5)} & \cs{\textsf{CS} (5/5)} & \cs{\textsf{CS} (5/5)} & \cs{\textsf{CS} (5/5)} & \cs{\textsf{CS} (5/5)} & \cs{\textsf{CS} (5/5)} & \cs{\textsf{CS} (5/5)}& \begin{tabular}{@{}c@{}}\begin{minipage}[c][0.5cm][c]{1.8cm}\centering \ac{\textsf{Ae} (2/5)}\ \end{minipage} \\\begin{minipage}[c][0.5cm][c]{1.8cm}\centering \cs{\textsf{CS} (3/5)}\ \end{minipage} \\\end{tabular} & \cs{\textsf{CS} (5/5)} & \cs{\textsf{CS} (5/5)} \\ 
\hline 
\ct{1.9} & \cs{\textsf{CS} (5/5)} & \cs{\textsf{CS} (5/5)} & \cs{\textsf{CS} (5/5)} & \cs{\textsf{CS} (5/5)} & \cs{\textsf{CS} (5/5)} & \cs{\textsf{CS} (5/5)} & \cs{\textsf{CS} (5/5)} & \cs{\textsf{CS} (5/5)} & \cs{\textsf{CS} (5/5)} & \cs{\textsf{CS} (5/5)} & \cs{\textsf{CS} (5/5)}& \begin{tabular}{@{}c@{}}\begin{minipage}[c][0.5cm][c]{1.8cm}\centering \ac{\textsf{Ae} (2/5)}\ \end{minipage} \\\begin{minipage}[c][0.5cm][c]{1.8cm}\centering \cs{\textsf{CS} (3/5)}\ \end{minipage} \\\end{tabular} & \cs{\textsf{CS} (5/5)} & \cs{\textsf{CS} (5/5)} \\ 
\hline 
5.0 & \cs{\textsf{CS} (5/5)} & \cs{\textsf{CS} (5/5)} & \cs{\textsf{CS} (5/5)} & \cs{\textsf{CS} (5/5)} & \cs{\textsf{CS} (5/5)} & \cs{\textsf{CS} (5/5)} & \cs{\textsf{CS} (5/5)} & \cs{\textsf{CS} (5/5)} & \cs{\textsf{CS} (5/5)} & \cs{\textsf{CS} (5/5)} & \cs{\textsf{CS} (5/5)}& \begin{tabular}{@{}c@{}}\begin{minipage}[c][0.5cm][c]{1.8cm}\centering \ac{\textsf{Ae} (3/5)}\ \end{minipage} \\\begin{minipage}[c][0.5cm][c]{1.8cm}\centering \cs{\textsf{CS} (2/5)}\ \end{minipage} \\\end{tabular} & \cs{\textsf{CS} (5/5)} & \cs{\textsf{CS} (5/5)} \\ 
\hline 
10.0 & \cs{\textsf{CS} (5/5)}& \begin{tabular}{@{}c@{}}\begin{minipage}[c][0.5cm][c]{1.8cm}\centering \ps{\textsf{PS} (1/5)}\ \end{minipage} \\\begin{minipage}[c][0.5cm][c]{1.8cm}\centering \ac{\textsf{Aa} (4/5)}\ \end{minipage} \\\end{tabular}& \begin{tabular}{@{}c@{}}\begin{minipage}[c][0.5cm][c]{1.8cm}\centering \ps{\textsf{PS} (1/5)}\ \end{minipage} \\\begin{minipage}[c][0.5cm][c]{1.8cm}\centering \ac{\textsf{Aa} (4/5)}\ \end{minipage} \\\end{tabular}& \begin{tabular}{@{}c@{}}\begin{minipage}[c][0.5cm][c]{1.8cm}\centering \ps{\textsf{PS} (1/5)}\ \end{minipage} \\\begin{minipage}[c][0.5cm][c]{1.8cm}\centering \ac{\textsf{Aa} (4/5)}\ \end{minipage} \\\end{tabular} & \ac{\textsf{Aa} (5/5)} & \ac{\textsf{Aa} (5/5)} & \ac{\textsf{Aa} (5/5)} & \ac{\textsf{Aa} (5/5)} & \ac{\textsf{Aa} (5/5)} & \ac{\textsf{Aa} (5/5)}& \begin{tabular}{@{}c@{}}\begin{minipage}[c][0.5cm][c]{1.8cm}\centering \ac{\textsf{Aa} (4/5)}\ \end{minipage} \\\begin{minipage}[c][0.5cm][c]{1.8cm}\centering \cs{\textsf{CS} (1/5)}\ \end{minipage} \\\end{tabular} & \ac{\textsf{Aa} (5/5)} & \ac{\textsf{Aa} (5/5)} & \ac{\textsf{Aa} (5/5)} \\ 
\hline 
15.0 & \cs{\textsf{CS} (5/5)} & \ps{\textsf{PS} (5/5)} & \ps{\textsf{PS} (5/5)} & \ps{\textsf{PS} (5/5)} & \ps{\textsf{PS} (5/5)} & \ps{\textsf{PS} (5/5)} & \ps{\textsf{PS} (5/5)} & \ps{\textsf{PS} (5/5)}& \begin{tabular}{@{}c@{}}\begin{minipage}[c][0.5cm][c]{1.8cm}\centering \ps{\textsf{PS} (4/5)}\ \end{minipage} \\\begin{minipage}[c][0.5cm][c]{1.8cm}\centering \ro{\textsf{DRe} (1/5)}\ \end{minipage} \\\end{tabular} & \ps{\textsf{PS} (5/5)} & \ps{\textsf{PS} (5/5)} & \ac{\textsf{Aa} (5/5)} & \ps{\textsf{PS} (5/5)} & \ps{\textsf{PS} (5/5)} \\ 
\hline 
20.0 & \cs{\textsf{CS} (5/5)} & \ps{\textsf{PS} (5/5)} & \ps{\textsf{PS} (5/5)} & \ps{\textsf{PS} (5/5)} & \ps{\textsf{PS} (5/5)} & \ps{\textsf{PS} (5/5)} & \ps{\textsf{PS} (5/5)} & \ps{\textsf{PS} (5/5)} & \ps{\textsf{PS} (5/5)} & \ps{\textsf{PS} (5/5)} & \ps{\textsf{PS} (5/5)}& \begin{tabular}{@{}c@{}}\begin{minipage}[c][0.5cm][c]{1.8cm}\centering \ac{\textsf{Aa} (4/5)}\ \end{minipage} \\\begin{minipage}[c][0.5cm][c]{1.8cm}\centering \ro{\textsf{DRe} (1/5)}\ \end{minipage} \\\end{tabular} & \ps{\textsf{PS} (5/5)} & \ps{\textsf{PS} (5/5)} \\ 
\hline 
25.0 & \cs{\textsf{CS} (5/5)} & \ps{\textsf{PS} (5/5)} & \ps{\textsf{PS} (5/5)} & \ps{\textsf{PS} (5/5)} & \ps{\textsf{PS} (5/5)} & \ps{\textsf{PS} (5/5)} & \ps{\textsf{PS} (5/5)} & \ps{\textsf{PS} (5/5)} & \ps{\textsf{PS} (5/5)} & \ps{\textsf{PS} (5/5)} & \ps{\textsf{PS} (5/5)} & \ac{\textsf{Aa} (5/5)} & \ps{\textsf{PS} (5/5)} & \ps{\textsf{PS} (5/5)} \\ 
\hline 
30.0 & \cs{\textsf{CS} (5/5)} & \ps{\textsf{PS} (5/5)} & \ps{\textsf{PS} (5/5)} & \ps{\textsf{PS} (5/5)} & \ps{\textsf{PS} (5/5)} & \ps{\textsf{PS} (5/5)}& \begin{tabular}{@{}c@{}}\begin{minipage}[c][0.5cm][c]{1.8cm}\centering \ps{\textsf{PS} (4/5)}\ \end{minipage} \\\begin{minipage}[c][0.5cm][c]{1.8cm}\centering \ro{\textsf{DRe} (1/5)}\ \end{minipage} \\\end{tabular} & \ps{\textsf{PS} (5/5)} & \ps{\textsf{PS} (5/5)} & \ps{\textsf{PS} (5/5)} & \ps{\textsf{PS} (5/5)}& \begin{tabular}{@{}c@{}}\begin{minipage}[c][0.5cm][c]{1.8cm}\centering \ps{\textsf{PS} (2/5)}\ \end{minipage} \\\begin{minipage}[c][0.5cm][c]{1.8cm}\centering \ac{\textsf{Aa} (3/5)}\ \end{minipage} \\\end{tabular}& \begin{tabular}{@{}c@{}}\begin{minipage}[c][0.5cm][c]{1.8cm}\centering \ps{\textsf{PS} (4/5)}\ \end{minipage} \\\begin{minipage}[c][0.5cm][c]{1.8cm}\centering \ro{\textsf{DRe} (1/5)}\ \end{minipage} \\\end{tabular} & \ps{\textsf{PS} (5/5)} \\ 
\hline 
35.0 & \cs{\textsf{CS} (5/5)} & \ps{\textsf{PS} (5/5)} & \ps{\textsf{PS} (5/5)} & \ps{\textsf{PS} (5/5)} & \ps{\textsf{PS} (5/5)} & \ps{\textsf{PS} (5/5)} & \ps{\textsf{PS} (5/5)} & \ps{\textsf{PS} (5/5)} & \ps{\textsf{PS} (5/5)} & \ps{\textsf{PS} (5/5)} & \ps{\textsf{PS} (5/5)} & \ps{\textsf{PS} (5/5)} & \ps{\textsf{PS} (5/5)} & \ps{\textsf{PS} (5/5)} \\ 
\hline 
40.0 & \cs{\textsf{CS} (5/5)} & \ps{\textsf{PS} (5/5)} & \ps{\textsf{PS} (5/5)} & \ps{\textsf{PS} (5/5)} & \ps{\textsf{PS} (5/5)} & \ps{\textsf{PS} (5/5)} & \ps{\textsf{PS} (5/5)} & \ps{\textsf{PS} (5/5)} & \ps{\textsf{PS} (5/5)} & \ps{\textsf{PS} (5/5)} & \ps{\textsf{PS} (5/5)} & \ps{\textsf{PS} (5/5)} & \ps{\textsf{PS} (5/5)} & \ps{\textsf{PS} (5/5)} \\ 
\hline 
\end{tabular} 

%% file: tables/transfuser-merging-4.tex
\begin{tabular}{|c| *{15}{c|}} 
\hline 
\diagbox{$\bm{x_a}$}{$\bm{x_f}$} & 0.0 & \ct{1.9} & 40.0 & 80.0 & 120.0 & 160.0 & 200.0 & 240.0 & 280.0 & 320.0 & 360.0 & 400.0 & 440.0 & 480.0 & 520.0 \\ 
\hline 
0.0 & - & \cs{\textsf{CS} (5/5)} & \cs{\textsf{CS} (5/5)} & \cs{\textsf{CS} (5/5)} & \cs{\textsf{CS} (5/5)} & \cs{\textsf{CS} (5/5)} & \cs{\textsf{CS} (5/5)} & \cs{\textsf{CS} (5/5)} & \cs{\textsf{CS} (5/5)} & \cs{\textsf{CS} (5/5)} & \cs{\textsf{CS} (5/5)} & \cs{\textsf{CS} (5/5)} & \ac{\textsf{Ae} (5/5)} & \cs{\textsf{CS} (5/5)} & \cs{\textsf{CS} (5/5)} \\ 
\hline 
\ct{3.8} & \cs{\textsf{CS} (5/5)} & \cs{\textsf{CS} (5/5)} & \cs{\textsf{CS} (5/5)} & \cs{\textsf{CS} (5/5)} & \cs{\textsf{CS} (5/5)} & \cs{\textsf{CS} (5/5)} & \cs{\textsf{CS} (5/5)} & \cs{\textsf{CS} (5/5)} & \cs{\textsf{CS} (5/5)} & \cs{\textsf{CS} (5/5)} & \cs{\textsf{CS} (5/5)} & \cs{\textsf{CS} (5/5)} & \ac{\textsf{Ae} (5/5)} & \cs{\textsf{CS} (5/5)} & \cs{\textsf{CS} (5/5)} \\ 
\hline 
5.0 & \cs{\textsf{CS} (5/5)} & \cs{\textsf{CS} (5/5)} & \cs{\textsf{CS} (5/5)} & \cs{\textsf{CS} (5/5)} & \cs{\textsf{CS} (5/5)} & \cs{\textsf{CS} (5/5)} & \cs{\textsf{CS} (5/5)} & \cs{\textsf{CS} (5/5)} & \cs{\textsf{CS} (5/5)} & \cs{\textsf{CS} (5/5)} & \cs{\textsf{CS} (5/5)} & \cs{\textsf{CS} (5/5)}& \begin{tabular}{@{}c@{}}\begin{minipage}[c][0.5cm][c]{1.2cm}\centering \ac{\textsf{Ae} (4/5)}\ \end{minipage} \\\begin{minipage}[c][0.5cm][c]{1.2cm}\centering \cs{\textsf{CS} (1/5)}\ \end{minipage} \\\end{tabular} & \cs{\textsf{CS} (5/5)} & \cs{\textsf{CS} (5/5)} \\ 
\hline 
7.5 & \cs{\textsf{CS} (5/5)} & \cs{\textsf{CS} (5/5)} & \ac{\textsf{Aa} (5/5)}& \begin{tabular}{@{}c@{}}\begin{minipage}[c][0.5cm][c]{1.2cm}\centering \ac{\textsf{Aa} (2/5)}\ \end{minipage} \\\begin{minipage}[c][0.5cm][c]{1.2cm}\centering \cs{\textsf{CS} (3/5)}\ \end{minipage} \\\end{tabular}& \begin{tabular}{@{}c@{}}\begin{minipage}[c][0.5cm][c]{1.2cm}\centering \ac{\textsf{Aa} (3/5)}\ \end{minipage} \\\begin{minipage}[c][0.5cm][c]{1.2cm}\centering \cs{\textsf{CS} (2/5)}\ \end{minipage} \\\end{tabular}& \begin{tabular}{@{}c@{}}\begin{minipage}[c][0.5cm][c]{1.2cm}\centering \ac{\textsf{Aa} (1/5)}\ \end{minipage} \\\begin{minipage}[c][0.5cm][c]{1.2cm}\centering \cs{\textsf{CS} (4/5)}\ \end{minipage} \\\end{tabular}& \begin{tabular}{@{}c@{}}\begin{minipage}[c][0.5cm][c]{1.2cm}\centering \ac{\textsf{Aa} (3/5)}\ \end{minipage} \\\begin{minipage}[c][0.5cm][c]{1.2cm}\centering \ac{\textsf{Ae} (2/5)}\ \end{minipage} \\\end{tabular} & \ac{\textsf{Aa} (5/5)}& \begin{tabular}{@{}c@{}}\begin{minipage}[c][0.5cm][c]{1.2cm}\centering \ac{\textsf{Aa} (2/5)}\ \end{minipage} \\\begin{minipage}[c][0.5cm][c]{1.2cm}\centering \cs{\textsf{CS} (3/5)}\ \end{minipage} \\\end{tabular} & \cs{\textsf{CS} (5/5)} & \ac{\textsf{Aa} (5/5)}& \begin{tabular}{@{}c@{}}\begin{minipage}[c][0.5cm][c]{1.2cm}\centering \ac{\textsf{Aa} (2/5)}\ \end{minipage} \\\begin{minipage}[c][0.5cm][c]{1.2cm}\centering \cs{\textsf{CS} (3/5)}\ \end{minipage} \\\end{tabular}& \begin{tabular}{@{}c@{}}\begin{minipage}[c][0.5cm][c]{1.2cm}\centering \ac{\textsf{Aa} (2/5)}\ \end{minipage} \\\begin{minipage}[c][0.5cm][c]{1.2cm}\centering \ac{\textsf{Ae} (3/5)}\ \end{minipage} \\\end{tabular}& \begin{tabular}{@{}c@{}}\begin{minipage}[c][0.5cm][c]{1.2cm}\centering \cs{\textsf{CS} (3/5)}\ \end{minipage} \\\begin{minipage}[c][0.5cm][c]{1.2cm}\centering \ac{\textsf{Aa} (2/5)}\ \end{minipage} \\\end{tabular}& \begin{tabular}{@{}c@{}}\begin{minipage}[c][0.5cm][c]{1.2cm}\centering \ac{\textsf{Aa} (2/5)}\ \end{minipage} \\\begin{minipage}[c][0.5cm][c]{1.2cm}\centering \ac{\textsf{Ae} (3/5)}\ \end{minipage} \\\end{tabular} \\
\hline
10.0 & \cs{\textsf{CS} (5/5)} & \cs{\textsf{CS} (5/5)} & \ps{\textsf{PS} (5/5)} & \ps{\textsf{PS} (5/5)} & \ps{\textsf{PS} (5/5)} & \ps{\textsf{PS} (5/5)} & \ps{\textsf{PS} (5/5)} & \ps{\textsf{PS} (5/5)} & \ps{\textsf{PS} (5/5)} & \ps{\textsf{PS} (5/5)} & \ps{\textsf{PS} (5/5)} & \ps{\textsf{PS} (5/5)} & \ac{\textsf{Aa} (5/5)} & \ps{\textsf{PS} (5/5)} & \ps{\textsf{PS} (5/5)} \\ 
\hline 
15.0 & \cs{\textsf{CS} (5/5)} & \cs{\textsf{CS} (5/5)} & \ps{\textsf{PS} (5/5)} & \ps{\textsf{PS} (5/5)} & \ps{\textsf{PS} (5/5)} & \ps{\textsf{PS} (5/5)} & \ps{\textsf{PS} (5/5)} & \ps{\textsf{PS} (5/5)} & \ps{\textsf{PS} (5/5)} & \ps{\textsf{PS} (5/5)} & \ps{\textsf{PS} (5/5)} & \ps{\textsf{PS} (4/4)} & \ac{\textsf{Aa} (5/5)} & \ps{\textsf{PS} (5/5)} & \ps{\textsf{PS} (5/5)} \\ 
\hline 
20.0 & \cs{\textsf{CS} (5/5)} & \cs{\textsf{CS} (5/5)} & \ps{\textsf{PS} (5/5)} & \ps{\textsf{PS} (5/5)} & \ps{\textsf{PS} (5/5)} & \ps{\textsf{PS} (5/5)} & \ps{\textsf{PS} (5/5)} & \ps{\textsf{PS} (5/5)} & \ps{\textsf{PS} (5/5)} & \ps{\textsf{PS} (5/5)} & \ps{\textsf{PS} (5/5)} & \ps{\textsf{PS} (5/5)} & \ac{\textsf{Aa} (5/5)} & \ps{\textsf{PS} (5/5)} & \ps{\textsf{PS} (5/5)} \\ 
\hline 
25.0 & \cs{\textsf{CS} (5/5)} & \cs{\textsf{CS} (5/5)} & \ps{\textsf{PS} (5/5)} & \ps{\textsf{PS} (5/5)} & \ps{\textsf{PS} (5/5)} & \ps{\textsf{PS} (5/5)} & \ps{\textsf{PS} (5/5)} & \ps{\textsf{PS} (5/5)} & \ps{\textsf{PS} (5/5)} & \ps{\textsf{PS} (5/5)} & \ps{\textsf{PS} (5/5)} & \ps{\textsf{PS} (5/5)}& \begin{tabular}{@{}c@{}}\begin{minipage}[c][0.5cm][c]{1.2cm}\centering \ps{\textsf{PS} (2/5)}\ \end{minipage} \\\begin{minipage}[c][0.5cm][c]{1.2cm}\centering \ac{\textsf{Aa} (3/5)}\ \end{minipage} \\\end{tabular} & \ps{\textsf{PS} (5/5)} & \ps{\textsf{PS} (5/5)} \\ 
\hline 
30.0 & \cs{\textsf{CS} (5/5)} & \cs{\textsf{CS} (5/5)} & \ps{\textsf{PS} (5/5)} & \ps{\textsf{PS} (5/5)} & \ps{\textsf{PS} (5/5)} & \ps{\textsf{PS} (5/5)} & \ps{\textsf{PS} (5/5)} & \ps{\textsf{PS} (5/5)} & \ps{\textsf{PS} (5/5)} & \ps{\textsf{PS} (5/5)} & \ps{\textsf{PS} (5/5)} & \ps{\textsf{PS} (5/5)}& \begin{tabular}{@{}c@{}}\begin{minipage}[c][0.5cm][c]{1.2cm}\centering \ac{\textsf{Aa} (4/5)}\ \end{minipage} \\\begin{minipage}[c][0.5cm][c]{1.2cm}\centering \ps{\textsf{PS} (1/5)}\ \end{minipage} \\\end{tabular} & \ps{\textsf{PS} (5/5)} & \ps{\textsf{PS} (5/5)} \\ 
\hline 
35.0 & \cs{\textsf{CS} (5/5)} & \cs{\textsf{CS} (5/5)} & \ps{\textsf{PS} (5/5)} & \ps{\textsf{PS} (5/5)} & \ps{\textsf{PS} (5/5)} & \ps{\textsf{PS} (5/5)} & \ps{\textsf{PS} (5/5)} & \ps{\textsf{PS} (5/5)} & \ps{\textsf{PS} (5/5)} & \ps{\textsf{PS} (5/5)} & \ps{\textsf{PS} (5/5)} & \ps{\textsf{PS} (5/5)} & \ps{\textsf{PS} (5/5)} & \ps{\textsf{PS} (5/5)} & \ps{\textsf{PS} (5/5)} \\ 
\hline 
40.0 & \cs{\textsf{CS} (5/5)} & \cs{\textsf{CS} (5/5)} & \ps{\textsf{PS} (5/5)} & \ps{\textsf{PS} (5/5)} & \ps{\textsf{PS} (5/5)} & \ps{\textsf{PS} (5/5)} & \ps{\textsf{PS} (5/5)} & \ps{\textsf{PS} (5/5)} & \ps{\textsf{PS} (5/5)} & \ps{\textsf{PS} (5/5)} & \ps{\textsf{PS} (5/5)} & \ps{\textsf{PS} (5/5)} & \ps{\textsf{PS} (5/5)} & \ps{\textsf{PS} (5/5)} & \ps{\textsf{PS} (5/5)} \\ 
\hline 
\end{tabular} 

%% file: tables/transfuser-lane_change-4.tex
\begin{tabular}{|c| *{15}{c|}}
\hline
\diagbox{$\bm{x_a}$}{$\bm{x_f}$} & 0.0 & \ct{1.9} & 40.0 & 80.0 & 120.0 & 160.0 & 200.0 & 240.0 & 280.0 & 320.0 & 360.0 & 400.0 & 440.0 & 480.0 & 520.0 \\
\hline
0.0 & - & \cs{\textsf{CS}} (5/5) & \cs{\textsf{CS}} (5/5) & \cs{\textsf{CS}} (5/5) & \cs{\textsf{CS}} (5/5) & \cs{\textsf{CS}} (5/5) & \cs{\textsf{CS}} (5/5) & \cs{\textsf{CS}} (5/5) & \cs{\textsf{CS}} (5/5) & \cs{\textsf{CS}} (5/5) & \cs{\textsf{CS}} (5/5) & \cs{\textsf{CS}} (5/5) & \cs{\textsf{CS}} (5/5) & \cs{\textsf{CS}} (5/5) & \cs{\textsf{CS}} (5/5) \\
\hline
\ct{15.4} & \cs{\textsf{CS}} (5/5) & \cs{\textsf{CS}} (5/5) & \cs{\textsf{CS}} (5/5) & \cs{\textsf{CS}} (5/5) & \cs{\textsf{CS}} (5/5) & \cs{\textsf{CS}} (5/5) & \cs{\textsf{CS}} (5/5) & \cs{\textsf{CS}} (5/5) & \cs{\textsf{CS}} (5/5) & \cs{\textsf{CS}} (5/5) & \cs{\textsf{CS}} (5/5) & \cs{\textsf{CS}} (5/5) & \cs{\textsf{CS}} (5/5) & \cs{\textsf{CS}} (5/5) & \cs{\textsf{CS}} (5/5)  \\
\hline
40.0 & \cs{\textsf{CS}} (5/5) & \cs{\textsf{CS}} (5/5) & \co{\textsf{CO}} (5/5) & \co{\textsf{CO}} (5/5) & \co{\textsf{CO}} (5/5) & \co{\textsf{CO}} (5/5) & \co{\textsf{CO}} (5/5) & \co{\textsf{CO}} (5/5) & \co{\textsf{CO}} (5/5) & \co{\textsf{CO}} (5/5) & \co{\textsf{CO}} (5/5) & \co{\textsf{CO}} (5/5) & \co{\textsf{CO}} (5/5) & \co{\textsf{CO}} (5/5) & \co{\textsf{CO}} (5/5) \\
\hline
80.0 & \cs{\textsf{CS}} (5/5) & \cs{\textsf{CS}} (5/5) & \co{\textsf{CO}} (5/5) & \co{\textsf{CO}} (5/5) & \co{\textsf{CO}} (5/5) & \co{\textsf{CO}} (5/5) & \co{\textsf{CO}} (5/5) & \co{\textsf{CO}} (5/5) & \co{\textsf{CO}} (5/5) & \co{\textsf{CO}} (5/5) & \co{\textsf{CO}} (5/5) & \co{\textsf{CO}} (5/5) & \co{\textsf{CO}} (5/5) & \co{\textsf{CO}} (5/5) & \co{\textsf{CO}} (5/5) \\
\hline
120.0 & \cs{\textsf{CS}} (5/5) & \cs{\textsf{CS}} (5/5) & \co{\textsf{CO}} (5/5) & \co{\textsf{CO}} (5/5) & \co{\textsf{CO}} (5/5) & \co{\textsf{CO}} (5/5) & \co{\textsf{CO}} (5/5) & \co{\textsf{CO}} (5/5) & \co{\textsf{CO}} (5/5) & \co{\textsf{CO}} (5/5) & \co{\textsf{CO}} (5/5) & \co{\textsf{CO}} (5/5) & \co{\textsf{CO}} (5/5) & \co{\textsf{CO}} (5/5) & \co{\textsf{CO}} (5/5) \\
\hline
160.0 & \cs{\textsf{CS}} (5/5) & \cs{\textsf{CS}} (5/5) & \co{\textsf{CO}} (5/5) & \co{\textsf{CO}} (5/5) & \co{\textsf{CO}} (5/5) & \co{\textsf{CO}} (5/5) & \co{\textsf{CO}} (5/5) & \co{\textsf{CO}} (5/5) & \co{\textsf{CO}} (5/5) & \co{\textsf{CO}} (5/5) & \co{\textsf{CO}} (5/5) & \co{\textsf{CO}} (5/5) & \co{\textsf{CO}} (5/5) & \co{\textsf{CO}} (5/5) & \co{\textsf{CO}} (5/5) \\
\hline
200.0 & \cs{\textsf{CS}} (5/5) & \cs{\textsf{CS}} (5/5) & \co{\textsf{CO}} (5/5) & \co{\textsf{CO}} (5/5) & \co{\textsf{CO}} (5/5) & \co{\textsf{CO}} (5/5) & \co{\textsf{CO}} (5/5) & \co{\textsf{CO}} (5/5) & \co{\textsf{CO}} (5/5) & \co{\textsf{CO}} (5/5) & \co{\textsf{CO}} (5/5) & \co{\textsf{CO}} (5/5) & \co{\textsf{CO}} (5/5) & \co{\textsf{CO}} (5/5) & \co{\textsf{CO}} (5/5) \\
\hline
240.0 & \cs{\textsf{CS}} (5/5) & \cs{\textsf{CS}} (5/5) & \co{\textsf{CO}} (5/5) & \co{\textsf{CO}} (5/5) & \co{\textsf{CO}} (5/5) & \co{\textsf{CO}} (5/5) & \co{\textsf{CO}} (5/5) & \co{\textsf{CO}} (5/5) & \co{\textsf{CO}} (5/5) & \co{\textsf{CO}} (5/5) & \co{\textsf{CO}} (5/5) & \co{\textsf{CO}} (5/5) & \co{\textsf{CO}} (5/5) & \co{\textsf{CO}} (5/5) & \co{\textsf{CO}} (5/5) \\
\hline
280.0 & \cs{\textsf{CS}} (5/5) & \cs{\textsf{CS}} (5/5) & \co{\textsf{CO}} (5/5) & \co{\textsf{CO}} (5/5) & \co{\textsf{CO}} (5/5) & \co{\textsf{CO}} (5/5) & \co{\textsf{CO}} (5/5) & \co{\textsf{CO}} (5/5) & \co{\textsf{CO}} (5/5) & \co{\textsf{CO}} (5/5) & \co{\textsf{CO}} (5/5) & \co{\textsf{CO}} (5/5) & \co{\textsf{CO}} (5/5) & \co{\textsf{CO}} (5/5) & \co{\textsf{CO}} (5/5) \\
\hline
320.0 & \cs{\textsf{CS}} (5/5) & \cs{\textsf{CS}} (5/5) & \co{\textsf{CO}} (5/5) & \co{\textsf{CO}} (5/5) & \co{\textsf{CO}} (5/5) & \co{\textsf{CO}} (5/5) & \co{\textsf{CO}} (5/5) & \co{\textsf{CO}} (5/5) & \co{\textsf{CO}} (5/5) & \co{\textsf{CO}} (5/5) & \co{\textsf{CO}} (5/5) & \co{\textsf{CO}} (5/5) & \co{\textsf{CO}} (5/5) & \co{\textsf{CO}} (5/5) & \co{\textsf{CO}} (5/5) \\
\hline
360.0 & \cs{\textsf{CS}} (5/5) & \cs{\textsf{CS}} (5/5) & \co{\textsf{CO}} (5/5) & \co{\textsf{CO}} (5/5) & \co{\textsf{CO}} (5/5) & \co{\textsf{CO}} (5/5) & \co{\textsf{CO}} (5/5) & \co{\textsf{CO}} (5/5) & \co{\textsf{CO}} (5/5) & \co{\textsf{CO}} (5/5) & \co{\textsf{CO}} (5/5) & \co{\textsf{CO}} (5/5) & \co{\textsf{CO}} (5/5) & \co{\textsf{CO}} (5/5) & \co{\textsf{CO}} (5/5) \\
\hline
400.0 & \cs{\textsf{CS}} (5/5) & \cs{\textsf{CS}} (5/5) & \co{\textsf{CO}} (5/5) & \co{\textsf{CO}} (5/5) & \co{\textsf{CO}} (5/5) & \co{\textsf{CO}} (5/5) & \co{\textsf{CO}} (5/5) & \co{\textsf{CO}} (5/5) & \co{\textsf{CO}} (5/5) & \co{\textsf{CO}} (5/5) & \co{\textsf{CO}} (5/5) & \co{\textsf{CO}} (5/5) & \co{\textsf{CO}} (5/5) & \co{\textsf{CO}} (5/5) & \co{\textsf{CO}} (5/5) \\
\hline
440.0 & \cs{\textsf{CS}} (5/5) & \cs{\textsf{CS}} (5/5) & \co{\textsf{CO}} (5/5) & \co{\textsf{CO}} (5/5) & \co{\textsf{CO}} (5/5) & \co{\textsf{CO}} (5/5) & \co{\textsf{CO}} (5/5) & \co{\textsf{CO}} (5/5) & \co{\textsf{CO}} (5/5) & \co{\textsf{CO}} (5/5) & \co{\textsf{CO}} (5/5) & \co{\textsf{CO}} (5/5) & \co{\textsf{CO}} (5/5) & \co{\textsf{CO}} (5/5) & \co{\textsf{CO}} (5/5) \\
\hline
480.0 & \cs{\textsf{CS}} (5/5) & \cs{\textsf{CS}} (5/5) & \co{\textsf{CO}} (5/5) & \co{\textsf{CO}} (5/5) & \co{\textsf{CO}} (5/5) & \co{\textsf{CO}} (5/5) & \co{\textsf{CO}} (5/5) & \co{\textsf{CO}} (5/5) & \co{\textsf{CO}} (5/5) & \co{\textsf{CO}} (5/5) & \co{\textsf{CO}} (5/5) & \co{\textsf{CO}} (5/5) & \co{\textsf{CO}} (5/5) & \co{\textsf{CO}} (5/5) & \co{\textsf{CO}} (5/5) \\
\hline
520.0 & \cs{\textsf{CS}} (5/5) & \cs{\textsf{CS}} (5/5) & \co{\textsf{CO}} (5/5) & \co{\textsf{CO}} (5/5) & \co{\textsf{CO}} (5/5) & \co{\textsf{CO}} (5/5) & \co{\textsf{CO}} (5/5) & \co{\textsf{CO}} (5/5) & \co{\textsf{CO}} (5/5) & \co{\textsf{CO}} (5/5) & \co{\textsf{CO}} (5/5) & \co{\textsf{CO}} (5/5) & \co{\textsf{CO}} (5/5) & \co{\textsf{CO}} (5/5) & \co{\textsf{CO}} (5/5) \\
\hline
\end{tabular}

%% file: tables/transfuser-crossing-0.tex
\begin{tabular}{|c| *{15}{c|}} 
\hline
\diagbox{$\bm{x_a}$}{$\bm{x_f}$} & 0.0 & \ct{1.9} & 40.0 & 80.0 & 120.0 & 160.0 & 200.0 & 240.0 & 280.0 & 320.0 & 360.0 & 400.0 & 440.0 & 480.0 & 520.0 \\
\hline
0.0& \begin{tabular}{@{}c@{}}\begin{minipage}[c][0.75cm][c]{1.8cm}\centering \ac{\textsf{Ae} (4/5)}\ \end{minipage} \\\begin{minipage}[c][0.75cm][c]{1.8cm}\centering \blk{\textsf{Blk} (1/5)}\ \end{minipage} \\\end{tabular}& \begin{tabular}{@{}c@{}}\begin{minipage}[c][0.75cm][c]{1.8cm}\centering \ac{\textsf{Ae} (4/5)}\ \end{minipage} \\\begin{minipage}[c][0.75cm][c]{1.8cm}\centering \blk{\textsf{Blk} (1/5)}\ \end{minipage} \\\end{tabular}& \begin{tabular}{@{}c@{}}\begin{minipage}[c][0.75cm][c]{1.8cm}\centering \ac{\textsf{Ae} (3/5)}\ \end{minipage} \\\begin{minipage}[c][0.75cm][c]{1.8cm}\centering \blk{\textsf{Blk} (2/5)}\ \end{minipage} \\\end{tabular}& \begin{tabular}{@{}c@{}}\begin{minipage}[c][0.5cm][c]{1.8cm}\centering \ac{\textsf{Ae} (1/5)}\ \end{minipage} \\\begin{minipage}[c][0.5cm][c]{1.8cm}\centering \blk{\textsf{Blk} (2/5)}\ \end{minipage} \\\begin{minipage}[c][0.5cm][c]{1.8cm}\centering \pu{\textsf{PU$p_1p_2$} (2/5)}\ \end{minipage} \\\end{tabular}& \begin{tabular}{@{}c@{}}\begin{minipage}[c][0.5cm][c]{1.8cm}\centering \ac{\textsf{Ae} (2/5)}\ \end{minipage} \\\begin{minipage}[c][0.5cm][c]{1.8cm}\centering \blk{\textsf{Blk} (2/5)}\ \end{minipage} \\\begin{minipage}[c][0.5cm][c]{1.8cm}\centering \pu{\textsf{PU$p_1p_2$} (1/5)}\ \end{minipage} \\\end{tabular}& \begin{tabular}{@{}c@{}}\begin{minipage}[c][0.75cm][c]{1.8cm}\centering \ac{\textsf{Ae} (3/5)}\ \end{minipage} \\\begin{minipage}[c][0.75cm][c]{1.8cm}\centering \blk{\textsf{Blk} (2/5)}\ \end{minipage} \\\end{tabular}& \begin{tabular}{@{}c@{}}\begin{minipage}[c][0.5cm][c]{1.8cm}\centering \ac{\textsf{Aa} (1/5)}\ \end{minipage} \\\begin{minipage}[c][0.5cm][c]{1.8cm}\centering \ac{\textsf{Ae} (3/5)}\ \end{minipage} \\\begin{minipage}[c][0.5cm][c]{1.8cm}\centering \pu{\textsf{PU$p_1p_2$} (1/5)}\ \end{minipage} \\\end{tabular}& \begin{tabular}{@{}c@{}}\begin{minipage}[c][0.75cm][c]{1.8cm}\centering \blk{\textsf{Blk} (4/5)}\ \end{minipage} \\\begin{minipage}[c][0.75cm][c]{1.8cm}\centering \pu{\textsf{PU$p_1p_2$} (1/5)}\ \end{minipage} \\\end{tabular}& \begin{tabular}{@{}c@{}}\begin{minipage}[c][0.5cm][c]{1.8cm}\centering \ac{\textsf{Ae} (2/5)}\ \end{minipage} \\\begin{minipage}[c][0.5cm][c]{1.8cm}\centering \blk{\textsf{Blk} (2/5)}\ \end{minipage} \\\begin{minipage}[c][0.5cm][c]{1.8cm}\centering \pu{\textsf{PU$p_1p_2$} (1/5)}\ \end{minipage} \\\end{tabular}& \begin{tabular}{@{}c@{}}\begin{minipage}[c][0.5cm][c]{1.8cm}\centering \blk{\textsf{Blk} (3/5)}\ \end{minipage} \\\begin{minipage}[c][0.5cm][c]{1.8cm}\centering \pu{\textsf{PU$p_1$} (1/5)}\ \end{minipage} \\\begin{minipage}[c][0.5cm][c]{1.8cm}\centering \pu{\textsf{PU$p_1p_2$} (1/5)}\ \end{minipage} \\\end{tabular}& \begin{tabular}{@{}c@{}}\begin{minipage}[c][0.75cm][c]{1.8cm}\centering \ac{\textsf{Ae} (1/5)}\ \end{minipage} \\\begin{minipage}[c][0.75cm][c]{1.8cm}\centering \blk{\textsf{Blk} (4/5)}\ \end{minipage} \\\end{tabular}& \begin{tabular}{@{}c@{}}\begin{minipage}[c][0.5cm][c]{1.8cm}\centering \ac{\textsf{Ae} (3/5)}\ \end{minipage} \\\begin{minipage}[c][0.5cm][c]{1.8cm}\centering \blk{\textsf{Blk} (1/5)}\ \end{minipage} \\\begin{minipage}[c][0.5cm][c]{1.8cm}\centering \pu{\textsf{PU$p_1p_2$} (1/5)}\ \end{minipage} \\\end{tabular}& \begin{tabular}{@{}c@{}}\begin{minipage}[c][0.5cm][c]{1.8cm}\centering \blk{\textsf{Blk} (2/5)}\ \end{minipage} \\\begin{minipage}[c][0.5cm][c]{1.8cm}\centering \pu{\textsf{PU$p_1$} (1/5)}\ \end{minipage} \\\begin{minipage}[c][0.5cm][c]{1.8cm}\centering \pu{\textsf{PU$p_1p_2$} (2/5)}\ \end{minipage} \\\end{tabular}& \begin{tabular}{@{}c@{}}\begin{minipage}[c][0.75cm][c]{1.8cm}\centering \ac{\textsf{Ae} (1/5)}\ \end{minipage} \\\begin{minipage}[c][0.75cm][c]{1.8cm}\centering \blk{\textsf{Blk} (4/5)}\ \end{minipage} \\\end{tabular}& \begin{tabular}{@{}c@{}}\begin{minipage}[c][0.5cm][c]{1.8cm}\centering \ac{\textsf{Ae} (2/5)}\ \end{minipage} \\\begin{minipage}[c][0.5cm][c]{1.8cm}\centering \blk{\textsf{Blk} (2/5)}\ \end{minipage} \\\begin{minipage}[c][0.5cm][c]{1.8cm}\centering \pu{\textsf{PU$p_1p_2$} (1/5)}\ \end{minipage} \\\end{tabular} \\
\hline
5.0 & \pu{\textsf{PU$p_1$} (5/5)} & \pu{\textsf{PU$p_1$} (5/5)} & \pu{\textsf{PU$p_1$} (5/5)} & \pu{\textsf{PU$p_1$} (5/5)} & \pu{\textsf{PU$p_1$} (5/5)} & \pu{\textsf{PU$p_1$} (5/5)} & \pu{\textsf{PU$p_1$} (5/5)} & \pu{\textsf{PU$p_1$} (5/5)} & \pu{\textsf{PU$p_1$} (5/5)} & \pu{\textsf{PU$p_1$} (5/5)} & \pu{\textsf{PU$p_1$} (5/5)} & \pu{\textsf{PU$p_1$} (5/5)} & \pu{\textsf{PU$p_1$} (5/5)} & \pu{\textsf{PU$p_1$} (5/5)} & \pu{\textsf{PU$p_1$} (5/5)} \\
\hline
10.0 & \begin{tabular}{@{}c@{}}\begin{minipage}[c][0.5cm][c]{1.8cm}\centering \pu{\textsf{PU$p_1$} (4/5)}\ \end{minipage} \\\begin{minipage}[c][0.5cm][c]{1.8cm}\centering \blk{\textsf{Blk} (1/5)}\ \end{minipage} \\\end{tabular}  & \pu{\textsf{PU$p_1$} (5/5)} & \pu{\textsf{PU$p_1$} (5/5)} & \pu{\textsf{PU$p_1$} (5/5)} & \pu{\textsf{PU$p_1$} (5/5)} & \pu{\textsf{PU$p_1$} (5/5)} & \pu{\textsf{PU$p_1$} (5/5)} & \pu{\textsf{PU$p_1$} (5/5)} & \pu{\textsf{PU$p_1$} (5/5)} & \pu{\textsf{PU$p_1$} (5/5)} & \pu{\textsf{PU$p_1$} (5/5)} & \pu{\textsf{PU$p_1$} (5/5)} & \pu{\textsf{PU$p_1$} (5/5)} & \pu{\textsf{PU$p_1$} (5/5)} & \pu{\textsf{PU$p_1$} (5/5)} \\
\hline
15.0 & \pu{\textsf{PU$p_1$} (5/5)} & \pu{\textsf{PU$p_1$} (5/5)} & \pu{\textsf{PU$p_1$} (5/5)} & \pu{\textsf{PU$p_1$} (5/5)} & \pu{\textsf{PU$p_1$} (5/5)} & \pu{\textsf{PU$p_1$} (5/5)} & \pu{\textsf{PU$p_1$} (5/5)} & \pu{\textsf{PU$p_1$} (5/5)} & \pu{\textsf{PU$p_1$} (5/5)} & \pu{\textsf{PU$p_1$} (5/5)} & \pu{\textsf{PU$p_1$} (5/5)} & \pu{\textsf{PU$p_1$} (5/5)} & \pu{\textsf{PU$p_1$} (5/5)} & \pu{\textsf{PU$p_1$} (5/5)} & \pu{\textsf{PU$p_1$} (5/5)} \\
\hline
20.0 & \pu{\textsf{PU$p_1$} (5/5)} & \pu{\textsf{PU$p_1$} (5/5)} & \pu{\textsf{PU$p_1$} (5/5)} & \pu{\textsf{PU$p_1$} (5/5)} & \pu{\textsf{PU$p_1$} (5/5)} & \pu{\textsf{PU$p_1$} (5/5)} & \pu{\textsf{PU$p_1$} (5/5)} & \pu{\textsf{PU$p_1$} (5/5)} & \pu{\textsf{PU$p_1$} (5/5)} & \pu{\textsf{PU$p_1$} (5/5)} & \pu{\textsf{PU$p_1$} (5/5)} & \pu{\textsf{PU$p_1$} (5/5)} & \pu{\textsf{PU$p_1$} (5/5)} & \pu{\textsf{PU$p_1$} (5/5)} & \pu{\textsf{PU$p_1$} (5/5)} \\
\hline
\ct{22.8} & \pu{\textsf{PU$p_1$} (5/5)} & \pu{\textsf{PU$p_1$} (5/5)} & \pu{\textsf{PU$p_1$} (5/5)} & \pu{\textsf{PU$p_1$} (5/5)} & \pu{\textsf{PU$p_1$} (5/5)} & \pu{\textsf{PU$p_1$} (5/5)} & \pu{\textsf{PU$p_1$} (5/5)} & \pu{\textsf{PU$p_1$} (5/5)} & \pu{\textsf{PU$p_1$} (5/5)} & \pu{\textsf{PU$p_1$} (5/5)} & \pu{\textsf{PU$p_1$} (5/5)} & \pu{\textsf{PU$p_1$} (5/5)} & \pu{\textsf{PU$p_1$} (5/5)} & \pu{\textsf{PU$p_1$} (5/5)} & \pu{\textsf{PU$p_1$} (5/5)} \\
\hline
25.0 & \begin{tabular}{@{}c@{}}\begin{minipage}[c][0.5cm][c]{1.8cm}\centering \pu{\textsf{PU$p_1$} (1/5)}\ \end{minipage} \\\begin{minipage}[c][0.5cm][c]{1.8cm}\centering \blk{\textsf{Blk} (4/5)}\ \end{minipage} \\\end{tabular}  & \pu{\textsf{PU$p_1$} (5/5)} & \pu{\textsf{PU$p_1$} (5/5)} & \pu{\textsf{PU$p_1$} (5/5)} & \pu{\textsf{PU$p_1$} (5/5)} & \pu{\textsf{PU$p_1$} (5/5)} & \pu{\textsf{PU$p_1$} (5/5)} & \pu{\textsf{PU$p_1$} (5/5)} & \pu{\textsf{PU$p_1$} (5/5)} & \pu{\textsf{PU$p_1$} (5/5)} & \pu{\textsf{PU$p_1$} (5/5)} & \pu{\textsf{PU$p_1$} (5/5)} & \pu{\textsf{PU$p_1$} (5/5)} & \pu{\textsf{PU$p_1$} (5/5)} & \pu{\textsf{PU$p_1$} (5/5)} \\
\hline
30.0& \begin{tabular}{@{}c@{}}\begin{minipage}[c][0.5cm][c]{1.8cm}\centering \pu{\textsf{PU$p_1$} (4/5)}\ \end{minipage} \\\begin{minipage}[c][0.5cm][c]{1.8cm}\centering \blk{\textsf{Blk} (1/5)}\ \end{minipage} \\\end{tabular} & \pu{\textsf{PU$p_1$} (5/5)} & \ps{\textsf{PS} (5/5)} & \ps{\textsf{PS} (5/5)} & \ps{\textsf{PS} (5/5)} & \ps{\textsf{PS} (5/5)} & \ps{\textsf{PS} (5/5)}& \begin{tabular}{@{}c@{}}\begin{minipage}[c][0.5cm][c]{1.8cm}\centering \pu{\textsf{PU$p_1$} (1/5)}\ \end{minipage} \\\begin{minipage}[c][0.5cm][c]{1.8cm}\centering \ps{\textsf{PS} (4/5)}\ \end{minipage} \\\end{tabular} & \ps{\textsf{PS} (5/5)}& \begin{tabular}{@{}c@{}}\begin{minipage}[c][0.5cm][c]{1.8cm}\centering \pu{\textsf{PU$p_1$} (1/5)}\ \end{minipage} \\\begin{minipage}[c][0.5cm][c]{1.8cm}\centering \ps{\textsf{PS} (4/5)}\ \end{minipage} \\\end{tabular} & \ps{\textsf{PS} (5/5)} & \ps{\textsf{PS} (5/5)} & \ps{\textsf{PS} (5/5)} & \ps{\textsf{PS} (5/5)} & \ps{\textsf{PS} (5/5)} \\ 
\hline 
35.0 & \ps{\textsf{PS} (5/5)} & \ps{\textsf{PS} (5/5)} & \ps{\textsf{PS} (5/5)} & \ps{\textsf{PS} (5/5)} & \ps{\textsf{PS} (5/5)} & \ps{\textsf{PS} (5/5)} & \ps{\textsf{PS} (5/5)} & \ps{\textsf{PS} (5/5)} & \ps{\textsf{PS} (5/5)} & \ps{\textsf{PS} (5/5)} & \ps{\textsf{PS} (5/5)} & \ps{\textsf{PS} (5/5)} & \ps{\textsf{PS} (5/5)} & \ps{\textsf{PS} (5/5)} & \ps{\textsf{PS} (5/5)} \\
\hline
40.0 & \ps{\textsf{PS} (5/5)} & \ps{\textsf{PS} (5/5)} & \ps{\textsf{PS} (5/5)} & \ps{\textsf{PS} (5/5)} & \ps{\textsf{PS} (5/5)} & \ps{\textsf{PS} (5/5)} & \ps{\textsf{PS} (5/5)} & \ps{\textsf{PS} (5/5)} & \ps{\textsf{PS} (5/5)} & \ps{\textsf{PS} (5/5)} & \ps{\textsf{PS} (5/5)} & \ps{\textsf{PS} (5/5)} & \ps{\textsf{PS} (5/5)} & \ps{\textsf{PS} (5/5)} & \ps{\textsf{PS} (5/5)} \\
\hline
\end{tabular}

%% file: tables/transfuser-crossing-4.tex
\begin{tabular}{|c| *{15}{c|}} 
\hline
\diagbox{$\bm{x_a}$}{$\bm{x_f}$} & 0.0 & \ct{1.9} & 40.0 & 80.0 & 120.0 & 160.0 & 200.0 & 240.0 & 280.0 & 320.0 & 360.0 & 400.0 & 440.0 & 480.0 & 520.0 \\
\hline
0.0& \begin{tabular}{@{}c@{}}\begin{minipage}[c][0.5cm][c]{1.8cm}\centering \pu{\textsf{PU$p_1$} (4/5)}\ \end{minipage} \\\begin{minipage}[c][0.5cm][c]{1.8cm}\centering \pu{\textsf{PU$p_1p_2$} (1/5)}\ \end{minipage} \\\end{tabular}& \begin{tabular}{@{}c@{}}\begin{minipage}[c][0.5cm][c]{1.8cm}\centering \pu{\textsf{PU$p_1$} (4/5)}\ \end{minipage} \\\begin{minipage}[c][0.5cm][c]{1.8cm}\centering \pu{\textsf{PU$p_1p_2$} (1/5)}\ \end{minipage} \\\end{tabular}& \begin{tabular}{@{}c@{}}\begin{minipage}[c][0.5cm][c]{1.8cm}\centering \pu{\textsf{PU$p_1$} (4/5)}\ \end{minipage} \\\begin{minipage}[c][0.5cm][c]{1.8cm}\centering \pu{\textsf{PU$p_1p_2$} (1/5)}\ \end{minipage} \\\end{tabular}& \begin{tabular}{@{}c@{}}\begin{minipage}[c][0.5cm][c]{1.8cm}\centering \pu{\textsf{PU$p_1$} (2/5)}\ \end{minipage} \\\begin{minipage}[c][0.5cm][c]{1.8cm}\centering \pu{\textsf{PU$p_1p_2$} (3/5)}\ \end{minipage} \\\end{tabular}& \begin{tabular}{@{}c@{}}\begin{minipage}[c][0.5cm][c]{1.8cm}\centering \pu{\textsf{PU$p_1p_2$} (3/5)}\ \end{minipage} \\\begin{minipage}[c][0.5cm][c]{1.8cm}\centering \pu{\textsf{PU$p_1$} (2/5)}\ \end{minipage} \\\end{tabular}& \begin{tabular}{@{}c@{}}\begin{minipage}[c][0.5cm][c]{1.8cm}\centering \pu{\textsf{PU$p_1p_2$} (3/5)}\ \end{minipage} \\\begin{minipage}[c][0.5cm][c]{1.8cm}\centering \pu{\textsf{PU$p_1$} (2/5)}\ \end{minipage} \\\end{tabular} & \pu{\textsf{PU$p_1p_2$} (5/5)}& \begin{tabular}{@{}c@{}}\begin{minipage}[c][0.5cm][c]{1.8cm}\centering \pu{\textsf{PU$p_1$} (2/5)}\ \end{minipage} \\\begin{minipage}[c][0.5cm][c]{1.8cm}\centering \pu{\textsf{PU$p_1p_2$} (3/5)}\ \end{minipage} \\\end{tabular} & \pu{\textsf{PU$p_1p_2$} (5/5)} & \pu{\textsf{PU$p_1p_2$} (5/5)}& \begin{tabular}{@{}c@{}}\begin{minipage}[c][0.5cm][c]{1.8cm}\centering \pu{\textsf{PU$p_1$} (3/5)}\ \end{minipage} \\\begin{minipage}[c][0.5cm][c]{1.8cm}\centering \pu{\textsf{PU$p_1p_2$} (2/5)}\ \end{minipage} \\\end{tabular}& \begin{tabular}{@{}c@{}}\begin{minipage}[c][0.5cm][c]{1.8cm}\centering \pu{\textsf{PU$p_1p_2$} (3/5)}\ \end{minipage} \\\begin{minipage}[c][0.5cm][c]{1.8cm}\centering \pu{\textsf{PU$p_1$} (2/5)}\ \end{minipage} \\\end{tabular}& \begin{tabular}{@{}c@{}}\begin{minipage}[c][0.5cm][c]{1.8cm}\centering \pu{\textsf{PU$p_1p_2$} (3/5)}\ \end{minipage} \\\begin{minipage}[c][0.5cm][c]{1.8cm}\centering \pu{\textsf{PU$p_1$} (2/5)}\ \end{minipage} \\\end{tabular}& \begin{tabular}{@{}c@{}}\begin{minipage}[c][0.5cm][c]{1.8cm}\centering \pu{\textsf{PU$p_1$} (4/5)}\ \end{minipage} \\\begin{minipage}[c][0.5cm][c]{1.8cm}\centering \pu{\textsf{PU$p_1p_2$} (1/5)}\ \end{minipage} \\\end{tabular}& \begin{tabular}{@{}c@{}}\begin{minipage}[c][0.5cm][c]{1.8cm}\centering \pu{\textsf{PU$p_1p_2$} (2/5)}\ \end{minipage} \\\begin{minipage}[c][0.5cm][c]{1.8cm}\centering \pu{\textsf{PU$p_1$} (3/5)}\ \end{minipage} \\\end{tabular} \\
\hline
5.0 & \pu{\textsf{PU$p_1$} (5/5)} & \pu{\textsf{PU$p_1$} (5/5)} & \pu{\textsf{PU$p_1$} (5/5)}& \begin{tabular}{@{}c@{}}\begin{minipage}[c][0.5cm][c]{1.8cm}\centering \pu{\textsf{PU$p_1$} (4/5)}\ \end{minipage} \\\begin{minipage}[c][0.5cm][c]{1.8cm}\centering \pu{\textsf{PU$p_1p_2$} (1/5)}\ \end{minipage} \\\end{tabular}& \begin{tabular}{@{}c@{}}\begin{minipage}[c][0.5cm][c]{1.8cm}\centering \pu{\textsf{PU$p_1$} (2/5)}\ \end{minipage} \\\begin{minipage}[c][0.5cm][c]{1.8cm}\centering \pu{\textsf{PU$p_1p_2$} (3/5)}\ \end{minipage} \\\end{tabular}& \begin{tabular}{@{}c@{}}\begin{minipage}[c][0.5cm][c]{1.8cm}\centering \pu{\textsf{PU$p_1p_2$} (3/5)}\ \end{minipage} \\\begin{minipage}[c][0.5cm][c]{1.8cm}\centering \pu{\textsf{PU$p_1$} (2/5)}\ \end{minipage} \\\end{tabular} & \pu{\textsf{PU$p_1$} (5/5)} & \pu{\textsf{PU$p_1p_2$} (5/5)} & \pu{\textsf{PU$p_1$} (5/5)} & \pu{\textsf{PU$p_1$} (5/5)} & \pu{\textsf{PU$p_1p_2$} (5/5)}& \begin{tabular}{@{}c@{}}\begin{minipage}[c][0.5cm][c]{1.8cm}\centering \pu{\textsf{PU$p_1$} (4/5)}\ \end{minipage} \\\begin{minipage}[c][0.5cm][c]{1.8cm}\centering \pu{\textsf{PU$p_1p_2$} (1/5)}\ \end{minipage} \\\end{tabular} & \pu{\textsf{PU$p_1$} (5/5)} & \pu{\textsf{PU$p_1$} (5/5)} & \pu{\textsf{PU$p_1$} (5/5)} \\
\hline
10.0 & \pu{\textsf{PU$p_1$} (5/5)} & \pu{\textsf{PU$p_1$} (5/5)} & \pu{\textsf{PU$p_1$} (5/5)} & \pu{\textsf{PU$p_1$} (5/5)} & \pu{\textsf{PU$p_1$} (5/5)} & \pu{\textsf{PU$p_1$} (5/5)} & \pu{\textsf{PU$p_1$} (5/5)} & \pu{\textsf{PU$p_1$} (5/5)} & \pu{\textsf{PU$p_1$} (5/5)} & \pu{\textsf{PU$p_1$} (5/5)} & \pu{\textsf{PU$p_1$} (5/5)} & \pu{\textsf{PU$p_1$} (5/5)} & \pu{\textsf{PU$p_1$} (5/5)} & \pu{\textsf{PU$p_1$} (5/5)} & \pu{\textsf{PU$p_1$} (5/5)} \\
\hline
15.0 & \pu{\textsf{PU$p_1$} (5/5)} & \pu{\textsf{PU$p_1$} (5/5)} & \pu{\textsf{PU$p_1$} (5/5)} & \pu{\textsf{PU$p_1$} (5/5)} & \pu{\textsf{PU$p_1$} (5/5)} & \pu{\textsf{PU$p_1$} (5/5)} & \pu{\textsf{PU$p_1$} (5/5)} & \pu{\textsf{PU$p_1$} (5/5)} & \pu{\textsf{PU$p_1$} (5/5)} & \pu{\textsf{PU$p_1$} (5/5)} & \pu{\textsf{PU$p_1$} (5/5)} & \pu{\textsf{PU$p_1$} (5/5)} & \pu{\textsf{PU$p_1$} (5/5)} & \pu{\textsf{PU$p_1$} (5/5)} & \pu{\textsf{PU$p_1$} (5/5)} \\
\hline
20.0 & \pu{\textsf{PU$p_1$} (5/5)} & \pu{\textsf{PU$p_1$} (5/5)} & \pu{\textsf{PU$p_1$} (5/5)} & \pu{\textsf{PU$p_1$} (5/5)} & \pu{\textsf{PU$p_1$} (5/5)} & \pu{\textsf{PU$p_1$} (5/5)} & \pu{\textsf{PU$p_1$} (5/5)} & \pu{\textsf{PU$p_1$} (5/5)} & \pu{\textsf{PU$p_1$} (5/5)} & \pu{\textsf{PU$p_1$} (5/5)} & \pu{\textsf{PU$p_1$} (5/5)} & \pu{\textsf{PU$p_1$} (5/5)} & \pu{\textsf{PU$p_1$} (5/5)} & \pu{\textsf{PU$p_1$} (5/5)} & \pu{\textsf{PU$p_1$} (5/5)} \\
\hline
\ct{21.9} & \pu{\textsf{PU$p_1$} (5/5)} & \pu{\textsf{PU$p_1$} (5/5)} & \pu{\textsf{PU$p_1$} (5/5)} & \pu{\textsf{PU$p_1$} (5/5)} & \pu{\textsf{PU$p_1$} (5/5)} & \pu{\textsf{PU$p_1$} (5/5)} & \pu{\textsf{PU$p_1$} (5/5)} & \pu{\textsf{PU$p_1$} (5/5)} & \pu{\textsf{PU$p_1$} (5/5)} & \pu{\textsf{PU$p_1$} (5/5)} & \pu{\textsf{PU$p_1$} (5/5)} & \pu{\textsf{PU$p_1$} (5/5)} & \pu{\textsf{PU$p_1$} (5/5)} & \pu{\textsf{PU$p_1$} (5/5)} & \pu{\textsf{PU$p_1$} (5/5)} \\
\hline
25.0 & \pu{\textsf{PU$p_1$} (5/5)} & \pu{\textsf{PU$p_1$} (5/5)} & \pu{\textsf{PU$p_1$} (5/5)} & \pu{\textsf{PU$p_1$} (5/5)} & \pu{\textsf{PU$p_1$} (5/5)} & \pu{\textsf{PU$p_1$} (5/5)} & \pu{\textsf{PU$p_1$} (5/5)} & \pu{\textsf{PU$p_1$} (5/5)} & \pu{\textsf{PU$p_1$} (5/5)} & \pu{\textsf{PU$p_1$} (5/5)} & \pu{\textsf{PU$p_1$} (5/5)} & \pu{\textsf{PU$p_1$} (5/5)} & \pu{\textsf{PU$p_1$} (5/5)} & \pu{\textsf{PU$p_1$} (5/5)} & \pu{\textsf{PU$p_1$} (5/5)} \\
\hline
30.0& \begin{tabular}{@{}c@{}}\begin{minipage}[c][0.5cm][c]{1.8cm}\centering \ps{\textsf{PS} (1/5)}\ \end{minipage} \\\begin{minipage}[c][0.5cm][c]{1.8cm}\centering \pu{\textsf{PU$p_1$} (4/5)}\ \end{minipage} \\\end{tabular}& \begin{tabular}{@{}c@{}}\begin{minipage}[c][0.5cm][c]{1.8cm}\centering \ps{\textsf{PS} (1/5)}\ \end{minipage} \\\begin{minipage}[c][0.5cm][c]{1.8cm}\centering \pu{\textsf{PU$p_1$} (4/5)}\ \end{minipage} \\\end{tabular} & \ps{\textsf{PS} (5/5)} & \ps{\textsf{PS} (5/5)} & \ps{\textsf{PS} (5/5)} & \ps{\textsf{PS} (5/5)} & \ps{\textsf{PS} (5/5)} & \ps{\textsf{PS} (5/5)} & \ps{\textsf{PS} (5/5)} & \ps{\textsf{PS} (5/5)} & \ps{\textsf{PS} (5/5)} & \ps{\textsf{PS} (5/5)} & \ps{\textsf{PS} (5/5)} & \ps{\textsf{PS} (5/5)} & \ps{\textsf{PS} (5/5)} \\
\hline
35.0& \begin{tabular}{@{}c@{}}\begin{minipage}[c][0.5cm][c]{1.8cm}\centering \ps{\textsf{PS} (4/5)}\ \end{minipage} \\\begin{minipage}[c][0.5cm][c]{1.8cm}\centering \pu{\textsf{PU$p_1$} (1/5)}\ \end{minipage} \\\end{tabular}& \begin{tabular}{@{}c@{}}\begin{minipage}[c][0.5cm][c]{1.8cm}\centering \ps{\textsf{PS} (4/5)}\ \end{minipage} \\\begin{minipage}[c][0.5cm][c]{1.8cm}\centering \pu{\textsf{PU$p_1$} (1/5)}\ \end{minipage} \\\end{tabular} & \ps{\textsf{PS} (5/5)} & \ps{\textsf{PS} (5/5)} & \ps{\textsf{PS} (5/5)} & \ps{\textsf{PS} (5/5)} & \ps{\textsf{PS} (5/5)} & \ps{\textsf{PS} (5/5)} & \ps{\textsf{PS} (5/5)} & \ps{\textsf{PS} (5/5)} & \ps{\textsf{PS} (5/5)} & \ps{\textsf{PS} (5/5)} & \ps{\textsf{PS} (5/5)} & \ps{\textsf{PS} (5/5)} & \ps{\textsf{PS} (5/5)} \\
\hline
40.0 & \ps{\textsf{PS} (5/5)} & \ps{\textsf{PS} (5/5)} & \ps{\textsf{PS} (5/5)} & \ps{\textsf{PS} (5/5)} & \ps{\textsf{PS} (5/5)} & \ps{\textsf{PS} (5/5)} & \ps{\textsf{PS} (5/5)} & \ps{\textsf{PS} (5/5)} & \ps{\textsf{PS} (5/5)} & \ps{\textsf{PS} (5/5)} & \ps{\textsf{PS} (5/5)} & \ps{\textsf{PS} (5/5)} & \ps{\textsf{PS} (5/5)} & \ps{\textsf{PS} (5/5)} & \ps{\textsf{PS} (5/5)} \\
\hline
\end{tabular}

%% file: tables/transfuser-traffic_light-0.tex
\begin{tabular}{|c|c|c|c|c|c|c|c|c|c|c|c|c|c|}
\hline
\multicolumn{14}{|c|}{$\bm{xf}$}\\
\hline
0.0 & 40.0 & 80.0 & 120.0 & 160.0 & 200.0 & 240.0 & 280.0 & 320.0 & 360.0 & 400.0 & 440.0 & 480.0 & 520.0 \\
\hline
\cs{\textsf{CS (5/5)}} & \cs{\textsf{CS (5/5)}} & \cs{\textsf{CS (5/5)}} & \cs{\textsf{CS (5/5)}} & \cs{\textsf{CS (5/5)}} & \cs{\textsf{CS (5/5)}} & \cs{\textsf{CS (5/5)}} & \cs{\textsf{CS (5/5)}} & \cs{\textsf{CS (5/5)}} & \cs{\textsf{CS (5/5)}} & \cs{\textsf{CS (5/5)}} & \cs{\textsf{CS (5/5)}} & \cs{\textsf{CS (5/5)}} & \cs{\textsf{CS (5/5)}} \\
\hline
\end{tabular}

%% file: tables/transfuser-traffic_light-4.tex
\begin{tabular}{|c|c|c|c|c|c|c|c|c|c|c|c|c|c|}
\hline
\multicolumn{14}{|c|}{$\bm{xf}$}\\
\hline
0.0 & 40.0 & 80.0 & 120.0 & 160.0 & 200.0 & 240.0 & 280.0 & 320.0 & 360.0 & 400.0 & 440.0 & 480.0 & 520.0 \\
\hline
\cs{\textsf{CS (5/5)}} & \cs{\textsf{CS (5/5)}} & \cs{\textsf{CS (5/5)}} & \cs{\textsf{CS (5/5)}} & \cs{\textsf{CS (5/5)}} & \cs{\textsf{CS (5/5)}} & \cs{\textsf{CS (5/5)}} & \cs{\textsf{CS (5/5)}} & \cs{\textsf{CS (5/5)}} & \cs{\textsf{CS (5/5)}} & \cs{\textsf{CS (5/5)}} & \cs{\textsf{CS (5/5)}} & \cs{\textsf{CS (5/5)}} & \cs{\textsf{CS (5/5)}} \\
\hline
\end{tabular}

%% file: tables/lmdrive-merging-0.tex
\begin{tabular}{|c| *{14}{c|}} 
\hline 
\diagbox{$\bm{x_a}$}{$\bm{x_f}$} & \ct{0.0} & 40.0 & 80.0 & 120.0 & 160.0 & 200.0 & 240.0 & 280.0 & 320.0 & 360.0 & 400.0 & 440.0 & 480.0 & 520.0 \\ 
\hline 
0.0 & - & \cs{\textsf{CS} (5/5)} & \cs{\textsf{CS} (5/5)} & \cs{\textsf{CS} (5/5)} & \cs{\textsf{CS} (5/5)} & \cs{\textsf{CS} (5/5)} & \cs{\textsf{CS} (5/5)} & \cs{\textsf{CS} (5/5)} & \cs{\textsf{CS} (5/5)}& \begin{tabular}{@{}c@{}}\begin{minipage}[c][0.75cm][c]{1.8cm}\centering \cs{\textsf{CS} (4/5)}\ \end{minipage} \\\begin{minipage}[c][0.75cm][c]{1.8cm}\centering \ro{\textsf{DRe} (1/5)}\ \end{minipage} \\\end{tabular}& \begin{tabular}{@{}c@{}}\begin{minipage}[c][0.75cm][c]{1.8cm}\centering \cs{\textsf{CS} (3/5)}\ \end{minipage} \\\begin{minipage}[c][0.75cm][c]{1.8cm}\centering \ro{\textsf{DRe} (2/5)}\ \end{minipage} \\\end{tabular}& \begin{tabular}{@{}c@{}}\begin{minipage}[c][0.5cm][c]{1.8cm}\centering \cs{\textsf{CS} (3/5)}\ \end{minipage} \\\begin{minipage}[c][0.5cm][c]{1.8cm}\centering \ro{\textsf{DRe} (1/5)}\ \end{minipage} \\\begin{minipage}[c][0.5cm][c]{1.8cm}\centering \ro{\textsf{DRa} (1/5)}\ \end{minipage} \\\end{tabular}& \begin{tabular}{@{}c@{}}\begin{minipage}[c][0.75cm][c]{1.8cm}\centering \cs{\textsf{CS} (4/5)}\ \end{minipage} \\\begin{minipage}[c][0.75cm][c]{1.8cm}\centering \ro{\textsf{DRa} (1/5)}\ \end{minipage} \\\end{tabular} & \cs{\textsf{CS} (5/5)} \\ 
\hline 
\ct{2.9} & \cs{\textsf{CS} (5/5)} & \cs{\textsf{CS} (5/5)} & \cs{\textsf{CS} (5/5)} & \cs{\textsf{CS} (5/5)}& \begin{tabular}{@{}c@{}}\begin{minipage}[c][0.75cm][c]{1.8cm}\centering \cs{\textsf{CS} (4/5)}\ \end{minipage} \\\begin{minipage}[c][0.75cm][c]{1.8cm}\centering \ro{\textsf{DRe} (1/5)}\ \end{minipage} \\\end{tabular} & \cs{\textsf{CS} (5/5)} & \cs{\textsf{CS} (5/5)} & \cs{\textsf{CS} (5/5)} & \cs{\textsf{CS} (5/5)}& \begin{tabular}{@{}c@{}}\begin{minipage}[c][0.75cm][c]{1.8cm}\centering \cs{\textsf{CS} (4/5)}\ \end{minipage} \\\begin{minipage}[c][0.75cm][c]{1.8cm}\centering \ro{\textsf{DRe} (1/5)}\ \end{minipage} \\\end{tabular}& \begin{tabular}{@{}c@{}}\begin{minipage}[c][0.75cm][c]{1.8cm}\centering \cs{\textsf{CS} (4/5)}\ \end{minipage} \\\begin{minipage}[c][0.75cm][c]{1.8cm}\centering \ro{\textsf{DRe} (1/5)}\ \end{minipage} \\\end{tabular} & \cs{\textsf{CS} (5/5)} & \cs{\textsf{CS} (5/5)}& \begin{tabular}{@{}c@{}}\begin{minipage}[c][0.5cm][c]{1.8cm}\centering \ro{\textsf{DRe} (1/5)}\ \end{minipage} \\\begin{minipage}[c][0.5cm][c]{1.8cm}\centering \ro{\textsf{DRa} (1/5)}\ \end{minipage} \\\begin{minipage}[c][0.5cm][c]{1.8cm}\centering \cs{\textsf{CS} (3/5)}\ \end{minipage} \\\end{tabular} \\ 
\hline
5.0 & \cs{\textsf{CS} (5/5)} & \cs{\textsf{CS} (5/5)} & \cs{\textsf{CS} (5/5)} & \cs{\textsf{CS} (5/5)} & \cs{\textsf{CS} (5/5)} & \cs{\textsf{CS} (5/5)} & \cs{\textsf{CS} (5/5)} & \cs{\textsf{CS} (5/5)}& \begin{tabular}{@{}c@{}}\begin{minipage}[c][0.5cm][c]{1.8cm}\centering \ro{\textsf{DRe} (1/5)}\ \end{minipage} \\\begin{minipage}[c][0.5cm][c]{1.8cm}\centering \cs{\textsf{CS} (4/5)}\ \end{minipage} \\\end{tabular} & \cs{\textsf{CS} (5/5)}& \begin{tabular}{@{}c@{}}\begin{minipage}[c][0.5cm][c]{1.8cm}\centering \cs{\textsf{CS} (4/5)}\ \end{minipage} \\\begin{minipage}[c][0.5cm][c]{1.8cm}\centering \ro{\textsf{DRa} (1/5)}\ \end{minipage} \\\end{tabular}& \begin{tabular}{@{}c@{}}\begin{minipage}[c][0.5cm][c]{1.8cm}\centering \ro{\textsf{DRa} (1/5)}\ \end{minipage} \\\begin{minipage}[c][0.5cm][c]{1.8cm}\centering \cs{\textsf{CS} (4/5)}\ \end{minipage} \\\end{tabular} & \cs{\textsf{CS} (5/5)}& \begin{tabular}{@{}c@{}}\begin{minipage}[c][0.5cm][c]{1.8cm}\centering \cs{\textsf{CS} (3/5)}\ \end{minipage} \\\begin{minipage}[c][0.5cm][c]{1.8cm}\centering \ro{\textsf{DRe} (2/5)}\ \end{minipage} \\\end{tabular} \\ 
\hline 
10.0 & \cs{\textsf{CS} (5/5)} & \cs{\textsf{CS} (5/5)} & \cs{\textsf{CS} (5/5)} & \cs{\textsf{CS} (5/5)} & \cs{\textsf{CS} (5/5)} & \cs{\textsf{CS} (5/5)} & \cs{\textsf{CS} (5/5)}& \begin{tabular}{@{}c@{}}\begin{minipage}[c][0.5cm][c]{1.8cm}\centering \cs{\textsf{CS} (4/6)}\ \end{minipage} \\\begin{minipage}[c][0.5cm][c]{1.8cm}\centering \ro{\textsf{DRe} (2/6)}\ \end{minipage} \\\end{tabular} & \cs{\textsf{CS} (5/5)}& \begin{tabular}{@{}c@{}}\begin{minipage}[c][0.5cm][c]{1.8cm}\centering \cs{\textsf{CS} (4/6)}\ \end{minipage} \\\begin{minipage}[c][0.5cm][c]{1.8cm}\centering \ro{\textsf{DRe} (2/6)}\ \end{minipage} \\\end{tabular}& \begin{tabular}{@{}c@{}}\begin{minipage}[c][0.5cm][c]{1.8cm}\centering \cs{\textsf{CS} (4/5)}\ \end{minipage} \\\begin{minipage}[c][0.5cm][c]{1.8cm}\centering \ro{\textsf{DRe} (1/5)}\ \end{minipage} \\\end{tabular}& \begin{tabular}{@{}c@{}}\begin{minipage}[c][0.5cm][c]{1.8cm}\centering \cs{\textsf{CS} (4/5)}\ \end{minipage} \\\begin{minipage}[c][0.5cm][c]{1.8cm}\centering \ro{\textsf{DRe} (1/5)}\ \end{minipage} \\\end{tabular} & \cs{\textsf{CS} (5/5)}& \begin{tabular}{@{}c@{}}\begin{minipage}[c][0.5cm][c]{1.8cm}\centering \cs{\textsf{CS} (4/5)}\ \end{minipage} \\\begin{minipage}[c][0.5cm][c]{1.8cm}\centering \ro{\textsf{DRe} (1/5)}\ \end{minipage} \\\end{tabular} \\ 
\hline 
15.0 & \cs{\textsf{CS} (5/5)} & \cs{\textsf{CS} (5/5)} & \cs{\textsf{CS} (5/5)} & \cs{\textsf{CS} (5/5)} & \cs{\textsf{CS} (5/5)} & \cs{\textsf{CS} (5/5)} & \cs{\textsf{CS} (5/5)} & \cs{\textsf{CS} (5/5)}& \begin{tabular}{@{}c@{}}\begin{minipage}[c][0.5cm][c]{1.8cm}\centering \cs{\textsf{CS} (4/5)}\ \end{minipage} \\\begin{minipage}[c][0.5cm][c]{1.8cm}\centering \ro{\textsf{DRe} (1/5)}\ \end{minipage} \\\end{tabular} & \cs{\textsf{CS} (5/5)}& \begin{tabular}{@{}c@{}}\begin{minipage}[c][0.5cm][c]{1.8cm}\centering \cs{\textsf{CS} (3/5)}\ \end{minipage} \\\begin{minipage}[c][0.5cm][c]{1.8cm}\centering \ro{\textsf{DRa} (2/5)}\ \end{minipage} \\\end{tabular} & \cs{\textsf{CS} (5/5)}& \begin{tabular}{@{}c@{}}\begin{minipage}[c][0.5cm][c]{1.8cm}\centering \cs{\textsf{CS} (4/5)}\ \end{minipage} \\\begin{minipage}[c][0.5cm][c]{1.8cm}\centering \ro{\textsf{DRa} (2/5)}\ \end{minipage} \\\end{tabular}& \begin{tabular}{@{}c@{}}\begin{minipage}[c][0.5cm][c]{1.8cm}\centering \cs{\textsf{CS} (4/5)}\ \end{minipage} \\\begin{minipage}[c][0.5cm][c]{1.8cm}\centering \ro{\textsf{DRe} (1/5)}\ \end{minipage} \\\end{tabular} \\ 
\hline 
20.0 & \cs{\textsf{CS} (5/5)} & \cs{\textsf{CS} (5/5)} & \cs{\textsf{CS} (5/5)} & \cs{\textsf{CS} (5/5)} & \cs{\textsf{CS} (5/5)} & \cs{\textsf{CS} (5/5)} & \cs{\textsf{CS} (5/5)} & \cs{\textsf{CS} (5/5)}& \begin{tabular}{@{}c@{}}\begin{minipage}[c][0.75cm][c]{1.8cm}\centering \ro{\textsf{DRe} (1/5)}\ \end{minipage} \\\begin{minipage}[c][0.75cm][c]{1.8cm}\centering \cs{\textsf{CS} (4/5)}\ \end{minipage} \\\end{tabular}& \begin{tabular}{@{}c@{}}\begin{minipage}[c][0.75cm][c]{1.8cm}\centering \cs{\textsf{CS} (4/5)}\ \end{minipage} \\\begin{minipage}[c][0.75cm][c]{1.8cm}\centering \ro{\textsf{DRe} (1/5)}\ \end{minipage} \\\end{tabular}& \begin{tabular}{@{}c@{}}\begin{minipage}[c][0.75cm][c]{1.8cm}\centering \cs{\textsf{CS} (4/5)}\ \end{minipage} \\\begin{minipage}[c][0.75cm][c]{1.8cm}\centering \ro{\textsf{DRe} (1/5)}\ \end{minipage} \\\end{tabular}& \begin{tabular}{@{}c@{}}\begin{minipage}[c][0.5cm][c]{1.8cm}\centering \ro{\textsf{DRe} (1/5)}\ \end{minipage} \\\begin{minipage}[c][0.5cm][c]{1.8cm}\centering \ro{\textsf{DRa} (2/5)}\ \end{minipage} \\\begin{minipage}[c][0.5cm][c]{1.8cm}\centering \cs{\textsf{CS} (2/5)}\ \end{minipage} \\\end{tabular} & \cs{\textsf{CS} (5/5)} & \cs{\textsf{CS} (5/5)} \\ 
\hline 
25.0 & \cs{\textsf{CS} (5/5)} & \cs{\textsf{CS} (5/5)}& \begin{tabular}{@{}c@{}}\begin{minipage}[c][0.5cm][c]{1.8cm}\centering \cs{\textsf{CS} (3/5)}\ \end{minipage} \\\begin{minipage}[c][0.5cm][c]{1.8cm}\centering \ac{\textsf{Ae} (1/5)}\ \end{minipage} \\\begin{minipage}[c][0.5cm][c]{1.8cm}\centering \ps{\textsf{PS} (1/5)}\ \end{minipage} \\\end{tabular}& \begin{tabular}{@{}c@{}}\begin{minipage}[c][0.75cm][c]{1.8cm}\centering \cs{\textsf{CS} (4/5)}\ \end{minipage} \\\begin{minipage}[c][0.75cm][c]{1.8cm}\centering \ac{\textsf{Aa} (1/5)}\ \end{minipage} \\\end{tabular}& \begin{tabular}{@{}c@{}}\begin{minipage}[c][0.75cm][c]{1.8cm}\centering \cs{\textsf{CS} (4/5)}\ \end{minipage} \\\begin{minipage}[c][0.75cm][c]{1.8cm}\centering \ro{\textsf{DRe} (1/5)}\ \end{minipage} \\\end{tabular}& \begin{tabular}{@{}c@{}}\begin{minipage}[c][0.75cm][c]{1.8cm}\centering \ps{\textsf{PS} (1/5)}\ \end{minipage} \\\begin{minipage}[c][0.75cm][c]{1.8cm}\centering \cs{\textsf{CS} (4/5)}\ \end{minipage} \\\end{tabular}& \begin{tabular}{@{}c@{}}\begin{minipage}[c][0.5cm][c]{1.8cm}\centering \ac{\textsf{Aa} (1/5)}\ \end{minipage} \\\begin{minipage}[c][0.5cm][c]{1.8cm}\centering \cs{\textsf{CS} (3/5)}\ \end{minipage} \\\begin{minipage}[c][0.5cm][c]{1.8cm}\centering \ro{\textsf{DRe} (1/5)}\ \end{minipage} \\\end{tabular}& \begin{tabular}{@{}c@{}}\begin{minipage}[c][0.75cm][c]{1.8cm}\centering \cs{\textsf{CS} (4/5)}\ \end{minipage} \\\begin{minipage}[c][0.75cm][c]{1.8cm}\centering \ac{\textsf{Aa} (1/5)}\ \end{minipage} \\\end{tabular}& \begin{tabular}{@{}c@{}}\begin{minipage}[c][0.75cm][c]{1.8cm}\centering \cs{\textsf{CS} (4/5)}\ \end{minipage} \\\begin{minipage}[c][0.75cm][c]{1.8cm}\centering \ps{\textsf{PS} (1/5)}\ \end{minipage} \\\end{tabular}& \begin{tabular}{@{}c@{}}\begin{minipage}[c][0.75cm][c]{1.8cm}\centering \cs{\textsf{CS} (4/5)}\ \end{minipage} \\\begin{minipage}[c][0.75cm][c]{1.8cm}\centering \ps{\textsf{PS} (1/5)}\ \end{minipage} \\\end{tabular}& \begin{tabular}{@{}c@{}}\begin{minipage}[c][0.5cm][c]{1.8cm}\centering \cs{\textsf{CS} (3/5)}\ \end{minipage} \\\begin{minipage}[c][0.5cm][c]{1.8cm}\centering \ac{\textsf{Aa} (1/5)}\ \end{minipage} \\\begin{minipage}[c][0.5cm][c]{1.8cm}\centering \ro{\textsf{DRe} (1/5)}\ \end{minipage} \\\end{tabular}& \begin{tabular}{@{}c@{}}\begin{minipage}[c][0.75cm][c]{1.8cm}\centering \ps{\textsf{PS} (1/5)}\ \end{minipage} \\\begin{minipage}[c][0.75cm][c]{1.8cm}\centering \cs{\textsf{CS} (4/5)}\ \end{minipage} \\\end{tabular}& \begin{tabular}{@{}c@{}}\begin{minipage}[c][0.5cm][c]{1.8cm}\centering \cs{\textsf{CS} (3/5)}\ \end{minipage} \\\begin{minipage}[c][0.5cm][c]{1.8cm}\centering \ps{\textsf{PS} (1/5)}\ \end{minipage} \\\begin{minipage}[c][0.5cm][c]{1.8cm}\centering \ac{\textsf{Aa} (1/5)}\ \end{minipage} \\\end{tabular}& \begin{tabular}{@{}c@{}}\begin{minipage}[c][0.5cm][c]{1.8cm}\centering \ps{\textsf{PS} (3/5)}\ \end{minipage} \\\begin{minipage}[c][0.5cm][c]{1.8cm}\centering \ac{\textsf{Aa} (1/5)}\ \end{minipage} \\\begin{minipage}[c][0.5cm][c]{1.8cm}\centering \cs{\textsf{CS} (1/5)}\ \end{minipage} \\\end{tabular} \\ 
\hline 
30.0 & \cs{\textsf{CS} (5/5)}& \begin{tabular}{@{}c@{}}\begin{minipage}[c][1.0cm][c]{1.2cm}\centering \ps{\textsf{PS} (3/5)}\ \end{minipage} \\\begin{minipage}[c][1.0cm][c]{1.2cm}\centering \ac{\textsf{Aa} (2/5)}\ \end{minipage} \\\end{tabular}& \begin{tabular}{@{}c@{}}\begin{minipage}[c][1.0cm][c]{1.8cm}\centering \ps{\textsf{PS} (3/5)}\ \end{minipage} \\\begin{minipage}[c][1.0cm][c]{1.8cm}\centering \ac{\textsf{Aa} (2/5)}\ \end{minipage} \\\end{tabular}& \begin{tabular}{@{}c@{}}\begin{minipage}[c][1.0cm][c]{1.8cm}\centering \ps{\textsf{PS} (2/5)}\ \end{minipage} \\\begin{minipage}[c][1.0cm][c]{1.8cm}\centering \cs{\textsf{CS} (3/5)}\ \end{minipage} \\\end{tabular} & \cs{\textsf{CS} (5/5)}& \begin{tabular}{@{}c@{}}\begin{minipage}[c][1.0cm][c]{1.8cm}\centering \cs{\textsf{CS} (4/5)}\ \end{minipage} \\\begin{minipage}[c][1.0cm][c]{1.8cm}\centering \ps{\textsf{PS} (1/5)}\ \end{minipage} \\\end{tabular}& \begin{tabular}{@{}c@{}}\begin{minipage}[c][1.0cm][c]{1.8cm}\centering \cs{\textsf{CS} (4/5)}\ \end{minipage} \\\begin{minipage}[c][1.0cm][c]{1.8cm}\centering \ps{\textsf{PS} (1/5)}\ \end{minipage} \\\end{tabular}& \begin{tabular}{@{}c@{}}\begin{minipage}[c][0.5cm][c]{1.8cm}\centering \ro{\textsf{DRe} (1/5)}\ \end{minipage} \\\begin{minipage}[c][0.5cm][c]{1.8cm}\centering \ps{\textsf{PS} (1/5)}\ \end{minipage} \\\begin{minipage}[c][0.5cm][c]{1.8cm}\centering \cs{\textsf{CS} (2/5)}\ \end{minipage} \\\begin{minipage}[c][0.5cm][c]{1.8cm}\centering \ac{\textsf{Aa} (1/5)}\ \end{minipage} \\\end{tabular}& \begin{tabular}{@{}c@{}}\begin{minipage}[c][0.6666666666666666cm][c]{1.8cm}\centering \cs{\textsf{CS} (3/5)}\ \end{minipage} \\\begin{minipage}[c][0.6666666666666666cm][c]{1.8cm}\centering \ps{\textsf{PS} (1/5)}\ \end{minipage} \\\begin{minipage}[c][0.6666666666666666cm][c]{1.8cm}\centering \ac{\textsf{Ae} (1/5)}\ \end{minipage} \\\end{tabular}& \begin{tabular}{@{}c@{}}\begin{minipage}[c][0.5cm][c]{1.8cm}\centering \ro{\textsf{DRe} (1/5)}\ \end{minipage} \\\begin{minipage}[c][0.5cm][c]{1.8cm}\centering \cs{\textsf{CS} (2/5)}\ \end{minipage} \\\begin{minipage}[c][0.5cm][c]{1.8cm}\centering \ac{\textsf{Aa} (1/5)}\ \end{minipage} \\\begin{minipage}[c][0.5cm][c]{1.8cm}\centering \ps{\textsf{PS} (1/5)}\ \end{minipage} \\\end{tabular}& \begin{tabular}{@{}c@{}}\begin{minipage}[c][0.6666666666666666cm][c]{1.8cm}\centering \cs{\textsf{CS} (3/5)}\ \end{minipage} \\\begin{minipage}[c][0.6666666666666666cm][c]{1.8cm}\centering \ps{\textsf{PS} (1/5)}\ \end{minipage} \\\begin{minipage}[c][0.6666666666666666cm][c]{1.8cm}\centering \ro{\textsf{DRe} (1/5)}\ \end{minipage} \\\end{tabular}& \begin{tabular}{@{}c@{}}\begin{minipage}[c][1.0cm][c]{1.8cm}\centering \cs{\textsf{CS} (4/5)}\ \end{minipage} \\\begin{minipage}[c][1.0cm][c]{1.8cm}\centering \ac{\textsf{Ae} (1/5)}\ \end{minipage} \\\end{tabular}& \begin{tabular}{@{}c@{}}\begin{minipage}[c][0.6666666666666666cm][c]{1.8cm}\centering \ro{\textsf{DRe} (1/5)}\ \end{minipage} \\\begin{minipage}[c][0.6666666666666666cm][c]{1.8cm}\centering \cs{\textsf{CS} (3/5)}\ \end{minipage} \\\begin{minipage}[c][0.6666666666666666cm][c]{1.8cm}\centering \ac{\textsf{Ae} (1/5)}\ \end{minipage} \\\end{tabular}& \begin{tabular}{@{}c@{}}\begin{minipage}[c][0.6666666666666666cm][c]{1.8cm}\centering \ro{\textsf{DRe} (3/5)}\ \end{minipage} \\\begin{minipage}[c][0.6666666666666666cm][c]{1.8cm}\centering \ps{\textsf{PS} (1/5)}\ \end{minipage} \\\begin{minipage}[c][0.6666666666666666cm][c]{1.8cm}\centering \cs{\textsf{CS} (1/5)}\ \end{minipage} \\\end{tabular} \\ 
\hline 
35.0 & \cs{\textsf{CS} (5/5)} & \ps{\textsf{PS} (5/5)} & \ps{\textsf{PS} (5/5)}& \begin{tabular}{@{}c@{}}\begin{minipage}[c][0.5cm][c]{1.8cm}\centering \ps{\textsf{PS} (4/5)}\ \end{minipage} \\\begin{minipage}[c][0.5cm][c]{1.8cm}\centering \ro{\textsf{DRe} (1/5)}\ \end{minipage} \\\end{tabular}& \begin{tabular}{@{}c@{}}\begin{minipage}[c][0.5cm][c]{1.8cm}\centering \ps{\textsf{PS} (1/5)}\ \end{minipage} \\\begin{minipage}[c][0.5cm][c]{1.8cm}\centering \ro{\textsf{DRe} (4/5)}\ \end{minipage} \\\end{tabular}& \begin{tabular}{@{}c@{}}\begin{minipage}[c][0.5cm][c]{1.8cm}\centering \ps{\textsf{PS} (2/5)}\ \end{minipage} \\\begin{minipage}[c][0.5cm][c]{1.8cm}\centering \ro{\textsf{DRe} (3/5)}\ \end{minipage} \\\end{tabular}& \begin{tabular}{@{}c@{}}\begin{minipage}[c][0.5cm][c]{1.8cm}\centering \ro{\textsf{DRe} (3/5)}\ \end{minipage} \\\begin{minipage}[c][0.5cm][c]{1.8cm}\centering \ps{\textsf{PS} (2/5)}\ \end{minipage} \\\end{tabular}& \begin{tabular}{@{}c@{}}\begin{minipage}[c][0.5cm][c]{1.8cm}\centering \ro{\textsf{DRe} (1/5)}\ \end{minipage} \\\begin{minipage}[c][0.5cm][c]{1.8cm}\centering \ps{\textsf{PS} (4/5)}\ \end{minipage} \\\end{tabular} & \ps{\textsf{PS} (5/5)}& \begin{tabular}{@{}c@{}}\begin{minipage}[c][0.5cm][c]{1.8cm}\centering \ps{\textsf{PS} (4/5)}\ \end{minipage} \\\begin{minipage}[c][0.5cm][c]{1.8cm}\centering \ro{\textsf{DRe} (1/5)}\ \end{minipage} \\\end{tabular}& \begin{tabular}{@{}c@{}}\begin{minipage}[c][0.5cm][c]{1.8cm}\centering \ro{\textsf{DRe} (2/5)}\ \end{minipage} \\\begin{minipage}[c][0.5cm][c]{1.8cm}\centering \ps{\textsf{PS} (3/5)}\ \end{minipage} \\\end{tabular}& \begin{tabular}{@{}c@{}}\begin{minipage}[c][0.5cm][c]{1.8cm}\centering \ro{\textsf{DRe} (1/5)}\ \end{minipage} \\\begin{minipage}[c][0.5cm][c]{1.8cm}\centering \ps{\textsf{PS} (4/5)}\ \end{minipage} \\\end{tabular}& \begin{tabular}{@{}c@{}}\begin{minipage}[c][0.5cm][c]{1.8cm}\centering \ps{\textsf{PS} (4/5)}\ \end{minipage} \\\begin{minipage}[c][0.5cm][c]{1.8cm}\centering \ro{\textsf{DRe} (1/5)}\ \end{minipage} \\\end{tabular}& \begin{tabular}{@{}c@{}}\begin{minipage}[c][0.5cm][c]{1.8cm}\centering \ps{\textsf{PS} (4/5)}\ \end{minipage} \\\begin{minipage}[c][0.5cm][c]{1.8cm}\centering \ro{\textsf{DRe} (1/5)}\ \end{minipage} \\\end{tabular} \\ 
\hline 
40.0 & \cs{\textsf{CS} (5/5)} & \ps{\textsf{PS} (5/5)}& \begin{tabular}{@{}c@{}}\begin{minipage}[c][0.5cm][c]{1.8cm}\centering \ro{\textsf{DRe} (2/5)}\ \end{minipage} \\\begin{minipage}[c][0.5cm][c]{1.8cm}\centering \ps{\textsf{PS} (3/5)}\ \end{minipage} \\\end{tabular} & \ps{\textsf{PS} (5/5)} & \ps{\textsf{PS} (5/5)} & \ps{\textsf{PS} (5/5)} & \ps{\textsf{PS} (5/5)} & \ps{\textsf{PS} (5/5)} & \ps{\textsf{PS} (5/5)} & \ps{\textsf{PS} (5/5)} & \ps{\textsf{PS} (5/5)}& \begin{tabular}{@{}c@{}}\begin{minipage}[c][0.5cm][c]{1.8cm}\centering \ps{\textsf{PS} (4/5)}\ \end{minipage} \\\begin{minipage}[c][0.5cm][c]{1.8cm}\centering \ro{\textsf{DRe} (1/5)}\ \end{minipage} \\\end{tabular}& \begin{tabular}{@{}c@{}}\begin{minipage}[c][0.5cm][c]{1.8cm}\centering \ro{\textsf{DRe} (2/5)}\ \end{minipage} \\\begin{minipage}[c][0.5cm][c]{1.8cm}\centering \ps{\textsf{PS} (3/5)}\ \end{minipage} \\\end{tabular}& \begin{tabular}{@{}c@{}}\begin{minipage}[c][0.5cm][c]{1.8cm}\centering \ps{\textsf{PS} (4/5)}\ \end{minipage} \\\begin{minipage}[c][0.5cm][c]{1.8cm}\centering \ro{\textsf{DRe} (1/5)}\ \end{minipage} \\\end{tabular} \\ 
\hline 
\end{tabular}

%% file: tables/lmdrive-merging-4.tex
\begin{tabular}{|c| *{15}{c|}} 
\hline 
\diagbox{$\bm{x_a}$}{$\bm{x_f}$} & 0.0 & \ct{2.4} & 40.0 & 80.0 & 120.0 & 160.0 & 200.0 & 240.0 & 280.0 & 320.0 & 360.0 & 400.0 & 440.0 & 480.0 & 520.0 \\ 
\hline 
0.0 & - & \cs{\textsf{CS} (5/5)} & \cs{\textsf{CS} (5/5)} & \cs{\textsf{CS} (5/5)} & \cs{\textsf{CS} (5/5)} & \cs{\textsf{CS} (5/5)} & \cs{\textsf{CS} (5/5)} & \cs{\textsf{CS} (5/5)} & \cs{\textsf{CS} (5/5)} & \cs{\textsf{CS} (5/5)} & \cs{\textsf{CS} (5/5)} & \cs{\textsf{CS} (5/5)} & \cs{\textsf{CS} (5/5)} & \cs{\textsf{CS} (5/5)} & \cs{\textsf{CS} (5/5)} \\ 
\hline 
5.0 & \cs{\textsf{CS} (5/5)} & \cs{\textsf{CS} (5/5)} & \cs{\textsf{CS} (5/5)} & \cs{\textsf{CS} (5/5)} & \cs{\textsf{CS} (5/5)} & \cs{\textsf{CS} (5/5)} & \cs{\textsf{CS} (5/5)} & \cs{\textsf{CS} (5/5)} & \cs{\textsf{CS} (5/5)}& \begin{tabular}{@{}c@{}}\begin{minipage}[c][0.5cm][c]{1.6cm}\centering \ro{\textsf{DRe} (1/5)}\ \end{minipage} \\\begin{minipage}[c][0.5cm][c]{1.6cm}\centering \cs{\textsf{CS} (4/5)}\ \end{minipage} \\\end{tabular} & \cs{\textsf{CS} (5/5)}& \begin{tabular}{@{}c@{}}\begin{minipage}[c][0.5cm][c]{1.6cm}\centering \ro{\textsf{DRe} (1/5)}\ \end{minipage} \\\begin{minipage}[c][0.5cm][c]{1.6cm}\centering \cs{\textsf{CS} (4/5)}\ \end{minipage} \\\end{tabular} & \cs{\textsf{CS} (5/5)}& \begin{tabular}{@{}c@{}}\begin{minipage}[c][0.5cm][c]{1.6cm}\centering \ro{\textsf{DRe} (1/5)}\ \end{minipage} \\\begin{minipage}[c][0.5cm][c]{1.6cm}\centering \cs{\textsf{CS} (4/5)}\ \end{minipage} \\\end{tabular} & \cs{\textsf{CS} (5/5)} \\ 
\hline 
\ct{5.4} & \cs{\textsf{CS} (5/5)} & \cs{\textsf{CS} (5/5)} & \cs{\textsf{CS} (5/5)} & \cs{\textsf{CS} (5/5)} & \cs{\textsf{CS} (5/5)} & \cs{\textsf{CS} (5/5)} & \cs{\textsf{CS} (5/5)} & \cs{\textsf{CS} (5/5)} & \cs{\textsf{CS} (5/5)} & \cs{\textsf{CS} (5/5)} & \begin{tabular}{@{}c@{}}\begin{minipage}[c][0.5cm][c]{1.6cm}\centering \cs{\textsf{CS} (4/5)}\ \end{minipage} \\\begin{minipage}[c][0.5cm][c]{1.6cm}\centering \ro{\textsf{DRa} (1/5)}\ \end{minipage} \\\end{tabular}  & \cs{\textsf{CS} (5/5)}& \begin{tabular}{@{}c@{}}\begin{minipage}[c][0.5cm][c]{1.6cm}\centering \cs{\textsf{CS} (4/5)}\ \end{minipage} \\\begin{minipage}[c][0.5cm][c]{1.6cm}\centering \ro{\textsf{DRa} (1/5)}\ \end{minipage} \\\end{tabular} & \cs{\textsf{CS} (5/5)} & \begin{tabular}{@{}c@{}}\begin{minipage}[c][0.5cm][c]{1.6cm}\centering \cs{\textsf{CS} (4/5)}\ \end{minipage} \\\begin{minipage}[c][0.5cm][c]{1.6cm}\centering \ro{\textsf{DRe} (1/5)}\ \end{minipage} \\\end{tabular}  \\ 
\hline 
10.0 & \cs{\textsf{CS} (5/5)} & \cs{\textsf{CS} (5/5)} & \cs{\textsf{CS} (5/5)} & \cs{\textsf{CS} (5/5)} & \cs{\textsf{CS} (5/5)} & \cs{\textsf{CS} (5/5)} & \cs{\textsf{CS} (5/5)} & \cs{\textsf{CS} (5/5)} & \cs{\textsf{CS} (5/5)} & \cs{\textsf{CS} (5/5)} & \cs{\textsf{CS} (5/5)}& \begin{tabular}{@{}c@{}}\begin{minipage}[c][0.5cm][c]{1.6cm}\centering \cs{\textsf{CS} (3/5)}\ \end{minipage} \\\begin{minipage}[c][0.5cm][c]{1.6cm}\centering \ro{\textsf{DRe} (1/5)}\ \end{minipage} \\\begin{minipage}[c][0.5cm][c]{1.6cm}\centering \ro{\textsf{DRa} (1/5)}\ \end{minipage} \\\end{tabular} & \cs{\textsf{CS} (4/4)}& \begin{tabular}{@{}c@{}}\begin{minipage}[c][0.5cm][c]{1.6cm}\centering \ro{\textsf{DRe} (2/5)}\ \end{minipage} \\\begin{minipage}[c][0.5cm][c]{1.6cm}\centering \cs{\textsf{CS} (2/5)}\ \end{minipage} \\\begin{minipage}[c][0.5cm][c]{1.6cm}\centering \ro{\textsf{DRa} (1/5)}\ \end{minipage} \\\end{tabular} & \cs{\textsf{CS} (4/4)} \\ 
\hline 
12.5 & \cs{\textsf{CS} (5/5)} & \cs{\textsf{CS} (5/5)} & \cs{\textsf{CS} (5/5)} & \cs{\textsf{CS} (5/5)} & \cs{\textsf{CS} (5/5)} & \cs{\textsf{CS} (5/5)} & \cs{\textsf{CS} (5/5)} & \cs{\textsf{CS} (5/5)} & \cs{\textsf{CS} (5/5)} & \cs{\textsf{CS} (5/5)}& \begin{tabular}{@{}c@{}}\begin{minipage}[c][0.5cm][c]{1.6cm}\centering \cs{\textsf{CS} (4/5)}\ \end{minipage} \\\begin{minipage}[c][0.5cm][c]{1.6cm}\centering \ro{\textsf{DRa} (1/5)}\ \end{minipage} \\\end{tabular}& \begin{tabular}{@{}c@{}}\begin{minipage}[c][0.5cm][c]{1.6cm}\centering \cs{\textsf{CS} (3/5)}\ \end{minipage} \\\begin{minipage}[c][0.5cm][c]{1.6cm}\centering \ro{\textsf{DRa} (2/5)}\ \end{minipage} \\\end{tabular}& \begin{tabular}{@{}c@{}}\begin{minipage}[c][0.5cm][c]{1.6cm}\centering \cs{\textsf{CS} (4/5)}\ \end{minipage} \\\begin{minipage}[c][0.5cm][c]{1.6cm}\centering \ro{\textsf{DRe} (1/5)}\ \end{minipage} \\\end{tabular}& \begin{tabular}{@{}c@{}}\begin{minipage}[c][0.5cm][c]{1.6cm}\centering \cs{\textsf{CS} (3/5)}\ \end{minipage} \\\begin{minipage}[c][0.5cm][c]{1.6cm}\centering \ro{\textsf{DRe} (2/5)}\ \end{minipage} \\\end{tabular}& \begin{tabular}{@{}c@{}}\begin{minipage}[c][0.5cm][c]{1.6cm}\centering \cs{\textsf{CS} (3/4)}\ \end{minipage} \\\begin{minipage}[c][0.5cm][c]{1.6cm}\centering \ro{\textsf{DRe} (1/4)}\ \end{minipage} \\\end{tabular} \\ 
\hline 
15.0 & \cs{\textsf{CS} (5/5)} & \cs{\textsf{CS} (5/5)}& \begin{tabular}{@{}c@{}}\begin{minipage}[c][0.5cm][c]{1.2cm}\centering \ps{\textsf{PS} (3/5)}\ \end{minipage} \\\begin{minipage}[c][0.5cm][c]{1.2cm}\centering \cs{\textsf{CS} (2/5)}\ \end{minipage} \\\end{tabular} & \ps{\textsf{PS} (5/5)} & \ps{\textsf{PS} (5/5)} & \ps{\textsf{PS} (5/5)} & \ps{\textsf{PS} (5/5)} & \ps{\textsf{PS} (5/5)} & \ps{\textsf{PS} (5/5)} & \ps{\textsf{PS} (5/5)}& \begin{tabular}{@{}c@{}}\begin{minipage}[c][0.5cm][c]{1.6cm}\centering \ro{\textsf{DRe} (1/5)}\ \end{minipage} \\\begin{minipage}[c][0.5cm][c]{1.6cm}\centering \ps{\textsf{PS} (4/5)}\ \end{minipage} \\\end{tabular}& \begin{tabular}{@{}c@{}}\begin{minipage}[c][0.5cm][c]{1.6cm}\centering \cs{\textsf{CS} (1/5)}\ \end{minipage} \\\begin{minipage}[c][0.5cm][c]{1.6cm}\centering \ps{\textsf{PS} (4/5)}\ \end{minipage} \\\end{tabular} & \ps{\textsf{PS} (5/5)}& \begin{tabular}{@{}c@{}}\begin{minipage}[c][0.5cm][c]{1.6cm}\centering \ps{\textsf{PS} (4/5)}\ \end{minipage} \\\begin{minipage}[c][0.5cm][c]{1.6cm}\centering \ro{\textsf{DRe} (1/5)}\ \end{minipage} \\\end{tabular} & \ps{\textsf{PS} (5/5)} \\ 
\hline 
20.0 & \cs{\textsf{CS} (5/5)} & \cs{\textsf{CS} (5/5)} & \ps{\textsf{PS} (5/5)} & \ps{\textsf{PS} (5/5)} & \ps{\textsf{PS} (5/5)} & \ps{\textsf{PS} (5/5)} & \ps{\textsf{PS} (5/5)} & \ps{\textsf{PS} (5/5)} & \ps{\textsf{PS} (5/5)} & \ps{\textsf{PS} (5/5)}& \begin{tabular}{@{}c@{}}\begin{minipage}[c][0.5cm][c]{1.6cm}\centering \ro{\textsf{DRe} (1/5)}\ \end{minipage} \\\begin{minipage}[c][0.5cm][c]{1.6cm}\centering \ps{\textsf{PS} (4/5)}\ \end{minipage} \\\end{tabular} & \ps{\textsf{PS} (5/5)} & \ps{\textsf{PS} (5/5)} & \ps{\textsf{PS} (5/5)} & \ps{\textsf{PS} (5/5)} \\ 
\hline 
25.0& \begin{tabular}{@{}c@{}}\begin{minipage}[c][0.5cm][c]{1.2cm}\centering \cs{\textsf{CS} (4/5)}\ \end{minipage} \\\begin{minipage}[c][0.5cm][c]{1.2cm}\centering \ac{\textsf{Af} (1/5)}\ \end{minipage} \\\end{tabular} & \cs{\textsf{CS} (5/5)} & \ps{\textsf{PS} (5/5)} & \ps{\textsf{PS} (5/5)} & \ps{\textsf{PS} (5/5)} & \ps{\textsf{PS} (5/5)}& \begin{tabular}{@{}c@{}}\begin{minipage}[c][0.5cm][c]{1.6cm}\centering \ps{\textsf{PS} (4/5)}\ \end{minipage} \\\begin{minipage}[c][0.5cm][c]{1.6cm}\centering \ro{\textsf{DRe} (1/5)}\ \end{minipage} \\\end{tabular} & \ps{\textsf{PS} (5/5)}& \begin{tabular}{@{}c@{}}\begin{minipage}[c][0.5cm][c]{1.6cm}\centering \ps{\textsf{PS} (4/5)}\ \end{minipage} \\\begin{minipage}[c][0.5cm][c]{1.6cm}\centering \ro{\textsf{DRe} (1/5)}\ \end{minipage} \\\end{tabular} & \ps{\textsf{PS} (5/5)} & \ps{\textsf{PS} (5/5)} & \ps{\textsf{PS} (5/5)}& \begin{tabular}{@{}c@{}}\begin{minipage}[c][0.5cm][c]{1.6cm}\centering \ps{\textsf{PS} (4/5)}\ \end{minipage} \\\begin{minipage}[c][0.5cm][c]{1.6cm}\centering \ro{\textsf{DRe} (1/5)}\ \end{minipage} \\\end{tabular}& \begin{tabular}{@{}c@{}}\begin{minipage}[c][0.5cm][c]{1.6cm}\centering \ps{\textsf{PS} (4/5)}\ \end{minipage} \\\begin{minipage}[c][0.5cm][c]{1.6cm}\centering \ro{\textsf{DRe} (1/5)}\ \end{minipage} \\\end{tabular}& \begin{tabular}{@{}c@{}}\begin{minipage}[c][0.5cm][c]{1.6cm}\centering \ps{\textsf{PS} (4/5)}\ \end{minipage} \\\begin{minipage}[c][0.5cm][c]{1.6cm}\centering \ro{\textsf{DRa} (1/5)}\ \end{minipage} \\\end{tabular} \\ 
\hline 
30.0 & \cs{\textsf{CS} (5/5)} & \cs{\textsf{CS} (5/5)} & \ps{\textsf{PS} (5/5)} & \ps{\textsf{PS} (5/5)}& \begin{tabular}{@{}c@{}}\begin{minipage}[c][0.5cm][c]{1.6cm}\centering \ps{\textsf{PS} (4/5)}\ \end{minipage} \\\begin{minipage}[c][0.5cm][c]{1.6cm}\centering \ro{\textsf{DRe} (1/5)}\ \end{minipage} \\\end{tabular} & \ps{\textsf{PS} (5/5)} & \ps{\textsf{PS} (5/5)} & \ps{\textsf{PS} (5/5)} & \ps{\textsf{PS} (5/5)} & \ps{\textsf{PS} (5/5)}& \begin{tabular}{@{}c@{}}\begin{minipage}[c][0.5cm][c]{1.6cm}\centering \ps{\textsf{PS} (4/5)}\ \end{minipage} \\\begin{minipage}[c][0.5cm][c]{1.6cm}\centering \ro{\textsf{DRe} (1/5)}\ \end{minipage} \\\end{tabular}& \begin{tabular}{@{}c@{}}\begin{minipage}[c][0.5cm][c]{1.6cm}\centering \ps{\textsf{PS} (4/5)}\ \end{minipage} \\\begin{minipage}[c][0.5cm][c]{1.6cm}\centering \ro{\textsf{DRe} (1/5)}\ \end{minipage} \\\end{tabular} & \ps{\textsf{PS} (5/5)} & \ps{\textsf{PS} (5/5)} & \ps{\textsf{PS} (5/5)} \\ 
\hline 
35.0 & \cs{\textsf{CS} (5/5)} & \cs{\textsf{CS} (5/5)} & \ps{\textsf{PS} (5/5)} & \ps{\textsf{PS} (5/5)} & \ps{\textsf{PS} (5/5)} & \ps{\textsf{PS} (5/5)} & \ps{\textsf{PS} (5/5)} & \ps{\textsf{PS} (5/5)} & \ps{\textsf{PS} (5/5)} & \ps{\textsf{PS} (5/5)} & \ps{\textsf{PS} (5/5)}& \begin{tabular}{@{}c@{}}\begin{minipage}[c][0.5cm][c]{1.6cm}\centering \ps{\textsf{PS} (3/5)}\ \end{minipage} \\\begin{minipage}[c][0.5cm][c]{1.6cm}\centering \ro{\textsf{DRe} (1/5)}\ \end{minipage} \\\begin{minipage}[c][0.5cm][c]{1.6cm}\centering \ro{\textsf{DRa} (1/5)}\ \end{minipage} \\\end{tabular} & \ps{\textsf{PS} (5/5)}& \begin{tabular}{@{}c@{}}\begin{minipage}[c][0.75cm][c]{1.6cm}\centering \ps{\textsf{PS} (4/5)}\ \end{minipage} \\\begin{minipage}[c][0.75cm][c]{1.6cm}\centering \ro{\textsf{DRe} (1/5)}\ \end{minipage} \\\end{tabular}& \begin{tabular}{@{}c@{}}\begin{minipage}[c][0.75cm][c]{1.6cm}\centering \ps{\textsf{PS} (4/5)}\ \end{minipage} \\\begin{minipage}[c][0.75cm][c]{1.6cm}\centering \ro{\textsf{DRe} (1/5)}\ \end{minipage} \\\end{tabular} \\ 
\hline 
40.0 & \cs{\textsf{CS} (5/5)} & \cs{\textsf{CS} (5/5)} & \ps{\textsf{PS} (5/5)} & \ps{\textsf{PS} (5/5)} & \ps{\textsf{PS} (5/5)} & \ps{\textsf{PS} (5/5)} & \ps{\textsf{PS} (5/5)} & \ps{\textsf{PS} (5/5)} & \ps{\textsf{PS} (5/5)} & \ps{\textsf{PS} (5/5)}& \begin{tabular}{@{}c@{}}\begin{minipage}[c][0.5cm][c]{1.6cm}\centering \ps{\textsf{PS} (4/5)}\ \end{minipage} \\\begin{minipage}[c][0.5cm][c]{1.6cm}\centering \ro{\textsf{DRe} (1/5)}\ \end{minipage} \\\end{tabular}& \begin{tabular}{@{}c@{}}\begin{minipage}[c][0.5cm][c]{1.6cm}\centering \ps{\textsf{PS} (4/5)}\ \end{minipage} \\\begin{minipage}[c][0.5cm][c]{1.6cm}\centering \ro{\textsf{DRe} (1/5)}\ \end{minipage} \\\end{tabular}& \begin{tabular}{@{}c@{}}\begin{minipage}[c][0.5cm][c]{1.6cm}\centering \ps{\textsf{PS} (4/5)}\ \end{minipage} \\\begin{minipage}[c][0.5cm][c]{1.6cm}\centering \ro{\textsf{DRe} (1/5)}\ \end{minipage} \\\end{tabular} & \ps{\textsf{PS} (4/4)}& \begin{tabular}{@{}c@{}}\begin{minipage}[c][0.5cm][c]{1.6cm}\centering \ps{\textsf{PS} (3/5)}\ \end{minipage} \\\begin{minipage}[c][0.5cm][c]{1.6cm}\centering \ro{\textsf{DRe} (2/5)}\ \end{minipage} \\\end{tabular} \\ 
\hline 
\end{tabular} 

%% file: tables/lmdrive-lane_change-4.tex
\begin{tabular}{|c| *{15}{c|}}
\hline
\diagbox{$\bm{x_a}$}{$\bm{x_f}$} & 0.0 & \ct{1.9} & 40.0 & 80.0 & 120.0 & 160.0 & 200.0 & 240.0 & 280.0 & 320.0 & 360.0 & 400.0 & 440.0 & 480.0 & 520.0 \\
\hline
0.0 & - & \cs{\textsf{CS}} (5/5) & \cs{\textsf{CS}} (5/5) & \cs{\textsf{CS}} (5/5) & \cs{\textsf{CS}} (5/5) & \cs{\textsf{CS}} (5/5) & \cs{\textsf{CS}} (5/5) & \cs{\textsf{CS}} (5/5) & \cs{\textsf{CS}} (5/5) & \cs{\textsf{CS}} (5/5) & \cs{\textsf{CS}} (5/5) & \cs{\textsf{CS}} (5/5) & \cs{\textsf{CS}} (5/5) & \cs{\textsf{CS}} (5/5) & \cs{\textsf{CS}} (5/5) \\
\hline
\ct{23.9} & \cs{\textsf{CS}} (5/5) & \cs{\textsf{CS}} (5/5) & \cs{\textsf{CS}} (5/5) & \cs{\textsf{CS}} (5/5) & \cs{\textsf{CS}} (5/5) & \cs{\textsf{CS}} (5/5) & \cs{\textsf{CS}} (5/5) & \cs{\textsf{CS}} (5/5) & \cs{\textsf{CS}} (5/5) & \cs{\textsf{CS}} (5/5) & \cs{\textsf{CS}} (5/5) & \cs{\textsf{CS}} (5/5) & \cs{\textsf{CS}} (5/5) & \cs{\textsf{CS}} (5/5) & \cs{\textsf{CS}} (5/5)  \\
\hline
40.0 & \cs{\textsf{CS}} (5/5) & \cs{\textsf{CS}} (5/5) & \co{\textsf{CO}} (5/5) & \co{\textsf{CO}} (5/5) & \co{\textsf{CO}} (5/5) & \co{\textsf{CO}} (5/5) & \co{\textsf{CO}} (5/5) & \co{\textsf{CO}} (5/5) & \co{\textsf{CO}} (5/5) & \co{\textsf{CO}} (5/5) & \co{\textsf{CO}} (5/5) & \co{\textsf{CO}} (5/5) & \co{\textsf{CO}} (5/5) & \co{\textsf{CO}} (5/5) & \co{\textsf{CO}} (5/5) \\
\hline
80.0 & \cs{\textsf{CS}} (5/5) & \cs{\textsf{CS}} (5/5) & \co{\textsf{CO}} (5/5) & \co{\textsf{CO}} (5/5) & \co{\textsf{CO}} (5/5) & \co{\textsf{CO}} (5/5) & \co{\textsf{CO}} (5/5) & \co{\textsf{CO}} (5/5) & \co{\textsf{CO}} (5/5) & \co{\textsf{CO}} (5/5) & \co{\textsf{CO}} (5/5) & \co{\textsf{CO}} (5/5) & \co{\textsf{CO}} (5/5) & \co{\textsf{CO}} (5/5) & \co{\textsf{CO}} (5/5) \\
\hline
120.0 & \cs{\textsf{CS}} (5/5) & \cs{\textsf{CS}} (5/5) & \co{\textsf{CO}} (5/5) & \co{\textsf{CO}} (5/5) & \co{\textsf{CO}} (5/5) & \co{\textsf{CO}} (5/5) & \co{\textsf{CO}} (5/5) & \co{\textsf{CO}} (5/5) & \co{\textsf{CO}} (5/5) & \co{\textsf{CO}} (5/5) & \co{\textsf{CO}} (5/5) & \co{\textsf{CO}} (5/5) & \co{\textsf{CO}} (5/5) & \co{\textsf{CO}} (5/5) & \co{\textsf{CO}} (5/5) \\
\hline
160.0 & \cs{\textsf{CS}} (5/5) & \cs{\textsf{CS}} (5/5) & \co{\textsf{CO}} (5/5) & \co{\textsf{CO}} (5/5) & \co{\textsf{CO}} (5/5) & \co{\textsf{CO}} (5/5) & \co{\textsf{CO}} (5/5) & \co{\textsf{CO}} (5/5) & \co{\textsf{CO}} (5/5) & \co{\textsf{CO}} (5/5) & \co{\textsf{CO}} (5/5) & \co{\textsf{CO}} (5/5) & \co{\textsf{CO}} (5/5) & \co{\textsf{CO}} (5/5) & \co{\textsf{CO}} (5/5) \\
\hline
200.0 & \cs{\textsf{CS}} (5/5) & \cs{\textsf{CS}} (5/5) & \co{\textsf{CO}} (5/5) & \co{\textsf{CO}} (5/5) & \co{\textsf{CO}} (5/5) & \co{\textsf{CO}} (5/5) & \co{\textsf{CO}} (5/5) & \co{\textsf{CO}} (5/5) & \co{\textsf{CO}} (5/5) & \co{\textsf{CO}} (5/5) & \co{\textsf{CO}} (5/5) & \co{\textsf{CO}} (5/5) & \co{\textsf{CO}} (5/5) & \co{\textsf{CO}} (5/5) & \co{\textsf{CO}} (5/5) \\
\hline
240.0 & \cs{\textsf{CS}} (5/5) & \cs{\textsf{CS}} (5/5) & \co{\textsf{CO}} (5/5) & \co{\textsf{CO}} (5/5) & \co{\textsf{CO}} (5/5) & \co{\textsf{CO}} (5/5) & \co{\textsf{CO}} (5/5) & \co{\textsf{CO}} (5/5) & \co{\textsf{CO}} (5/5) & \co{\textsf{CO}} (5/5) & \co{\textsf{CO}} (5/5) & \co{\textsf{CO}} (5/5) & \co{\textsf{CO}} (5/5) & \co{\textsf{CO}} (5/5) & \co{\textsf{CO}} (5/5) \\
\hline
280.0 & \cs{\textsf{CS}} (5/5) & \cs{\textsf{CS}} (5/5) & \co{\textsf{CO}} (5/5) & \co{\textsf{CO}} (5/5) & \co{\textsf{CO}} (5/5) & \co{\textsf{CO}} (5/5) & \co{\textsf{CO}} (5/5) & \co{\textsf{CO}} (5/5) & \co{\textsf{CO}} (5/5) & \co{\textsf{CO}} (5/5) & \co{\textsf{CO}} (5/5) & \co{\textsf{CO}} (5/5) & \co{\textsf{CO}} (5/5) & \co{\textsf{CO}} (5/5) & \co{\textsf{CO}} (5/5) \\
\hline
320.0 & \cs{\textsf{CS}} (5/5) & \cs{\textsf{CS}} (5/5) & \co{\textsf{CO}} (5/5) & \co{\textsf{CO}} (5/5) & \co{\textsf{CO}} (5/5) & \co{\textsf{CO}} (5/5) & \co{\textsf{CO}} (5/5) & \co{\textsf{CO}} (5/5) & \co{\textsf{CO}} (5/5) & \co{\textsf{CO}} (5/5) & \co{\textsf{CO}} (5/5) & \co{\textsf{CO}} (5/5) & \co{\textsf{CO}} (5/5) & \co{\textsf{CO}} (5/5) & \co{\textsf{CO}} (5/5) \\
\hline
360.0 & \cs{\textsf{CS}} (5/5) & \cs{\textsf{CS}} (5/5) & \co{\textsf{CO}} (5/5) & \co{\textsf{CO}} (5/5) & \co{\textsf{CO}} (5/5) & \co{\textsf{CO}} (5/5) & \co{\textsf{CO}} (5/5) & \co{\textsf{CO}} (5/5) & \co{\textsf{CO}} (5/5) & \co{\textsf{CO}} (5/5) & \co{\textsf{CO}} (5/5) & \co{\textsf{CO}} (5/5) & \co{\textsf{CO}} (5/5) & \co{\textsf{CO}} (5/5) & \co{\textsf{CO}} (5/5) \\
\hline
400.0 & \cs{\textsf{CS}} (5/5) & \cs{\textsf{CS}} (5/5) & \co{\textsf{CO}} (5/5) & \co{\textsf{CO}} (5/5) & \co{\textsf{CO}} (5/5) & \co{\textsf{CO}} (5/5) & \co{\textsf{CO}} (5/5) & \co{\textsf{CO}} (5/5) & \co{\textsf{CO}} (5/5) & \co{\textsf{CO}} (5/5) & \co{\textsf{CO}} (5/5) & \co{\textsf{CO}} (5/5) & \co{\textsf{CO}} (5/5) & \co{\textsf{CO}} (5/5) & \co{\textsf{CO}} (5/5) \\
\hline
440.0 & \cs{\textsf{CS}} (5/5) & \cs{\textsf{CS}} (5/5) & \co{\textsf{CO}} (5/5) & \co{\textsf{CO}} (5/5) & \co{\textsf{CO}} (5/5) & \co{\textsf{CO}} (5/5) & \co{\textsf{CO}} (5/5) & \co{\textsf{CO}} (5/5) & \co{\textsf{CO}} (5/5) & \co{\textsf{CO}} (5/5) & \co{\textsf{CO}} (5/5) & \co{\textsf{CO}} (5/5) & \co{\textsf{CO}} (5/5) & \co{\textsf{CO}} (5/5) & \co{\textsf{CO}} (5/5) \\
\hline
480.0 & \cs{\textsf{CS}} (5/5) & \cs{\textsf{CS}} (5/5) & \co{\textsf{CO}} (5/5) & \co{\textsf{CO}} (5/5) & \co{\textsf{CO}} (5/5) & \co{\textsf{CO}} (5/5) & \co{\textsf{CO}} (5/5) & \co{\textsf{CO}} (5/5) & \co{\textsf{CO}} (5/5) & \co{\textsf{CO}} (5/5) & \co{\textsf{CO}} (5/5) & \co{\textsf{CO}} (5/5) & \co{\textsf{CO}} (5/5) & \co{\textsf{CO}} (5/5) & \co{\textsf{CO}} (5/5) \\
\hline
520.0 & \cs{\textsf{CS}} (5/5) & \cs{\textsf{CS}} (5/5) & \co{\textsf{CO}} (5/5) & \co{\textsf{CO}} (5/5) & \co{\textsf{CO}} (5/5) & \co{\textsf{CO}} (5/5) & \co{\textsf{CO}} (5/5) & \co{\textsf{CO}} (5/5) & \co{\textsf{CO}} (5/5) & \co{\textsf{CO}} (5/5) & \co{\textsf{CO}} (5/5) & \co{\textsf{CO}} (5/5) & \co{\textsf{CO}} (5/5) & \co{\textsf{CO}} (5/5) & \co{\textsf{CO}} (5/5) \\
\hline
\end{tabular}

%% file: tables/lmdrive-crossing-0.tex
\begin{tabular}{|c| *{15}{c|}} 
\hline
\diagbox{$\bm{x_a}$}{$\bm{x_f}$} & 0.0 & \ct{2.9} & 40.0 & 80.0 & 120.0 & 160.0 & 200.0 & 240.0 & 280.0 & 320.0 & 360.0 & 400.0 & 440.0 & 480.0 & 520.0 \\
\hline
0.0 & \blk{\textsf{Blk} (5/5)} & \begin{tabular}{@{}c@{}}\begin{minipage}[c][0.5cm][c]{1.8cm}\centering \ro{\textsf{CDRea} (2/5)}\ \end{minipage} \\\begin{minipage}[c][0.5cm][c]{1.8cm}\centering \blk{\textsf{Blk} (3/5)}\ \end{minipage} \\\end{tabular}& \begin{tabular}{@{}c@{}}\begin{minipage}[c][0.5cm][c]{1.8cm}\centering \ro{\textsf{CRea} (1/5)}\ \end{minipage} \\\begin{minipage}[c][0.5cm][c]{1.8cm}\centering \blk{\textsf{Blk} (4/5)}\ \end{minipage} \\\end{tabular}& \begin{tabular}{@{}c@{}}\begin{minipage}[c][0.5cm][c]{1.8cm}\centering \ro{\textsf{CRea} (1/5)}\ \end{minipage} \\\begin{minipage}[c][0.5cm][c]{1.8cm}\centering \blk{\textsf{Blk} (4/5)}\ \end{minipage} \\\end{tabular}& \begin{tabular}{@{}c@{}}\begin{minipage}[c][0.5cm][c]{1.8cm}\centering \ro{\textsf{CRea} (1/5)}\ \end{minipage} \\\begin{minipage}[c][0.5cm][c]{1.8cm}\centering \blk{\textsf{Blk} (4/5)}\ \end{minipage} \\\end{tabular}& \begin{tabular}{@{}c@{}}\begin{minipage}[c][0.5cm][c]{1.8cm}\centering \ro{\textsf{CDRea} (1/5)}\ \end{minipage} \\\begin{minipage}[c][0.5cm][c]{1.8cm}\centering \blk{\textsf{Blk} (4/5)}\ \end{minipage} \\\end{tabular}& \begin{tabular}{@{}c@{}}\begin{minipage}[c][0.5cm][c]{1.8cm}\centering \ro{\textsf{CRea} (1/5)}\ \end{minipage} \\\begin{minipage}[c][0.5cm][c]{1.8cm}\centering \blk{\textsf{Blk} (4/5)}\ \end{minipage} \\\end{tabular} & \blk{\textsf{Blk} (5/5)}& \blk{\textsf{Blk} (5/5)} & \blk{\textsf{Blk} (5/5)}& \blk{\textsf{Blk} (5/5)}& \blk{\textsf{Blk} (5/5)} & \blk{\textsf{Blk} (5/5)} & \blk{\textsf{Blk} (5/5)} & \blk{\textsf{Blk} (5/5)} \\
\hline
5.0 & \ro{\textsf{CDRea} (5/5)} & \ro{\textsf{CDRea} (5/5)} & \ro{\textsf{CDRea} (5/5)} & \ro{\textsf{CDRea} (5/5)} & \ro{\textsf{CDRea} (5/5)} & \ro{\textsf{CDRea} (5/5)} & \ro{\textsf{CDRea} (5/5)} & \ro{\textsf{CDRea} (5/5)} & \ro{\textsf{CDRea} (5/5)}& \begin{tabular}{@{}c@{}}\begin{minipage}[c][0.5cm][c]{1.8cm}\centering \ro{\textsf{CDRea} (3/5)}\ \end{minipage} \\\begin{minipage}[c][0.5cm][c]{1.8cm}\centering \ro{\textsf{CDRa} (2/5)}\ \end{minipage} \\\end{tabular}& \begin{tabular}{@{}c@{}}\begin{minipage}[c][0.5cm][c]{1.8cm}\centering \ro{\textsf{CDRa} (2/5)}\ \end{minipage} \\\begin{minipage}[c][0.5cm][c]{1.8cm}\centering \ro{\textsf{CDRea} (3/5)}\ \end{minipage} \\\end{tabular} & \ro{\textsf{CDRea} (5/5)} & \ro{\textsf{CDRea} (5/5)} & \ro{\textsf{CDRea} (5/5)} & \ro{\textsf{CDRea} (5/5)} \\
\hline
10.0& \begin{tabular}{@{}c@{}}\begin{minipage}[c][0.5cm][c]{1.8cm}\centering \ro{\textsf{CDRea} (3/5)}\ \end{minipage} \\\begin{minipage}[c][0.5cm][c]{1.8cm}\centering \ro{\textsf{CDRe} (1/5)}\ \end{minipage} \\\begin{minipage}[c][0.5cm][c]{1.8cm}\centering \blk{\textsf{Blk} (1/5)}\ \end{minipage} \\\end{tabular}& \begin{tabular}{@{}c@{}}\begin{minipage}[c][0.75cm][c]{1.8cm}\centering \ro{\textsf{CDRea} (4/5)}\ \end{minipage} \\\begin{minipage}[c][0.75cm][c]{1.8cm}\centering \ro{\textsf{CRea} (1/5)}\ \end{minipage} \\\end{tabular} & \ro{\textsf{CDRea} (5/5)} & \ro{\textsf{CDRea} (5/5)} & \ro{\textsf{CDRea} (5/5)}& \begin{tabular}{@{}c@{}}\begin{minipage}[c][0.75cm][c]{1.8cm}\centering \ro{\textsf{CDRe} (1/5)}\ \end{minipage} \\\begin{minipage}[c][0.75cm][c]{1.8cm}\centering \ro{\textsf{CDRea} (4/5)}\ \end{minipage} \\\end{tabular} & \ro{\textsf{CDRea} (5/5)} & \ro{\textsf{CDRea} (5/5)} & \ro{\textsf{CDRea} (5/5)}& \begin{tabular}{@{}c@{}}\begin{minipage}[c][0.75cm][c]{1.8cm}\centering \ro{\textsf{CDRea} (4/5)}\ \end{minipage} \\\begin{minipage}[c][0.75cm][c]{1.8cm}\centering \ro{\textsf{CRea} (1/5)}\ \end{minipage} \\\end{tabular} & \ro{\textsf{CDRea} (5/5)} & \ro{\textsf{CDRea} (5/5)} & \ro{\textsf{CDRea} (5/5)} & \ro{\textsf{CDRea} (5/5)} & \ro{\textsf{CDRea} (5/5)} \\
\hline
15.0& \begin{tabular}{@{}c@{}}\begin{minipage}[c][0.5cm][c]{1.8cm}\centering \blk{\textsf{Blk} (1/5)}\ \end{minipage} \\\begin{minipage}[c][0.5cm][c]{1.8cm}\centering \ro{\textsf{CDRea} (3/5)}\ \end{minipage} \\\begin{minipage}[c][0.5cm][c]{1.8cm}\centering \ro{\textsf{CRea} (1/5)}\ \end{minipage} \\\end{tabular}&\ro{\textsf{CDRa} (5/5)} & \ro{\textsf{CDRa} (5/5)}& \begin{tabular}{@{}c@{}}\begin{minipage}[c][0.75cm][c]{1.8cm}\centering \ro{\textsf{CDRea} (1/5)}\ \end{minipage} \\\begin{minipage}[c][0.75cm][c]{1.8cm}\centering \ro{\textsf{CDRa} (4/5)}\ \end{minipage} \\\end{tabular}& \begin{tabular}{@{}c@{}}\begin{minipage}[c][0.75cm][c]{1.8cm}\centering \ro{\textsf{CDRa} (4/5)}\ \end{minipage} \\\begin{minipage}[c][0.75cm][c]{1.8cm}\centering \ro{\textsf{CDRea} (1/5)}\ \end{minipage} \\\end{tabular}& \begin{tabular}{@{}c@{}}\begin{minipage}[c][0.75cm][c]{1.8cm}\centering \ro{\textsf{CDRa} (4/5)}\ \end{minipage} \\\begin{minipage}[c][0.75cm][c]{1.8cm}\centering \ro{\textsf{CDRea} (1/5)}\ \end{minipage} \\\end{tabular}& \begin{tabular}{@{}c@{}}\begin{minipage}[c][0.75cm][c]{1.8cm}\centering \ro{\textsf{CDRea} (1/5)}\ \end{minipage} \\\begin{minipage}[c][0.75cm][c]{1.8cm}\centering \ro{\textsf{CDRa} (4/5)}\ \end{minipage} \\\end{tabular}& \begin{tabular}{@{}c@{}}\begin{minipage}[c][0.75cm][c]{1.8cm}\centering \ro{\textsf{CDRea} (4/5)}\ \end{minipage} \\\begin{minipage}[c][0.75cm][c]{1.8cm}\centering \ro{\textsf{CDRa} (1/5)}\ \end{minipage} \\\end{tabular}& \begin{tabular}{@{}c@{}}\begin{minipage}[c][0.75cm][c]{1.8cm}\centering \ro{\textsf{CDRea} (3/5)}\ \end{minipage} \\\begin{minipage}[c][0.75cm][c]{1.8cm}\centering \ro{\textsf{CDRa} (2/5)}\ \end{minipage} \\\end{tabular}& \begin{tabular}{@{}c@{}}\begin{minipage}[c][0.75cm][c]{1.8cm}\centering \ro{\textsf{CDRa} (3/5)}\ \end{minipage} \\\begin{minipage}[c][0.75cm][c]{1.8cm}\centering \ro{\textsf{CDRea} (2/5)}\ \end{minipage} \\\end{tabular}& \begin{tabular}{@{}c@{}}\begin{minipage}[c][0.75cm][c]{1.8cm}\centering \ro{\textsf{CDRe} (2/5)}\ \end{minipage} \\\begin{minipage}[c][0.75cm][c]{1.8cm}\centering \ro{\textsf{CDRa} (3/5)}\ \end{minipage} \\\end{tabular}& \begin{tabular}{@{}c@{}}\begin{minipage}[c][0.75cm][c]{1.8cm}\centering \ro{\textsf{CDRa} (2/3)}\ \end{minipage} \\\begin{minipage}[c][0.75cm][c]{1.8cm}\centering \ro{\textsf{CDRea} (1/3)}\ \end{minipage} \\\end{tabular}& \begin{tabular}{@{}c@{}}\begin{minipage}[c][0.75cm][c]{1.8cm}\centering \ro{\textsf{CDRe} (2/5)}\ \end{minipage} \\\begin{minipage}[c][0.75cm][c]{1.8cm}\centering \ro{\textsf{CDRa} (3/5)}\ \end{minipage} \\\end{tabular}& \begin{tabular}{@{}c@{}}\begin{minipage}[c][0.5cm][c]{1.8cm}\centering \ro{\textsf{CDRea} (1/5)}\ \end{minipage} \\\begin{minipage}[c][0.5cm][c]{1.8cm}\centering \ro{\textsf{CDRa} (2/5)}\ \end{minipage} \\\begin{minipage}[c][0.5cm][c]{1.8cm}\centering \ro{\textsf{CDRe} (2/5)}\ \end{minipage} \\\end{tabular}& \begin{tabular}{@{}c@{}}\begin{minipage}[c][0.75cm][c]{1.8cm}\centering \ro{\textsf{CDRea} (3/5)}\ \end{minipage} \\\begin{minipage}[c][0.75cm][c]{1.8cm}\centering \ro{\textsf{CDRa} (2/5)}\ \end{minipage} \\\end{tabular} \\
\hline
20.0 & \blk{\textsf{Blk} (5/5)} & \pu{\textsf{PU$p_1$} (5/5)}& \begin{tabular}{@{}c@{}}\begin{minipage}[c][0.5cm][c]{1.8cm}\centering \ro{\textsf{CDRe} (4/5)}\ \end{minipage} \\\begin{minipage}[c][0.5cm][c]{1.8cm}\centering \ro{\textsf{CRe} (1/5)}\ \end{minipage} \\\end{tabular} & \ro{\textsf{CDRe} (5/5)}& \begin{tabular}{@{}c@{}}\begin{minipage}[c][0.5cm][c]{1.8cm}\centering \ro{\textsf{CDRe} (3/5)}\ \end{minipage} \\\begin{minipage}[c][0.5cm][c]{1.8cm}\centering \ro{\textsf{CRe} (2/5)}\ \end{minipage} \\\end{tabular} & \ro{\textsf{CRe} (5/5)}& \begin{tabular}{@{}c@{}}\begin{minipage}[c][0.5cm][c]{1.8cm}\centering \ro{\textsf{CRe} (3/5)}\ \end{minipage} \\\begin{minipage}[c][0.5cm][c]{1.8cm}\centering \pu{\textsf{PU$p_1$} (2/5)}\ \end{minipage} \\\end{tabular}& \begin{tabular}{@{}c@{}}\begin{minipage}[c][0.5cm][c]{1.8cm}\centering \ro{\textsf{CRe} (3/5)}\ \end{minipage} \\\begin{minipage}[c][0.5cm][c]{1.8cm}\centering \pu{\textsf{PU$p_1$} (2/5)}\ \end{minipage} \\\end{tabular}& \begin{tabular}{@{}c@{}}\begin{minipage}[c][0.5cm][c]{1.8cm}\centering \pu{\textsf{PU$p_1$} (1/5)}\ \end{minipage} \\\begin{minipage}[c][0.5cm][c]{1.8cm}\centering \ro{\textsf{CRe} (4/5)}\ \end{minipage} \\\end{tabular}& \begin{tabular}{@{}c@{}}\begin{minipage}[c][0.5cm][c]{1.8cm}\centering \ro{\textsf{CRe} (3/5)}\ \end{minipage} \\\begin{minipage}[c][0.5cm][c]{1.8cm}\centering \pu{\textsf{PU$p_1$} (2/5)}\ \end{minipage} \\\end{tabular} & \ro{\textsf{CRe} (5/5)} & \pu{\textsf{PU$p_1$} (5/5)}& \begin{tabular}{@{}c@{}}\begin{minipage}[c][0.5cm][c]{1.8cm}\centering \pu{\textsf{PU$p_1$} (3/5)}\ \end{minipage} \\\begin{minipage}[c][0.5cm][c]{1.8cm}\centering \ro{\textsf{CRe} (2/5)}\ \end{minipage} \\\end{tabular}& \begin{tabular}{@{}c@{}}\begin{minipage}[c][0.5cm][c]{1.8cm}\centering \ro{\textsf{CRe} (1/5)}\ \end{minipage} \\\begin{minipage}[c][0.5cm][c]{1.8cm}\centering \pu{\textsf{PU$p_1$} (4/5)}\ \end{minipage} \\\end{tabular}& \begin{tabular}{@{}c@{}}\begin{minipage}[c][0.5cm][c]{1.8cm}\centering \ro{\textsf{CRe} (4/5)}\ \end{minipage} \\\begin{minipage}[c][0.5cm][c]{1.8cm}\centering \pu{\textsf{PU$p_1$} (1/5)}\ \end{minipage} \\\end{tabular} \\
\hline
25.0 & \blk{\textsf{Blk} (5/5)} & \pu{\textsf{PU$p_1$} (5/5)}& \begin{tabular}{@{}c@{}}\begin{minipage}[c][0.5cm][c]{1.8cm}\centering \ro{\textsf{CDRe} (4/5)}\ \end{minipage} \\\begin{minipage}[c][0.5cm][c]{1.8cm}\centering \ro{\textsf{CRe} (1/5)}\ \end{minipage} \\\end{tabular} & \ro{\textsf{CDRe} (5/5)} & \ro{\textsf{CDRe} (5/5)} & \ro{\textsf{CDRe} (5/5)}& \begin{tabular}{@{}c@{}}\begin{minipage}[c][0.5cm][c]{1.8cm}\centering \ro{\textsf{CDRe} (4/5)}\ \end{minipage} \\\begin{minipage}[c][0.5cm][c]{1.8cm}\centering \ac{\textsf{Ae} (1/5)}\ \end{minipage} \\\end{tabular} & \ro{\textsf{CDRe} (5/5)} & \ro{\textsf{CDRe} (5/5)}& \begin{tabular}{@{}c@{}}\begin{minipage}[c][0.5cm][c]{1.8cm}\centering \ro{\textsf{CDRe} (3/5)}\ \end{minipage} \\\begin{minipage}[c][0.5cm][c]{1.8cm}\centering \ac{\textsf{Ae} (2/5)}\ \end{minipage} \\\end{tabular}& \begin{tabular}{@{}c@{}}\begin{minipage}[c][0.5cm][c]{1.8cm}\centering \ro{\textsf{CDRe} (3/5)}\ \end{minipage} \\\begin{minipage}[c][0.5cm][c]{1.8cm}\centering \ro{\textsf{CRe} (2/5)}\ \end{minipage} \\\end{tabular} & \ro{\textsf{CDRe} (5/5)} & \ro{\textsf{CRe} (5/5)}& \begin{tabular}{@{}c@{}}\begin{minipage}[c][0.5cm][c]{1.8cm}\centering \ro{\textsf{CDRe} (2/5)}\ \end{minipage} \\\begin{minipage}[c][0.5cm][c]{1.8cm}\centering \ro{\textsf{CRe} (3/5)}\ \end{minipage} \\\end{tabular}& \begin{tabular}{@{}c@{}}\begin{minipage}[c][0.5cm][c]{1.8cm}\centering \ro{\textsf{CDRe} (1/5)}\ \end{minipage} \\\begin{minipage}[c][0.5cm][c]{1.8cm}\centering \ro{\textsf{CRe} (4/5)}\ \end{minipage} \\\end{tabular} \\
\hline
\ct{27.0} & \blk{\textsf{Blk} (5/5)} & \pu{\textsf{PU$p_1$} (5/5)} & \ro{\textsf{CDRea} (5/5)}& \begin{tabular}{@{}c@{}}\begin{minipage}[c][0.5cm][c]{1.8cm}\centering \ro{\textsf{CDRe} (2/5)}\ \end{minipage} \\\begin{minipage}[c][0.5cm][c]{1.8cm}\centering \ro{\textsf{CDRea} (3/5)}\ \end{minipage} \\\end{tabular}& \begin{tabular}{@{}c@{}}\begin{minipage}[c][0.5cm][c]{1.8cm}\centering \ro{\textsf{CDRea} (3/5)}\ \end{minipage} \\\begin{minipage}[c][0.5cm][c]{1.8cm}\centering \ro{\textsf{CDRe} (2/5)}\ \end{minipage} \\\end{tabular}& \begin{tabular}{@{}c@{}}\begin{minipage}[c][0.5cm][c]{1.8cm}\centering \ro{\textsf{CDRe} (2/5)}\ \end{minipage} \\\begin{minipage}[c][0.5cm][c]{1.8cm}\centering \ro{\textsf{CDRea} (3/5)}\ \end{minipage} \\\end{tabular}& \begin{tabular}{@{}c@{}}\begin{minipage}[c][0.5cm][c]{1.8cm}\centering \ro{\textsf{CDRea} (3/5)}\ \end{minipage} \\\begin{minipage}[c][0.5cm][c]{1.8cm}\centering \ro{\textsf{CDRe} (2/5)}\ \end{minipage} \\\end{tabular}& \begin{tabular}{@{}c@{}}\begin{minipage}[c][0.5cm][c]{1.8cm}\centering \ro{\textsf{CDRe} (2/5)}\ \end{minipage} \\\begin{minipage}[c][0.5cm][c]{1.8cm}\centering \ro{\textsf{CDRea} (3/5)}\ \end{minipage} \\\end{tabular}& \begin{tabular}{@{}c@{}}\begin{minipage}[c][0.5cm][c]{1.8cm}\centering \ro{\textsf{CDRea} (4/5)}\ \end{minipage} \\\begin{minipage}[c][0.5cm][c]{1.8cm}\centering \ro{\textsf{CDRe} (1/5)}\ \end{minipage} \\\end{tabular}& \begin{tabular}{@{}c@{}}\begin{minipage}[c][0.5cm][c]{1.8cm}\centering \ro{\textsf{CDRea} (4/5)}\ \end{minipage} \\\begin{minipage}[c][0.5cm][c]{1.8cm}\centering \ro{\textsf{CDRe} (1/5)}\ \end{minipage} \\\end{tabular}& \begin{tabular}{@{}c@{}}\begin{minipage}[c][0.5cm][c]{1.8cm}\centering \ro{\textsf{CDRe} (2/5)}\ \end{minipage} \\\begin{minipage}[c][0.5cm][c]{1.8cm}\centering \ro{\textsf{CDRea} (3/5)}\ \end{minipage} \\\end{tabular}& \begin{tabular}{@{}c@{}}\begin{minipage}[c][0.5cm][c]{1.8cm}\centering \ro{\textsf{CDRe} (1/5)}\ \end{minipage} \\\begin{minipage}[c][0.5cm][c]{1.8cm}\centering \ro{\textsf{CDRea} (4/5)}\ \end{minipage} \\\end{tabular} & \ro{\textsf{CDRea} (5/5)} & \ro{\textsf{CDRea} (5/5)} & \ro{\textsf{CDRea} (5/5)} \\ 
\hline 
30.0 & \blk{\textsf{Blk} (5/5)} & \ps{\textsf{PS} (5/5)}& \begin{tabular}{@{}c@{}}\begin{minipage}[c][0.75cm][c]{1.8cm}\centering \ro{\textsf{CDRe} (4/5)}\ \end{minipage} \\\begin{minipage}[c][0.75cm][c]{1.8cm}\centering \ro{\textsf{CDRea} (1/5)}\ \end{minipage} \\\end{tabular} & \ro{\textsf{CDRe} (5/5)}& \begin{tabular}{@{}c@{}}\begin{minipage}[c][0.75cm][c]{1.8cm}\centering \ro{\textsf{CDRe} (4/5)}\ \end{minipage} \\\begin{minipage}[c][0.75cm][c]{1.8cm}\centering \ro{\textsf{CDRea} (1/5)}\ \end{minipage} \\\end{tabular}& \begin{tabular}{@{}c@{}}\begin{minipage}[c][0.5cm][c]{1.8cm}\centering \ro{\textsf{CDRe} (3/5)}\ \end{minipage} \\\begin{minipage}[c][0.5cm][c]{1.8cm}\centering \ro{\textsf{CDRea} (1/5)}\ \end{minipage} \\\begin{minipage}[c][0.5cm][c]{1.8cm}\centering \ro{\textsf{CRe} (1/5)}\ \end{minipage} \\\end{tabular}& \begin{tabular}{@{}c@{}}\begin{minipage}[c][0.75cm][c]{1.8cm}\centering \ro{\textsf{CDRe} (4/5)}\ \end{minipage} \\\begin{minipage}[c][0.75cm][c]{1.8cm}\centering \ro{\textsf{CDRea} (1/5)}\ \end{minipage} \\\end{tabular}& \begin{tabular}{@{}c@{}}\begin{minipage}[c][0.75cm][c]{1.8cm}\centering \ro{\textsf{CDRe} (2/5)}\ \end{minipage} \\\begin{minipage}[c][0.75cm][c]{1.8cm}\centering \ro{\textsf{CDRea} (3/5)}\ \end{minipage} \\\end{tabular}& \begin{tabular}{@{}c@{}}\begin{minipage}[c][0.75cm][c]{1.8cm}\centering \ro{\textsf{CDRe} (4/5)}\ \end{minipage} \\\begin{minipage}[c][0.75cm][c]{1.8cm}\centering \ro{\textsf{CDRea} (1/5)}\ \end{minipage} \\\end{tabular} & \ro{\textsf{CDRe} (5/5)}& \begin{tabular}{@{}c@{}}\begin{minipage}[c][0.75cm][c]{1.8cm}\centering \ro{\textsf{CDRe} (4/5)}\ \end{minipage} \\\begin{minipage}[c][0.75cm][c]{1.8cm}\centering \ro{\textsf{CDRea} (1/5)}\ \end{minipage} \\\end{tabular}& \begin{tabular}{@{}c@{}}\begin{minipage}[c][0.75cm][c]{1.8cm}\centering \ro{\textsf{CDRea} (3/5)}\ \end{minipage} \\\begin{minipage}[c][0.75cm][c]{1.8cm}\centering \ro{\textsf{CDRe} (2/5)}\ \end{minipage} \\\end{tabular}& \begin{tabular}{@{}c@{}}\begin{minipage}[c][0.75cm][c]{1.8cm}\centering \ro{\textsf{CDRea} (3/5)}\ \end{minipage} \\\begin{minipage}[c][0.75cm][c]{1.8cm}\centering \ro{\textsf{CDRe} (2/5)}\ \end{minipage} \\\end{tabular} & \ro{\textsf{CDRe} (5/5)}& \begin{tabular}{@{}c@{}}\begin{minipage}[c][0.75cm][c]{1.8cm}\centering \ro{\textsf{CDRe} (4/5)}\ \end{minipage} \\\begin{minipage}[c][0.75cm][c]{1.8cm}\centering \ro{\textsf{CDRea} (1/5)}\ \end{minipage} \\\end{tabular} \\
\hline
35.0 & \blk{\textsf{Blk} (5/5)} & \ps{\textsf{PS} (5/5)}& \begin{tabular}{@{}c@{}}\begin{minipage}[c][0.5cm][c]{1.8cm}\centering \ro{\textsf{CDRea} (3/5)}\ \end{minipage} \\\begin{minipage}[c][0.5cm][c]{1.8cm}\centering \ro{\textsf{CDRe} (2/5)}\ \end{minipage} \\\end{tabular}& \begin{tabular}{@{}c@{}}\begin{minipage}[c][0.5cm][c]{1.8cm}\centering \ro{\textsf{CDRea} (4/5)}\ \end{minipage} \\\begin{minipage}[c][0.5cm][c]{1.8cm}\centering \ro{\textsf{CDRe} (1/5)}\ \end{minipage} \\\end{tabular}& \begin{tabular}{@{}c@{}}\begin{minipage}[c][0.5cm][c]{1.8cm}\centering \ro{\textsf{CDRea} (3/5)}\ \end{minipage} \\\begin{minipage}[c][0.5cm][c]{1.8cm}\centering \ro{\textsf{CDRe} (2/5)}\ \end{minipage} \\\end{tabular}& \begin{tabular}{@{}c@{}}\begin{minipage}[c][0.5cm][c]{1.8cm}\centering \ro{\textsf{CDRea} (3/5)}\ \end{minipage} \\\begin{minipage}[c][0.5cm][c]{1.8cm}\centering \ro{\textsf{CDRe} (2/5)}\ \end{minipage} \\\end{tabular}& \begin{tabular}{@{}c@{}}\begin{minipage}[c][0.5cm][c]{1.8cm}\centering \ro{\textsf{CDRea} (3/5)}\ \end{minipage} \\\begin{minipage}[c][0.5cm][c]{1.8cm}\centering \ro{\textsf{CDRe} (2/5)}\ \end{minipage} \\\end{tabular} & \ro{\textsf{CDRe} (5/5)} & \ro{\textsf{CDRe} (5/5)}& \begin{tabular}{@{}c@{}}\begin{minipage}[c][0.5cm][c]{1.8cm}\centering \ro{\textsf{CDRe} (3/5)}\ \end{minipage} \\\begin{minipage}[c][0.5cm][c]{1.8cm}\centering \ro{\textsf{CDRea} (2/5)}\ \end{minipage} \\\end{tabular} & \ro{\textsf{CDRe} (5/5)}& \begin{tabular}{@{}c@{}}\begin{minipage}[c][0.5cm][c]{1.8cm}\centering \ro{\textsf{CDRe} (3/5)}\ \end{minipage} \\\begin{minipage}[c][0.5cm][c]{1.8cm}\centering \ro{\textsf{CDRea} (2/5)}\ \end{minipage} \\\end{tabular} & \ro{\textsf{CDRe} (5/5)} & \ro{\textsf{CDRe} (5/5)} & \ro{\textsf{CDRe} (5/5)} \\
\hline
40.0 & \blk{\textsf{Blk} (5/5)} & \ps{\textsf{PS} (5/5)}& \begin{tabular}{@{}c@{}}\begin{minipage}[c][0.5cm][c]{1.8cm}\centering \ro{\textsf{CDRe} (1/5)}\ \end{minipage} \\\begin{minipage}[c][0.5cm][c]{1.8cm}\centering \ro{\textsf{CDRea} (4/5)}\ \end{minipage} \\\end{tabular} & \ro{\textsf{CDRea} (5/5)}& \begin{tabular}{@{}c@{}}\begin{minipage}[c][0.5cm][c]{1.8cm}\centering \ro{\textsf{CRe} (1/5)}\ \end{minipage} \\\begin{minipage}[c][0.5cm][c]{1.8cm}\centering \ro{\textsf{CDRea} (4/5)}\ \end{minipage} \\\end{tabular} & \ro{\textsf{CDRea} (5/5)} & \ro{\textsf{CDRea} (5/5)} & \ro{\textsf{CDRea} (5/5)} & \ro{\textsf{CDRea} (5/5)} & \ro{\textsf{CDRea} (5/5)} & \ro{\textsf{CDRea} (5/5)} & \ro{\textsf{CDRea} (5/5)} & \ro{\textsf{CDRea} (5/5)} & \ro{\textsf{CDRea} (5/5)} & \ro{\textsf{CDRea} (5/5)} \\
\hline
\end{tabular}

%% file: tables/lmdrive-crossing-4.tex
\begin{tabular}{|c| *{15}{c|}}
\hline
\diagbox{$\bm{x_a}$}{$\bm{x_f}$} & 0.0 & \ct{2.9} & 40.0 & 80.0 & 120.0 & 160.0 & 200.0 & 240.0 & 280.0 & 320.0 & 360.0 & 400.0 & 440.0 & 480.0 & 520.0 \\
\hline
0.0& \begin{tabular}{@{}c@{}}\begin{minipage}[c][0.5cm][c]{1.8cm}\centering \ro{\textsf{CDRea} (1/5)}\ \end{minipage} \\\begin{minipage}[c][0.5cm][c]{1.8cm}\centering \blk{\textsf{Blk} (4/5)}\ \end{minipage} \\\end{tabular}& \ro{\textsf{CDRea} (5/5)}& \begin{tabular}{@{}c@{}}\begin{minipage}[c][0.5cm][c]{1.8cm}\centering \ro{\textsf{CDRea} (1/5)}\ \end{minipage} \\\begin{minipage}[c][0.5cm][c]{1.8cm}\centering \blk{\textsf{Blk} (4/5)}\ \end{minipage} \\\end{tabular}& \begin{tabular}{@{}c@{}}\begin{minipage}[c][0.5cm][c]{1.8cm}\centering \ro{\textsf{CDRea} (1/5)}\ \end{minipage} \\\begin{minipage}[c][0.5cm][c]{1.8cm}\centering \blk{\textsf{Blk} (4/5)}\ \end{minipage} \\\end{tabular}& \begin{tabular}{@{}c@{}}\begin{minipage}[c][0.5cm][c]{1.8cm}\centering \ro{\textsf{CRea} (1/5)}\ \end{minipage} \\\begin{minipage}[c][0.5cm][c]{1.8cm}\centering \blk{\textsf{Blk} (4/5)}\ \end{minipage} \\\end{tabular}& \begin{tabular}{@{}c@{}}\begin{minipage}[c][0.5cm][c]{1.8cm}\centering \ro{\textsf{CRea} (1/5)}\ \end{minipage} \\\begin{minipage}[c][0.5cm][c]{1.8cm}\centering \blk{\textsf{Blk} (4/5)}\ \end{minipage} \\\end{tabular}& \begin{tabular}{@{}c@{}}\begin{minipage}[c][0.5cm][c]{1.8cm}\centering \ro{\textsf{CRea} (1/5)}\ \end{minipage} \\\begin{minipage}[c][0.5cm][c]{1.8cm}\centering \blk{\textsf{Blk} (4/5)}\ \end{minipage} \\\end{tabular} & \blk{\textsf{Blk} (5/5)} & \blk{\textsf{Blk} (5/5)} & \blk{\textsf{Blk} (5/5)} & \blk{\textsf{Blk} (5/5)} & \blk{\textsf{Blk} (5/5)} & \blk{\textsf{Blk} (5/5)} & \blk{\textsf{Blk} (5/5)} & \blk{\textsf{Blk} (5/5)} \\
\hline
5.0& \begin{tabular}{@{}c@{}}\begin{minipage}[c][0.5cm][c]{1.8cm}\centering \ro{\textsf{CDRe} (1/5)}\ \end{minipage} \\\begin{minipage}[c][0.5cm][c]{1.8cm}\centering \blk{\textsf{Blk} (1/5)}\ \end{minipage} \\\begin{minipage}[c][0.5cm][c]{1.8cm}\centering \ro{\textsf{CDRea} (3/5)}\ \end{minipage} \\\end{tabular}& \ro{\textsf{CDRea} (5/5)} & \ro{\textsf{CDRa} (5/5)} & \ro{\textsf{CDRea} (5/5)} & \ro{\textsf{CDRa} (5/5)}& \begin{tabular}{@{}c@{}}\begin{minipage}[c][0.75cm][c]{1.8cm}\centering \ro{\textsf{CDRea} (1/5)}\ \end{minipage} \\\begin{minipage}[c][0.75cm][c]{1.8cm}\centering \ro{\textsf{CDRa} (4/5)}\ \end{minipage} \\\end{tabular}& \begin{tabular}{@{}c@{}}\begin{minipage}[c][0.75cm][c]{1.8cm}\centering \ro{\textsf{CDRea} (2/5)}\ \end{minipage} \\\begin{minipage}[c][0.75cm][c]{1.8cm}\centering \ro{\textsf{CDRa} (3/5)}\ \end{minipage} \\\end{tabular}& \begin{tabular}{@{}c@{}}\begin{minipage}[c][0.75cm][c]{1.8cm}\centering \ro{\textsf{CDRea} (4/5)}\ \end{minipage} \\\begin{minipage}[c][0.75cm][c]{1.8cm}\centering \ro{\textsf{CDRa} (1/5)}\ \end{minipage} \\\end{tabular} & \ro{\textsf{CDRea} (5/5)} & \ro{\textsf{CDRea} (5/5)}& \begin{tabular}{@{}c@{}}\begin{minipage}[c][0.75cm][c]{1.8cm}\centering \ro{\textsf{CDRa} (1/5)}\ \end{minipage} \\\begin{minipage}[c][0.75cm][c]{1.8cm}\centering \ro{\textsf{CDRea} (4/5)}\ \end{minipage} \\\end{tabular}& \begin{tabular}{@{}c@{}}\begin{minipage}[c][0.5cm][c]{1.8cm}\centering \ro{\textsf{CDRea} (3/5)}\ \end{minipage} \\\begin{minipage}[c][0.5cm][c]{1.8cm}\centering \ro{\textsf{CRea} (1/5)}\ \end{minipage} \\\begin{minipage}[c][0.5cm][c]{1.8cm}\centering \ro{\textsf{CDRa} (1/5)}\ \end{minipage} \\\end{tabular} & \ro{\textsf{CDRea} (5/5)}& \begin{tabular}{@{}c@{}}\begin{minipage}[c][0.75cm][c]{1.8cm}\centering \ro{\textsf{CDRa} (2/5)}\ \end{minipage} \\\begin{minipage}[c][0.75cm][c]{1.8cm}\centering \ro{\textsf{CDRea} (3/5)}\ \end{minipage} \\\end{tabular}& \begin{tabular}{@{}c@{}}\begin{minipage}[c][0.75cm][c]{1.8cm}\centering \ro{\textsf{CDRea} (4/5)}\ \end{minipage} \\\begin{minipage}[c][0.75cm][c]{1.8cm}\centering \ro{\textsf{CDRa} (1/5)}\ \end{minipage} \\\end{tabular} \\
\hline
10.0& \begin{tabular}{@{}c@{}}\begin{minipage}[c][0.75cm][c]{1.8cm}\centering \blk{\textsf{Blk} (1/5)}\ \end{minipage} \\\begin{minipage}[c][0.75cm][c]{1.8cm}\centering \ro{\textsf{CRe} (4/5)}\ \end{minipage} \\\end{tabular}& \ro{\textsf{CDRe} (5/5)}& \begin{tabular}{@{}c@{}}\begin{minipage}[c][0.75cm][c]{1.8cm}\centering \blk{\textsf{Blk} (1/5)}\ \end{minipage} \\\begin{minipage}[c][0.75cm][c]{1.8cm}\centering \ro{\textsf{CDRea} (4/5)}\ \end{minipage} \\\end{tabular}& \begin{tabular}{@{}c@{}}\begin{minipage}[c][0.75cm][c]{1.8cm}\centering \blk{\textsf{Blk} (1/5)}\ \end{minipage} \\\begin{minipage}[c][0.75cm][c]{1.8cm}\centering \ro{\textsf{CDRea} (4/5)}\ \end{minipage} \\\end{tabular} & \ro{\textsf{CDRea} (5/5)} & \ro{\textsf{CDRea} (5/5)} & \ro{\textsf{CDRea} (5/5)} & \ro{\textsf{CDRea} (5/5)}& \begin{tabular}{@{}c@{}}\begin{minipage}[c][0.75cm][c]{1.8cm}\centering \ro{\textsf{CRea} (1/5)}\ \end{minipage} \\\begin{minipage}[c][0.75cm][c]{1.8cm}\centering \ro{\textsf{CDRea} (4/5)}\ \end{minipage} \\\end{tabular}& \begin{tabular}{@{}c@{}}\begin{minipage}[c][0.75cm][c]{1.8cm}\centering \ro{\textsf{CDRea} (4/5)}\ \end{minipage} \\\begin{minipage}[c][0.75cm][c]{1.8cm}\centering \ro{\textsf{CDRe} (1/5)}\ \end{minipage} \\\end{tabular} & \ro{\textsf{CDRea} (5/5)}& \begin{tabular}{@{}c@{}}\begin{minipage}[c][0.5cm][c]{1.8cm}\centering \ro{\textsf{CDRea} (3/5)}\ \end{minipage} \\\begin{minipage}[c][0.5cm][c]{1.8cm}\centering \ro{\textsf{CDRa} (1/5)}\ \end{minipage} \\\begin{minipage}[c][0.5cm][c]{1.8cm}\centering \ro{\textsf{CDRe} (1/5)}\ \end{minipage} \\\end{tabular}& \begin{tabular}{@{}c@{}}\begin{minipage}[c][0.75cm][c]{1.8cm}\centering \ro{\textsf{CDRea} (4/5)}\ \end{minipage} \\\begin{minipage}[c][0.75cm][c]{1.8cm}\centering \ro{\textsf{CDRe} (1/5)}\ \end{minipage} \\\end{tabular} & \ro{\textsf{CDRea} (5/5)}& \begin{tabular}{@{}c@{}}\begin{minipage}[c][0.75cm][c]{1.8cm}\centering \ro{\textsf{CDRea} (4/5)}\ \end{minipage} \\\begin{minipage}[c][0.75cm][c]{1.8cm}\centering \ro{\textsf{CDRe} (1/5)}\ \end{minipage} \\\end{tabular} \\
\hline
15.0 & \blk{\textsf{Blk} (5/5)} & \ro{\textsf{CDRe} (5/5)} & \ro{\textsf{CDRe} (5/5)} & \ro{\textsf{CDRe} (5/5)} & \ro{\textsf{CDRe} (5/5)} & \ro{\textsf{CDRe} (5/5)} & \ro{\textsf{CDRe} (5/5)}& \begin{tabular}{@{}c@{}}\begin{minipage}[c][0.5cm][c]{1.8cm}\centering \ro{\textsf{CRe} (1/5)}\ \end{minipage} \\\begin{minipage}[c][0.5cm][c]{1.8cm}\centering \ro{\textsf{CDRe} (4/5)}\ \end{minipage} \\\end{tabular} & \ro{\textsf{CDRe} (5/5)} & \ro{\textsf{CDRe} (5/5)}& \begin{tabular}{@{}c@{}}\begin{minipage}[c][0.5cm][c]{1.8cm}\centering \ro{\textsf{CDRe} (4/5)}\ \end{minipage} \\\begin{minipage}[c][0.5cm][c]{1.8cm}\centering \ro{\textsf{CDRea} (1/5)}\ \end{minipage} \\\end{tabular} & \ro{\textsf{CDRe} (5/5)}& \begin{tabular}{@{}c@{}}\begin{minipage}[c][0.5cm][c]{1.8cm}\centering \ro{\textsf{CDRe} (4/5)}\ \end{minipage} \\\begin{minipage}[c][0.5cm][c]{1.8cm}\centering \ro{\textsf{CDRea} (1/5)}\ \end{minipage} \\\end{tabular} & \ro{\textsf{CDRe} (5/5)} & \ro{\textsf{CDRe} (5/5)} \\
\hline
20.0 & \blk{\textsf{Blk} (5/5)} & \ro{\textsf{CDRe} (5/5)}& \begin{tabular}{@{}c@{}}\begin{minipage}[c][0.75cm][c]{1.8cm}\centering \ro{\textsf{CDRe} (3/5)}\ \end{minipage} \\\begin{minipage}[c][0.75cm][c]{1.8cm}\centering \ro{\textsf{CDRea} (2/5)}\ \end{minipage} \\\end{tabular}& \begin{tabular}{@{}c@{}}\begin{minipage}[c][0.75cm][c]{1.8cm}\centering \ro{\textsf{CDRe} (2/5)}\ \end{minipage} \\\begin{minipage}[c][0.75cm][c]{1.8cm}\centering \ro{\textsf{CDRea} (3/5)}\ \end{minipage} \\\end{tabular}& \begin{tabular}{@{}c@{}}\begin{minipage}[c][0.75cm][c]{1.8cm}\centering \ro{\textsf{CDRe} (2/5)}\ \end{minipage} \\\begin{minipage}[c][0.75cm][c]{1.8cm}\centering \ro{\textsf{CDRea} (3/5)}\ \end{minipage} \\\end{tabular}& \begin{tabular}{@{}c@{}}\begin{minipage}[c][0.75cm][c]{1.8cm}\centering \ro{\textsf{CDRe} (2/5)}\ \end{minipage} \\\begin{minipage}[c][0.75cm][c]{1.8cm}\centering \ro{\textsf{CDRea} (3/5)}\ \end{minipage} \\\end{tabular}& \begin{tabular}{@{}c@{}}\begin{minipage}[c][0.75cm][c]{1.8cm}\centering \ro{\textsf{CDRe} (4/5)}\ \end{minipage} \\\begin{minipage}[c][0.75cm][c]{1.8cm}\centering \ro{\textsf{CDRea} (1/5)}\ \end{minipage} \\\end{tabular}& \begin{tabular}{@{}c@{}}\begin{minipage}[c][0.75cm][c]{1.8cm}\centering \ro{\textsf{CDRea} (2/5)}\ \end{minipage} \\\begin{minipage}[c][0.75cm][c]{1.8cm}\centering \ro{\textsf{CDRe} (3/5)}\ \end{minipage} \\\end{tabular}& \begin{tabular}{@{}c@{}}\begin{minipage}[c][0.5cm][c]{1.8cm}\centering \ro{\textsf{CDRea} (1/5)}\ \end{minipage} \\\begin{minipage}[c][0.5cm][c]{1.8cm}\centering \ro{\textsf{CDRe} (3/5)}\ \end{minipage} \\\begin{minipage}[c][0.5cm][c]{1.8cm}\centering \ro{\textsf{CRe} (1/5)}\ \end{minipage} \\\end{tabular}& \begin{tabular}{@{}c@{}}\begin{minipage}[c][0.75cm][c]{1.8cm}\centering \ro{\textsf{CDRea} (3/5)}\ \end{minipage} \\\begin{minipage}[c][0.75cm][c]{1.8cm}\centering \ro{\textsf{CDRe} (2/5)}\ \end{minipage} \\\end{tabular}& \begin{tabular}{@{}c@{}}\begin{minipage}[c][0.5cm][c]{1.8cm}\centering \ro{\textsf{CDRea} (3/5)}\ \end{minipage} \\\begin{minipage}[c][0.5cm][c]{1.8cm}\centering \ps{\textsf{PS} (1/5)}\ \end{minipage} \\\begin{minipage}[c][0.5cm][c]{1.8cm}\centering \ro{\textsf{CDRe} (1/5)}\ \end{minipage} \\\end{tabular}& \begin{tabular}{@{}c@{}}\begin{minipage}[c][0.5cm][c]{1.8cm}\centering \ps{\textsf{PS} (2/5)}\ \end{minipage} \\\begin{minipage}[c][0.5cm][c]{1.8cm}\centering \ro{\textsf{CDRea} (2/5)}\ \end{minipage} \\\begin{minipage}[c][0.5cm][c]{1.8cm}\centering \ro{\textsf{CDRe} (1/5)}\ \end{minipage} \\\end{tabular}& \begin{tabular}{@{}c@{}}\begin{minipage}[c][0.75cm][c]{1.8cm}\centering \ro{\textsf{CDRea} (3/5)}\ \end{minipage} \\\begin{minipage}[c][0.75cm][c]{1.8cm}\centering \ro{\textsf{CDRe} (2/5)}\ \end{minipage} \\\end{tabular}& \begin{tabular}{@{}c@{}}\begin{minipage}[c][0.5cm][c]{1.8cm}\centering \ps{\textsf{PS} (1/5)}\ \end{minipage} \\\begin{minipage}[c][0.5cm][c]{1.8cm}\centering \ro{\textsf{CDRea} (3/5)}\ \end{minipage} \\\begin{minipage}[c][0.5cm][c]{1.8cm}\centering \ro{\textsf{CDRe} (1/5)}\ \end{minipage} \\\end{tabular}& \begin{tabular}{@{}c@{}}\begin{minipage}[c][0.75cm][c]{1.8cm}\centering \ro{\textsf{CDRe} (2/5)}\ \end{minipage} \\\begin{minipage}[c][0.75cm][c]{1.8cm}\centering \ro{\textsf{CDRea} (3/5)}\ \end{minipage} \\\end{tabular} \\
\hline
\ct{23.1} & \ro{\textsf{CDRea} (5/5)} & \ro{\textsf{CDRea} (5/5)} & \ro{\textsf{CDRea} (5/5)} & \ro{\textsf{CDRea} (5/5)} & \ro{\textsf{CDRe} (5/5)}& \begin{tabular}{@{}c@{}}\begin{minipage}[c][0.5cm][c]{1.8cm}\centering \ro{\textsf{CDRea} (3/5)}\ \end{minipage} \\\begin{minipage}[c][0.5cm][c]{1.8cm}\centering \ro{\textsf{CDRe} (2/5)}\ \end{minipage} \\\end{tabular} & \ro{\textsf{CDRea} (5/5)}& \begin{tabular}{@{}c@{}}\begin{minipage}[c][0.5cm][c]{1.8cm}\centering \ro{\textsf{CDRea} (4/5)}\ \end{minipage} \\\begin{minipage}[c][0.5cm][c]{1.8cm}\centering \ro{\textsf{CDRe} (1/5)}\ \end{minipage} \\\end{tabular}& \begin{tabular}{@{}c@{}}\begin{minipage}[c][0.5cm][c]{1.8cm}\centering \ro{\textsf{CDRea} (3/5)}\ \end{minipage} \\\begin{minipage}[c][0.5cm][c]{1.8cm}\centering \ro{\textsf{CDRe} (2/5)}\ \end{minipage} \\\end{tabular}& \begin{tabular}{@{}c@{}}\begin{minipage}[c][0.5cm][c]{1.8cm}\centering \ro{\textsf{CDRe} (4/5)}\ \end{minipage} \\\begin{minipage}[c][0.5cm][c]{1.8cm}\centering \ro{\textsf{CDRea} (1/5)}\ \end{minipage} \\\end{tabular}& \begin{tabular}{@{}c@{}}\begin{minipage}[c][0.5cm][c]{1.8cm}\centering \ro{\textsf{CDRe} (4/5)}\ \end{minipage} \\\begin{minipage}[c][0.5cm][c]{1.8cm}\centering \ro{\textsf{CDRea} (1/5)}\ \end{minipage} \\\end{tabular} & \ro{\textsf{CDRe} (5/5)}& \begin{tabular}{@{}c@{}}\begin{minipage}[c][0.5cm][c]{1.8cm}\centering \ro{\textsf{CDRe} (2/5)}\ \end{minipage} \\\begin{minipage}[c][0.5cm][c]{1.8cm}\centering \ro{\textsf{CDRea} (3/5)}\ \end{minipage} \\\end{tabular}& \begin{tabular}{@{}c@{}}\begin{minipage}[c][0.5cm][c]{1.8cm}\centering \ro{\textsf{CDRe} (4/5)}\ \end{minipage} \\\begin{minipage}[c][0.5cm][c]{1.8cm}\centering \ro{\textsf{CDRea} (1/5)}\ \end{minipage} \\\end{tabular}& \begin{tabular}{@{}c@{}}\begin{minipage}[c][0.5cm][c]{1.8cm}\centering \ro{\textsf{CDRe} (3/5)}\ \end{minipage} \\\begin{minipage}[c][0.5cm][c]{1.8cm}\centering \ro{\textsf{CDRea} (2/5)}\ \end{minipage} \\\end{tabular} \\
\hline
25.0& \begin{tabular}{@{}c@{}}\begin{minipage}[c][0.5cm][c]{1.8cm}\centering \ro{\textsf{CDRe} (2/5)}\ \end{minipage} \\\begin{minipage}[c][0.5cm][c]{1.8cm}\centering \ro{\textsf{CDRea} (3/5)}\ \end{minipage} \\\end{tabular}& \begin{tabular}{@{}c@{}}\begin{minipage}[c][0.5cm][c]{1.8cm}\centering \ro{\textsf{CRe} (2/5)}\ \end{minipage} \\\begin{minipage}[c][0.5cm][c]{1.8cm}\centering \ro{\textsf{CDRe} (3/5)}\ \end{minipage} \\\end{tabular} & \ro{\textsf{CDRe} (5/5)} & \ro{\textsf{CDRe} (5/5)} & \ro{\textsf{CDRe} (5/5)} & \ro{\textsf{CDRe} (5/5)} & \ro{\textsf{CDRe} (5/5)} & \ro{\textsf{CDRe} (5/5)} & \ro{\textsf{CDRe} (5/5)} & \ro{\textsf{CDRe} (5/5)} & \ro{\textsf{CDRe} (5/5)} & \ro{\textsf{CDRe} (5/5)} & \ro{\textsf{CDRe} (5/5)} & \ro{\textsf{CDRe} (5/5)} & \ro{\textsf{CDRe} (5/5)} \\
\hline
30.0 & \ro{\textsf{CDRea} (5/5)} & \begin{tabular}{@{}c@{}}\begin{minipage}[c][0.5cm][c]{1.8cm}\centering \ro{\textsf{CDRea} (4/5)}\ \end{minipage} \\\begin{minipage}[c][0.5cm][c]{1.8cm}\centering \ro{\textsf{CRea} (1/5)}\ \end{minipage} \\\end{tabular} & \ro{\textsf{CDRea} (5/5)} & \ro{\textsf{CDRea} (5/5)} & \ro{\textsf{CDRea} (5/5)}& \begin{tabular}{@{}c@{}}\begin{minipage}[c][0.5cm][c]{1.8cm}\centering \ro{\textsf{CDRe} (1/5)}\ \end{minipage} \\\begin{minipage}[c][0.5cm][c]{1.8cm}\centering \ro{\textsf{CDRea} (4/5)}\ \end{minipage} \\\end{tabular} & \ro{\textsf{CDRea} (5/5)} & \ro{\textsf{CDRea} (5/5)} & \ro{\textsf{CDRea} (5/5)} & \ro{\textsf{CDRea} (5/5)} & \ro{\textsf{CDRea} (5/5)} & \ro{\textsf{CDRea} (5/5)} & \ro{\textsf{CDRea} (5/5)}& \begin{tabular}{@{}c@{}}\begin{minipage}[c][0.5cm][c]{1.8cm}\centering \ro{\textsf{CDRea} (4/5)}\ \end{minipage} \\\begin{minipage}[c][0.5cm][c]{1.8cm}\centering \ro{\textsf{CDRa} (1/5)}\ \end{minipage} \\\end{tabular} & \ro{\textsf{CDRea} (5/5)} \\
\hline
35.0& \begin{tabular}{@{}c@{}}\begin{minipage}[c][0.5cm][c]{1.8cm}\centering \ro{\textsf{CDRea} (2/5)}\ \end{minipage} \\\begin{minipage}[c][0.5cm][c]{1.8cm}\centering \ro{\textsf{CDRe} (2/5)}\ \end{minipage} \\\begin{minipage}[c][0.5cm][c]{1.8cm}\centering \ro{\textsf{CRea} (1/5)}\ \end{minipage} \\\end{tabular}& \ro{\textsf{CDRea} (5/5)} & \ro{\textsf{CDRea} (5/5)}& \begin{tabular}{@{}c@{}}\begin{minipage}[c][0.75cm][c]{1.8cm}\centering \ro{\textsf{CDRea} (3/5)}\ \end{minipage} \\\begin{minipage}[c][0.75cm][c]{1.8cm}\centering \ro{\textsf{CDRe} (2/5)}\ \end{minipage} \\\end{tabular}& \begin{tabular}{@{}c@{}}\begin{minipage}[c][0.75cm][c]{1.8cm}\centering \ro{\textsf{CDRea} (2/5)}\ \end{minipage} \\\begin{minipage}[c][0.75cm][c]{1.8cm}\centering \ro{\textsf{CDRe} (3/5)}\ \end{minipage} \\\end{tabular}& \begin{tabular}{@{}c@{}}\begin{minipage}[c][0.75cm][c]{1.8cm}\centering \ro{\textsf{CDRea} (3/5)}\ \end{minipage} \\\begin{minipage}[c][0.75cm][c]{1.8cm}\centering \ro{\textsf{CDRe} (2/5)}\ \end{minipage} \\\end{tabular}& \begin{tabular}{@{}c@{}}\begin{minipage}[c][0.75cm][c]{1.8cm}\centering \ro{\textsf{CDRea} (1/5)}\ \end{minipage} \\\begin{minipage}[c][0.75cm][c]{1.8cm}\centering \ro{\textsf{CDRe} (4/5)}\ \end{minipage} \\\end{tabular}& \begin{tabular}{@{}c@{}}\begin{minipage}[c][0.75cm][c]{1.8cm}\centering \ro{\textsf{CDRea} (3/5)}\ \end{minipage} \\\begin{minipage}[c][0.75cm][c]{1.8cm}\centering \ro{\textsf{CDRe} (2/5)}\ \end{minipage} \\\end{tabular}& \begin{tabular}{@{}c@{}}\begin{minipage}[c][0.75cm][c]{1.8cm}\centering \ro{\textsf{CDRe} (4/5)}\ \end{minipage} \\\begin{minipage}[c][0.75cm][c]{1.8cm}\centering \ro{\textsf{CDRea} (1/5)}\ \end{minipage} \\\end{tabular}& \begin{tabular}{@{}c@{}}\begin{minipage}[c][0.75cm][c]{1.8cm}\centering \ro{\textsf{CDRea} (4/5)}\ \end{minipage} \\\begin{minipage}[c][0.75cm][c]{1.8cm}\centering \ro{\textsf{CDRe} (1/5)}\ \end{minipage} \\\end{tabular}& \begin{tabular}{@{}c@{}}\begin{minipage}[c][0.5cm][c]{1.8cm}\centering \ro{\textsf{CDRea} (3/5)}\ \end{minipage} \\\begin{minipage}[c][0.5cm][c]{1.8cm}\centering \ro{\textsf{CRa} (1/5)}\ \end{minipage} \\\begin{minipage}[c][0.5cm][c]{1.8cm}\centering \ro{\textsf{CDRe} (1/5)}\ \end{minipage} \\\end{tabular}& \begin{tabular}{@{}c@{}}\begin{minipage}[c][0.75cm][c]{1.8cm}\centering \ro{\textsf{CDRe} (3/5)}\ \end{minipage} \\\begin{minipage}[c][0.75cm][c]{1.8cm}\centering \ro{\textsf{CDRea} (2/5)}\ \end{minipage} \\\end{tabular}& \begin{tabular}{@{}c@{}}\begin{minipage}[c][0.5cm][c]{1.8cm}\centering \ro{\textsf{CDRea} (2/5)}\ \end{minipage} \\\begin{minipage}[c][0.5cm][c]{1.8cm}\centering \ro{\textsf{CDRa} (2/5)}\ \end{minipage} \\\begin{minipage}[c][0.5cm][c]{1.8cm}\centering \ro{\textsf{CDRe} (1/5)}\ \end{minipage} \\\end{tabular}& \begin{tabular}{@{}c@{}}\begin{minipage}[c][0.75cm][c]{1.8cm}\centering \ro{\textsf{CDRea} (2/5)}\ \end{minipage} \\\begin{minipage}[c][0.75cm][c]{1.8cm}\centering \ro{\textsf{CDRe} (3/5)}\ \end{minipage} \\\end{tabular}& \begin{tabular}{@{}c@{}}\begin{minipage}[c][0.75cm][c]{1.8cm}\centering \ps{\textsf{PS} (2/5)}\ \end{minipage} \\\begin{minipage}[c][0.75cm][c]{1.8cm}\centering \ro{\textsf{CDRa} (3/5)}\ \end{minipage} \\\end{tabular} \\
\hline
40.0& \begin{tabular}{@{}c@{}}\begin{minipage}[c][0.5cm][c]{1.8cm}\centering \ro{\textsf{CDRe} (1/5)}\ \end{minipage} \\\begin{minipage}[c][0.5cm][c]{1.8cm}\centering \ro{\textsf{CDRea} (4/5)}\ \end{minipage} \\\end{tabular}& \begin{tabular}{@{}c@{}}\begin{minipage}[c][0.5cm][c]{1.8cm}\centering \ro{\textsf{CDRea} (4/5)}\ \end{minipage} \\\begin{minipage}[c][0.5cm][c]{1.8cm}\centering \ro{\textsf{CRea} (1/5)}\ \end{minipage} \\\end{tabular} & \ro{\textsf{CDRea} (5/5)} & \ro{\textsf{CDRea} (5/5)} & \ro{\textsf{CDRea} (5/5)} & \ro{\textsf{CDRea} (5/5)} & \ro{\textsf{CDRea} (5/5)} & \ro{\textsf{CDRea} (5/5)} & \ro{\textsf{CDRea} (5/5)} & \ro{\textsf{CDRea} (5/5)} & \ro{\textsf{CDRea} (5/5)} & \ro{\textsf{CDRea} (5/5)} & \ro{\textsf{CDRea} (5/5)} & \ro{\textsf{CDRea} (5/5)} & \ro{\textsf{CDRea} (5/5)} \\
\hline
\end{tabular}

%% file: tables/lmdrive-traffic_light-0.tex
\begin{tabular}{|c|c|c|c|c|c|c|c|c|c|c|c|c|c|c|}
\hline
\multicolumn{15}{|c|}{$\bm{xf}$}\\
\hline
0.0 & \ct{2.9} & 40.0 & 80.0 & 120.0 & 160.0 & 200.0 & 240.0 & 280.0 & 320.0 & 360.0 & 400.0 & 440.0 & 480.0 & 520.0 \\
\hline
\blk{\textsf{Blk (5/5)}} & \pu{\textsf{PU$p_4$ (5/5)}} & \pu{\textsf{PU$p_4$ (5/5)}} & \pu{\textsf{PU$p_4$ (5/5)}} & \pu{\textsf{PU$p_4$ (5/5)}} & \pu{\textsf{PU$p_4$ (5/5)}} & \pu{\textsf{PU$p_4$ (5/5)}} & \pu{\textsf{PU$p_4$ (5/5)}} & \pu{\textsf{PU$p_4$ (5/5)}} & \pu{\textsf{PU$p_4$ (5/5)}} & \pu{\textsf{PU$p_4$ (5/5)}} & \pu{\textsf{PU$p_4$ (5/5)}} & \pu{\textsf{PU$p_4$ (5/5)}} & \pu{\textsf{PU$p_4$ (5/5)}} & \pu{\textsf{PU$p_4$ (5/5)}} \\
\hline
\end{tabular}

%% file: tables/lmdrive-traffic_light-4.tex
\begin{tabular}{|c|c|c|c|c|c|c|c|c|c|c|c|c|c|c|}
\hline
\multicolumn{15}{|c|}{$\bm{xf}$}\\
\hline
0.0 & \ct{2.9} & 40.0 & 80.0 & 120.0 & 160.0 & 200.0 & 240.0 & 280.0 & 320.0 & 360.0 & 400.0 & 440.0 & 480.0 & 520.0 \\
\hline
\ro{\textsf{CRe (5/5)}} & \ro{\textsf{CRe (5/5)}} & \ro{\textsf{CRe (5/5)}} & \ro{\textsf{CRe (5/5)}} & \ro{\textsf{CRe (5/5)}} & \ro{\textsf{CRe (5/5)}} & \ro{\textsf{CRe (5/5)}} & \ro{\textsf{CRe (5/5)}} & \ro{\textsf{CRe (5/5)}} & \ro{\textsf{CRe (5/5)}} & \ro{\textsf{CRe (5/5)}} & \ro{\textsf{CRe (5/5)}} & \ro{\textsf{CRe (5/5)}} & \ro{\textsf{CRe (5/5)}} & \ro{\textsf{CRe (5/5)}} \\
\hline
\end{tabular}